\documentclass[10pt,a4paper,twoside]{report}
\usepackage[latin1]{inputenc}
\usepackage{amsmath}
\usepackage{amsfonts}
\usepackage{amssymb}
\usepackage[inner=3cm,outer=2cm,top=2cm,bottom=2cm,marginparsep=0cm,marginparwidth=3cm]{geometry}
\usepackage{setspace} 
\usepackage{booktabs} 
\usepackage{array} 
\usepackage{xcolor}
\definecolor{darkblue}{rgb}{0,0,0.6}
\definecolor{MMA1}{rgb}{0.368417, 0.506779, 0.709798} 
\definecolor{MMA2}{rgb}{0.880722, 0.611041, 0.142051}
\definecolor{MMA3}{rgb}{0.560181, 0.691569, 0.194885}
\definecolor{MMA4}{rgb}{0.922526, 0.385626, 0.209179}
\definecolor{MMA5}{rgb}{0.528488, 0.470624, 0.701351}
\definecolor{MMA6}{rgb}{0.772079, 0.431554, 0.102387}
\definecolor{MMA7}{rgb}{0.363898, 0.618501, 0.782349}
\definecolor{MMA8}{rgb}{1, 0.75, 0}
\definecolor{MMA9}{rgb}{0.647624, 0.37816, 0.614037}
\definecolor{MMA10}{rgb}{0.571589, 0.586483, 0}
\definecolor{MMA11}{rgb}{0.915, 0.3325, 0.2125}
\definecolor{MAT1}{rgb}{0,     0.447, 0.741} 
\definecolor{MAT2}{rgb}{0.85,  0.325, 0.098}
\definecolor{MAT3}{rgb}{0.929, 0.694, 0.125}
\definecolor{MAT4}{rgb}{0.494, 0.184, 0.556}
\definecolor{MAT5}{rgb}{0.466, 0.674, 0.188}
\definecolor{MAT6}{rgb}{0.301, 0.745, 0.933}
\definecolor{MAT7}{rgb}{0.635, 0.078, 0.184}

\usepackage{tensor} 
\usepackage[backend=bibtex8, backref=true, doi=false, url=false, isbn=false, style=phys, eprint=true, biblabel=brackets]{biblatex} 
\DeclareFieldFormat{titlecase}{#1} 
\bibliography{Full}
\DefineBibliographyStrings{english}{%
  backrefpage = {\hspace{-1.2mm}},
  backrefpages = {\hspace{-1.2mm}},
}
\usepackage[colorlinks, linkcolor=darkblue, citecolor=teal, pdfstartview={XYZ null null 1}]{hyperref} 
\usepackage{graphicx} 
\usepackage{animate} 
\usepackage{tikz}
\usepackage{mathrsfs} 
\usepackage{multirow}
\usepackage[safe]{tipa} 

\newcommand{\dx}[2][]{\,{#1}\mathrm{d}{#2}}
\newcommand{\dy}{\mathrm{d}}
\newcommand{\dydx}[2]{\frac{\mathrm{d}{#1}}{\mathrm{d}{#2}}}
\newcommand{\pypx}[2]{\frac{\partial{#1}}{\partial{#2}}}

\DeclareMathOperator{\tr}{tr}
\DeclareMathOperator{\co}{co}
\DeclareMathOperator{\sgn}{sgn}

\DeclareMathOperator{\arcosh}{arcosh}

\title{A Numerical Exploration of the Spherically Symmetric SU(2) Einstein-Yang-Mills Equations}
\author{Daniel Jackson\\\\\\
A thesis submitted for the degree of Doctor of Philosophy\\
School of Mathematical Sciences\\
Monash University}
\date{February 13, 2018}

\begin{document}
\pagenumbering{roman}
\maketitle

\section*{Copyright notice}
\copyright\hspace{0pt} The author (2018).

\begin{abstract}
\setcounter{page}{3}
\thispagestyle{plain}
\phantomsection
\addcontentsline{toc}{section}{Abstract}
The Einstein-Yang-Mills equations are the source of many interesting solutions within general relativity, including families of particle-like and black hole solutions, and critical phenomena of more than one type.
These solutions, discovered in the last thirty years, all assume a restricted form for the Yang-Mills gauge potential known as the ``magnetic" ansatz.
In this thesis we relax that assumption and investigate the most general solutions of the Einstein-Yang-Mills system assuming spherically symmetry, a Yang-Mills gauge group of SU(2), and zero cosmological constant.
We proceed primarily by numerically integrating the equations and find new static solutions, for both regular and black hole boundary conditions, which are not asymptotically flat, and attempt to classify the possible static behaviours.
We develop a code to solve the dynamic equations that uses a novel adaptive mesh refinement algorithm making full use of double-null coordinates.
We find that the ``type II" critical behaviour in the general case exhibits non-universal critical solutions, in contrast to the magnetic case and to all previously observed type II critical behaviour.
\end{abstract}

\section*{Declaration}
\addcontentsline{toc}{section}{Declaration}
This thesis contains no material which has been accepted for the award of any other degree or diploma at any university or equivalent institution and that, to the best of my knowledge and belief, this thesis contains no material previously published or written by another person, except where due reference is made in the text of the thesis.
\vspace{16pt}

\noindent
\includegraphics[scale=0.1]{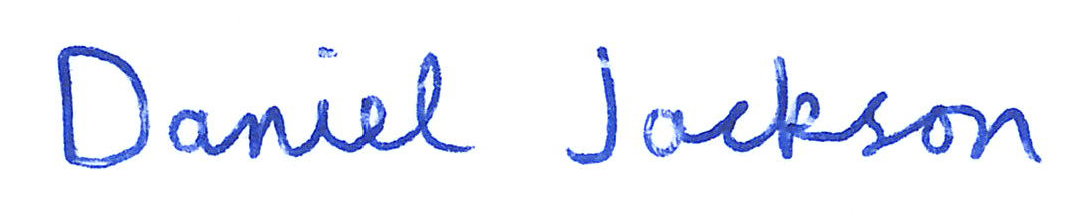}

\noindent
Daniel Jackson

\noindent
February 13, 2018

\newpage
\section*{Acknowledgements}
\addcontentsline{toc}{section}{Acknowledgements}
I began this project under the supervision of Robert Bartnik, who unfortunately became ill during this time.
I regret that our time working together was shortened due to this and I wish to thank him for his persistence, advice, and for never giving up on me.
There are many people, too many to name, who helped me in some way. To those from the School of Mathematical Sciences, my family, church family, and friends, who offered support and encouragement, thank you.
I'd like to specifically thank Todd Oliynyk, Leo Brewin, and Yann Bernard for their helpful suggestions and feedback as I came close to completion.
Thanks to Matt Choptuik for his hospitality, advice, and encouragement; he was most helpful during my brief visit.
Special thanks to my fellow students David Hartley and Stephen McCormick for making the time that much more enjoyable.

\newpage
\pdfbookmark[1]{Contents}{contents}
\tableofcontents
\cleardoublepage

\chapter{Introduction}
\setcounter{page}{0}
\pagenumbering{arabic}
\usetikzlibrary{calc}
The Einstein-Yang-Mills (EYM) equations consist of the Einstein field equations of general relativity and the Yang-Mills equations of non-Abelian gauge theory. These two theories are motivated by physics, and both can be considered from a purely geometric point of view. We will begin by giving a brief non-technical overview of the geometries involved in these physical theories. We will then review the relevant known results, state the scope of the thesis, and outline how this is achieved.

\section{The geometry and physical motivation of the equations}
Developed over one hundred years ago by Albert Einstein, the general theory of relativity provided an empirical and conceptual improvement upon the Newtonian description of gravity. It proposed that the force of gravity can be attributed to the curvature of spacetime, which is caused by the matter and energy distribution within the spacetime.
Space and time were conceptually first unified in Einstein's special theory of relativity, which resolved contradictions between Newtonian mechanics and Maxwell's theory of electromagnetism.
The three dimensions of space and the one dimension of time are in these theories considered inseparable aspects of four-dimensional spacetime.
The concept of curvature is made precise in the mathematical field of differential geometry, which provides the language for general relativity. The relevant object is called a manifold, whose geometry is described by a metric. The curvature of the manifold is determined by taking derivatives of the metric.
The Einstein equations then equate the curvature of the spacetime manifold with the matter and energy content of the spacetime. A solution to these equations means determining the spacetime metric, that is, being able to describe the geometry of the spacetime.

There are many kinds of matter that it is possible to couple to the Einstein equations in an attempt to solve them.
In the simplest case we can consider an absence of matter; the manifold is then called a vacuum spacetime.
If the matter is electromagnetic (described by Maxwell's equations), then the spacetime can contain electromagnetic phenomena such as electric charge. If there are no electromagnetic sources, then the spacetime can contain electromagnetic fields and it is called an electrovacuum spacetime.
We will only be considering spherically symmetric spacetimes, where the solutions in the vacuum and electrovacuum cases have been understood for over 50 years. Non-trivial solutions require a more complicated matter model and we will use a Yang-Mills field, which generalises the electromagnetic field in a certain sense.

Electromagnetism is the theory of electric and magnetic fields, unified by Maxwell in the 19th century. By the 20th century, this was recognised as a gauge theory. It can be described in geometric terms by considering an additional dimension of spacetime; then the electric and magnetic fields correspond to the curvature or twisting this additional dimension allows.
Again using the precise language of differential geometry, the object in question is a principal fibre bundle, whose geometry is described by a connection. A principal fibre bundle consists of a base space (the spacetime manifold) and a fibre (circle) at every point. The connection determines how adjacent fibres relate to each other, and the curvature is found by differentiating the connection.
The connection can be written purely in terms of objects on the base manifold (spacetime) once a section is chosen, which is a continuous choice of ``height" on the fibre for each point in the base space. The connection written in this way is called the gauge potential.
Maxwell's equations can then be written as a derivative of the curvature equated to the source of the electromagnetic field, called the four-current. A solution to these equations means determining the gauge potential, that is, being able to describe the geometry of the principal bundle.

The additional dimensions must have a structure known as a compact Lie group, referred to as the gauge group. Electromagnetism is called an Abelian gauge theory, where Abelian refers to the simplistic structure of the gauge group in this case, known as U(1), whose geometry is simply a circle.
A non-Abelian gauge theory allows for a more complicated gauge group, and this is how the Yang-Mills equations generalise Maxwell's equations.
The Yang-Mills equation we will consider will have gauge group SU(2), which has the geometry of a three-dimensional sphere.
While the structure of the Yang-Mills equations is the same as Maxwell's equations (derivative of curvature equals source), the equations are significantly more complicated.
We will only consider sourceless Yang-Mills fields (analogous to electrovacuum), and the language of electric and magnetic components of the curvature is borrowed directly from electromagnetism.
The Yang-Mills equations are named for Chen-Ning Yang and Robert Mills who first used them to describe a phenomenon of another of the four fundamental forces, the strong force. Today they are used in the standard model of particle physics, however we are not considering either the Higgs field or the quantisation of the equations that are applied in the physical model, but only the underlying classical Yang-Mills equations.

Solving the Einstein-Yang-Mills equations means determining the metric and the gauge potential, indicating what geometries are possible with this physical model. There are many more interesting solutions than in the electrovacuum case, and we review the known results in the next section.

\section{Previous work}
In this section we review the known solutions and other results of the Einstein-Yang-Mills system, beginning with the static equations. The properties discussed here and some of the more relevant solutions will be developed more fully in the coming chapters.

The earliest solutions to the Einstein-Yang-Mills equations were found by recognising U(1) as a subgroup of any compact Lie group, and that, therefore, all solutions to the Einstein-Maxwell equations have corresponding EYM solutions, termed ``embedded Abelian solutions" \cite{Yasskin75}. These include the Reissner-Nordstr\"{o}m and Kerr-Newman families of solutions.

The recent interest in the Einstein-Yang-Mills equations began in 1988 when Bartnik and McKinnon published their numerical discovery of the first essentially non-Abelian solutions \cite{BK88}. They are regular, static, spherically symmetric, asymptotically flat (``particle-like") solutions of the SU(2) EYM equations. There are a countably infinite number of such solutions, labelled by $ k $, the number of times a component of the gauge potential crosses zero. Each solution exhibits a high-density interior region and a Schwarzschild-like exterior, with the higher $ k $ solutions also containing an extremal Reissner-Nordstr\"{o}m-like intermediate region. They are purely magnetic, have zero Yang-Mills charge, and they were later termed ``BK solitons" \cite{VG99}.

Soon afterwards, analogous black hole solution were found numerically by a number of authors \cite{VG89,KM90,Bizon90}. These are also static, spherically symmetric, and asymptotically flat, but now with a regular event horizon. For each horizon radius $ r_h \in (0,\infty) $ there are again a countably infinite number, all of which are purely magnetic with zero Yang-Mills charge.
The interior of the EYM black hole solutions was investigated by Donets, Gal'tsov, and Zotov \cite{DGZ97} and Breitenlohner, Lavrelashvili, and Maison \cite{BLM98}.
They found that the generic behaviour near the singularity was not like the Schwarzschild or Reissner-Nordstr\"{o}m solutions, but instead, a metric function exhibited infinite oscillations as the singularity was approached; there were repeated cycles of mass inflation.

These numerical discoveries were complemented by analytic proofs of existence by Smoller, Wasserman, Yau, and McLeod \cite{SWYL91,SW93-1,SWY93}.
The (linear mode) stability of both families of solutions was investigated by Straumann and Zhou \cite{SZ90-1,SZ90-2,Zhou92}.
Even within the magnetic, spherically symmetric sector the BK solitons and the EYM black holes were found to be unstable.

In 1992, Bizon and Popp investigated whether there exist similar solutions with non-zero Yang-Mills charge.
They proved that there are no globally regular, static, spherically symmetric solutions with a net electric or magnetic charge and that the only such black hole solutions belong to the Reissner-Nordstr\"{o}m family \cite{BP92}.
This non-existence result probably accounts for much of the neglect of the electric (that is, not purely magnetic) solutions. Indeed, due to an apparent misunderstanding of the gauge freedom, Zhou \cite{Zhou92} erroneously claims that this result extends to the dynamic system.

In 1994, Breitenlohner, Forg\'{a}cs and Maison classified the static, spherically symmetric, magnetic solutions of the SU(2) EYM equations \cite{BFM94}. This classification included the asymptotically flat regular and black hole cases mentioned above, as well as non-asymptotically flat solutions, some with compact $ t = \mathrm{constant} $ hypersurfaces that occur generically and some that have the gauge potential forever oscillating.

While we will only be concerned in this thesis with the spherically symmetric, SU(2) EYM equations with zero cosmological constant, we briefly note some results of investigations that relax one of these assumptions.
Regular, static, spherically symmetric, purely magnetic solutions have been found with higher gauge groups such as SU(3) \cite{Kunzle94} and SO(5) \cite{BFO10}. Fisher and Oliynyk proved that for all higher gauge groups, there are no magnetically charged particle-like solutions for ``Abelian models" \cite{FO11}.
Static, spherically symmetric, black hole solutions with gauge groups SU($ n $) are also known \cite{GV92,KKS98}, including solutions with non-zero Yang-Mills charge that have non-trivial non-Abelian components.
Static, asymptotically flat, purely magnetic SU(2) solutions have been found with axial symmetry, both regular \cite{KK97-1,KK98-1,IKKS05} and black hole \cite{KK97-2,KK98-2,IKKW05}, as well as globally regular, non-asymptotically flat solutions \cite{GDV07}.
The non-zero cosmological constant case has been investigated by many authors, with the static SU(2) EYM equations producing both particle-like and black hole solutions.
Purely magnetic, spherically symmetric solutions have been found with either sign of the cosmological constant $ \Lambda $ \cite{TMT95,VSLHB96,Winstanley99,HB00}, with the $ \Lambda > 0 $ solutions found to be similar to the $ \Lambda = 0 $ case, while the $ \Lambda < 0 $ solutions show different behaviour such as a continuous spectrum of solutions and linear stability.
For the $ \Lambda < 0 $ case, there have been solutions found that include the electric field; spherically symmetric \cite{BH00-1,BH00-2} and axially symmetric \cite{Radu02-1,RW04}.
There has also been much work with additional matter fields such as a Higgs and dilaton fields; see the 1999 review \cite{VG99} and the references therein.

The dynamics of the spherically symmetric SU(2) EYM equations have also been previously studied.
After determining the linear instability of the first BK soliton, Zhou and Straumann performed numerical evolutions of perturbations of that solution \cite{ZS91,Zhou92}. They found that, depending on the sign of the perturbation, it would either collapse to a Schwarzschild black hole or disperse to flat space.
In 1996, Choptuik, Chmaj, and Bizon observed two distinct critical behaviours in the gravitational collapse of regular initial data \cite{CCB96}, building on Choptuik's previous work on the scalar field \cite{Choptuik93}.
Their investigation proceeded by evolving initial data with a free parameter such that small values produce dispersion and large values produce collapse to a black hole, and tuning this parameter to the boundary of these different end-states (Minkowski and Schwarzschild).
For some initial data families the results were similar to the scalar field case, where a self-similar critical solution \cite{Gundlach97-2} was found and the super-critical black holes exhibited universal mass scaling (termed ``Type II" behaviour).
They also found some families of initial data that would produce the first BK soliton as the critical solution, and the super-critical black holes would have finite mass (``Type I" behaviour).
In 1999, Choptuik, Hirschmann, and Marsa \cite{CHM99} determined that the first EYM black hole solution is also a critical solution, separating Schwarzschild black hole end-states with different values of the Yang-Mills gauge potential (``Type III" behaviour).
In fact the Yang-Mills field of this critical solution can take either sign and in 2014, Rinne found by tuning a second parameter to the value separating these two critical solutions that a magnetic Reissner-Nordstr\"{o}m solution can be evolved as an approximate co-dimension two attractor \cite{Rinne14}.
Finally, the behaviour of the Yang-Mills field for the dispersing solutions seen at great distances (well approximated by their behaviour on future null infinity) has been investigated by P\"{u}rrer and Aichelburg \cite{PA09}. Their numerical results found that the Yang-Mills fields decay as fast or faster than than the nonlinear tails on the Minkowski background.

As far as the author is aware\footnote{During the final editing stages of this thesis, Maliborski and Rinne released a preprint \cite{MR18} discussing critical phenomena in the general case. The relation to this work will be discussed in chapter \ref{ch:C}.}, the only previous time the electric field has been considered in dynamics is by Rinne and Moncrief in 2013, who investigated the decay of the Yang-Mills fields on future null infinity \cite{RM13}.
However, the electric perturbations on the purely magnetic BK and EYM black hole solution have been considered by authors investigating the linear mode stability of those solutions \cite{BS94-2,LM95,VG95,VBLS95}.

\section{Outline}
In this thesis we are concerned only with spherically symmetric solutions of the SU(2) EYM equations, primarily in the most general (electric as well as magnetic, dynamic) case that has not been thoroughly investigated previously. We therefore start in chapter \ref{ch:SSE} by deriving the equations and all the relevant quantities in this case.
The EYM equations are coordinate-invariant and gauge-invariant, and choosing appropriate coordinates and an appropriate gauge are crucial to being able to efficiently solve the equations. Most of chapter \ref{ch:SSE} will be devoted to establishing the coordinate and gauge choices to be used throughout the thesis. The previously mentioned embedded Abelian solutions will be referred to throughout the thesis and are discussed here as well.
We will then consider the static solutions, including a review of known results, in chapter \ref{ch:S}. These are interesting in their own right but will also be important when we consider the EYM dynamics in later chapters.
We construct dynamic solutions by numerically integrating the EYM equations. We develop a new code to do this efficiently by using double-null coordinates and an adaptive mesh refinement algorithm, and this is detailed in chapter \ref{ch:code}.
We consider dynamic magnetic solutions in chapter \ref{ch:Dm}, which by comparing to previous work will form a useful confirmation of our code as well as provide points of comparison for the results of evolving the full equations in chapter \ref{ch:Df}.
Our conclusions are in chapter \ref{ch:C}.
A number of the figures in this document are animated and will respond to a mouse click (tested in Adobe Reader), so the preferred viewing method is on a digital device. These figures are marked ``digital only".

\chapter{The equations in spherical symmetry} \label{ch:SSE}
In this chapter we specify a number of coordinate systems that will be used to write the Einstein-Yang-Mills equations, each exploiting the spherical symmetry we assume for the spacetime.
In each case we write out all of the equations and specify a gauge choice that allows further simplification.
We also introduce new variables that allow us to write the equations in a first-order form, and then briefly indicate how they may be solved numerically.
Finally, we state the known trivial solutions that are direct generalisations of the Einstein-Maxwell solutions.
We begin by defining a spherically symmetric spacetime and writing out the metric and gauge potential in the most general terms, before considering fully specified coordinates.

\section{The Einstein-Yang-Mills equations}
We begin by defining our terms and fixing our notation. Useful references are \cite{H&E,Wald} for general relativity and \cite{GTP,AMP} more generally and in particular for the geometry of fibre bundles.

The geometry of general relativity is described by the pair $ (M,g) $, where $ M $ is the spacetime manifold and $ g $ is a Lorentzian metric.
The fundamental variable is the metric $ g_{\mu\nu} $, whose final form depends on the coordinate choice.
The geometry of non-Abelian gauge theory is described by the principal fibre bundle $ P(M,G) $ and a connection $ \omega $, where $ M $ is the base manifold (spacetime) and $ G $ is the fibre and gauge group, with $ \mathfrak{g} $ its associated Lie algebra.
The fundamental variable is the local form of the connection, the gauge potential $ A_\mu^a $; a $ \mathfrak{g} $-valued one-form that depends on the gauge choice as well as the coordinate choice.
Here we use the abstract index notation, and throughout we will use Greek letters for the spacetime coordinate indices, and Latin letters for the Lie algebra indices.
We will also use the metric signature $ (-,+,+,+) $, and the Einstein summation convention.

The content of the combined EYM theory can be written compactly using the principle of stationary action.
The Einstein-Hilbert action, where we assume zero cosmological constant ($ \Lambda = 0 $) and a sourceless Yang-Mills field, is
\begin{equation} \label{action}
S_\mathrm{EYM} = \int_\mathcal{V} \left(\frac{c^4}{16\pi G}R -\frac{1}{2\text{\textg}^2}|F|^2\right) \sqrt{-g} \dx{^4x} \;,
\end{equation}
where $ R $ is the scalar curvature of $ (M,g) $ and $ |F|^2 $ is the square of the Yang-Mills field strength (the local form of the curvature of $ P $).
These are calculated from the metric $ g $ and the gauge potential $ A $ through the usual intermediate variables and formulas:
\begin{subequations}
\begin{align}
&\text{the Christoffel symbols} & \Gamma\indices{^\kappa_{\mu\nu}} &= \frac{1}{2}g^{\kappa\lambda}\left(g_{\lambda\nu,\mu}+g_{\mu\lambda,\nu}-g_{\mu\nu,\lambda}\right), \\
&\text{the Riemann curvature tensor} & R\indices{^\kappa_{\mu\lambda\nu}} &= \Gamma\indices{^\kappa_{\mu\nu,\lambda}}-\Gamma\indices{^\kappa_{\mu\lambda,\nu}} + \Gamma\indices{^\eta_{\mu\nu}}\Gamma\indices{^\kappa_{\lambda\eta}} - \Gamma\indices{^\eta_{\mu\lambda}}\Gamma\indices{^\kappa_{\eta\nu}} \;, \\
&\text{the Ricci curvature tensor} & R_{\mu\nu} &= R\indices{^\kappa_{\mu\kappa\nu}} \;, \\
&\text{the scalar curvature} & R &= g^{\mu\nu}R_{\mu\nu} \;, \label{R} \\
&\text{the field strength} & F_{\mu\nu}^a &= A_{\nu,\mu}^a - A_{\mu,\nu}^a + f_{abc} A_\mu^b A_\nu^c \;, \label{F} \\
&\text{and the field strength squared} & |F|^2 &= \frac{1}{2}g^{\mu\kappa}g^{\nu\lambda}\delta_{ab}F_{\mu\nu}^a F_{\kappa\lambda}^b \;, \label{F2}
\end{align}
\end{subequations}
where $ \delta_{ab} $ is the metric on $ \mathfrak{g} $ given by a constant multiple of the Killing form, and $ f_{abc} $ are the structure constants of the gauge group $ G $.
The field strength formula (\ref{F}) can be written purely in terms of geometric operations as $ F = \dy A + A \wedge A $, where d is the exterior derivative and $ \wedge $ is the wedge product.
The $ \frac{1}{2} $ in the definition of $ |F|^2 $ (\ref{F2}) accounts for the antisymmetry of $ F^a_{\mu\nu} $.

We slightly simplify the equations by using geometrised units so that Newton's gravitational constant $ G $ and the speed of light $ c $ are both equal to one, and units such that the coupling constant satisfies $ \text{\textg}^2 = 4\pi $.
Requiring the action (\ref{action}) to be stationary with respect to variations of the (inverse) metric gives Einstein's equations
\begin{equation} \label{EE}
G_{\mu\nu} := R_{\mu\nu} - \frac{1}{2} R g_{\mu\nu} = 8\pi T_{\mu\nu} \;,
\end{equation}
where $ G_{\mu\nu} $ is the Einstein tensor, and $ T_{\mu\nu} $ is the Yang-Mills stress-energy tensor
\begin{equation} \label{T}
T_{\mu\nu} := \frac{1}{4\pi}\left(g^{\alpha\beta}\delta_{ab}F_{\mu\alpha}^b F_{\nu\beta}^a - \frac{1}{2}g_{\mu\nu}|F|^2 \right).
\end{equation}
Requiring the action (\ref{action}) to be stationary with respect to the variations of the gauge potential gives the curved-space sourceless Yang-Mills equations
\begin{equation} \label{YME}
g^{\kappa\mu}\left(F_{\mu\nu;\kappa}^a + f_{abc} A_\kappa^b F_{\mu\nu}^c\right) = 0 \;.
\end{equation}
The Yang-Mills equations (\ref{YME}) can be written geometrically as $ *\left(\mathrm{d}{*F} + A \wedge *F - *F \wedge A\right) = 0 $ and the stress-energy tensor (\ref{T}) can be written more symmetrically as $ T_{\mu\nu} = \frac{1}{8\pi}g^{\alpha\beta}\delta_{ab}\left(F_{\mu\alpha}^b F_{\nu\beta}^a + *F_{\mu\alpha}^b *F_{\nu\beta}^a \right)$, where $ * $ is the Hodge star operator.

We remark here that taking the trace of (\ref{EE}), using (\ref{T}), shows us that the scalar curvature $ R $ (\ref{R}) is zero for any solution to the EYM equations.

Given an element $ g \in G $, the gauge potential transforms by
\begin{equation} \label{gauge transform def}
\tilde{A} = g^{-1}Ag +g^{-1}\dy{g} \;.
\end{equation}
A Yang-Mills theory can have any compact Lie group as the gauge group $ G $, but we consider only the group SU(2).
The Lie algebra indices will hence correspond to the three basis vectors $ \tau_1 $, $ \tau_2 $, $ \tau_3 $ of $ \mathfrak{su}(2) $, and the structure constants $ f_{abc} $ are given by the Levi-Civita symbol, and so we have $ [\tau_a,\tau_b] = \epsilon_{abc} \tau_c $.
If we write $ \tau_a = -\frac{i}{2}\sigma_a $, where $ \sigma_a $ are the Pauli matrices, then
\begin{align*}
\tau_1 &= \begin{pmatrix} 0 & -\frac{i}{2} \\ -\frac{i}{2} & 0 \end{pmatrix}, & \tau_2 &= \begin{pmatrix} 0 & -\frac{1}{2} \\ \frac{1}{2} & 0 \end{pmatrix}, & \tau_3 &= \begin{pmatrix} -\frac{i}{2} & 0 \\ 0 & \frac{i}{2} \end{pmatrix},
\end{align*}
and the metric on $ \mathfrak{g} $ is given explicitly by $ \delta_{ab} = -2\tr(\tau_a\tau_b) $.

\section{Spherical symmetry}\label{s:SS}
A spacetime $ (M,g) $ is spherically symmetric if its group of isometries includes a subgroup isomorphic to SO(3) and the orbits are spacelike, two-dimensional spheres (or points, which can lie on, at most, two lines) \cite{H&E}. 
Let $ A(p) $ be the area of the orbit 2-sphere of $ p \in M $; then the areal function is $ r = \sqrt{\frac{A}{4\pi}} $. The metric on these spheres is then $ r^2\dy{\theta^2} + r^2\sin^2\theta\dy{\phi^2} $ in standard spherical coordinates. The quotient manifold $ U = M/\text{SO}(3) $ is a 2-dimensional Lorentzian manifold for $ r > 0 $, which is orthogonal to each sphere. 
There still remains the freedom to choose coordinates on $ U $, which will be exploited in the following sections to determine several useful coordinate systems. For now we use $ x $ and $ y $ as arbitrary coordinates on $ U $.
The metric is then
\begin{equation} \label{metric}
g = g_{xx}\dy{x}^2 + 2g_{xy}\dy{x}\dy{y} +  g_{yy}\dy{y}^2 + r^2\dy{\theta^2} + r^2\sin^2\theta\dy{\phi^2} \;,
\end{equation}
where $ g_{xx}, g_{xy}, g_{yy} $, and $ r $ are functions of $ x $ and $ y $ only, and the Lorentzian condition is $ g_{xx}g_{yy}-g_{xy}^2 < 0 $.
For any choice of the coordinates $ (x,y) $ on $ U $, the Einstein equations (\ref{EE}) will be non-zero only for $ \mu\nu = xx $, $ xy $, $ yy $, $ \theta\theta $, and $ \phi\phi $, with the $ \phi\phi $ component equal to $ \sin^2\theta $ times the $ \theta\theta $ component, due to the spherical symmetry.

In spherical symmetry we can make use of the Misner-Sharp mass \cite{MS64,Hayward96}, which is defined by
\begin{equation} \label{MS}
m = \frac{r}{2}\left(1-\nabla_\mu r \nabla^\mu r\right) \;.
\end{equation}
We will often make use of the function $ N(x,y) = 1-\frac{2m(x,y)}{r(x,y)} = |\nabla r|^2 $, which has a useful physical interpretation. Consider the null geodesics that are perpendicular to the metric spheres; there are two future-directed congruences of null geodesics. $ N $ is proportional to the negative of the product of the expansions of these two congruences \cite{Hayward96}. A sphere on which $ N > 0 $ is called untrapped, and since the expansions have opposite signs we can assign the directions of their null lines as outwards or inwards. A sphere on which $ N < 0 $ is called trapped; when both expansions are negative it is a future trapped surface. A sphere on which $ N = 0 $ and $ \nabla_\mu N \neq 0 $ (one expansion is non-zero) is called a marginally trapped surface (MTS), and future MTSs (the non-zero expansion is negative) occur in evolution as part of a marginally trapped tube (MTT) \cite{AG05-2,BBGvdB06}. Note we use this terminology rather than that of a trapping horizon \cite{Hayward94-2} as we will see that the future horizons we form can have both spacelike (outer) and timelike (inner) components.
We will be concerned primarily with asymptotically flat spacetimes, which in spherical symmetry we characterise by a finite Misner-Sharp mass as $ r $ goes to infinity. This implies that $ \lim\limits_{r \to \infty} N = 1 $.
In this case the Misner-Sharp mass converges to the Arnowitt-Deser-Misner (ADM) mass at spatial infinity and the Bondi mass on null infinity \cite{Hayward96}.

The assumptions of spherical symmetry and asymptotic flatness imply that the bundle is trivial and hence a global gauge exists.
Clearly $ \mathbb{R} \times \mathbb{R}^3 $ can be covered by a single coordinate chart, and hence needs no transition functions. The one-point compactification of the spatial slices produces a spacetime manifold with topology $ \mathbb{R} \times \mathbb{S}^3 $, and the transition functions on $ \mathbb{S}^3 $ are homeomorphic to maps from $ \mathbb{S}^2 $ to SU(2). Since the second fundamental group of SU(2) is trivial, the transition functions must be trivial, hence the bundle is trivial.

A principal bundle $ P(M,G) $ is spherically symmetric if there is a left action of SO(3) on $ P $ that projects to $ M $, and a connection $ \omega $ on $ P $ is spherically symmetric if it is invariant under the action of SO(3).
Bartnik showed the form of a spherically symmetric SU(2) gauge potential depends on an integer $ k $ corresponding to the homotopy class of the isotropy homomorphism \cite{Bartnik91}.
When $ k = 1 $ we obtain the usual form for the gauge potential \cite{Witten77,BK88,VG99}, which will be used for the majority of this thesis. Written in the Abelian gauge, it is
\begin{equation} \label{potential}
A = a\tau_3\dy{x} + b\tau_3\dy{y} + (w\tau_1-d\tau_2)\dy{\theta} + \left(d\tau_1 + w\tau_2 + \cot\theta\tau_3\right)\sin\theta\dy{\phi} \;,
\end{equation}
where $ a $, $ b $, $ w $, and $ d $ are functions of $ x $ and $ y $ only.
When $ k = -1 $ we find a potential related to (\ref{potential}) by a gauge transformation (\ref{gauge transform def}) with $ g = e^{\pi \tau_1} $.
When $ k \neq \pm 1 $, the functions $ w $ and $ d $ must be zero and the gauge potential is
\begin{equation} \label{potentialAbel}
A = \left(a\dy{x} + b\dy{y} + k\cos\theta\dy{\phi}\right)\tau_3 \;.
\end{equation}
This is in an embedded Abelian form since it is proportional to a single Lie algebra element, and the corresponding solutions are completely considered in section \ref{s:EA}.

The gauge is not completely fixed by (\ref{potential}); the form is preserved by a gauge transformation (\ref{gauge transform def}) with a group element $ g = e^{\lambda\tau_3} $, where $ \lambda(x,y) $ is an arbitrary function. In that case the gauge functions transform as;
\begin{equation} \label{gauge transform}
\begin{aligned}
a &\mapsto a + \lambda_x \;, \qquad & w &\mapsto w\cos\lambda - d\sin\lambda \;, \\
b &\mapsto b + \lambda_y \;, & d &\mapsto d\cos\lambda + w\sin\lambda \;,
\end{aligned}
\end{equation}
where a subscript denotes a partial derivative.
We see the following combinations that are invariant under this residual gauge transformation will appear throughout; $ a_y - b_x $, $ w^2 + d^2 $, $ (w_x+ad)^2+(d_x-aw)^2 $, and $ (w_y+bd)^2+(d_y-bw)^2 $. These appear in the stress-energy tensor and thus the Einstein equations, for example.

We note that the gauge choice (\ref{potential}) is singular, at $ \theta = 0,\pi $, however a (singular) gauge change can transform it into a regular gauge \cite{BH00-2}.
A gauge transformation (\ref{gauge transform def}) with $ g = e^{-\frac{\pi}{2}\tau_3} e^{-\theta\tau_2} e^{-\phi\tau_3} $ produces
\begin{equation} \label{potential regular}
\tilde{A} = a \tau_r \dy{x} + b \tau_r \dy{y} +(d \tau_\theta -(1-w)\tau_\phi)\dy{\theta} +((1-w)\tau_\theta +d\tau_\phi)\sin\theta \dy{\phi} \;,
\end{equation}
where $ \tau_r $, $ \tau_\theta $ and $ \tau_\phi $ are a Lie algebra basis suited to spherical coordinates:
\[ \begin{bmatrix} \tau_r \\ \tau_\theta \\ \tau_\phi \end{bmatrix} = \begin{bmatrix} \sin\theta\cos\phi & \sin\theta\sin\phi & \cos\theta \\ \cos\theta\cos\phi & \cos\theta\sin\phi & -\sin\theta \\ -\sin\phi & \cos\phi & 0 \end{bmatrix} \begin{bmatrix} \tau_1 \\ \tau_2 \\ \tau_3 \end{bmatrix}. \]
They satisfy the same commutation relations; $ [\tau_r, \tau_\theta] = \tau_\phi $, $ [\tau_\theta, \tau_\phi] = \tau_r $, and $ [\tau_\phi, \tau_r] = \tau_\theta $.
In this regular gauge, the residual gauge freedom is expressed as $ g = e^{\lambda(x,y)\tau_r} $.
The potential (\ref{potential regular}) has no singularities and is defined on all of $ M $, and therefore the singularities in (\ref{potential}) are merely coordinate (gauge) singularities.
We choose to write everything in the computationally simpler Abelian gauge.
The field strength in either gauge is manifestly not singular. In the Abelian gauge, (\ref{F}) is written
\begin{align*}
F &= -(a_y-b_x)\tau_3\dx{x} \wedge \dx{y} + (w^2+d^2-1)\tau_3\dx{\theta} \wedge \sin\theta\dx{\phi} \\
 & \qquad + ((w_x+ad)\tau_1-(d_x-aw)\tau_2)\dx{x} \wedge \dx{\theta} + ((d_x-aw)\tau_1+(w_x+ad)\tau_2)\dx{x} \wedge \sin\theta\dx{\phi} \\
 & \qquad + ((w_y+bd)\tau_1-(d_y-bw)\tau_2)\dx{y} \wedge \dx{\theta} + ((d_y-bw)\tau_1+(w_y+bd)\tau_2)\dx{y} \wedge \sin\theta\dx{\phi} \;.
\end{align*}

Using the ``straightforward definition" \cite{VG99} we can write the total electric and magnetic charges as
\begin{align*}
\frac{1}{4\pi}\oint_{S^2_\infty} *F &= \lim_{r \to \infty}\frac{r^2(a_y-b_x)}{\sqrt{g_{xy}^2-g_{xx}g_{yy}}}\tau_3 \;, & \frac{1}{4\pi} \oint_{S^2_\infty} F &= \lim_{r \to \infty}(w^2+d^2-1)\tau_3 \;.
\end{align*}
These Lie algebra-valued charges are invariant with respect to the residual gauge, however they are gauge dependent under general gauge transformations where $ F $ transforms according to $ \tilde{F} = g^{-1}Fg $.
The effect of such a gauge transformation is only to rotate the charges in the Lie algebra; their length (with respect to $ \delta_{ab} $) remains constant.
Therefore we define the total Yang-Mills charges as the scalars
\begin{align} \label{charge}
Q &= \lim_{r \to \infty}\frac{r^2(a_y-b_x)}{\sqrt{g_{xy}^2-g_{xx}g_{yy}}} \;, & P &= \lim_{r \to \infty}(w^2+d^2-1) \;,
\end{align}
where we have preserved their ability to change sign, as in the paper by Sudarsky and Wald \cite{SW92}.

There are three rather natural choices for the coordinates on $ U $ which we find useful, and we describe them here, along with the resulting equations.

\section{Polar-areal coordinates}
Where $ \nabla_\mu r $ is spacelike ($ N > 0 $), the areal radius $ r $ may be used as a spatial coordinate.
In this case, the Misner-Sharp mass can be used as a metric variable, and we choose $ t $ to be an orthogonal time coordinate (the ``polar" condition \cite{BP83});
\begin{equation} \label{metric tr}
g = -S(t,r)^2N(t,r)\dy{t}^2 + \frac{1}{N(t,r)}\dy{r}^2 + r^2\dy{\theta^2} + r^2\sin^2\theta\dy{\phi^2} \;,
\end{equation}
where $ N(t,r) = 1 - \frac{2m(t,r)}{r} $, and we include it in the coefficient of $ \dy{t}^2 $ to simplify the equations.

There remains some coordinate freedom; a new time coordinate can be chosen as any increasing function of the old: $ \tilde{t}(t) $ with $ \tilde{t}' > 0 $. In the metric this has the effect of $ S = \tilde{t}'\tilde{S} $, so we can fix the remaining coordinate freedom by specifying $ S $ at a constant radius. Since $ N = 1 $ at a regular origin (see (\ref{regular tr metric}) below) and at asymptotically flat infinity, we can easily use this freedom to set $ t $ to be the proper time at either place.
One choice is $ \lim\limits_{r \to \infty} S(t,r) = 1 $, however in practice we will typically set $ S(t,0) = 1 $ as we are more concerned with behaviour at the origin, where $ t $ will indicate the proper time. If we can calculate $ \lim\limits_{r \to \infty} S(t,r) $ it is easy to divide $ S $ by this to transform the coordinates so that $ t $ is the proper time at infinity.
Polar-areal coordinates are also known as Schwarzschild coordinates, and are a common choice in the literature.
For these coordinates, we use a prime to indicate $ \pypx{}{r} $ and an overdot to indicate $ \pypx{}{t} $.

These coordinates will not cover the whole spacetime in general, however they are appropriate in any untrapped region ($ N > 0 $), and do cover the manifold in the globally regular case.
They are most useful for discussing static solutions, for which $ \pypx{}{t} $ is a Killing vector.

The definitions of $ a $ and $ b $ in (\ref{potential}) depend on the coordinate choice.
To distinguish their definition in polar areal coordinates, we use a hat above them.
This is meant to remind of the orthogonal nature of these coordinates.
The gauge potential (\ref{potential}) is therefore written
\begin{equation} \label{potential tr}
A = \hat{a}\tau_3\dy{t} + \hat{b}\tau_3\dy{r} + (w\tau_1-d\tau_2)\dy{\theta} + \left(d\tau_1 + w\tau_2 + \cot\theta\tau_3\right)\sin\theta\dy{\phi} \;,
\end{equation}
and the non-zero components of the Yang-Mills stress-energy tensor are
\begin{equation} \label{T pa}
\begin{aligned}
8\pi T_{tt} &= S^2N \frac{(w^2+d^2-1)^2}{r^4} + N(\hat{a}'-\dot{\hat{b}})^2 \\
 & \qquad + \frac{2S^2N^2}{r^2}\left(\frac{(\dot{w}+\hat{a}d)^2+(\dot{d}-\hat{a}w)^2}{S^2N^2}+(w'+\hat{b}d)^2+(d'-\hat{b}w)^2\right), \\
8\pi T_{tr} &= \frac{4}{r^2}\left((\dot{w}+\hat{a}d)(w'+\hat{b}d)+(\dot{d}-\hat{a}w)(d'-\hat{b}w)\right), \\
8\pi T_{rr} &= -\frac{1}{N}\frac{(w^2+d^2-1)^2}{r^4} -\frac{(\hat{a}'-\dot{\hat{b}})^2}{S^2N} \\
 & \qquad + \frac{2}{r^2}\left(\frac{(\dot{w}+\hat{a}d)^2+(\dot{d}-\hat{a}w)^2}{S^2N^2}+(w'+\hat{b}d)^2+(d'-\hat{b}w)^2\right), \\
8\pi T_{\theta\theta} &= \frac{(w^2+d^2-1)^2}{r^2} + \frac{r^2(\hat{a}'-\dot{\hat{b}})^2}{S^2} \;, \\
8\pi T_{\phi\phi} &= \left(\frac{(w^2+d^2-1)^2}{r^2} + \frac{r^2(\hat{a}'-\dot{\hat{b}})^2}{S^2}\right)\sin^2\theta \;.
\end{aligned}
\end{equation}

Given a unit timelike vector $ n^\mu $, the energy density is given by $ E = T_{\mu\nu}n^\mu n^\nu $.
We can obtain a unit timelike vector simply from the orthonormal frame found by normalising the coordinate one-forms, in which case the energy density is
\begin{equation} \label{energy density}
8\pi E = \frac{(w^2+d^2-1)^2}{r^4} + \frac{\left(\hat{a}'-\dot{\hat{b}}\right)^2}{S^2} + \frac{2|N|}{r^2}\left(\frac{(\dot{w}+\hat{a}d)^2+(\dot{d}-\hat{a}w)^2}{S^2N^2} +(w'+\hat{b}d)^2 +(d'-\hat{b}w)^2\right).
\end{equation}

The Einstein equations (\ref{EE}) reduce to
\begin{subequations} \label{EEtr}
\begin{align}
m' &= \frac{(w^2+d^2-1)^2}{2r^2} + \frac{r^2\left(\hat{a}'-\dot{\hat{b}}\right)^2}{2S^2} + N\left(\frac{(\dot{w}+\hat{a}d)^2+(\dot{d}-\hat{a}w)^2}{S^2N^2} + (w'+\hat{b}d)^2 + (d'-\hat{b}w)^2\right), \label{mr}\\
\dot{m} &= 2N\left((\dot{w}+\hat{a}d)(w'+\hat{b}d)+(\dot{d}-\hat{a}w)(d'-\hat{b}w)\right), \label{mt}\\
S' &= \frac{2S}{r}\left(\frac{(\dot{w}+\hat{a}d)^2+(\dot{d}-\hat{a}w)^2}{S^2N^2} + (w'+\hat{b}d)^2 + (d'-\hat{b}w)^2\right), \label{Sr}
\end{align}
and
\begin{equation}
\frac{r^2}{2S}\left(\left(\frac{\dot{N}}{SN^2}\right)^. +\left(\frac{(S^2N)'}{S}\right)' +\frac{2(SN)'}{r}\right) = \frac{r^2\left(\hat{a}'-\dot{\hat{b}}\right)^2}{S^2} + \frac{(w^2+d^2-1)^2}{r^2} \;, \label{Etheta}
\end{equation}
\end{subequations}
where (\ref{mr}-\ref{Sr}) are a slight rearrangement of the $ \mu\nu = tt $, $ tr $, and $ rr $ components, and (\ref{Etheta}) is the $ \theta\theta $ component.

The non-zero Yang-Mills equations (\ref{YME}), respectively the $ a\nu = 3t $, $ 3r $, $ 1\theta $, and $ 2\theta $ components, are
\begin{subequations} \label{YMEtr}
\begin{align}
\partial_r \left(\frac{r^2\left(\hat{a}'-\dot{\hat{b}}\right)}{2S}\right) &= \frac{d(\dot{w}+\hat{a}d)-w(\dot{d}-\hat{a}w)}{SN}\;, \label{a} \\
\partial_t \left(\frac{r^2\left(\hat{a}'-\dot{\hat{b}}\right)}{2S}\right) &= SN(d(w'+\hat{b}d)-w(d'-\hat{b}w))\;, \label{b} \\
\partial_t \left(\frac{\dot{w}+\hat{a}d}{SN}\right) - \partial_r \left(SN(w'+\hat{b}d)\right) &= -Sw\frac{w^2+d^2-1}{r^2} - \hat{a}\frac{\dot{d}-\hat{a}w}{SN} + SN\hat{b}(d'-\hat{b}w)\;, \label{w} \\
\partial_t \left(\frac{\dot{d}-\hat{a}w}{SN}\right) - \partial_r \left(SN(d'-\hat{b}w)\right) &= -Sd\frac{w^2+d^2-1}{r^2} + \hat{a}\frac{\dot{w}+\hat{a}d}{SN} - SN\hat{b}(w'+\hat{b}d)\;. \label{d}
\end{align}
\end{subequations}
Due to the spherical symmetry, the $ a\nu = 2\phi $ and $ 1\phi $ components also give equations (\ref{w},\ref{d}), respectively.
Equations (\ref{a}--\ref{d}) are not independent due to the identity
\[ \partial_t (\ref{a}) - \partial_r (\ref{b}) + d(\ref{w}) - w(\ref{d}) \equiv 0\;, \]
where here and henceforth, we use an equation number to refer to all the terms of the equation, transposed to the left-hand side (for example, $ (\ref{a}) = \partial_r \left(\frac{r^2\left(\hat{a}'-\dot{\hat{b}}\right)}{2S}\right) - \frac{d(\dot{w}+\hat{a}d)-w(\dot{d}-\hat{a}w)}{SN} $).
This dependency corresponds to the freedom to choose a gauge (\ref{gauge transform}), which can be used to specify one of the four gauge functions.

The twice contracted second Bianchi identity, $ \nabla^\mu G_{\mu\nu} = 0 $ and the Einstein equations (\ref{EE}) imply that the divergence of the stress-energy tensor is zero; $ \nabla^\mu T_{\mu\nu} = 0 $.
The left-hand side is not identically zero only for $ \nu = t $ and $ \nu = r $ and these equations give the following combinations of the Yang-Mills equations:
\begin{subequations} \label{Bianchi}
\begin{align}
\left(\hat{a}'-\dot{\hat{b}}\right)(\ref{b}) + (\dot{w}+\hat{a}d)(\ref{w}) + (\dot{d}-\hat{a}w)(\ref{d}) &= 0 \;, \label{Bianchit}\\
\left(\hat{a}'-\dot{\hat{b}}\right)(\ref{a}) + (w'+\hat{b}d)(\ref{w}) + (d'-\hat{b}w)(\ref{d}) &= 0 \;. \label{Bianchir}
\end{align}
\end{subequations}
It is precisely these combinations of the Yang-Mills equations (\ref{YMEtr}) that can be used to show the interdependency of the Einstein equations (\ref{EEtr});
\begin{align*}
\partial_t(\ref{mr}) - \partial_r(\ref{mt}) + \frac{2}{S}(\ref{Bianchit}) &\equiv 0 \;, \\
(\ref{Etheta})+\frac{2r}{S}(\ref{Bianchir}) &\equiv 0 \;,
\end{align*}
where use has been made of (\ref{mr}-\ref{Sr}). We remark that these results hold for any matter model.

We therefore conclude that (\ref{Etheta}) is (analytically) satisfied if (\ref{mr}-\ref{Sr},\ref{a},\ref{w},\ref{d}) are satisfied.
Hence we may use (\ref{YMEtr}) and (\ref{mr},\ref{mt},\ref{Sr}), remembering that one Yang-Mills equation can be satisfied by completely fixing the gauge, and one of (\ref{mr}) and (\ref{mt}) is redundant and can be used as a check of the numerical method along with (\ref{Etheta}).

There is a well-known subcase of these equations we call purely magnetic, where $ \hat{a} \equiv \hat{b} \equiv d \equiv 0 $.
We can define the electric $ E $ and magnetic field $ B $ analogously to the Maxwell case, by contracting the field strength and its Hodge dual with a unit timelike normal $ n^\mu = \frac{1}{S\sqrt{N}}\pypx{}{t} $, so that $ E_\nu = n^\mu F_{\mu\nu} $ and $ B_\nu = n^\mu {*F}_{\mu\nu} $. Then we find
\begin{align*}
E &= -\frac{\hat{a}'-\dot{\hat{b}}}{S\sqrt{N}}\tau_3\dy{r} +\frac{(\dot{w}+\hat{a}d)\tau_1-(\dot{d}-\hat{a}w)\tau_2}{S\sqrt{N}}\dy{\theta} +\frac{(\dot{d}-\hat{a}w)\tau_1+(\dot{w}+\hat{a}d)\tau_2}{S\sqrt{N}}\sin\theta\dy{\phi} \;, \\
B &= \frac{w^2+d^2-1}{r^2\sqrt{N}}\tau_3\dy{r} -\sqrt{N}\left((d'-\hat{b}w)\tau_1+(w'+\hat{b}d)\tau_2\right)\dy{\theta} +\sqrt{N}\left((w'+\hat{b}d)\tau_1-(d'-\hat{b}w)\tau_2\right)\sin\theta\dy{\phi} \;.
\end{align*}
We see that in the static case ($ \dot{w} = 0 $), this terminology is precise and the electric field is zero (analogously to the Maxwell case). We will nevertheless continue to use the term purely magnetic when considering dynamic solutions to refer to the subcase where $ w $ is the only non-zero gauge function.
If we were to consider a purely electric solution, we would require $ w^2+d^2 = 1 $ and $ \hat{b} \equiv 0 $. Using the gauge freedom (\ref{gauge transform}) we can set $ w \equiv 1 $ and $ d \equiv 0 $. Then equation (\ref{w}) implies $ \hat{a} = 0 $, and so there is no electric field either. Therefore any non-trivial solution has a non-zero magnetic field.
In the purely magnetic case, (\ref{Bianchi}) completely specifies the (single) Yang-Mills equation (\ref{w})
\[ \partial_t \left(\frac{\dot{w}}{SN}\right) - \partial_r \left(SNw'\right) = -Sw\frac{w^2-1}{r^2} \;, \]
similarly to the real massless scalar field case \cite{Christodoulou91}. In general, the twice contracted second Bianchi identity (\ref{Bianchi}) lacks the full information of the matter equations. A similar result was noted in \cite{Smolic14}.

\section{Isothermal coordinates}
When $ N $ goes to zero, polar-areal coordinates exhibit a coordinate singularity.
In the case of a static spacetime with a regular event horizon ($ S $ bounded, $ N' \geqslant 0 $), the coordinates are still useful on the interior although the $ t = \text{constant} $ surfaces are now timelike.
While studying the static solutions we will observe a second possibility, when $ N $ goes to zero and doesn't change sign, coinciding with a maximum of $ r $. The sphere on which this occurs is called an equator and $ r $ can no longer be used as a global coordinate.
It is particularly for this case that we choose isothermal coordinates on $ U $:
\begin{equation} \label{metric TR}
g = -\alpha(T,R)^2\dy{T}^2 +\alpha(T,R)^2\dy{R}^2 + r(T,R)^2\dy{\theta}^2 +r(T,R)^2\sin^2\theta\dy{\phi}^2 \;.
\end{equation}
The Misner-Sharp mass (\ref{MS}) and $ N $ in these coordinates are
\begin{align} \label{mN TR}
N(T,R) &= \frac{r'^2-\dot{r}^2}{\alpha^2} \;, & m(T,R) &= \frac{r}{2}\left(1-\frac{r'^2-\dot{r}^2}{\alpha^2}\right).
\end{align}
Where there's little chance of confusion, we use a prime to indicate $ \pypx{}{R} $ and an overdot to indicate $ \pypx{}{T} $ in these coordinates.

There remains some coordinate freedom; the metric (\ref{metric TR}) retains the same form for new coordinates given by
\begin{align} \label{coord freedom TR}
T &= f(\tilde{T}+\tilde{R})+g(\tilde{T}-\tilde{R}) \;, & R &= f(\tilde{T}+\tilde{R})-g(\tilde{T}-\tilde{R}) \;,
\end{align}
for arbitrary functions $ f $ and $ g $, with $ f' > 0 $ and $ g' > 0 $ to preserve the directions of the coordinates in both time and space.
If we have $ r = 0 $ at $ R = 0 $ and desire it to be at $ \tilde{R} = 0 $ in the new coordinates, then this demands the functions $ f $ and $ g $ be identical.
If the spacetime is static with Killing vector $ \pypx{}{T} $ and we demand that $ \pypx{}{\tilde{T}} $ also be a Killing vector, then in general the only possible transformations are
\begin{align} \label{TR stat free}
T &= C\tilde{T}+D \;, & R &= C\tilde{R}+E \;,
\end{align}
for some constants $ C > 0 $, $ D $, and $ E $.
There is one exceptional case, when the areal radius $ r $ is constant. Then there are further possible coordinate transformations that will preserve the coordinate staticity (see section \ref{s:EA}).
Assuming a regular origin, the coordinates can be completely fixed by demanding the origin remains at $ R = 0 $ ($ r(T,0) = 0 $) and $ T $ is the proper time at the origin, so $ \alpha(T,0) = 1 $.
Regularity then implies $ r'(T,0) = 1 $.

To distinguish the gauge functions $ a $ and $ b $ in isothermal coordinates from the other coordinate choices, we use an overbar. This is to remind of the constant time slices, which are horizontal when viewed in double-null coordinates.
The gauge potential (\ref{potential}) is therefore written
\begin{equation} \label{potential isothermal}
A = \bar{a}\tau_3\dy{T} + \bar{b}\tau_3\dy{R} + (w\tau_1-d\tau_2)\dy{\theta} + \left(d\tau_1 + w\tau_2 + \cot\theta\tau_3\right)\sin\theta\dy{\phi} \;,
\end{equation}
and the energy density $ E $ calculated in the orthonormal frame found by normalising the coordinate one-forms is
\begin{equation} \label{energy density isothermal}
8\pi E = \frac{(w^2+d^2-1)^2}{r^4} + \frac{\left(\bar{a}'-\dot{\bar{b}}\right)^2}{\alpha^4} + 2\frac{(\dot{w}+\bar{a}d)^2 +(w'+\bar{b}d)^2 +(\dot{d}-\bar{a}w)^2 +(d'-\bar{b}w)^2}{r^2\alpha^2} \;.
\end{equation}

The Einstein equations (\ref{EE}) reduce to
\begin{subequations} \label{EETR}
\begin{align}
\left(\frac{r'}{\alpha^2}\right)^. +\left(\frac{\dot{r}}{\alpha^2}\right)' &= -4\frac{(\dot{w}+\bar{a}d)(w'+\bar{b}d)+(\dot{d}-\bar{a}w)(d'-\bar{b}w)}{r\alpha^2} \;, \label{ralphaTR} \\
\left(\frac{\dot{r}}{\alpha^2}\right)^. +\left(\frac{r'}{\alpha^2}\right)' &= -2\frac{(\dot{w}+\bar{a}d)^2+(w'+\bar{b}d)^2+(\dot{d}-\bar{a}w)^2+(d'-\bar{b}w)^2}{r\alpha^2} \;, \label{ralphaTT} \\
\left(r^2\right)^{..}-\left(r^2\right)'' &= -2\alpha^2\left(1-\frac{(w^2+d^2-1)^2}{r^2}-\frac{r^2\left(\bar{a}'-\dot{\bar{b}}\right)^2}{\alpha^4}\right), \label{rTT} \\
\left(\ln\alpha\right)^{..}-\left(\ln\alpha\right)'' +\frac{r'^2-\dot{r}^2}{r^2} &= \frac{\alpha^2}{r^2}\left(1-2\frac{(w^2+d^2-1)^2}{r^2}-2\frac{r^2\left(\bar{a}'-\dot{\bar{b}}\right)^2}{\alpha^4}\right), \label{alphaTT}
\end{align}
\end{subequations}
where (\ref{ralphaTT}-\ref{alphaTT}) are a slight rearrangement of the $ \mu\nu = TT $, $ RR $, and $ \theta\theta $ components, and (\ref{ralphaTR}) is the $ TR $ component.

The Yang-Mills equations (\ref{YME}), respectively the $ a\nu = 3T $, $ 3R $, $ 1\theta $, and $ 2\theta $ components, are
\begin{subequations} \label{YMETR}
\begin{align}
\partial_R \left(\frac{r^2\left(\bar{a}'-\dot{\bar{b}}\right)}{2\alpha^2}\right) &= d(\dot{w}+\bar{a}d)-w(\dot{d}-\bar{a}w) \;, \label{aTR} \\
\partial_T \left(\frac{r^2\left(\bar{a}'-\dot{\bar{b}}\right)}{2\alpha^2}\right) &= d(w'+\bar{b}d)-w(d'-\bar{b}w) \;, \label{bTR} \\
\partial_T \left(\dot{w}+\bar{a}d\right) -\partial_R \left(w'+\bar{b}d\right) &= -\alpha^2w\frac{w^2+d^2-1}{r^2} - \bar{a}\left(\dot{d}-\bar{a}w\right) + \bar{b}\left(d'-\bar{b}w\right) \;, \label{wTR} \\
\partial_T \left(\dot{d}-\bar{a}w\right) -\partial_R \left(d'-\bar{b}w\right) &= -\alpha^2d\frac{w^2+d^2-1}{r^2} + \bar{a}\left(\dot{w}+\bar{a}d\right) - \bar{b}\left(w'+\bar{b}d\right) \;. \label{dTR}
\end{align}
\end{subequations}
Due to the spherical symmetry, the $ a\nu = 2\phi $ and $ 1\phi $ components also give equations (\ref{wTR},\ref{dTR}), respectively.
These equations are not independent;
\[ \partial_T (\ref{aTR}) - \partial_R (\ref{bTR}) + d(\ref{wTR}) - w(\ref{dTR}) \equiv 0 \;. \]
This dependency corresponds to the freedom to choose a gauge (\ref{gauge transform}), which can be used to specify one of the four gauge functions.

The twice contracted second Bianchi identity again gives two combinations of the Yang-Mills equations:
\begin{subequations} \label{BianchiTR}
\begin{align}
\left(\bar{a}'-\dot{\bar{b}}\right)(\ref{bTR}) + (\dot{w}+\bar{a}d)(\ref{wTR}) + (\dot{d}-\bar{a}w)(\ref{dTR}) &= 0 \;, \label{BianchiT}\\
\left(\bar{a}'-\dot{\bar{b}}\right)(\ref{aTR}) + (w'+\bar{b}d)(\ref{wTR}) + (d'-\bar{b}w)(\ref{dTR}) &= 0 \;. \label{BianchiR}
\end{align}
\end{subequations}
It is precisely these combinations of the Yang-Mills equations (\ref{YMETR}) that can be used to show the interdependency of the Einstein equations (\ref{EETR}):
\begin{align*}
2r\alpha^2\left(\partial_R(\ref{ralphaTR}) - \partial_T(\ref{ralphaTT})\right) +\partial_T(\ref{rTT}) +8(\ref{BianchiT}) &\equiv 0 \;, \\
2r\alpha^2\left(\partial_T(\ref{ralphaTR}) - \partial_R(\ref{ralphaTT})\right) -\partial_R(\ref{rTT}) -8(\ref{BianchiR}) &\equiv 0 \;,
\end{align*}
where use has been made of (\ref{EETR}).
As before there are two independent Einstein equations and three independent Yang-Mills equations

We also note here that in the purely magnetic case, (\ref{BianchiTR}) completely specifies the (single) Yang-Mills equation (\ref{wTR})
\[ \ddot{w} -w'' = -\alpha^2w\frac{w^2-1}{r^2} \;. \]

\section{Double-null coordinates}
The third natural choice is to use double-null coordinates, $ u $ and $ v $, where the $ u = \text{constant} $ lines (in the direction $ \pypx{}{v} $) are outgoing null lines ($ r $ is increasing);
\begin{equation} \label{metric dn}
g = -4\alpha(u,v)^2\dx{u}\dx{v} + r(u,v)^2\dy{\theta^2} + r(u,v)^2\sin^2\theta\dy{\phi^2} \;.
\end{equation}
We choose a factor of four as it's the smallest even square, in contrast with the choice of $ 1 $ \cite{Christodoulou93,HS96} and $ 2 $ \cite{Hayward96}, but similarly to \cite{GP97}.
We use the variable $ \alpha $ here as well as in the metric in isothermal coordinates (\ref{metric TR}) because with the appropriate choice of coordinate transformation (see appendix \ref{ss:pa dn}), they represent the same quantity.
Double-null coordinates will be most useful when considering the dynamic equations, and by appropriate scaling of the coordinates (see (\ref{coord freedom dn})), will cover the entire manifold.

\begin{figure}[!ht]
  \centering
  \begin{tikzpicture}
    \draw[->] (0,0)--(-1,1) node[above left]{$ u $};
    \draw[->] (0,0)--(1,1) node[above right]{$ v $};
    \draw (0,0)--node[above right]{$ r > 0 $}(0,3) node[above]{$ r = 0 $};
  \end{tikzpicture}
  \caption{The geometry of the double-null coordinates.}
  \label{fig:dn geometry}
\end{figure}
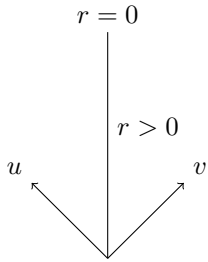

There remains the coordinate freedom
\begin{align} \label{coord freedom dn}
\tilde{u} &= \tilde{u}(u) \;, & \tilde{v} &= \tilde{v}(v) \;,
\end{align}
where $ \tilde{u}' > 0 $ and $ \tilde{v}' > 0 $.
In the metric this has the effect of $ \alpha(u,v)^2 = \tilde{u}'(u)\tilde{v}'(v)\tilde\alpha(\tilde{u}(u),\tilde{v}(v))^2 $.
Somewhat analogously to the polar-areal case, the double-null coordinate freedom (\ref{coord freedom dn}) allows us practically to specify the value of the metric function $ \alpha $ on two (non-parallel) lines (as opposed to one for $ S $).
We use up part of the freedom by choosing to have the regular origin $ r = 0 $ on the line $ u = v $, as in \cite{HS96} and shown in figure \ref{fig:dn geometry}. This reduces the available freedom by requiring the functions $ \tilde{u} $ and $ \tilde{v} $ to be equal.
We can use up the remaining freedom by specifying $ \alpha $ on a line of constant $ u $.
In fact we will do this by specifying $ \alpha(0,0) $ and $ \alpha_v(0,v) $ as part of the initial conditions; this will be done in section \ref{s:IBCs}.
This completely fixes the coordinates.

With this regular origin we see that in the untrapped region we have $ r_v > 0 $ and $ r_u < 0 $.
In the scalar field case it was shown that a regular origin necessarily implies $ r_u < 0 $ everywhere (\cite{Christodoulou93} prop. 1.1), and we note that with a Yang-Mills field this simple result doesn't hold.
In these coordinates we can write the Misner-Sharp mass (\ref{MS}) and $ N $ as
\begin{align} \label{mN dn}
m &= \frac{r}{2}\left(1+\frac{r_u r_v}{\alpha^2}\right) \;, & N &= -\frac{r_u r_v}{\alpha^2} \;.
\end{align}
The null expansions are proportional to $ r_u $ and $ r_v $ \cite{HS96}, and in particular a future MTS is indicated by $ r_u < 0 $ and $ r_v = 0 $.
We also note that it is possible to use the Misner-Sharp mass as a metric variable in place of $ \alpha $ (at the expense of introducing derivatives of $ r $ into the metric), but we have not found it as useful as it is in the polar-areal case.

As calculated in \cite{HS96}, the proper time along a line of constant $ r $ is
\begin{equation} \label{tau dn def}
\tau = \int_{u_0}^u 2\alpha \sqrt{\dydx{v}{u}} \dx{u} \;,
\end{equation}
where the line of constant $ r $ is parameterised by $ (u,v(u)) $, beginning at $ u_0 $ (usually 0). Note that (\ref{tau dn def}) simplifies to
\begin{equation} \label{tau dn}
\tau = \int_{u_0}^u 2\alpha \sqrt{-\frac{r_u}{r_v}} \dx{u} \;,
\end{equation}
which is useful because, as in \cite{HS96}, we will use $ r_u $ and $ r_v $ as variables in the evolution (see section \ref{ss:dn choices}).
On $ r = 0 $ this further simplifies to
\[ \tau_0 = \int_{u_0}^u 2\alpha \dx{u} \;. \]

For the double-null coordinates the first two gauge potential components will have no hat or bar over the top;
\begin{equation*} \label{potential dn}
A = a\tau_3\dy{u} + b\tau_3\dy{v} + (w\tau_1-d\tau_2)\dy{\theta} + \left(d\tau_1 + w\tau_2 + \cot\theta\tau_3\right)\sin\theta\dy{\phi} \;.
\end{equation*}
The non-zero components of the Yang-Mills stress-energy tensor are
\begin{equation} \label{T dn}
\begin{aligned}
8\pi T_{uu} &= 4\frac{(w_u+ad)^2+(d_u-aw)^2}{r^2} \;, \\
8\pi T_{uv} &= 2\frac{\alpha^2(w^2+d^2-1)^2}{r^4} +\frac{(a_v-b_u)^2}{2\alpha^2} \;, \\
8\pi T_{vv} &= 4\frac{(w_v+bd)^2+(d_v-bw)^2}{r^2} \;, \\
8\pi T_{\theta\theta} &= \frac{(w^2+d^2-1)^2}{r^2} +\frac{r^2(a_v-b_u)^2}{4\alpha^4} \;, \\
8\pi T_{\phi\phi} &= \left(\frac{(w^2+d^2-1)^2}{r^2} +\frac{r^2(a_v-b_u)^2}{4\alpha^4}\right)\sin^2\theta \;.
\end{aligned}
\end{equation}
and the energy density $ E $ in the non-coordinate basis formed by normalising $ \dy{v}+\dy{u} $, $ \dy{v}-\dy{u} $, $ \dy\theta $, and $ \dy\phi $ (cf. (\ref{energy density isothermal})), is
\begin{equation} \label{energy density dn}
8\pi E = \frac{(w^2+d^2-1)^2}{r^4} + \frac{(a_v-b_u)^2}{4\alpha^4} + \frac{(w_u+ad)^2+(w_v+bd)^2+(d_u-aw)^2+(d_v-bw)^2}{r^2\alpha^2} \;.
\end{equation}

In double-null coordinates, the non-zero components of the Einstein equations (\ref{EE}) reduce to
\begin{subequations} \label{EEuv}
\begin{align}
\left(\frac{r_u}{\alpha^2}\right)_u &= -2\frac{(w_u+ad)^2+(d_u-aw)^2}{r\alpha^2} \;, \label{ruu} \\
\frac{rr_{uv}+r_ur_v+\alpha^2}{r^2} &= \frac{\alpha^2(w^2+d^2-1)^2}{r^4} +\frac{(a_v-b_u)^2}{4\alpha^2} \;, \label{ruv} \\
\left(\frac{r_v}{\alpha^2}\right)_v &= -2\frac{(w_v+bd)^2+(d_v-bw)^2}{r\alpha^2} \;, \label{rvv} \\
(\ln\alpha)_{uv} - \frac{\alpha^2+r_ur_v}{r^2} &= -2\frac{\alpha^2(w^2+d^2-1)^2}{r^4} -\frac{(a_v-b_u)^2}{2\alpha^2} \;, \label{EEuvtheta}
\end{align}
\end{subequations}
where (\ref{ruu}-\ref{rvv}) are essentially the $ \mu\nu = uu $, $ uv $, and $ vv $ components, and (\ref{EEuvtheta}) is a combination of the $ \mu\nu = uv $ and $ \theta\theta $ components.

The non-zero Yang-Mills equations (\ref{YME}), respectively the $ a\nu = 3u $, $ 3v $, $ 1\theta $ and $ 2\theta $ components, are
\begin{subequations} \label{YMEuv}
\begin{align}
\partial_u \left(\frac{r^2(a_v-b_u)}{4\alpha^2}\right) &= -d(w_u+ad)+w(d_u-aw) \;, \label{abu} \\
\partial_v \left(\frac{r^2(a_v-b_u)}{4\alpha^2}\right) &= d(w_v+bd)-w(d_v-bw) \;, \label{abv} \\
(w_u+ad)_v +(w_v+bd)_u +a(d_v-bw) +b(d_u-aw) &= -\frac{2\alpha^2 w(w^2+d^2-1)}{r^2} \;, \label{wuv} \\
(d_u-aw)_v +(d_v-bw)_u -a(w_v+bd) -b(w_u+ad) &= -\frac{2\alpha^2 d(w^2+d^2-1)}{r^2} \;. \label{duv}
\end{align}
\end{subequations}
Due to the spherical symmetry, the $ a\nu =  2\phi $ and $ 1\phi $ components also give equations (\ref{wuv},\ref{duv}), respectively.
These equations are not independent;
\[  \partial_v (\ref{abu}) - \partial_u (\ref{abv}) -d(\ref{wuv}) +w(\ref{duv}) \equiv 0 \;. \]
This dependency corresponds to the freedom to choose a gauge (\ref{gauge transform}).

The twice contracted second Bianchi identity, for $ \nu = u $ and $ \nu = v $, gives the following combinations of the Yang-Mills equations:
\begin{subequations} \label{Bianchiuv}
\begin{align}
(a_v-b_u)(\ref{abu}) +(w_u+ad)(\ref{wuv}) +(d_u-aw)(\ref{duv}) &= 0 \;, \label{Bianchiu}\\
(a_v-b_u)(\ref{abv}) +(w_v+bd)(\ref{wuv}) +(d_v-bw)(\ref{duv}) &= 0 \;. \label{Bianchiv}
\end{align}
\end{subequations}
It is precisely these combinations of the Yang-Mills equations (\ref{YMEuv}) that can be used to show the interdependency of the Einstein equations (\ref{EEuv});
\begin{align*}
\partial_u(\ref{ruv}) -2\partial_v(\ref{ruu}) +\frac{2}{r^2}(\ref{Bianchiu}) &\equiv 0 \;, \\
\partial_v(\ref{ruv}) -2\partial_u(\ref{rvv}) +\frac{2}{r^2}(\ref{Bianchiv}) &\equiv 0 \;,
\end{align*}
where use has been made of (\ref{EEuv}).
As before there are two independent Einstein equations and three independent Yang-Mills equations in general.

We also note here that in the purely magnetic case, (\ref{Bianchiuv}) completely specifies the (single) Yang-Mills equation (\ref{wuv})
\[ w_{uv} = -\frac{\alpha^2 w(w^2-1)}{r^2} \;. \]

\section{Gauge choices and initial and boundary conditions} \label{s:GC}
In this section we detail how a final choice of gauge and appropriate initial and boundary conditions can be used to solve the equations in polar-areal (\ref{EEtr},\ref{YMEtr}) and double-null coordinates (\ref{EEuv},\ref{YMEuv}) numerically.
While only double-null coordinates will be used here for numerical evolution, we will describe the remaining gauge choices and the initial and boundary conditions required to evolve the equations in polar-areal coordinates, as they are fairly straightforward and will make a useful comparison with the double-null case.
We assume throughout that the solution is regular at the origin.
\subsection{Polar-areal coordinates}
We demand the metric to be regular at the origin, and we can see the restrictions this puts on the metric functions by writing (\ref{metric tr}) in Cartesian coordinates $ (x_1,x_2,x_3) = (r\sin\theta\cos\phi,r\sin\theta\sin\phi,r\cos\theta) $, as in \cite{BP92}:
\[ g = -S^2N\dy{t}^2 + \frac{\left(\frac{1}{N}-1\right)}{r^2}\left(x_1\dy{x_1}+x_2\dy{x_2}+x_3\dy{x_3}\right)^2 +\dy{x_1}^2+\dy{x_2}^2+\dy{x_3}^2 \;. \]
Thus if
\begin{align} \label{regular tr metric}
S(t,r) &= \tilde{S}(t,r^2) \;, & N(t,r) &= 1-r^2\tilde{N}(t,r^2) \;,
\end{align}
and so $ m(t,r) = \frac{r^3}{2}\tilde{N}(t,r^2) $, where $ \tilde{S} $ and $ \tilde{N} $ are smooth on $ \mathbb{R} \times [0,\infty) $, then the metric is regular across $ r = 0 $.

The remaining gauge freedom (\ref{gauge transform}) requires $ \lambda(t,r) $ to be completely specified.
We will demand the gauge potential to be smooth in the regular gauge (\ref{potential regular}), and this will restrict the possible choices of $ \lambda $.
Note that the $ (w,d) $ symmetry of the Abelian gauge (\ref{potential}) is broken in the regular gauge due to the particular gauge transformation chosen and thus the restrictions on the functions we find are not the only possible ones.

Writing the gauge potential (\ref{potential tr}) in the regular gauge and Cartesian coordinates we find
\begin{align*}
\tilde{A} &= \left(x_1\tau_1+x_2\tau_2+x_3\tau_3\right)\left(\frac{\hat{a}}{r}\dy{t} +\left(\frac{\hat{b}}{r^2}-\frac{d}{r^3}\right)\left(x_1\dy{x_1}+x_2\dy{x_2}+x_3\dy{x_3}\right)\right) \\
 &+\frac{1-w}{r^2}\left((x_2\tau_3-x_3\tau_2)\dy{x_1}+(x_3\tau_1-x_1\tau_3)\dy{x_2}+(x_1\tau_2-x_2\tau_1)\dy{x_3}\right) +\frac{d}{r}\left(\tau_1\dy{x_1}+\tau_2\dy{x_2}+\tau_3\dy{x_3}\right) \;.
\end{align*}
Noting that $ \{\tau_i\dy{t}, \tau_i\dy{x_j}, 1 \leqslant i,j \leqslant 3 \} $ form a global basis on the bundle $ SU(2) \otimes T^*\mathbb{R}^{3,1} $, if we have
\begin{equation} \label{regular tr}
\begin{aligned}
\hat{a}(t,r) &= r \tilde{\hat{a}}(t,r^2) \;, \\
\hat{b}(t,r) &= \tilde{d}(t,r^2) +r^2\tilde{\hat{b}}(t,r^2) \;, \\
w(t,r) &= 1-r^2\tilde{w}(t,r^2) \;, \\
d(t,r) &= r\tilde{d}(t,r^2) \;,
\end{aligned}
\end{equation}
where $ \tilde{\hat{a}} $, $ \tilde{\hat{b}} $, $ \tilde{w} $, and $ \tilde{d} $ are smooth on $ \mathbb{R} \times [0,\infty) $, then the connection $ \tilde{A} $ is explicitly globally defined and smooth. The converse was proven by Bartnik \cite{Bartnik95}.

In polar-areal coordinates the gauge freedom (\ref{gauge transform}) is
\begin{equation*}
\begin{aligned}
\hat{a} &\mapsto \hat{a} + \dot{\lambda} \;, \qquad & w &\mapsto w\cos\lambda - d\sin\lambda \;, \\
\hat{b} &\mapsto \hat{b} + \lambda' \;, & d &\mapsto d\cos\lambda + w\sin\lambda \;.
\end{aligned}
\end{equation*}
Thus to preserve (\ref{regular tr}) we will require $ \lambda(t,r) $ to be an odd function of $ r $, and we will make this same requirement when using double-null coordinates.

We find it useful to specify the polar gauge $ \hat{b} \equiv 0 $, particularly as this can be achieved by specifying $ \lambda' $ only. This means in the static case, the gauge functions remain explicitly time-independent in this gauge.
In (\ref{regular tr}) this requires $ \tilde{d} = -r^2\tilde{\hat{b}} $, thus $ d = O(r^3) $ as $ r \to 0 $.
While there still remains the possibility of setting $ \lambda $ on a line of constant $ r $, the demand that $ \lambda$ is odd requires $ \lambda(t,0) = 0 $, and thus this completely uses up the residual gauge freedom.

To clearly write the initial conditions we introduce the new variables
\begin{equation} \label{pa first}
\begin{aligned}
\hat{p} &:= w'+\hat{b}d \;, \qquad & \hat{x} &:= d'-\hat{b}w \;, \qquad & z &:= \frac{r^2\left(\hat{a}'-\dot{\hat{b}}\right)}{S} \;, \\
\hat{q} &:= \frac{\dot{w}+\hat{a}d}{SN} \;, & \hat{y} &:= \frac{\dot{d}-\hat{a}w}{SN} \;.
\end{aligned}
\end{equation}
Note that these are independent of the final coordinate fixing; if we change time coordinates to $ \tilde{t}(t) $, then $ S = \tilde{t}'\tilde{S} $, $ \hat{a} = \tilde{t}'\tilde{\hat{a}} $, and $ \pypx{}{t} = \tilde{t}'\pypx{}{\tilde{t}} $.
We also note that the electric charge $ Q $ (\ref{charge}) is given by $ \lim\limits_{r \to \infty} z $.
Then the equations (\ref{EEtr},\ref{YMEtr}) with the polar gauge choice become
\begin{equation} \label{tr evo}
\begin{aligned}
\dot{\hat{p}} &= (SN\hat{q})' -Sd\frac{z}{r^2} -\hat{a}\hat{x} \;, \\
\dot{\hat{q}} &= (SN\hat{p})' -Sw\frac{w^2+d^2-1}{r^2} -\hat{a}\hat{y} \;, \\
\dot{\hat{x}} &= (SN\hat{y})' +Sw\frac{z}{r^2} +\hat{a}\hat{p} \;, \\
\dot{\hat{y}} &= (SN\hat{x})' -Sd\frac{w^2+d^2-1}{r^2} +\hat{a}\hat{q} \;,
\end{aligned}
\end{equation}
and
\begin{subequations} \label{tr int}
\begin{align}
w' &= \hat{p} \;, \\
d' &= \hat{x} \;, \\
\hat{a}' &= \frac{Sz}{r^2} \;, \\
z' &= 2(\hat{q}d-w\hat{y}) \;, \\
S' &= \frac{2S}{r}\left(\hat{p}^2+\hat{x}^2+\hat{q}^2+\hat{y}^2\right), \\
m' &= \frac{z^2+(w^2+d^2-1)^2}{2r^2} +N\left(\hat{p}^2+\hat{x}^2+\hat{q}^2+\hat{y}^2\right), \label{tr m}
\end{align}
\end{subequations}
along with
\begin{equation} \label{tr check}
\begin{aligned}
\dot{m} &= 2SN^2\left(\hat{p}\hat{q}+\hat{x}\hat{y}\right), \\
\dot{z} &= 2SN\left(d\hat{p}-w\hat{x}\right), \\
\frac{r^2}{2S}\left(\left(\frac{\dot{N}}{SN^2}\right)^. +\left(\frac{(S^2N)'}{S}\right)' +\frac{2(SN)'}{r}\right) &= \frac{z^2+(w^2+d^2-1)^2}{r^2} \;.
\end{aligned}
\end{equation}
We can set initial conditions by specifying the four functions $ \hat{p}(0,r) $, $ \hat{q}(0,r) $, $ \hat{x}(0,r) $, and $ \hat{y}(0,r) $ so that $ \hat{p} $ and $ \hat{y} $ are odd, and $ \hat{q} $ and $ \hat{x} $ are even and zero at the origin, but are otherwise unconstrained.

The boundary conditions at the origin are then
\begin{align} \label{BCs 0 pa}
S(t,0) &= 1 \;, & m(t,0) &= 0 \;, & \hat{a}(t,0) &= 0 \;, & w(t,0) &= 1 \;, & d(t,0) &= 0 \;.
\end{align}
From (\ref{regular tr}) we see that the boundary conditions for these variables due to regularity are
\begin{align*}
\hat{p}(t,0) &= 0 \;, & \hat{x}(t,0) &= 0 \;, & z(t,0) &= 0 \;, \\
\hat{q}(t,0) &= 0 \;, & \hat{y}(t,0) &= 0 \;.
\end{align*}

Writing out the energy density (\ref{energy density}) in these variables;
\[ 8\pi E = \frac{z^2+(w^2+d^2-1)^2}{r^4} +\frac{2|N|}{r^2}\left(\hat{p}^2+\hat{x}^2+\hat{q}^2+\hat{y}^2\right) \;, \]
we see that the boundary conditions and (\ref{regular tr}) are consistent with having finite energy density at the origin. Demanding regularity purely through finite energy density (as in \cite{BP92} for example) allows a broader range of gauge choices ($ \lambda(t,r) $ no longer has to be odd with respect to $ r $, and $ w^2+d^2 = 1 $ is the only restriction on $ w $ and $ d $ at the origin), but we here consistently use the full gauge specification above.

In the following order, equations (\ref{tr int}) can be integrated to determine the remaining functions on $ t = 0 $:
\begin{align*}
w &= 1+\int_0^r \hat{p}\dx{r} \;, & d &= \int_0^r \hat{x}\dx{r} \;, \\
z &= 2\int_0^r \hat{q}d-w\hat{y}\dx{r} \;, & S &= e^{\int_0^r \frac{2}{r}\left(\hat{p}^2+\hat{x}^2+\hat{q}^2+\hat{y}^2\right)\dx{r}} \;, \\
\hat{a} &= \int_0^r \frac{Sz}{r^2}\dx{r} \;, & m &= \frac{1}{S}\int_0^r S\left(\frac{z^2+(w^2+d^2-1)^2}{2r^2} +\hat{p}^2+\hat{x}^2+\hat{q}^2+\hat{y}^2\right)\dx{r} \;.
\end{align*}
This completes the initial data specification.

The evolution can then begin using (\ref{tr evo}) with an appropriate numerical method such as (one- or two-step) Lax-Wendroff, leap-frog, or the method of lines, to find $ \hat{p} $, $ \hat{q} $, $ \hat{x} $, and $ \hat{y} $ on a future $ t = \text{constant} $ line.
Since the system is not purely hyperbolic, each of these methods requires the supplemental integration of the equations (\ref{tr int}) for the remaining variables at each new time step.
The equations (\ref{tr check}) are not needed in the evolution and can be used as a check on the numerical results.
The time derivatives of $ m $ and $ z $ are merely consistent with their respective spatial derivatives, while the last equation is an identity when all the other equations are satisfied.

Lastly we discuss the boundary conditions at (spacelike) infinity.
$ S $ and $ m $ require only one boundary condition each, which we have already specified at the origin. Since we are mostly interested in asymptotically flat spacetimes, we here note what restrictions this places on the various functions.
We enforce asymptotic flatness by having finite ADM mass $ M = \lim\limits_{r \to \infty} m(t,r) $.
This also implies $ \lim\limits_{r \to \infty} S(t,r) < \infty $.
From (\ref{tr m}) we see that for any $ t $, $ \int_0^\infty N \left(\hat{p}^2+\hat{x}^2+\hat{q}^2+\hat{y}^2\right) \dy{r} < \infty $, and so
\[ \frac{S(t,\infty)}{S(t,0)} = \int_0^\infty \frac{2}{r} \left(\hat{p}^2+\hat{x}^2+\hat{q}^2+\hat{y}^2\right) \dy{r} < \infty \;, \]
since $ \frac{2}{r} $ is eventually bounded by $ N $.

This condition also puts restrictions on the asymptotic behaviour of the Yang-Mills fields. We make the further simplifying assumption that all our functions have a power series expansion in $ \frac{1}{r} $ at infinity, as did \cite{GE89,EG90,Zhou92,BH00-2}.
The more general asymptotics were considered in the static case in \cite{BP92}.
Since $ M = \int\limits_0^\infty \frac{z^2+(w^2+d^2-1)^2}{2r^2} +N\left(\hat{p}^2+\hat{x}^2+\hat{q}^2+\hat{y}^2\right) \dx{r} $, we find that finite ADM mass requires
\begin{align} \label{BCs inf pa}
w^2+d^2 &= O(1) \;, & \hat{p} &= O\mathopen{}\left(\frac{1}{r}\right)\mathclose{}, & \hat{q} &= O\mathopen{}\left(\frac{1}{r}\right)\mathclose{}, \nonumber \\
z &= O(1) \;, & \hat{x} &= O\mathopen{}\left(\frac{1}{r}\right)\mathclose{}, & \hat{y} &= O\mathopen{}\left(\frac{1}{r}\right)\mathclose{},  \\
\hat{a}w &= O\mathopen{}\left(\frac{1}{r}\right)\mathclose{} \;. \nonumber
\end{align}

\subsection{Double-null coordinates} \label{ss:dn choices}
To speak of the regularity requirements in double-null coordinates, we find it useful to consider isothermal coordinates, which relate simply  (see (\ref{uvTR})).
We see from (\ref{metric TR}) that regularity at the origin demands $ \alpha $ is an even function of $ R $.
Evaluating (\ref{mN TR}) at the origin tells us $ r' = \alpha $ there.
Furthermore, writing (\ref{metric TR}) in Cartesian coordinates (with $ x_1^2+x_2^2+x_3^2 = R^2 $) tells us that $ r $ is an odd function of $ R $.
We will use the terms even and odd when referring to functions of any of $ r $, $ R $, or $ v-u $ equivalently.

The gauge freedom (\ref{gauge transform}) allows us to choose $ \lambda(u,v) $ to affect the gauge functions via:
\begin{equation*}
\begin{aligned}
a &\mapsto a + \lambda_u \;, \qquad & w &\mapsto w\cos\lambda - d\sin\lambda \;, \\
b &\mapsto b + \lambda_v \;, & d &\mapsto d\cos\lambda + w\sin\lambda \;.
\end{aligned}
\end{equation*}
It is tempting to try to achieve a similar simplification as in polar-areal coordinates, and set $ \lambda $ to make $ b \equiv 0 $.
However, this is not compatible with the simple $ w = 1 $ and $ d = 0 $ boundary conditions as it generically involves a $ \lambda $ that is not purely odd.
Another sensible option might be to enforce the polar gauge $ \hat{b} \equiv 0 $, which we know is compatible with these boundary conditions.
From equations (\ref{ab dn to pa}) we see that this means that $ a $ and $ b $ are related by
\begin{equation} \label{dn polar gauge}
-\frac{a}{r_u} = \frac{b}{r_v} \;.
\end{equation}
We can use this choice with (\ref{EEuv},\ref{YMEuv}) and form a first-order system by introducing new variables (see (\ref{dn first}) below). However, the resulting first-order equation for $ a $, which comes from inserting (\ref{dn polar gauge}) into the definition of $ z $, is an advection equation (which involves both derivatives of $ a $) as opposed to the simple first-order equations that exist for the remaining variables.

Therefore we take a different approach and look for a relationship between $ a $ and $ b $ involving only the derivatives $ a_v $ and $ b_u $ that can be enforced using an odd-in-$ r $ $ \lambda $.
We find that the restriction
\begin{equation} \label{dn gauge}
a_v + b_u = 0
\end{equation}
fits these requirements well.
This requires $ \lambda_{uv} = -\frac{a_v+b_u}{2} $.
Using (\ref{dn derivs in pa}) and (\ref{c}) we write these terms with respect to polar-areal coordinates and find that
\begin{align*}
\lambda_{uv} &= r_ur_v\left(\lambda''-\frac{1}{c^2}\ddot\lambda +\frac{c'}{c}\lambda'+\frac{\dot{c}}{c^3}\dot\lambda\right), \\
a_v-b_u &= 2r_ur_v\left(\hat{b}'-\frac{1}{c^2}\dot{\hat{a}} +\frac{c'}{c}\hat{b}+\frac{\dot{c}}{c^3}\hat{a}\right).
\end{align*}
Recalling $ S $, $ N $, and $ \alpha $ are even, the equations (\ref{SN from dn}) tell us that $ c $ and $ r_ur_v $ are also even.
Using (\ref{regular tr}), we therefore see that this gauge choice is compatible with an odd $ \lambda $ and the $ (w,d) = (1,0) $ boundary conditions.

Since only $ \lambda_{uv} $ has been specified, there is still some gauge freedom remaining. We can still specify $ \lambda $ on two non-parallel lines, for example $ u = 0 $ and $ u = v $. Indeed on $ u = 0 $ we will set $ \lambda_v(0,v) $ (and thus $ b(0,v) $) as part of the initial conditions, which is unrestricted. But due to the odd requirement on $ \lambda $, it must remain $ 0 $ on $ u = v $, therefore this completes the gauge fixing.

To write the Einstein-Yang-Mills equations (\ref{EEuv},\ref{YMEuv}) in a first-order form, we introduce the new variables
\begin{equation} \label{dn first}
\begin{aligned}
\beta &= \frac{\alpha_u}{\alpha} \;, \qquad & f &= r_u \;, \qquad & p &= w_u+ad \;, \qquad & x &= d_u-aw \;, \qquad & z &= \frac{r^2(a_v-b_u)}{2\alpha^2} \;, \\
\gamma &= \frac{\alpha_v}{\alpha} \;, & g &= r_v \;, & q &= \frac{w_v+bd}{r} \;, & y &= \frac{d_v-bw}{r} \;.
\end{aligned}
\end{equation}
These variables in general depend on the final coordinate choice, but $ z $ does not. If we choose new coordinates $ \tilde{u}(u) $ and $ \tilde{v}(v) $, then $ \alpha^2 = \tilde{u}'\tilde{v}'\tilde\alpha^2 $, $ a = \tilde{u}'\tilde{a} $, $ b = \tilde{v}'\tilde{b} $, $ \pypx{}{u} = \tilde{u}'\pypx{}{\tilde{u}} $, and $ \pypx{}{v} = \tilde{v}'\pypx{}{\tilde{v}} $.
We use the variable name $ z $ here as well as in (\ref{pa first}) because they represent the same coordinate-independent function (see appendix \ref{ss:pa dn}).

Applying the gauge choice (\ref{dn gauge}), the commutation of mixed partial derivatives, and the equations (\ref{EEuv},\ref{YMEuv}) we find four variables that need to be evolved with $ u $-derivatives;
\begin{subequations} \label{F4}
\begin{align}
q_u &= -\frac{fq}{r} -\frac{\alpha^2\left(w(w^2+d^2-1)+dz\right)}{r^3} -ay \;, \label{gau} \\
y_u &= -\frac{fy}{r} +\frac{\alpha^2\left(d(w^2+d^2-1)+wz\right)}{r^3} +aq \;, \\
\gamma_u &= \frac{\alpha^2+fg}{r^2} -\frac{2\alpha^2\left((w^2+d^2-1)^2+z^2\right)}{r^4} \;, \\
b_u &= -\frac{\alpha^2 z}{r^2} \;.
\end{align}
\end{subequations}
These will be referred to as $ u $-direction variables.
There are 11 remaining variables that can be integrated with $ v $-derivative equations, which will be referred to as $ v $-direction variables;
\begin{subequations} \label{G11}
\begin{align}
(\ln\alpha)_v &= \gamma \;, \label{alphav} \\
r_v &= g \;, \label{rv} \\
\left(\frac{g}{\alpha^2}\right)_v &= -2(q^2+y^2)\frac{r}{\alpha^2} \;, \label{gv} \\
w_v &= rq-bd \;, \label{wv} \\
d_v &= ry+bw \;, \label{dv} \\
z_v &= 2r(dq-wy) \;, \label{zv} \\
a_v &= \frac{\alpha^2 z}{r^2} \;, \label{av} \\
f_v &= -\frac{\alpha^2+fg}{r} +\frac{\alpha^2\left((w^2+d^2-1)^2+z^2\right)}{r^3} \;, \label{fv} \\
p_v &= -bx -\frac{\alpha^2\left(w(w^2+d^2-1)-dz\right)}{r^2} \;, \label{pv} \\
x_v &= bp -\frac{\alpha^2\left(d(w^2+d^2-1)+wz\right)}{r^2} \;, \label{xv} \\
\beta_v &= \frac{\alpha^2+fg}{r^2} -\frac{2\alpha^2\left((w^2+d^2-1)^2+z^2\right)}{r^4} \;. \label{betav}
\end{align}
\end{subequations}
We note that it is possible to evolve the remaining variables without $ p $, $ x $, and $ \beta $.
There are 7 remaining $ u $-derivative equations that are not required in the evolution and can be used as checks on the numerical method:
\begin{equation} \label{H7}
\begin{aligned}
(\ln\alpha)_u &= \beta \;, \\
r_u &= f \;, \\
w_u &= p-ad \;, \\
d_u &= x+aw \;, \\
g_u &= -\frac{\alpha^2+fg}{r} +\frac{\alpha^2\left((w^2+d^2-1)^2+z^2\right)}{r^3} \;, \\
\left(\frac{f}{\alpha^2}\right)_u &= -\frac{2(p^2+x^2)}{r\alpha^2} \;, \\
z_u &= 2(wx-dp) \;.
\end{aligned}
\end{equation}

While the gauge variable dependencies of $ p $, $ q $, $ x $, and $ y $ in (\ref{dn first}) are suggested by the appearance of these combinations in (\ref{EEuv},\ref{YMEuv}), the $ r $ in the denominators of $ q $ and $ y $ are necessary to keep (\ref{gv}) linear in $ r $.
The importance of this is explained in section \ref{s:HS}.
While formally singular, $ q $ and $ y $ are in fact regular functions, which follows from requiring finite energy density (\ref{energy density dn}) at the origin.
With these variables the Misner-Sharp mass (\ref{mN dn}) can now be written
\begin{equation} \label{MS dn}
m = \frac{r}{2}\left(1+\frac{fg}{\alpha^2}\right).
\end{equation}

The initial conditions required for this system are $ \gamma(0,v) $, $ q(0,v) $, $ y(0,v) $, and $ b(0,v) $. $ \gamma(0,v) $ (along with $ \alpha(0,0) $) completes the coordinate specification while $ b(0,v) $ completes the gauge specification.
We wish to disentangle the physically meaningful initial conditions from purely coordinate conditions. Therefore we do not specify $ q(0,v) $ and $ y(0,v) $ which involve $ v $ derivatives of the gauge variables $ w $ and $ d $, but rather the derivatives of $ w $ and $ d $ with respect to the areal radius $ r $. These are geometric quantities and are independent of the coordinate condition chosen by specifying $ \gamma(0,v) $.
On the line $ u = 0 $ we relate these to $ q $ and $ y $ via $ w_v = w'(r(v))r_v(v) $ and $ d_v = d'(r(v))r_v(v) $. To set $ q(0,v) $ from $ w'(r) $ we need to find $ r(v) $ which we do by solving (\ref{rvv}) (or equivalently (\ref{rv},\ref{gv})).

We note that although $ w $ needs to be have an even expansion at $ r = 0 $ on any constant $ t $ slice, on a constant $ u $ slice there is no restriction on the form of $ w'(r) $ other than $ w'(0) = 0 $ and that $ w $ remains finite at future null infinity (\ref{BCs inf dn}).
There is then as much information in the initial conditions for $ w $ along $ u = 0 $ as there is in polar-areal coordinates along $ t = 0 $, which consisted of the odd $ w'(0,r) $ and the even $ \dot{w}(0,r) $.
Similarly, although $ d $ needs to have an odd expansion at $ r = 0 $ on any constant $ t $ slice, $ d' $ is unrestricted on $ u = 0 $ other than $ d $ remains finite at future null infinity.

The boundary conditions that (\ref{G11}) requires are each of the $ v $-direction variables at the origin.
The boundary conditions are simply those that keep $ r = 0 $ on the line $ u = v $ as well as those that follow from regularity.
Therefore on $ u = v $, we set
\begin{subequations} \label{BCs 0 dn}
\begin{align}
r &= 0 \;,& w &= 1 \;,& d &= 0 \;,& a &= 0 \;,& z &= 0 \;.
\end{align}
Since $ r $, $ w $, $ d $ are constant along $ u = v $, we have $ r_u + r_v = 0 $, etc., thus
\begin{align}
f &= -g \;,& p &= 0 \;,& x &= 0 \;.
\end{align}
Furthermore since $ \pypx{\alpha}{r} = 0 $ at the origin, we conclude
\begin{equation}
\beta = \gamma \;,
\end{equation}
on $ u = v $.
We use $ N = -\frac{fg}{\alpha^2} = 1 $ to set
\begin{equation}
g = \alpha \;.
\end{equation}
\end{subequations}

We follow \cite{HS96} and set $ \alpha(u,u) $ by requiring $ \pypx{\alpha}{r} = 0 $, that is, $ \pypx{\alpha}{v}-\pypx{\alpha}{u} = 0 $.
Such a boundary condition can be achieved numerically by approximating $ \pypx{\alpha}{v}-\pypx{\alpha}{u} = 0 $ at the point $ (u,u) $ with a second-order approximation to the first derivative using additional values of $ \alpha $ at $ (u-h,u+h) $ and $ (u-2h,u+2h) $.
In \cite{HS96}, the same technique is used to set the value of the scalar field on the origin boundary. It is not necessary to set the gauge variables on the origin with this method due to the gauge choice made.

Differentiating the mass (\ref{MS dn}) with respect to $ v $, and using (\ref{G11}) we find
\begin{equation} \label{mv}
m_v = \left(\frac{(w^2+d^2-1)^2+z^2}{2r^2} +N\frac{r^2(q^2+y^2)}{g^2}\right)r_v \;.
\end{equation}
Then integrating this to future null infinity, changing coordinates to $ r $, and demanding the mass is finite tells us
\begin{align} \label{BCs inf dn}
w^2+d^2 &= O(1) \;, & z &= O(1) \;, & \frac{q}{g} &= O\mathopen{}\left(\frac{1}{r^2}\right)\mathclose{}, & \frac{y}{g} &= O\mathopen{}\left(\frac{1}{r^2}\right)\mathclose{}.
\end{align}
This is consistent with the results in polar-areal coordinates (\ref{BCs inf pa}).

\section{Embedded Abelian solutions} \label{s:EA}
In this section we derive the solutions obtainable from the embedded Abelian gauge potential (\ref{potentialAbel}), and how they may be realised in the general case (\ref{potential}). These are essentially the electrovacuum (source-free Einstein-Maxwell) solutions.
By an extension of the Jebsen-Birkhoff theorem by Hoffmann to the electrovacuum case \cite{Hoffmann32,Hoffmann62}, the only solutions possible are the Reissner-Nordstr\"{o}m family and the Bertotti-Robinson family.
This becomes quite clear in isothermal coordinates.

With the potential (\ref{potentialAbel}), the Yang-Mills equations (\ref{YMETR}) become
\begin{align*}
\partial_R \left(\frac{r^2\left(\bar{a}'-\dot{\bar{b}}\right)}{2\alpha^2}\right) &= 0 \;, & \partial_T \left(\frac{r^2\left(\bar{a}'-\dot{\bar{b}}\right)}{2\alpha^2}\right) &= 0 \;,
\end{align*}
and we write $ \bar{a}'-\dot{\bar{b}} = \frac{\alpha^2 Q}{r^2} $, where $ Q $ is an arbitrary constant, and retain for the moment the residual gauge freedom.
The Einstein equations (\ref{EETR}) then become
\begin{subequations} \label{EETRAbel}
\begin{align}
\left(\frac{r'}{\alpha^2}\right)^. +\left(\frac{\dot{r}}{\alpha^2}\right)' &= 0 \;, \label{ralphaTRAbel} \\
\left(\frac{\dot{r}}{\alpha^2}\right)^. +\left(\frac{r'}{\alpha^2}\right)' &= 0 \;, \label{ralphaTTAbel} \\
\left(r^2\right)^{..}-\left(r^2\right)'' &= -2\alpha^2\left(1-\frac{Q^2+k^2}{r^2}\right) \;, \label{rTTAbel} \\
\left(\ln\alpha\right)^{..}-\left(\ln\alpha\right)'' +\frac{r'^2-\dot{r}^2}{r^2} &= \frac{\alpha^2}{r^2}\left(1-2\frac{Q^2+k^2}{r^2}\right) \;. \label{alphaTTAbel}
\end{align}
\end{subequations}

Turning briefly to double-null coordinates, we can write (\ref{ralphaTRAbel}-\ref{rTTAbel}) as
\begin{align*}
\left(\frac{r_u}{\alpha^2}\right)_u &= 0 \;, & \left(\frac{r_v}{\alpha^2}\right)_v &= 0 \;, & rr_{uv} +r_ur_v &= -\alpha^2\left(1-\frac{Q^2+k^2}{r^2}\right) \;,
\end{align*}
and so $ r_u = G(v)\alpha^2 $ and $ r_v = F(u)\alpha^2 $ where $ F $ and $ G $ are arbitrary functions.
Now considering the Misner-Sharp mass (\ref{mN dn}) $ m = \frac{r}{2}\left(1+FG\alpha^2\right) $, we find that $ m_u = r_u \frac{Q^2+k^2}{2r^2} $ and $ m_v = r_v \frac{Q^2+k^2}{2r^2} $, thus $ m = M -\frac{Q^2+k^2}{2r} $, where $ M $ is another arbitrary constant.
We now have
\[ N = 1-\frac{2M}{r}+\frac{Q^2+k^2}{r^2} = -Fr_u = -Gr_v \;. \]

If both $ F $ and $ G $ are non-zero, we can solve the above equations for $ r $.
Letting $ F \equiv -1 $ and $ G \equiv 1 $, we return to isothermal coordinates and find $ \pypx{r}{T} = 0 $ and $ \dydx{r}{R} = 1-\frac{2M}{r}+\frac{Q^2+k^2}{r^2} $, which gives the Reissner-Nordstr\"{o}m solution \cite{Reissner16}. Other non-zero choices of $ F $ and $ G $ correspond to coordinate transformations (\ref{coord freedom dn}) of this solution.
For the gauge potential we again choose the polar gauge $ \bar{b} \equiv 0 $, with $ \lim\limits_{r \to \infty} \bar{a} = 0 $ (which is not compatible with our standard gauge choice that ensures regularity at the origin (\ref{regular tr})), so we have:
\begin{subequations} \label{RN TR}
\begin{align}
g &= -\left(1-\frac{2M}{r}+\frac{Q^2+k^2}{r^2}\right)\dy{T}^2 +\left(1-\frac{2M}{r}+\frac{Q^2+k^2}{r^2}\right)\dy{R}^2 + r^2\dy{\theta}^2 + r^2\sin^2\theta\dy{\phi}^2 \;, \\
A &= -\frac{Q}{r}\tau_3\dy{T} +k\tau_3\cos\theta\dy{\phi} \;, \label{RN TR A}
\end{align}
\end{subequations}
where $ M $ and $ Q $ are two free parameters corresponding to the ADM mass and the electric charge, and $ -k $ corresponds to the magnetic charge.
For $ M = Q = k = 0 $ this is Minkowski space.
The choice $ Q = k = 0 $ gives the Schwarzschild family, with event horizon $ r_h = 2M $ for $ M > 0 $.
In general, the Reissner-Nordstr\"{o}m family has an event horizon $ r_h = r_+ $ when $ M^2 \geqslant Q^2 + k^2 $, where $ r_\pm = M \pm \sqrt{M^2-Q^2-k^2} $.
This can be written in the familiar polar-areal coordinates:
\begin{subequations} \label{RN pa}
\begin{align}
g &= -\left(1-\frac{2M}{r}+\frac{Q^2+k^2}{r^2}\right)\dy{t}^2 +\frac{1}{1-\frac{2M}{r}+\frac{Q^2+k^2}{r^2}}\dy{r}^2 + r^2\dy{\theta}^2 + r^2\sin^2\theta\dy{\phi}^2 \;, \\
A &= -\frac{Q}{r}\tau_3\dy{t} +k\tau_3\cos\theta\dy{\phi} \;,
\end{align}
\end{subequations}
and in double-null coordinates, where the coordinate freedom (equivalently the choice of $ F $ and $ G $ above) can be utilised to cover the range $ r_- < r < \infty $ with a single coordinate patch.
For example,
\begin{subequations} \label{RN dn}
\begin{align*}
g &= -\frac{4r_+^4}{(r_+-r_-)^2}\frac{(r-r_-)^{1+\frac{r_-^2}{r_+^2}}}{r^2\cos^2u\cos^2v}e^{-\frac{r_+-r_-}{r_+^2}r}\dy{u}\dy{v} + r^2\dy{\theta}^2 + r^2\sin^2\theta\dy{\phi}^2 \;, \\
A &= -\frac{2r_+^2}{r_+-r_-}\frac{Q}{r}\tau_3\left(\frac{1}{\sin(2v)}\dy{v}-\frac{1}{\sin(2u)}\dy{u}\right) +k\tau_3\cos\theta\dy{\phi} \;,
\end{align*}
\end{subequations}
where $ r(u,v) $ satisfies $ e^{\frac{r_+-r_-}{r_+^2}r}\frac{r-r_+}{(r-r_-)^{r_-^2 r_+^{-2}}} = -\tan(u)\tan(v) $ and $ -\frac{\pi}{2} < u < \frac{\pi}{2} $ and $ -\frac{\pi}{2} < v < \frac{\pi}{2} $.

If, on the other hand, either $ F $ or $ G $ are chosen zero, we find that $ r = r_\pm $, and hence both $ F $ and $ G $ must be zero to ensure that $ \alpha^2 > 0 $. Turning to (\ref{rTTAbel}), we see that $ r = \sqrt{Q^2+k^2} $, which in turn means (\ref{alphaTTAbel}) reduces to
\[ \left(\ln\alpha\right)^{..}-\left(\ln\alpha\right)'' = -\frac{\alpha^2}{r^2} \;. \]
This equation has a special solution $ \alpha = \frac{r}{R} $, and because there are two arbitrary functions (\ref{coord freedom TR}) associated to the coordinate freedom of $ \alpha $, any other solution is equivalent to this one by choosing the two initial conditions of the above wave equation.
We can then choose $ \bar{a} = -\frac{Q}{R} $, $ \bar{b} = 0 $ as before.
Other special solutions include $ \alpha = \frac{r}{\sinh{R}} $ and $ \bar{a} = -Q\coth{R} $, $ \alpha = \frac{r}{\cos{R}} $ and $ \bar{a} = Q\tan{R} $, $ \alpha = \frac{r}{\cosh{T}} $ and $ \bar{b} = -Q\tanh{T} $, as well as simple transformations of these by $ R \mapsto CR+D $ or $ T \mapsto CT+D $.
These are all the solutions that depend on only one of the variables.
This is the Bertotti-Robinson solution \cite{Bertotti59}, where the $ \alpha = \frac{r}{R} $ choice gives
\[ g = -\frac{Q^2+k^2}{R^2}\dy{T}^2 +\frac{Q^2+k^2}{R^2}\dy{R}^2 + (Q^2+k^2)\dy{\theta}^2 +(Q^2+k^2)\sin^2\theta\dy{\phi}^2 \;, \]
which is static.

We now consider how the above solutions may appear when considering the non-Abelian potential (\ref{potential}).
The $ k = 1 $ Abelian solutions are achieved when $ w \equiv d \equiv 0 $. Thus a Reissner-Nordstr\"{o}m solution may appear, but only with magnetic charge equal to $ -1 $.
We note here that these Reissner-Nordstr\"{o}m solutions differ slightly from their electromagnetic counterparts; the principal bundles are trivial, and they are unstable \cite{LNW92}.

The Schwarzschild solution cannot be achieved in this way because it requires $ k = 0 $. However, the vacuum ($ T_{\mu\nu} = 0 $) configuration for (\ref{potential isothermal}) is $ w \equiv 1 $, $ \bar{a} \equiv \bar{b} \equiv d \equiv 0 $, up to an arbitrary residual gauge transformation.
It is straightforward to see that the Einstein equations (\ref{EEtr}) give $ m = M $, and thus the Schwarzschild solution
\[ g = -\left(1-\frac{2M}{r}\right)\dy{T}^2 +\left(1-\frac{2M}{r}\right)\dy{R}^2 + r^2\dy{\theta}^2 +r^2\sin^2\theta\dy{\phi}^2 \;. \]
The gauge potential $ A = \tau_1\dx{\theta} + (\sin\theta\tau_2 + \cos\theta\tau_3)\dx{\phi} $ is pure gauge, and can be written as $ A = g^{-1}\dx{g} $ with $ g = e^{\phi\tau_3}e^{\theta\tau_1} $, and thus is equivalent to (\ref{RN TR A}) with $ Q = k = 0 $.

If we consider an $ r = $ constant solution with (\ref{potential isothermal}), we see that (\ref{ralphaTT}) implies
\[ \dot{w}+\bar{a}d = w'+\bar{b}d = \dot{d}-\bar{a}w = d'-\bar{b}w = 0 \;, \]
and then the Yang-Mills equations (\ref{YMETR}) imply that $ w \equiv d \equiv 0 $.
Thus we recover the Bertotti-Robinson solution with $ k = 1 $, and so there are no new $ r = \text{constant} $ dynamics in the non-Abelian case.
We will henceforth only be considering the fully non-Abelian gauge potential (\ref{potential}) with $ r $ non-constant.

\chapter{Static solutions} \label{ch:S}
In this chapter we assume $ \pypx{}{t} $ is a Killing vector so that the metric and gauge variables depend only on the other quotient manifold coordinate (for example, $ r $).
The spacetime is therefore static where $ N > 0 $ and spatially homogeneous where $ N < 0 $.
We begin by writing out the full equations in this case, which simplify considerably due to the additional symmetry. We will use a number of coordinate systems, as different properties of the solutions lend themselves to be analysed most easily in different coordinates.
To investigate the static solutions, we consider in each section one possible property; a regular origin, a singular origin, a regular horizon, an equator (a maximum of $ r $), Bertotti-Robinson asymptotics, and asymptotic flatness. A single solution will exhibit at least two of these behaviours, and our results are summarised after each of these properties is considered.

We make use of the polar gauge $ \hat{b} \equiv 0 $ ($ \bar{b} \equiv 0 $), which can be implemented without affecting the explicit time-independence of the gauge functions.
With this choice the equations (\ref{mt}) and (\ref{b}) both imply $ d = Cw $ for some constant $ C $, which by the regularity conditions (\ref{regular tr}) gives us $ d \equiv 0 $.
Therefore the resulting static Einstein-Yang-Mills equations in polar-areal coordinates (\ref{EEtr},\ref{YMEtr}) are
\begin{subequations} \label{EYME stat}
\begin{align}
m' &= \frac{(w^2-1)^2}{2r^2} + \frac{r^2\hat{a}'^2}{2S^2} + N\left(w'^2 + \frac{\hat{a}^2w^2}{S^2N^2}\right), \label{m stat}\\
S' &= \frac{2S}{r}\left(w'^2 + \frac{\hat{a}^2w^2}{S^2N^2}\right), \label{S stat} \\
\left(\frac{r^2\hat{a}'}{S}\right)' &= \frac{2w^2\hat{a}}{SN} \;, \label{a stat} \\
\left(SNw'\right)' &= \left(S\frac{w^2-1}{r^2} - \frac{\hat{a}^2}{SN}\right)w \;, \label{w stat}
\end{align}
with the additional second-order Einstein equation (\ref{Etheta}), which can be used as a check of the numerical results:
\begin{equation}
\frac{r^2}{2S}\left(\left(\frac{(S^2N)'}{S}\right)' +\frac{2(SN)'}{r}\right) = \frac{r^2\hat{a}'^2}{S^2} + \frac{(w^2-1)^2}{r^2} \;.
\end{equation}
\end{subequations}
It is possible to remove $ S $ from the above system of equations and solve (\ref{S stat}) separately, for example by defining $ \hat\Phi := \frac{\hat{a}}{S} $. This was commonly done in the purely magnetic case, for example in \cite{BFM94, SW98, BP92}. 
Here we choose to keep $ S $ in the system to retain the simplicity of (\ref{a stat}).

Isothermal coordinates will prove to be very useful in examining the static solutions. Since we will use different coordinates in this chapter, we will denote $ \pypx{}{R} $ with a subscript $ R $, and a prime will only refer to a derivative with respect to $ r $.
In the static case the equations (\ref{EETR},\ref{YMETR}) can be written:
\begin{subequations} \label{EYME stat TR}
\begin{align}
\left(\frac{r_R}{\alpha^2}\right)_R &= -2\frac{w\indices{_R^2}+\bar{a}^2w^2}{r\alpha^2} \;, \label{rRR+} \\
\left(\frac{r^2\alpha_R}{\alpha}\right)_R &= \frac{\alpha^2(w^2-1)^2}{r^2}+\frac{r^2\bar{a}\indices{_R^2}}{\alpha^2} +2\left(w\indices{_R^2}+\bar{a}^2w^2\right), \label{alphaRR+} \\
\left(\frac{r^2\bar{a}_R}{\alpha^2}\right)_R &= 2w^2\bar{a} \;, \label{aRR} \\
w_{RR} +\bar{a}^2w &= \alpha^2w\frac{w^2-1}{r^2} \;. \label{wRR}
\end{align}

There are three algebraically independent Einstein equations (which are of course not differentially independent when including the Yang-Mills equations as shown in chapter \ref{ch:SSE}) and various combinations of these can be used to solve for $ r $ and $ \alpha $. Above we have chosen combinations that are second-order in the metric variables and have single-signed right-hand sides, resulting in monotonic quantities.
For the third algebraically independent equation, we choose the $ RR $ component of Einstein's equations, which contains only first-order derivatives:
\begin{equation}
2\frac{rr_R\alpha_R}{\alpha^3} +\frac{r\indices{_R^2}}{\alpha^2}-1 =-\frac{(w^2-1)^2}{r^2} -\frac{r^2\bar{a}\indices{_R^2}}{\alpha^4}  +\frac{2}{\alpha^2}\left(w\indices{_R^2}+\bar{a}^2w^2\right). \label{ralpha1}
\end{equation}
This is equivalent to the Hamiltonian that results from the Legendre transform of the reduced Lagrangian, which is preserved because the equations are autonomous in these coordinates.
We do not use this equation (or (\ref{rRR+})) to solve for $ \alpha $ because it becomes singular when $ r_R = 0 $, which occurs on an equator. We do not use it to solve for $ r $ because it is not linear in $ r_R $.
There are some further potentially useful combinations of the Einstein equations that we list here:
\begin{align}
\left(r^2\right)_{RR} &= 2\alpha^2\left(1-\frac{(w^2-1)^2}{r^2}-\frac{r^2\bar{a}\indices{_R^2}}{\alpha^4}\right), \label{rRR} \\
\left(\ln\alpha\right)_{RR} -\frac{r\indices{_R^2}}{r^2} &= -\frac{\alpha^2}{r^2}\left(1-2\frac{(w^2-1)^2}{r^2}-2\frac{r^2\bar{a}\indices{_R^2}}{\alpha^4}\right), \label{alphaRR} \\
\left(\frac{r}{\alpha}(r\alpha)_R\right)_R &= \alpha^2 +2\left(w\indices{_R^2}+\bar{a}^2w^2\right). \label{ralphaRR}
\end{align}
\end{subequations}

We will also make use of the Kretschmann scalar in these coordinates:
\begin{equation*} \label{K iso}
K = R_{\mu\nu\kappa\lambda}R^{\mu\nu\kappa\lambda} = 4\left(\frac{(\ln\alpha)\indices{_{RR}^2}}{\alpha^4} +\frac{1}{r^2}\left(\frac{r_R}{\alpha^2}\right)\indices{_R^2} +\frac{r\indices{_{RR}^2}}{r^2\alpha^4} + \frac{1}{r^4}\left(1-\frac{r\indices{_R^2}}{\alpha^2}\right)^2\right).
\end{equation*}

Having written the equations in two different coordinate systems, we now make note of the dimensionality of the space of solutions and ensure that it is consistent.
We can in general determine a unique solution in polar-areal coordinates by specifying finite values of $ r = r_0 $, $ m(r_0) $, $ w(r_0) $, $ w_R(r_0) $, $ \hat{a}(r_0) $, and $ \hat{a}_R(r_0) $. Note this requires $ r_0 > 0 $ and $ m(r_0) \neq \frac{r_0}{2} $, for which the equations are singular (these special cases will be studied in detail in sections \ref{s:reg} and \ref{s:EH} respectively).
The further choice of $ S(r_0) > 0 $ determines $ S $ but trivially affects the other variables, so we write this as 6+1 dimensional. In the purely magnetic case it is 4+1 dimensional.
This independence of the solution from $ S(r_0) $ is tied to the freedom inherent in the orthogonal time coordinate that allows us to choose the value of $ S $ at any point.

To uniquely determine a solution in isothermal coordinates, we can specify the variables at any value of the coordinate $ R $ because the system is autonomous. Thus in general we can specify finite values for $ r(R_0) $, $ r_R(R_0) $, $ w(R_0) $, $ w_R(R_0) $, $ \bar{a}(R_0) $, and $ \bar{a}_R(R_0) $. Just as the coordinate freedom allows us to choose $ S $ above, the coordinate freedom in isothermal coordinates allows us to choose $ \alpha(R_0) > 0 $ while only trivially affecting the remaining variables. In general, (\ref{ralpha1}) then gives $ \alpha_R(R_0) $ and the solution is completely specified. Note that this requires $ r(R_0) > 0 $ and $ r_R(R_0) \neq 0 $ (these special cases will be considered further in sections \ref{s:reg} and \ref{sec:s:equators} respectively).
We thus recover the 6+1 dimensionality of the system.
Any series solution to the static Einstein-Yang-Mills equations that has 6 essential parameters (or 4 in the purely magnetic case) will be said to be ``generic", otherwise it will be ``special".

Since the ``polar" coordinate $ t $ is used in each case, the definition of $ a $ in (\ref{potential}) is the same and will be denoted by $ \bar{a} $. We also use the polar gauge in each case so the gauge potential is
\[ A = \bar{a}\tau_3\dy{t} + w\tau_1\dy{\theta} + \left(w\tau_2 + \cot\theta\tau_3\right)\sin\theta\dy{\phi} \;. \]

\section{Alternative coordinates for static spacetimes}
For the static case there is an additional coordinate choice that will simplify the analysis of certain properties.
Here we choose a metric of the form
\begin{equation} \label{metric ttau}
g = -\alpha(\tau)^2\dy{t}^2 +r(\tau)^2\left(\dy{\tau}^2 + \dy{\theta}^2 +\sin^2\theta\dy{\phi}^2\right) \;.
\end{equation}
The only coordinate freedom is the ability to scale $ t $; that is, choosing $ \alpha $ at one point.
The further transformation $ X = e^\tau $ converts (\ref{metric ttau}) into standard isotropic coordinates, and so we will refer to these as semi-isotropic coordinates, and denote $ \dydx{}{\tau} $ by a subscript $ \tau $.
These coordinates were used extensively by \cite{BFM94}.

The Misner-Sharp mass (\ref{MS}) and $ N $ in these coordinates are
\begin{align*}
N(\tau) &= \frac{r\indices{_\tau^2}}{r^2} \;, & m(\tau) &= \frac{r}{2}\left(1-\frac{r\indices{_\tau^2}}{r^2}\right).
\end{align*}
The Yang-Mills equations are
\begin{subequations} \label{YME tau}
\begin{align}
\frac{r}{\alpha}\left(\frac{\alpha}{r}w_\tau\right)_\tau &= \left((w^2-1)-\frac{r^2}{\alpha^2}\bar{a}^2\right)w \;, \label{wtautau} \\
\frac{\alpha}{r}\left(\frac{r}{\alpha}\bar{a}_\tau\right)_\tau &= 2w^2\bar{a} \;. \label{atautau}
\end{align}
Again, there are a number of ways of arranging the Einstein equations into two independent equations for $ r $ and $ \alpha $. As we did in isothermal coordinates, we choose combinations that are second order for $ r $ and $ \alpha $, and are manifestly monotonic:
\begin{align}
\alpha\left(\frac{r_\tau}{r\alpha}\right)_\tau &= -2\left(\frac{w\indices{_\tau^2}}{r^2} +\frac{\bar{a}^2w^2}{\alpha^2}\right), \label{rtautau} \\
\frac{(r\alpha_\tau)_\tau}{r\alpha} &= \frac{(w^2-1)^2}{r^2} +\frac{\bar{a}\indices{_\tau^2}}{\alpha^2} +2\left(\frac{w\indices{_\tau^2}}{r^2} +\frac{\bar{a}^2w^2}{\alpha^2}\right).
\end{align}
We will later also make use of the $ \tau\tau $ component of the Einstein equations, which is first order in $ r $ and $ \alpha $:
\begin{equation} \label{alphatau}
2\frac{r_\tau\alpha_\tau}{r\alpha} +\frac{r\indices{_\tau^2}}{r^2} -1 = -\frac{(w^2-1)^2}{r^2} -\frac{\bar{a}\indices{_\tau^2}}{\alpha^2} +2\left(\frac{w\indices{_\tau^2}}{r^2} +\frac{\bar{a}^2w^2}{\alpha^2}\right).
\end{equation}
\end{subequations}
Note that just like for isothermal coordinates this equation is not well suited to numerical implementation in general because of the quadratic term in $ r_\tau $ and the $ r_\tau $ factor in front of $ \alpha_\tau $ that can go to zero.

The transformations between the different coordinates are given by
\begin{align*}
\dydx{r}{R} &= SN \;, & \dydx{r}{\tau} &= r\sqrt{N} \;, & \dydx{R}{\tau} &= \frac{r}{\alpha} \;,
\end{align*}
and
\begin{align*}
S^2N &= \alpha^2 \;, & N &= 1-\frac{2m}{r} = \frac{r\indices{_R^2}}{\alpha^2} = \frac{r\indices{_\tau^2}}{r^2} \;, & z &= \frac{r^2\bar{a}'}{S} = \frac{r^2\bar{a}_R}{\alpha^2} = \frac{r\bar{a}_\tau}{\alpha}\;.
\end{align*}

We summarise the features of each of these coordinate choices in table \ref{table:stat coords} by listing the behaviour of the metric functions for various solution properties. The details are in the following sections.
At a regular origin only the semi-isotropic coordinate is not finite.
At a singular origin one of the metric functions blows up in each case and only the isothermal coordinates produce fractional powers.
Only the polar-areal coordinates are capable of describing the interior of black holes, and only the isothermal coordinates become infinite as a horizon is approached from the exterior.
Only the isothermal and semi-isotropic coordinates are capable of crossing an equator.
The metric functions have series in powers of $ \frac{1}{r} $ at a regular, asymptotically flat infinity.

\begin{table}[!ht]
  \centering
  \begin{tabular}{clllll}
    \toprule
    Coordinates & Regular origin & Singular origin & Horizon & Equator & Infinity \\
    \midrule
    \multirow{4}{*}{Polar-areal} & $ r = 0 $ & $ r = 0 $ & $ r = r_h $ & $ r = r_e $ & $ r \to \infty $ \\
     & $ S = O(1) $ & $ S = O(1) $ & $ S = O(1) $ & $ S =O((r_e-r)^{-\frac{1}{2}}) $ & $ S = O(1) $ \\
     & $ N = 1 $ & $ N = O(\frac{1}{r^n}) $ & $ N = O(r-r_h) $ & $ N = O(r_e-r) $ & $ N = 1 $ \\
     & & $ n = 1 \text{ or } 2 $ & Coord. sing. & Coord. sing. \\
     \midrule
    \multirow{4}{*}{Isothermal} & $ R = 0 $ & $ R = 0 $ & $ R \to -\infty $ & $ R = 0 $ & $ R \to \infty $ \\
     & $ r = O(R) $ & $ r = O(R^\frac{1}{n}) $ & $ r = r_h $ & $ r_R = 0 $ & $ r = O(R) $ \\
     & $ \alpha = O(1) $ & $ \alpha = O(R^{-\frac{1}{n}}) $ & $ \alpha = O(e^\frac{R}{2}) $ & $ \alpha = O(1) $ & $ \alpha = O(1) $ \\
     & & $ n = 2 \text{ or } 3 $ & Exterior only & & \\ 
     \midrule
    \multirow{4}{*}{Semi-isotropic} & $ \tau \to -\infty $ & $ \tau = 0 $ & $ \tau = 0 $ & $ \tau = 0 $ & $ \tau \to \infty $ \\
     & $ r = O(e^\tau) $ & $ r = O(\tau^n) $ & $ r = r_h $ & $ r_\tau = 0 $ & $ r = O(e^{\tau}) $ \\
     & $ \alpha = O(1) $ & $ \alpha = O(\frac{1}{\tau^2}) $ & $ \alpha = O(\tau) $ & $ \alpha = O(1) $ & $ \alpha = O(1) $ \\
     & & $ n = 2 \text{ or } 1 $ & Exterior only & \\
    \bottomrule
  \end{tabular}
  \caption{A comparison of coordinate choices for static spherically symmetric spacetimes (Coord. sing. indicates a coordinate singularity).}
  \label{table:stat coords}
\end{table}

We note that we are able to find coordinates that can be used inside a horizon and across an equator; by taking
\[ g = -\beta(\rho)\dy{t}^2 +\frac{1}{\beta(\rho)}\dy{\rho}^2 +r(\rho)^2\left(\dy{\theta}^2 +\sin^2\theta\dy{\phi}^2\right), \]
where $ \beta = \alpha^2 $ outside any horizon and $ \dydx{R}{\rho} = \frac{1}{\alpha^2} $, $ \dydx{r}{\rho} = \frac{1}{S} $.
However, we do not find these coordinates as useful as the three that we will use in this chapter.

Polar-areal coordinates have the advantage of explicitly being able to decouple the equation for $ S $, however this comes at the cost of being unable to cover an equator.
In the case of semi-isotropic coordinates it is also possible to explicitly decouple $ \alpha $ from the equations using (\ref{rRR+}), however this then makes the resulting equations singular when $ r_\tau = 0 $.

\section{Solutions regular at the origin} \label{s:reg}
In this section we assume that all the unknown variables have convergent power series expansions at $ r = 0 $, and begin working with polar-areal coordinates.
Inserting the power series ansatz into the equations (\ref{EYME stat}) we find the following expansions:
\begin{align} \label{seriesO stat}
w(r) &= 1 + br^2 + \frac{3b^2-d^2+8b^3+2bd^2}{10}r^4 + O(r^6) \;,  \nonumber\\
\bar{a}(r) &= S_0d\left(r + \frac{2b+8b^2+2d^2}{5}r^3 + \frac{12b^2-d^2+136b^3+34bd^2+272b^4+136b^2d^2+17d^4}{70}r^5 + O(r^7)\right), \nonumber \\
m(r) &= \frac{4b^2+d^2}{2}r^3 + \frac{8b^3+2bd^2}{5}r^5 + O(r^7) \;, \\
S(r) &= S_0\left(1 + (4b^2+d^2)r^2 + \frac{3(8b^3+2bd^2+48b^4+24b^2d^2+3d^4)}{10}r^4 + O(r^6)\right). \nonumber 
\end{align}
There are two essential parameters in this case, $ b = \frac{w''(0)}{2} $ and $ d = \frac{\bar{a}'(0)}{S(0)} $.
There is a further parameter $ S_0 = S(0) $, which scales $ \bar{a}(r) $ and $ S(r) $, and corresponds to the freedom to scale the time coordinate $ t $ by an arbitrary factor.
We choose $ S_0 = 1 $, which makes $ t $ the proper time at the origin.
Note that the invariance of the equations (\ref{EYME stat}) when swapping the sign of $ \bar{a} $ here manifests itself as the sign of $ d $ only affecting the sign of $ \bar{a} $. We will therefore only consider $ d > 0 $.
That only every second power of $ r $ appears in these series is expected from the regularity conditions (\ref{regular tr metric}) and (\ref{regular tr}).
For $ d = 0 $, these expansions equal those found previously by \cite{BK88}, and they correspond to those found by \cite{BH00-2} in the case of vanishing cosmological constant.

We can also find power series expansions in terms of isothermal coordinates. We set $ r = 0 $ to be at $ R = 0 $ to use up the freedom of the autonomous system, and use (\ref{EYME stat TR}) to find
\begin{equation} \label{seriesO stat TR}
\begin{aligned}
r(R) &= \alpha_0R -\frac{8b^3+16b^4+2bd^2+8b^2d^2+d^4}{50}(\alpha_0R)^5 +O(\alpha_0R)^7 \;, \\
\alpha(R) &= \alpha_0\left(1 +\frac{4b^2+d^2}{2}(\alpha_0R)^2 +\frac{32b^3+176b^4+8bd^2+88b^2d^2+11d^4}{40}(\alpha_0R)^4 + O(\alpha_0R)^6\right), \\
w(R) &= 1 +b(\alpha_0R)^2 +\frac{3b^2+8b^3-d^2+2bd^2}{10}(\alpha_0R)^4 + O(\alpha_0R)^6 \;, \\
\bar{a}(R) &= \alpha_0d\left(\alpha_0R +\frac{2b+8b^2+2d^2}{5}(\alpha_0R)^3 + O(\alpha_0R)^5\right). 
\end{aligned}
\end{equation}
Again there are two essential parameters, $ b = \frac{w_{RR}(0)}{2\alpha(0)^2} $ and $ d = \frac{\bar{a}_R(0)}{\alpha(0)^2} $.
The further parameter $ \alpha_0 = \alpha(0) $ corresponds to the coordinate freedom to scale $ T $ and $ R $ together (\ref{TR stat free}).
We choose $ \alpha_0 = 1 $ so the free parameters $ (b,d) $ are equal to those in the polar-areal coordinates.

\subsection{Purely magnetic solutions}
We begin our investigation of the solutions to the equations by looking at the well-known magnetic case, $ \bar{a} \equiv 0 $ ($ d = 0 $), where the equations in polar-areal coordinates are
\begin{subequations} \label{EYME stat mag}
\begin{align}
m' &= \frac{(w^2-1)^2}{2r^2} + Nw'^2 \;, \label{m stat mag}\\
S' &= \frac{2Sw'^2}{r} \;, \label{S stat mag} \\
\left(SNw'\right)' &= S\frac{w^2-1}{r^2}w \;. \label{w stat mag}
\end{align}
\end{subequations}

In this case there is one parameter, $ b = \frac{w''(0)}{2} $, and for generic values of this parameter the equations (\ref{EYME stat}) develop a singularity where $ N \to 0 $, $ S \to \infty $ and $ |w'| \to \infty $. This is however only a coordinate singularity; $ S^2N $ remains finite and the areal radius $ r $ can no longer be used as a spatial coordinate because it reaches a maximum when $ N = 0 $ and decreases beyond that.
This property is known as an equator and will be discussed in more detail using the isothermal coordinates in section \ref{sec:s:full}.

For a countably infinite number of values of $ b $ in the range $ (-0.707,0] $ we recover the Bartnik-McKinnon solutions \cite{BK88} (labelled by the number of times $ k $ the gauge potential $ w $ crosses zero), an infinitely oscillating solution \cite{BFM94} and Minkowski space for $ b = 0 $.
For these solutions, $ r $ is a good coordinate, and we discuss some of the details and properties of the solutions here.

To find these solutions numerically we use Mathematica \cite{Mathematica} for the integration and also to solve for the first few terms in the power series (\ref{seriesO stat}), which can be used to set initial conditions to the system of ordinary differential equations (\ref{EYME stat}) at a small value of $ r $.
We find that calculating the terms up to $ O(r^{14}) $ is sufficient to have the series be accurate to at least 60 significant figures at $ r = 10^{-5} $.

We note that it is possible to change variables, in particular to use $ W = \frac{1-w}{r^2} $ as in \cite{BFM94}, to desingularise the equations at $ r = 0 $ in order to start integration precisely from $ r = 0 $.
However, we choose not to do this because small errors in the calculation of $ W $ can result in large errors in $ w $ when $ r $ is large. In the magnetic case this is not very noticeable, and the method works rather well (see section \ref{s:PM instability}).
However, in the general case these errors have a significant impact on the solution. This is probably because in the general case we will see that $ w $ is not monotonic for large $ r $, but rather oscillating.
An alternative but still desingularising variable $ w = 1-\frac{r^2}{1+r^2}W $ was also considered, but this time the equations in the full case became too complicated to easily handle. So we find a power series approach to be best overall.

We use Mathematica to solve the system (\ref{EYME stat}), making use of its inbuilt function \verb_NDSolve_, the method \verb_"StiffnessSwitching"_, and its ability to use arbitrary precision arithmetic to present highly accurate solutions beyond machine precision for the first time.

Let us briefly consider the Yang-Mills equation in isothermal coordinates (\ref{wRR}), which in the magnetic case is
\[ w_{RR} = \alpha^2w\frac{w^2-1}{r^2} \;. \]
We see that when $ |w| > 1 $, the second derivative of $ w $ has the same sign as $ w $, so if $ |w| $ crosses one from below it remains greater, and in particular the solution can never become asymptotically flat. We use this fact to find the precise values of $ b $ corresponding to the various Bartnik-McKinnon solutions. We integrate equations (\ref{EYME stat mag}) using the series solution (\ref{seriesO stat}) to set initial conditions at $ r = 10^{-5} $, and cease integrating when $ |w| > 1.1 $.
Figure \ref{fig:stat mag b ani} shows how the solution changes as the parameter $ b $ varies.
\begin{figure}[!ht]
  \centering
  \animategraphics[scale=1]{12}{SSSEYMpara}{0}{600}
  \caption{An animation of the dependence of $ w(r) $ on the initial conditions determined by the parameter $ b $ (digital only). Note that $ b $ is not varied linearly, see equation (\ref{b asy}).}
  \label{fig:stat mag b ani}
\end{figure}
Each asymptotically flat solution is between solutions that pass through $ +1 $ and $ -1 $ respectively. To find the precise values we first bracket each solution and use a bisection search down to an accuracy (at least $ 2^{-85} $) required to be sure of the 24th decimal place.
The parameter values for the first 20 Bartnik-McKinnon solutions are shown in table \ref{table:BKs}, and plots of the solutions are shown in figure \ref{fig:BKs}.
\begin{table}[!ht]
  \centering
  \begin{tabular}{cll}
    \toprule
    $ k $ & $ b_k $ \\
    \midrule
    0 & 0 \\
    1 & $-$0.453 716 272 705 877 156 120 850 \\
    2 & $-$0.651 725 525 595 250 171 108 639 \\
    3 & $-$0.697 040 050 306 902 022 539 489 \\
    4 & $-$0.704 878 477 943 515 118 601 339 \\
    5 & $-$0.706 168 660 867 561 437 188 341 \\
    6 & $-$0.706 379 329 970 121 599 474 987 \\
    7 & $-$0.706 413 684 761 743 764 021 418 \\
    8 & $-$0.706 419 285 975 101 207 815 240 \\
    9 & $-$0.706 420 199 166 833 789 429 545 \\
    10 & $-$0.706 420 348 047 872 580 355 742 \\
    11 & $-$0.706 420 372 320 478 925 790 677 \\ 
    12 & $-$0.706 420 376 277 727 852 536 414 \\
    13 & $-$0.706 420 376 922 892 136 287 157 \\
    14 & $-$0.706 420 377 028 075 550 088 862 \\
    15 & $-$0.706 420 377 045 223 973 847 713 \\
    16 & $-$0.706 420 377 048 019 741 989 794 \\
    17 & $-$0.706 420 377 048 475 545 952 540 \\
    18 & $-$0.706 420 377 048 549 857 283 773 \\
    19 & $-$0.706 420 377 048 561 972 522 781 \\
    20 & $-$0.706 420 377 048 563 947 713 021 \\
    \vdots & \vdots \\
    $ \infty $ & $-$0.706 420 377 048 564 332 462 304 \\
    \bottomrule
  \end{tabular}
  \caption{The calculated parameter $ b $ for the first 20 Bartnik-McKinnon solutions. The decimal expansions are truncated, not rounded.}
  \label{table:BKs}
\end{table}
\begin{figure}[!ht]
  \centering
  \animategraphics[scale=0.75,step]{}{BK20_pa}{00}{20}
  \caption{A plot of $ w $ for the first 20 Bartnik-McKinnon solutions in polar-areal coordinates. Click through to see them individually plotted with $ m $, $ N $, and $ S $ scaled so $ \lim\limits_{r \to \infty} S = 1 $ (digital only).}
  \label{fig:BKs}
\end{figure}

With these results we can confirm the result in \cite{BFM94} that the values of $ b_k $ are asymptotically given by
\begin{equation} \label{b asy}
b_k = b_\infty + 2.185943445 e^{-1.8137993642 k} \;.
\end{equation}
We find that the exponent coefficient agrees with the expected value $ -\frac{\pi}{\sqrt{3}} $ to at least the 11 significant figures found.

The parameter value $ b_\infty $ produces a solution with infinite oscillations, and its value was found using isothermal coordinates.
The equations in the magnetic case are
\begin{subequations} \label{EYME stat mag TR}
\begin{align}
\left(\frac{r_R}{\alpha^2}\right)_R &= -2\frac{w\indices{_R^2}}{r\alpha^2} \;, \label{rRR+ mag} \\
\left(\frac{r^2\alpha_R}{\alpha}\right)_R &= \frac{\alpha^2(w^2-1)^2}{r^2} +2w\indices{_R^2} \;, \label{alphaRR+ mag} \\
w_{RR} &= \alpha^2w\frac{w^2-1}{r^2} \;. \label{wRR mag}
\end{align}
\end{subequations}
We use the series (\ref{seriesO stat TR}) to set initial conditions at $ R = 10^{-5} $, and integrate equations (\ref{EYME stat mag TR}).
A bisection search was again used since for $ b > b_\infty $, we stop the integration when $ |w| > 1 $, where $ N > 0 $, while for $ b < b_\infty $, we stop the integration when $ N = 0 $ with $ |w| < 1 $.

Although the limit of the Bartnik-McKinnon solutions for large $ r $ is the magnetic extremal Reissner-Nordstr\"{o}m solution, the oscillating solution is not asymptotically flat \cite{BFM94}.
In fact, the oscillating solution approaches the magnetic Bertotti-Robinson solution as $ R \to \infty $, as will be discussed in section \ref{sec:s:ABR}.

\begin{figure}[!ht]
  \centering
  \includegraphics[scale=0.4]{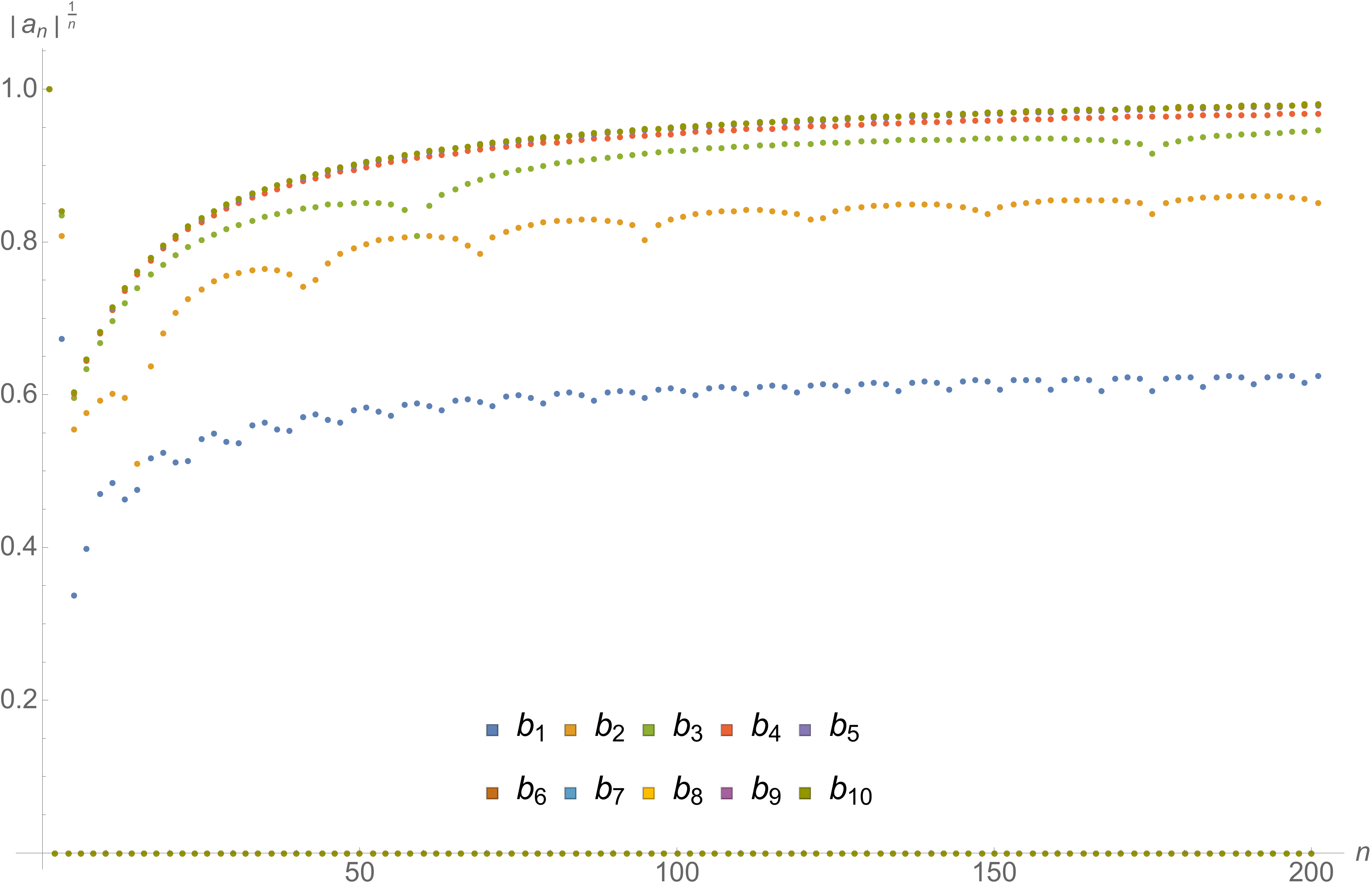}
  \caption{The coefficients of the power series for $ w $ for the first 10 Bartnik-McKinnon solutions (the odd coefficients are all zero).}
  \label{fig:BKs roc}
\end{figure}

In 1993, Schunck considered the coefficients of the power series $ w(r) = \sum\limits_{n=0}^\infty a_n r^n $, and used them to estimate the radius of convergence $ R_k $ for the $ k $th Bartnik-McKinnon solution ($ a_2 = b_k $) with $ \frac{1}{R_k} = \limsup\limits_{n \to \infty} \sqrt[n]{|a_n|} $ \cite{Schunck93}.
We here briefly note that the arching pattern observed for small values of $ n $ in their figure 3 in fact continues to large values of $ n $ (as shown in figure \ref{fig:BKs roc}) and their incorrect later values are due to numerical error.

\subsection{General solutions} \label{sec:s:full}
Before considering the solutions with non-zero electric field, we recall the main relevant result \cite{BP92}, that solutions regular at the origin and asymptotically flat must have $ \bar{a} \equiv 0 $.
With our assumptions about regularity at infinity this is easy to see.
Firstly, (\ref{BCs inf pa}) tells us that either $ \bar{a} $ or $ w $ go to zero as $ r $ goes to infinity. We will show in section \ref{s:AF} that $ \lim\limits_{r \to \infty} w = 0 $ implies the Reissner-Nordstr\"{o}m solution, which is not regular at the origin, thus we assume $ \lim\limits_{r \to \infty} \bar{a} = 0 $.
We can observe (like \cite{BP92}) that equation (\ref{aRR}) does not allow maxima with $ \bar{a} > 0 $ or minima for $ \bar{a} < 0 $, therefore using either boundary condition ($ \bar{a} = 0 $ at the origin (\ref{BCs 0 pa}) or infinity (\ref{BCs inf pa})) implies $ \bar{a} $ is monotonic.
Both boundary conditions imply $ \bar{a} \equiv 0 $.

Alternatively, we can multiply (\ref{a stat}) by $ \bar{a} $ and integrate by parts (like \cite{EG90,Zhou92,BH00-2}) to find
\[ \int_0^\infty \frac{\bar{a}^2w^2}{SN} +\frac{r^2\bar{a}'^2}{2S} \dx{r} = \lim_{r \to \infty} \frac{r^2\bar{a}\bar{a}'}{2S} - \lim_{r \to 0} \frac{r^2\bar{a}\bar{a}'}{2S} \;. \]
Regularity at the origin (\ref{regular tr}) requires $ \bar{a} = O(r) $ and $ \bar{a}' = O(1) $, and asymptotic flatness (\ref{BCs inf pa}) requires $ \bar{a} = O\mathopen{}\left(\frac{1}{r}\right)\mathclose{} $ and $ \bar{a}' = O\mathopen{}\left(\frac{1}{r^2}\right)\mathclose{} $ so the right hand side is zero. The terms in the integral are non-negative and thus must vanish. Since $ w \equiv 0 $ is incompatible with regularity at the origin, we conclude $ \bar{a} \equiv 0 $.

We now look numerically at what happens to solutions that are regular at the origin and have non-zero electric field, and see in what way they fail to become asymptotically flat. For this purpose the isothermal coordinates are most useful, and we calculate one additional equation.
By differentiating (\ref{mN TR}) in the static case (or transforming (\ref{m stat})), we find
\begin{equation} \label{mR}
m_R = \frac{r_R}{2}\left(\frac{(w^2-1)^2}{r^2}+\frac{r^2\bar{a}\indices{_R^2}}{\alpha^4}+2\frac{w\indices{_R^2}+a^2w^2}{\alpha^2}\right),
\end{equation}
and we see that extrema of $ m $ and $ r $ coincide.

\begin{figure}[!ht]
  \centering
  \includegraphics[scale=0.75]{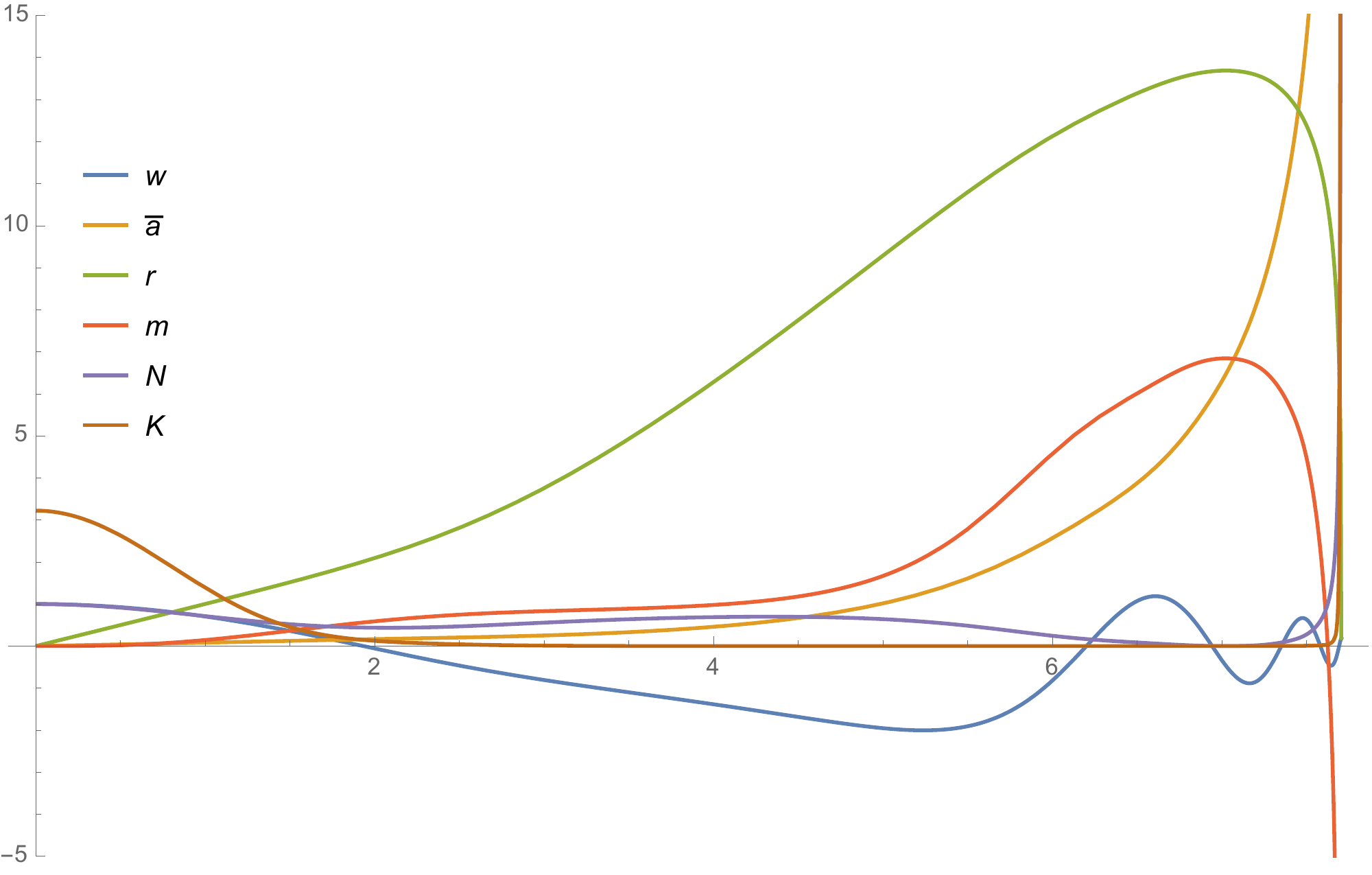}
  \caption{A typical short-lived solution, with $ (b,d) = (-0.3,0.08) $. The plot includes $ w $, $ \bar{a} $, $ r $, $ m $, $ N $, and the Kretschmann scalar $ K $ as functions of $ R $.}
  \label{fig:stat elec S3 typ}
\end{figure}

We solve the equations (\ref{EYME stat TR}) as in the magnetic case; by calculating the power series expansions (\ref{seriesO stat TR}) and setting initial conditions at small $ R $ by choosing the values of the two parameters $ b $ and $ d $.
We find two generic behaviours as the parameters are varied.
The first is similar to the $ b > 0 $ or $ b < b_\infty $ behaviour in the magnetic case, and a fairly typical singular solution is shown in figure \ref{fig:stat elec S3 typ}.
An equator (maximum of $ r $ and $ m $, $ N = 0 $) forms, after which the areal radius reduces back to zero where the solution is singular (the Kretschmann scalar $ K = R_{\mu\nu\kappa\lambda}R^{\mu\nu\kappa\lambda} $ blows up).
One difference to the magnetic case is that due to the non-zero $ \bar{a} $ term in (\ref{wRR}), the magnetic potential $ w $ generically oscillates.
The behaviour of these solutions with equators are analysed further in section \ref{sec:s:equators}.

\begin{figure}[!ht]
  \centering
  \includegraphics[scale=0.7]{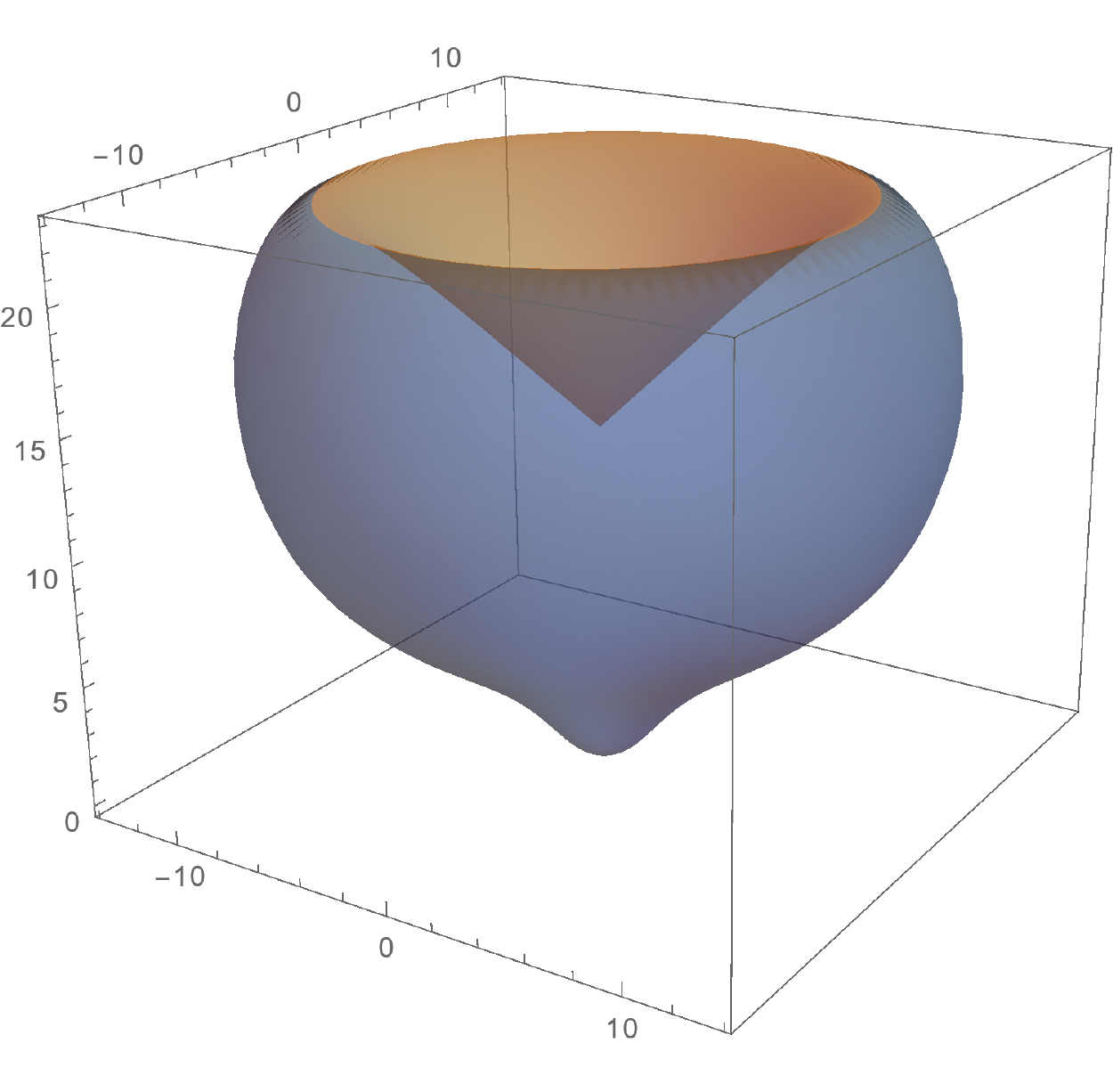}
  \includegraphics[scale=0.5]{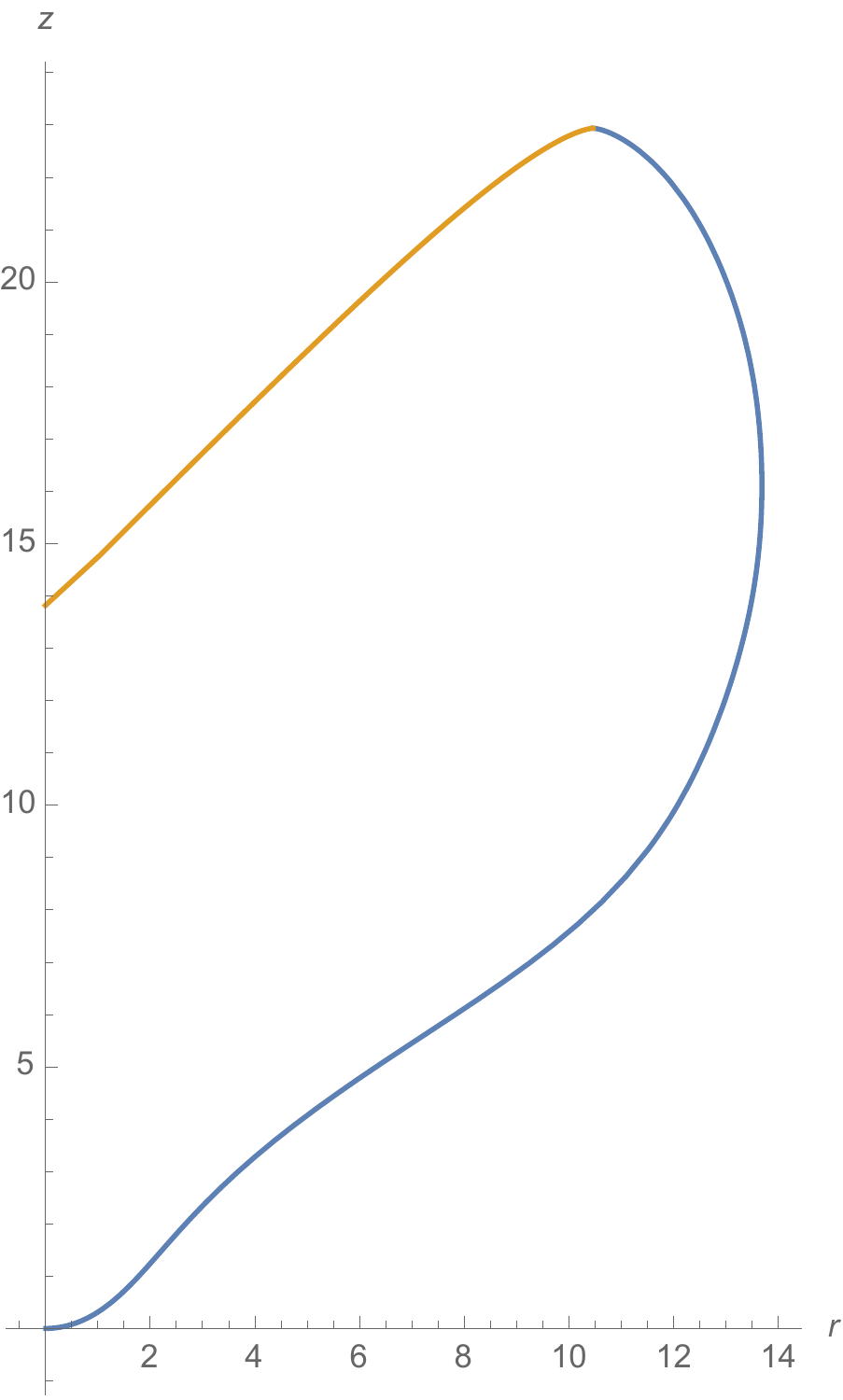}
  \caption{An embedding diagram of a typical short-lived solution, with $ (b,d) = (-0.3,0.08) $. The Euclidean embedding section is in blue while the Minkowskian embedding section is in yellow.}
  \label{fig:stat elec S3 emb}
\end{figure}

In order to visualise the solution with an equator, we make use of an embedding diagram.
Its spatial topology is $ \mathbb{S}^3 $.
We equate the $ t = 0 $, $ \theta = \frac{\pi}{2} $ slice of the metric $ \alpha^2\dy{R}^2 + r^2\dy{\phi}^2 $ with an arbitrary surface of revolution with the $ z $-axis being the axis of symmetry $ \left(r\indices{_R^2}+z\indices{_R^2}\right)\dy{R}^2 +r^2\dy{\phi}^2 $.
Note that these can only be equated when $ \alpha^2 -r\indices{_R^2} \geqslant 0 $, which is precisely when $ m \geqslant 0 $. Our solution has $ m < 0 $ near the singularity, and so this section cannot be isometrically embedded in Euclidean space.
For this section we instead embed into Minkwoski space $ \mathbb{R}^{3,1} $, where the condition becomes $ \alpha^2 -r\indices{_R^2} \leqslant 0 $.
The result for the example of figure \ref{fig:stat elec S3 typ} is shown in figure \ref{fig:stat elec S3 emb}, where the separation of the Euclidean and Minkowskian embedding is indicated.
Note the gradient of the Minkowski section is near $ 45^\circ $, indicating that the radius decreases rapidly over a metrically small part of the space.

For the second kind of generic behaviour, the solution exists for very large $ r $, with apparently infinite oscillations of $ w $, while decreasing in amplitude and increasing in frequency. This corresponds with both the electric potential $ \bar{a} $ and the Misner-Sharp mass $ m $ increasing without bound, so that if they were to continue in this fashion they would not be asymptotically flat.
Some typical examples are shown in figure \ref{fig:stat elec R3 typ}.
Such solutions we will refer to as ``long-lived".

\begin{figure}[!ht]
  \centering
  \animategraphics[scale=0.75,step]{}{bd_long_}{1}{6}
  \caption{Six long-lived solutions as functions of $ R $, with various values of $ (b,d) $. Click to step through (digital only).}
  \label{fig:stat elec R3 typ}
\end{figure}

\begin{figure}[!ht]
  \centering
  \includegraphics[scale=0.75]{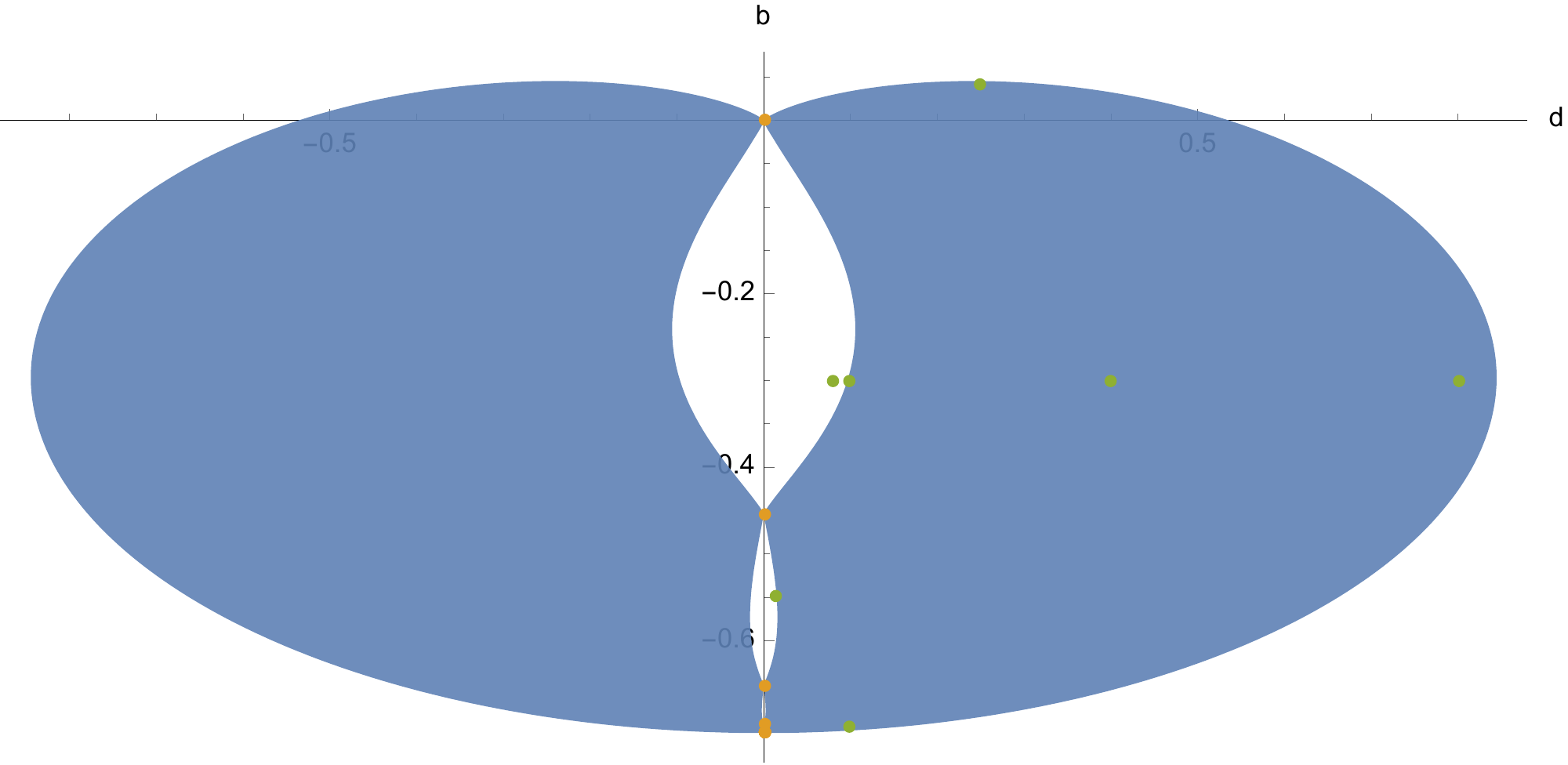}
  \caption{A plot of the static, regular origin parameter space $ (b,d) $, with the shaded region indicating the long-lived solutions. The BK solutions are marked in yellow and the example short-lived and long-lived solutions in green.}
  \label{fig:stat elec space}
\end{figure}

We find the boundary of the long-lived region in the parameter space by using a bisection search. An equator is indicated by $ r_R < 0 $, at which point the integration can be stopped. We find a useful way of determining a long-lived solution by testing if $ N $ increases above the amplitude of $ w $. This is approximated by $ \sqrt{w^2+\frac{w\indices{_R^2}}{\bar{a}^2}} $ (see below).
Recall that the equations (\ref{EYME stat TR}) are invariant under the change $ \bar{a} \to -\bar{a} $, and we see that the solutions with negative $ d $ will be equal to those with positive $ d $ except for the sign of $ \bar{a} $.
A plot of the region in the $ (b,d) $ plane is shown in figure \ref{fig:stat elec space}.
We see that the solution space contracts to the BK solutions as $ d $ goes to zero.
Unlike the magnetic case, solutions apparently without an equator occur generically, in a bounded region of the parameter space.
This is reminiscent of the generalisation of the magnetic EYM solutions to those with negative cosmological constant shown in \cite{BML04} figures 1 to 5.

Outside the long-lived region, $ N $ goes to zero for an equator, while just inside $ N $ has a positive minimum. We see that near the outer boundary this is the first minimum of $ N $, while for the inner boundary it is typically the second minimum of $ N $ that approaches zero as the boundary is approached.

\begin{figure}[!ht]
  \centering
  \includegraphics[scale=0.75]{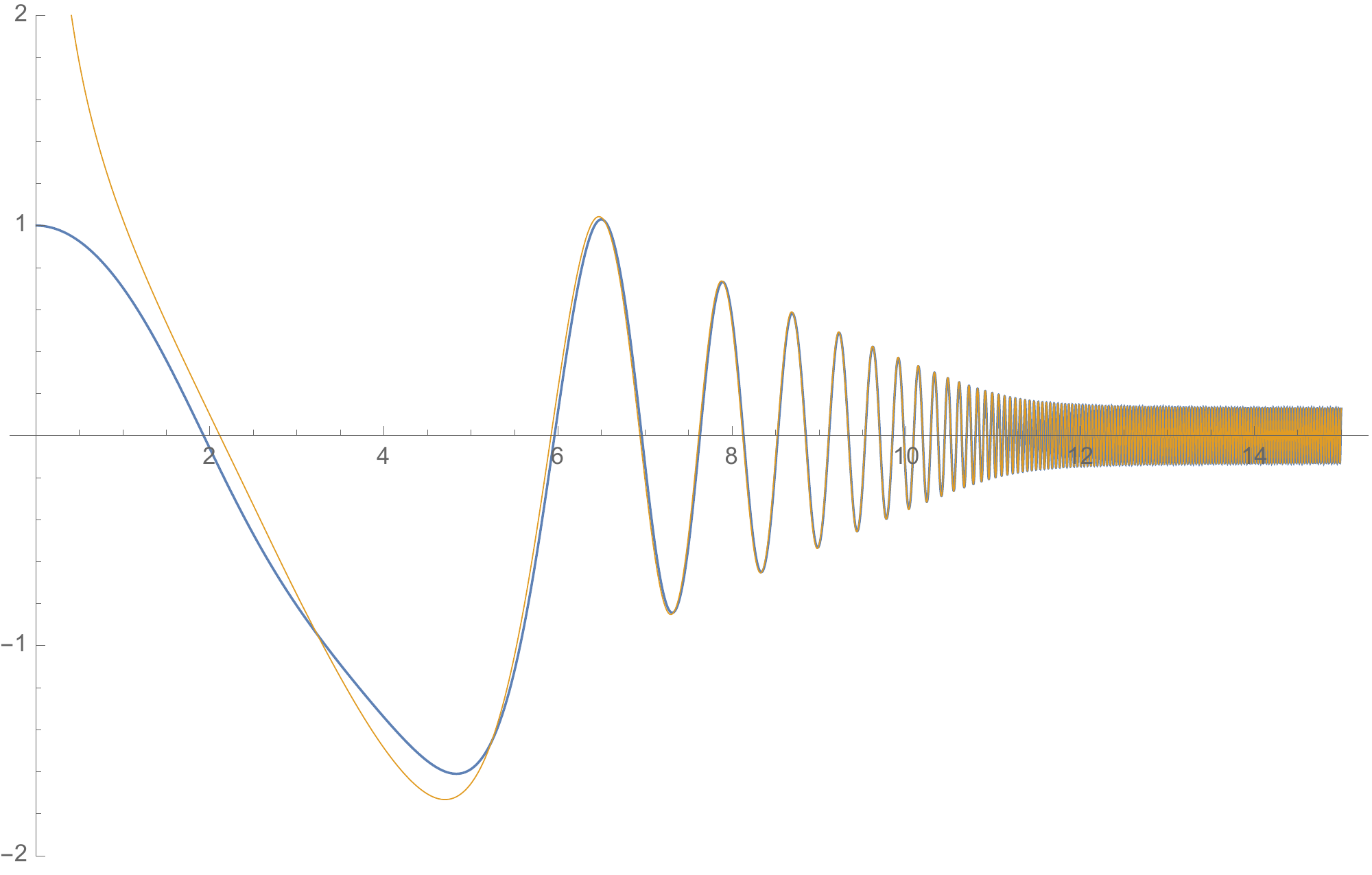}
  \caption{A typical long-lived solution, with $ (b,d) = (-0.3,0.1) $, showing $ w $ (blue) and $ \frac{A}{\sqrt{\bar{a}}}\cos\lambda $ (orange) with respect to $ R $, with $ A $ and the phase of $ \lambda $ tuned appropriately.}
  \label{fig:stat elec w asy}
\end{figure}

We now turn to the static equations to attempt to understand the exhibited behaviour. Again, the equations in isothermal coordinates (\ref{EYME stat TR}) are most useful. We wish to find the long-term behaviour using asymptotic analysis \cite{B&O}.
Let us begin by analysing the $ w $ equation (\ref{wRR}). We see numerically that the term on the right-hand side is negligible compared to the others, as $ w $ is bounded, $ \bar{a} $ increases, and $ r $ increases much faster than $ \alpha $. Therefore we analyse the approximate equation
\begin{equation} \label{w asy R}
w_{RR} + \bar{a}^2w = 0 \;.
\end{equation}
We find a slightly different but successful alternative to the techniques presented in \cite{B&O} to analyse this and the other static Yang-Mills equation.
To capture the oscillatory nature of $ w $, we assume it has the form
\[ w = h\cos\lambda \;, \]
where $ h $ and $ \lambda $ are monotonic functions of $ R $. The frequency of the oscillations is then $ \lambda_R $, which should also be monotonically increasing. Inserting this ansatz into (\ref{w asy R}) gives
\[ \left(\bar{a}^2-\lambda\indices{_R^2}+\frac{h_{RR}}{h}\right)h\cos\lambda -\left(\frac{\lambda_{RR}}{\lambda_R}+\frac{2h_R}{h}\right)\lambda_R h \sin\lambda = 0 \;. \]
We solve for $ h $ and $ \lambda $ by setting the coefficients of the trigonometric terms each to zero.
For the first, we note $ \bar{a} $ and $ \lambda_R $ are increasing, while the amplitude $ h $ is decreasing, so we assume $ \bar{a}^2 \sim \lambda\indices{_R^2} $ and conclude $ \lambda_R \sim \bar{a} $.
The second coefficient can be solved exactly; $ h = \frac{A}{\sqrt{\lambda_R}} $ for some constant $ A $, and so we find $ h \sim \frac{A}{\sqrt{\bar{a}}} $.
These results determine the leading behaviour of $ w $ for large $ R $, and we are able to also confirm that corrections to the leading behaviour decay as expected.
This asymptotic behaviour of $ w $ agrees very well with the numerical solutions, as shown in figure \ref{fig:stat elec w asy}.
With our approximate $ w $, we find $ w\indices{_R^2}+\bar{a}^2w^2 \sim A^2 \bar{a} $, which justifies our use of $ \sqrt{w^2+\frac{w\indices{_R^2}}{\bar{a}^2}} $ to estimate the amplitude of $ w $.

We now turn to the $ \bar{a} $ equation (\ref{aRR}), inserting this result for $ w $;
\begin{equation} \label{a asy R}
\left(\frac{r^2 \bar{a}_R}{\alpha^2}\right)_R = A^2(1+\cos(2\lambda)) \;.
\end{equation}
We see that while $ \bar{a} $ grows monotonically, it must also have an oscillatory component with double the frequency of $ w $.
We will use a similar technique here and assume that $ \bar{a} = o+n\cos(2\lambda)+p\sin(2\lambda) $, where $ o $ is a monotonically increasing function of $ R $, $ o \gg n $, and $ o \gg p $.
If we insert this ansatz into (\ref{a asy R}) we find
\begin{align} \label{onp asy R}
0 &= \left(\left(\frac{r^2}{\alpha^2}o_R\right)_R-A^2\right) \nonumber \\
 &\quad+\left(\left(\frac{r^2}{\alpha^2}n_R\right)_R +2p\lambda_R\left(\frac{r^2}{\alpha^2}\right)_R +\frac{r^2}{\alpha^2}\left(4p_R\lambda_R+2p\lambda_{RR}\right)-4\frac{r^2}{\alpha^2}\lambda\indices{_R^2}n-A^2\right)\cos(2\lambda) \nonumber \\
 &\quad+\left(\left(\frac{r^2}{\alpha^2}p_R\right)_R -2n\lambda_R\left(\frac{r^2}{\alpha^2}\right)_R -\frac{r^2}{\alpha^2}\left(4n_R\lambda_R+2n\lambda_{RR}\right)-4\frac{r^2}{\alpha^2}\lambda\indices{_R^2}p\right)\sin(2\lambda) \;.
\end{align}
We then attempt to approximately solve the equations formed by setting each of the terms grouped in parentheses to zero.

The first can be solved exactly for $ o $, and we find
\[ o = A^2\left(\int R\frac{\alpha^2}{r^2}\dx{R} +B\int\frac{\alpha^2}{r^2}\dx{R}+C\right), \]
where $ B $ and $ C $ are additional integration constants, and the integrals are indefinite integrals (although in practice begin at the same small value of $ R $ as the numerical solutions).
Since $ n \ll o $, we have $ \bar{a} \sim o $ and we can use this and $ \lambda_R \sim \bar{a} $ to determine $ \lambda $;
\begin{equation} \label{lambda dom}
\lambda \sim A^2\left(\iint R\frac{\alpha^2}{r^2}\dx{R}\dx{R} +B\iint\frac{\alpha^2}{r^2}\dx{R}\dx{R}+CR+D\right),
\end{equation}
so we now have a description of the $ w $ asymptotics purely in terms of the metric variables $ r $ and $ \alpha $.
Note that we also have $ \bar{a}_R \sim o_R $, but for further derivatives the oscillatory terms due to $ n $ and $ p $ are no longer subdominant.

Now considering the second term in (\ref{onp asy R}), we find a suitable balance with $ 4\frac{r^2}{\alpha^2}\lambda\indices{_R^2}n \sim -A^2 $, and all other terms being subdominant. We thus have
\begin{equation} \label{n dom}
n \sim -\frac{A^2\alpha^2}{4o^2r^2} \;.
\end{equation}
For the final term in (\ref{onp asy R}), we assume $ \left(\frac{r^2}{\alpha^2}p_R\right)_R $ is subdominant to the remaining terms and find $ p = -\frac{n}{\lambda_R}\left(\frac{r_R}{r}-\frac{\alpha_R}{\alpha}+\frac{\lambda_{RR}}{2\lambda_R}+\frac{n_R}{n}\right) $.
Using (\ref{lambda dom}) and (\ref{n dom}) we find
\[ p \sim -\frac{A^2\alpha^2}{4o^3r^2}\left(\frac{r_R}{r}-\frac{\alpha_R}{\alpha}+\frac{3o_R}{2o}\right). \]
Inserting these into the remaining terms of (\ref{onp asy R}) confirms our assumptions. Furthermore, analysis of the subsequent terms for $ n $ and $ p $ reveals they decay quicker.
Therefore $ o $ fully determines the leading behaviour of $ \bar{a} $.

In summary, we now have the asymptotics of both Yang-Mills variables in terms of the metric functions, and four constants $ A $, $ B $, $ C $, and $ D $;
\begin{align*}
w &\sim \frac{A}{\sqrt{o}}\cos\lambda \;, \\
\bar{a} &\sim o -\frac{A^2\alpha^2}{4o^2r^2}\left(\cos\left(2\lambda\right) +\frac{1}{o}\left(\frac{r_R}{r}-\frac{\alpha_R}{\alpha}+\frac{3o_R}{2o}\right)\sin\left(2\lambda\right)\right),
\end{align*}
where $ o $ and $ \lambda $ are given by
\begin{align*}
o &= A^2\left(\int R\frac{\alpha^2}{r^2}\dx{R} +B\int\frac{\alpha^2}{r^2}\dx{R}+C\right), \\
\lambda &\sim A^2\left(\iint R\frac{\alpha^2}{r^2}\dx{R}\dx{R} +B\iint\frac{\alpha^2}{r^2}\dx{R}\dx{R}+CR+D\right).
\end{align*}

Using the numerical solutions for $ (b,d) = (-0.3,0.1) $, we are able to determine the value of the constants. We estimate $ A $ by plotting the function $ \sqrt{\frac{w\indices{_R^2}+\bar{a}^2w^2}{\bar{a}}} $, which we see approaches a constant very quickly, and $ A \approx 1.836634 $. We use this to determine $ B $ by plotting $ \frac{r^2}{A^2\alpha^2}\bar{a}_R -R $ and find $ B \approx -3.2487 $. $ C $ is approximated by $ \frac{a}{A^2} -\int R\frac{\alpha^2}{r^2}\dx{R} -B\int\frac{\alpha^2}{r^2}\dx{R} $, and we find $ C \approx 320.4836 $. $ D $ (essentially the phase of $ w $) is found to be approximately 0.79.
\begin{figure}[!ht]
  \centering
  \includegraphics[scale=0.5]{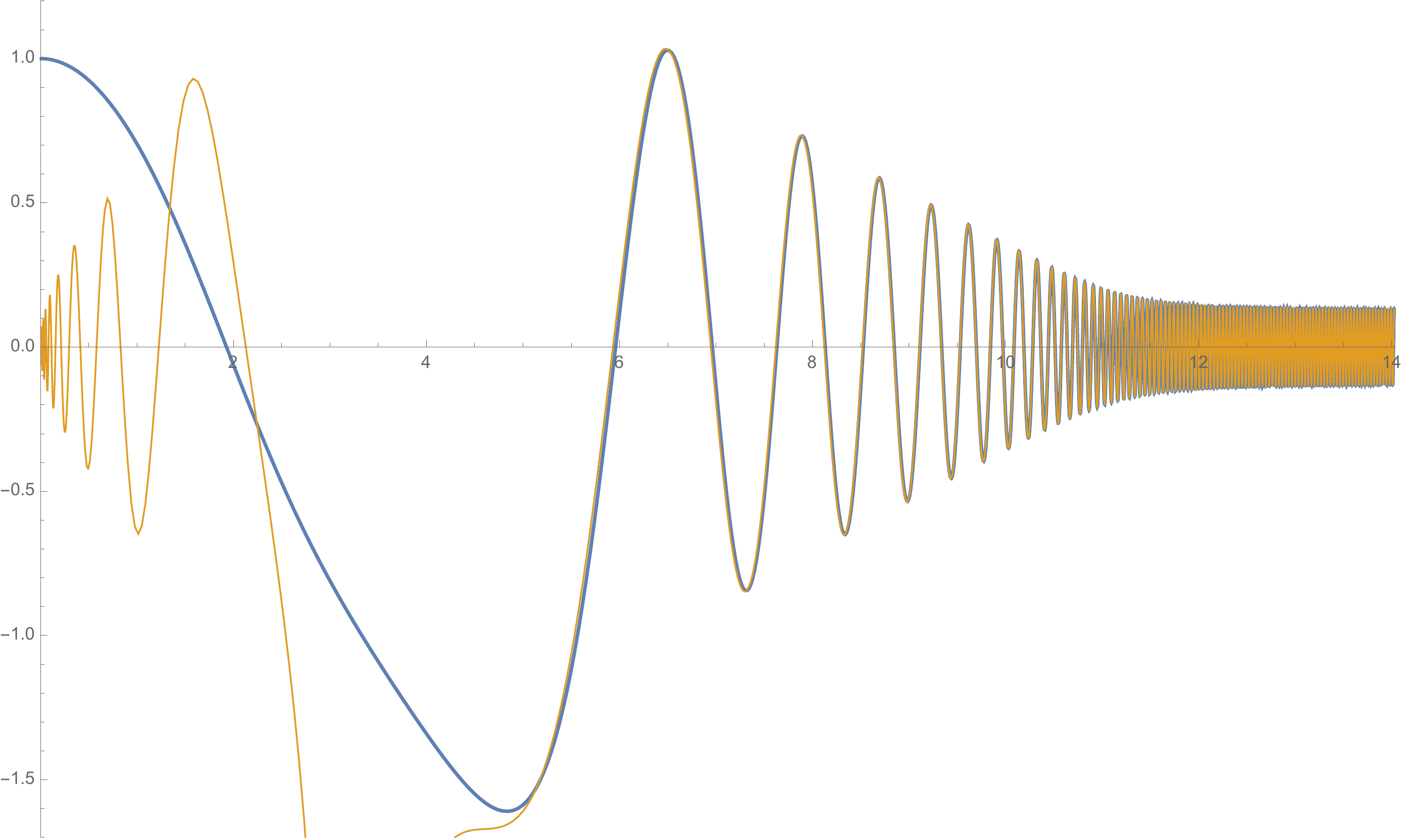}
  \caption{The long-lived solution with $ (b,d) = (-0.3,0.1) $, showing $ w $ (blue) and $ \frac{A}{\sqrt{o}}\cos\lambda $ (orange) with respect to $ R $.}
  \label{fig:stat elec ABCD w}
\end{figure}
\begin{figure}[!ht]
  \centering
  \includegraphics[scale=0.5]{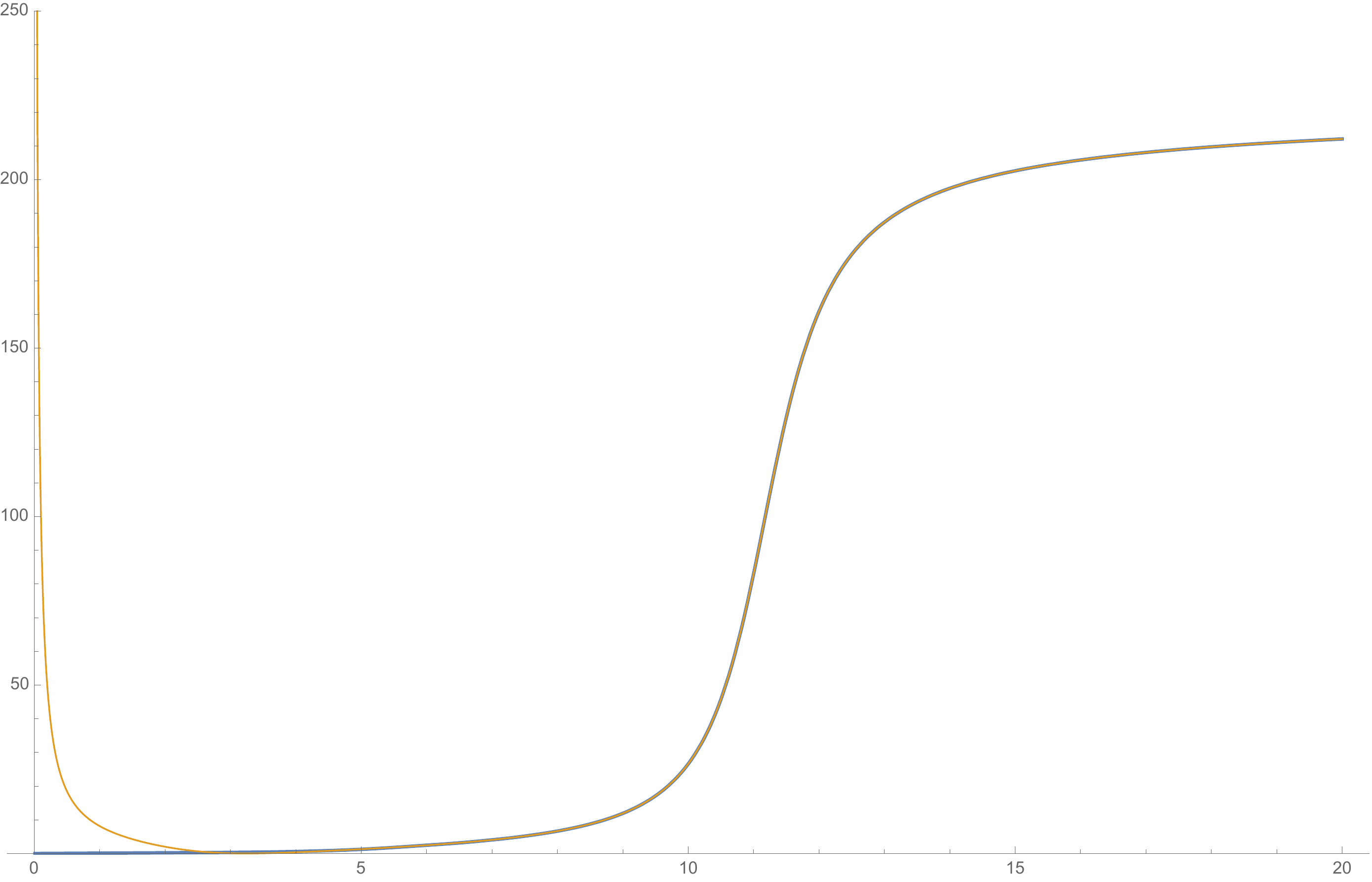}
  \caption{The long-lived solution with $ (b,d) = (-0.3,0.1) $, showing $ \bar{a} $ (blue) and $ o $ (orange) with respect to $ R $.}
  \label{fig:stat elec ABCD a}
\end{figure}
In figures \ref{fig:stat elec ABCD w} and \ref{fig:stat elec ABCD a} we plot the numerical and the asymptotic solutions for $ w $ and $ \bar{a} $, and we see they agree very closely after $ R \approx 5 $.

We can use (\ref{rRR},\ref{alphaRR}) to obtain a second-order system for which the oscillatory terms are all subdominant, unless they're differentiated:
\begin{align*}
r r_{RR} +r\indices{_R^2} &\sim \alpha^2\left(1-A^4\frac{(R+B)^2}{r^2}\right), \\
\frac{\alpha_{RR}}{\alpha}-\frac{\alpha\indices{_R^2}}{\alpha^2} &\sim \frac{r\indices{_R^2}}{r^2}-\frac{\alpha^2}{r^2}\left(1-2A^4\frac{(R+B)^2}{r^2}\right).
\end{align*}

At the moment we have very good approximations of $ w $ and $ \bar{a} $, mostly because their reduced equations were linear in their unknowns and able to be integrated. The above equations for $ r $ and $ \alpha $ are highly nonlinear and their asymptotic behaviour is therefore much more difficult to determine.
Since (\ref{EYME stat TR}) is essentially a six-dimensional dynamical system, it is not surprising that the behaviour is difficult to determine.
Without knowledge of the metric variables' behaviour, we can't say for certain that the observed behaviour continues indefinitely. There could, for example, be an equator with a very large radius.
We nevertheless conjecture that the observed numerical behaviour continues indefinitely and $ r $ increases up to infinity.

We will analyse solutions with an equator in section \ref{sec:s:equators} and the solutions on the boundary of these regions in section \ref{sec:s:ABR}.

\section{Solutions singular at the origin} \label{sec:s:stat sing}
In this section we consider solutions that are singular at the origin, since these behaviours occur in the interior of the black holes, as well as quite generically as will be seen shortly.

Inserting a power series ansatz into (\ref{EYME stat}) we find three separate singular behaviours.
Firstly, there is a Schwarzschild-like (SL) singularity;
\begin{align*}
w(r) &= \pm\left(1 +br^2 -\frac{d^2}{8}r^4 -\frac{8b^2+(1+8b)(4b^2+d^2)}{120a}r^5 +O(r^6)\right), \\
\bar{a}(r) &= 2S_0ad\left(1 -\frac{1}{2a}r -\frac{b+4b^2+d^2}{6a}r^3 +\frac{b+4b^2+d^2}{120a^2}r^3 +O(r^5)\right), \\
m(r) &= a -a(4b^2+d^2)r^2 +\frac{4b^2+d^2}{2}r^3 +a\frac{(4b^2+d^2)^2}{2}r^4 +O(r^5) \;, \\
S(r) &= S_0\left(1 +\left(4b^2+d^2\right)r^2 +\frac{(4b^2+d^2)^2}{2}r^4 +O(r^5)\right),
\end{align*}
where the Misner-sharp mass $ m $ is bounded and $ N = O\left(\frac{1}{r}\right) $ and negative.
There are three essential parameters; $ b = \pm \frac{w''(0)}{2} $, $ a = m(0) $, and $ d = \left.\frac{\bar{a}}{2Sm}\right\vert_{r=0} $, as well as $ S_0 = S(0) $ which can be freely chosen to complete the specification of the $ t $ coordinate (which at the origin parameterises a spacelike direction).
Note that this series requires $ a \neq 0 $, and that the regular origin expansion (\ref{seriesO stat}) can be considered as the limiting $ a = 0 $ case.
The Schwarzschild solution with mass $ a $ is recovered when $ b = d = 0 $.
This generalises the purely magnetic SL series found in \cite{DGZ97}, and the following series likewise generalise the other series stated there.

Secondly there are series with $ N = O\left(\frac{1}{r^2}\right) $ and negative, called pseudo-Reissner-Nordstr\"{o}m-like (PRNL).
We find the most generic series in this case to be:
\begin{align*}
w(r) &= b +\sin(c)r -\frac{2(b^2-1)^2+\left((b^2-1)^2+d^2\right)\cos^2c}{2b\left((b^2-1)^2+d^2\right)}r^2 +O(r^3) \;, \\
\bar{a}(r) &= S_2\left(\frac{(b^2-1)^2+d^2}{b}\cos{c} +dr -\frac{b\left(\left((b^2-1)^2-d^2\right)\cos{c}+2d(b^2-1)\sin{c}\right)}{(b^2-1)^2+d^2}r^2 +O(r^3)\right), \\
m(r) &= \frac{(b^2-1)^2+d^2}{2r} +2b\left((b^2-1)\sin{c}-d\cos{c}\right) +O(r) \;, \\
S(r) &= S_2\left(r^2 -\frac{4b\left((b^2-1)\sin{c}-d\cos{c}\right)}{(b^2-1)^2+d^2}r^3 +O(r^4)\right).
\end{align*}
This has three essential parameters; $ b = w(0) \neq 0 $, $ d = \frac{2\bar{a}'(0)}{S''(0)} $, and $ \sin{c} = w'(0) $, where $ |w'(0)| \leqslant 1 $ and there are two choices for $ c $ in the range $ [0,2\pi) $, giving effectively a choice of sign of $ \cos c $.
This series is singular when $ b = 0 $. When $ w(0) = 0 $, the following series is found:
\begin{align*}
w(r) &= \pm\left(r +\frac{9(1+d^2)-5e^2}{30(1+d^2)^2}r^3 +O(r^4)\right), \\
\bar{a}(r) &= S_2\left(e +dr +\frac{3d}{5(1+d^2)}r^3 +O(r^4)\right), \\
m(r) &= \frac{1+d^2}{2r} -\frac{3}{5}r -\frac{5ed}{12(1+d^2)}r^2 +O(r^3) \;, \\
S(r) &= S_2\left(r^2 +\frac{9}{5(1+d^2)}r^4 +O(r^5)\right),
\end{align*}
where here there are two essential parameters; $ e = \frac{2\bar{a}(0)}{S''(0)} $ and $ d = \frac{2\bar{a}'(0)}{S''(0)} $.
In both of the above PRNL series, the additional parameter $ S_2 = \frac{S''(0)}{2} $ can be freely chosen to complete the specification of the $ t $ coordinate.
The purely magnetic series found in \cite{DGZ97} is recovered with $ c = \pm \frac{\pi}{2} $ and $ d = 0 $ in the generic case, and $ d = e = 0 $ in the $ w(0) = 0 $ case.

Thirdly and finally, there are singular solutions with $ N = O\left(\frac{1}{r^2}\right) $ and positive, called Reissner-Nordstr\"{o}m-like (RNL);
\begin{align*}
w(r) &= b +\frac{b(1-b^2)}{2\left((b^2-1)^2+d^2\right)}r^2 +\frac{c}{6\left((b^2-1)^2+d^2\right)}r^3 +\frac{2ac+b\left((b^2+3)(b^2-1)-2d^2\right)}{8\left((b^2-1)^2+d^2\right)^2}r^4 +O(r^5) \;, \\
\bar{a}(r) &= S_0\left(\frac{d}{r} +\frac{e}{(b^2-1)^2+d^2} \vphantom{\frac{b\left(cd(b^2-1)\left((b^2-1)^2+d^2\right)+b(e+2ad)\left((b^2-1)^2-d^2\right)\right)}{3\left((b^2-1)^2+d^2\right)^3}r^2}\right. \\
&\qquad\qquad\qquad \left.+\frac{b\left(cd(b^2-1)\left((b^2-1)^2+d^2\right)+b(e+2ad)\left((b^2-1)^2-d^2\right)\right)}{3\left((b^2-1)^2+d^2\right)^3}r^2 +O(r^3)\right), \\
m(r) &= -\frac{(b^2-1)^2+d^2}{2r} + a -\frac{b\left(c(b^2-1)\left((b^2-1)^2+d^2\right)+ab\left(3(b^2-1)^2-d^2\right)-2bde\right)}{3\left((b^2-1)^2+d^2\right)^2}r^2 +O(r^3) \;, \\
S(r) &= S_0\left(1 +\frac{b^2}{(b^2-1)^2+d^2}r^2 -\frac{2b\left(c(b^2-1)\left((b^2-1)^2+d^2\right)-2bd(e+2ad)\right)}{3\left((b^2-1)^2+d^2\right)^3}r^3 +O(r^4)\right),
\end{align*}
where there are five essential parameters; $ b = w(0) $, $ c = \left.w'''\left((w^2-1)^2+\left(\frac{r\bar{a}}{S}\right)^2\right)\right\vert_{r=0} $, $ d = \left.\frac{r\bar{a}}{S}\right\vert_{r=0} $, $ e = \left.\frac{1}{S}\dydx{}{r}\left(r\bar{a}\right)\left((w^2-1)^2+\left(\frac{r\bar{a}}{S}\right)^2\right)\right\vert_{r=0} $, and $ a = \left.\dydx{}{r}(rm)\right\vert_{r=0} $. There is the usual additional parameter $ S_0 = S(0) $.
This series requires $ (b,d) \neq (\pm 1,0) $. The Schwarzschild-like singularity can be thought of as the limiting case.
The Reissner-Nordstr\"{o}m family is recovered when $ b = c = 0 $, with ADM mass $ a $ and electric charge $ d $.

\begin{figure}[!ht]
  \centering
  \includegraphics[scale=0.75]{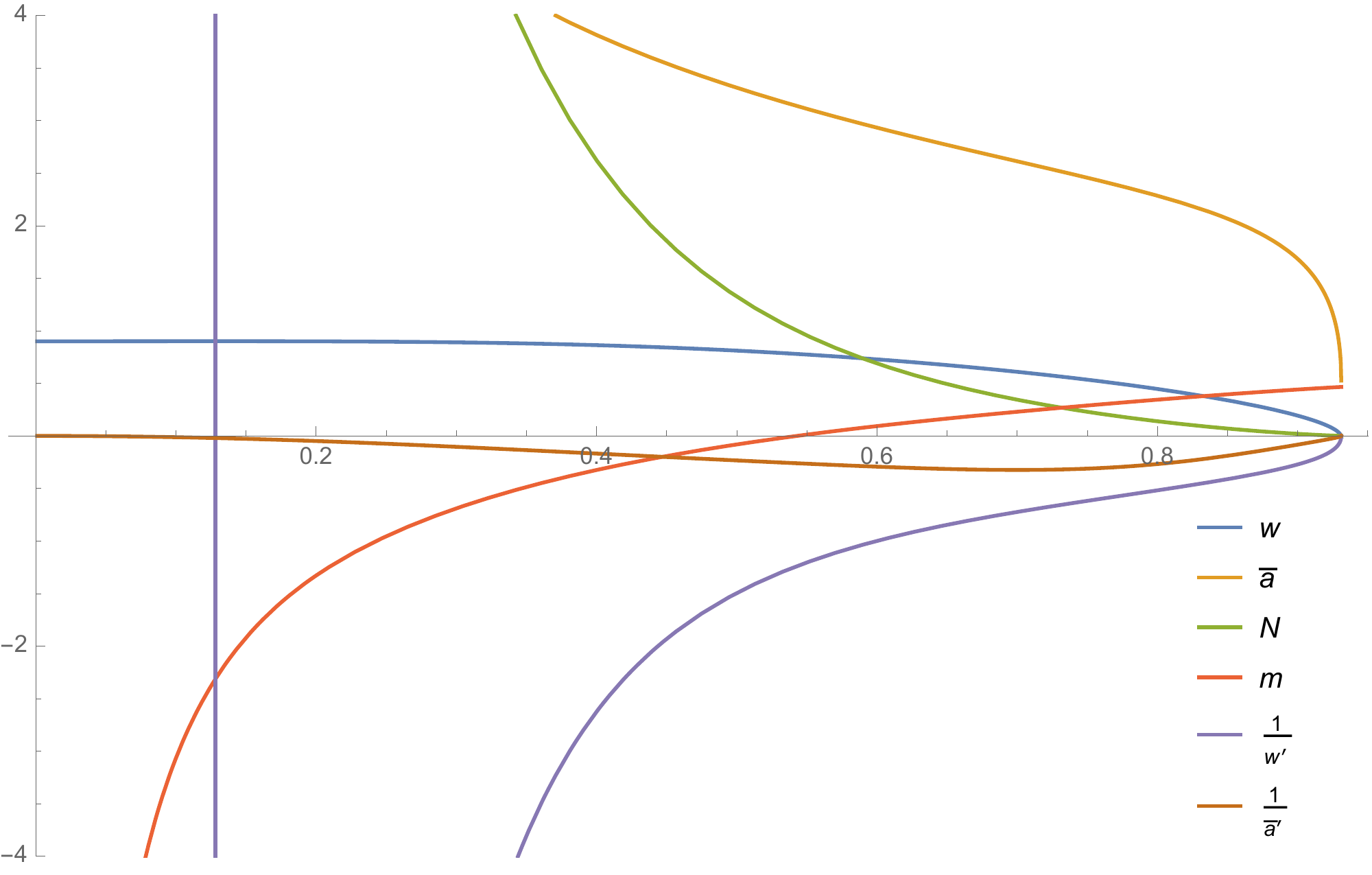}
  \caption{A plot of the RNL solution against $ r $ with parameters $ (a,b,c,d,e,S_0) $ equal to $ (0.3,0.9,-0.5,2,0.8,1) $.}
  \label{fig:RNLs}
\end{figure}

Polar-areal coordinates are most useful to consider RNL singularities; isothermal coordinates produce a series in powers of $ R^\frac{1}{3} $. 
The RNL singularity has the full number of parameters (five plus one trivial at $ r = 0 $) and so we expect to see this behaviour occur generically.
An example integration of an RNL singularity is shown in figure \ref{fig:RNLs}.
We observe that when generic values of the parameters are chosen in the RNL expansion, the solution proceeds to an equator (section \ref{sec:s:equators}) and then returns to a new RNL expansion. This is not the only possible behaviour for an RNL singularity as will be seen in section \ref{s:AF}.

\section{Solutions with a regular horizon} \label{s:EH}
We here consider solutions to (\ref{EYME stat}) that have a regular horizon at $ r = r_h $, that is, $ N(r_h) = 0 $, $ N'(r_h) > 0 $.
If a solution extends to an asymptotically flat infinity then the horizon may be referred to as an event horizon, while in general it will be a bifurcate Killing horizon.
We let $ \rho = r-r_h $ and find the local power series solution;
\begin{align*}
w(r_h+\rho) &= w_h + \frac{r_hw_h(w_h^2-1)}{r_h^2-(w_h^2-1)^2-z_h^2}\rho + O(\rho^2) \;, \\
\bar{a}(r_h+\rho) &= \frac{S_hz_h}{r_h^2}\left(\rho + \frac{(r_h^2-(w_h^2-1)^2-z_h^2)^2-r_h^4w_h^2}{r_h(r_h^2-(w_h^2-1)^2-z_h^2)^2}\rho^2 + O(\rho^3)\right), \\
m(r_h+\rho) &= \frac{r_h}{2} + \frac{(w_h^2-1)^2+z_h^2}{2r_h^2}\rho +O(\rho^2) \;, \\
S(r_h+\rho) &= S_h\left(1 + \frac{2r_hw_h^2((w_h^2-1)^2+z_h^2)}{(r_h^2-(w_h^2-1)^2-z_h^2)^2}\rho +O(\rho^2)\right),
\end{align*}
which depends essentially on three parameters, $ r_h $, $ w_h = w(r_h) $, and $ z_h = \frac{r_h^2\bar{a}(r_h)}{S(r_h)} $, and trivially on $ S_h = S(r_h) $.
We see that the Reissner-Nordstr\"{o}m family is recovered when $ w_h = 0 $, with horizon radius $ r_h > 1 $, electric charge $ z_h $, and ADM mass $ \frac{r_h}{2}+\frac{1+z_h^2}{2r_h} $.
It is also possible to obtain power series for the extremal Reissner-Nordstr\"{o}m solution ($ N'(r_h) = 0 $) and singular ($ N \equiv 0 $) series that correspond to Bertotti-Robinson solutions with $ r_h \geqslant 1 $, but we here concentrate on the generic $ N'(r_h) > 0 $ case.
We choose $ S_h = 1 $ to use up the coordinate freedom in $ t $.

Since $ N(r_h+\rho) = \frac{1}{r_h}\left(1-\frac{(w_h^2-1)^2+z_h^2}{r_h^2}\right)\rho +O(\rho^2) $, the $ N'(r_h) > 0 $ condition means the parameter space is restricted to $ \{(r_h,w_h,z_h) \mid r_h^2 > (w_h^2-1)^2+z_h^2 \} $.
Since the signs of $ w_h $ and $ z_h $ correspond to the signs of $ w $ and $ \bar{a} $ respectively, we further restrict to $ w_h \geqslant 0 $ and $ z_h \geqslant 0 $ without loss of generality.

\subsection{Purely magnetic solutions} \label{s:EH:m}
We first consider the purely magnetic case $ z_h = 0 $. As with the case of a regular origin, we use the power series to create initial data for the equations at a non-singular point. We find a fifth order power series evaluated at $ \rho = 10^{-4} $ is sufficient for better than machine precision (double precision).
Numerically integrating to large $ r $ produced results analogous to the regular origin case for each horizon radius.
For generic values of $ w_h $, an equator forms, at which the polar-areal coordinates become singular. However, for each $ r_h \in (0,\infty) $, $ w_h $ could be tuned to find asymptotically flat solutions by again ensuring $ |w| < 1 $. There are again a countably infinite number of solutions, labelled by the number of times $ w $ crosses zero.
This results in the values $ (w_h)_n $, for each $ n \in \mathbb{N} $ and $ r_h \in (0,\infty) $. Such black hole solutions were first found in \cite{VG89,KM90,Bizon90}.
The exterior behaviour of the first black hole solutions (single $ w $ zero crossing) for varying horizon radii are shown in figure \ref{fig:BHs}.
In the figure we rescale $ S $ so that $ \lim\limits_{r \to \infty} S = 1 $.
\begin{figure}[!ht]
  \centering
  \animategraphics[scale=1,timeline=BH1_r_timeline.txt]{8}{BH1_r}{1}{51}
  \caption{An animated plot of $ w $, $ N $, $ m $, and $ S $ versus $ r $ for the first black hole solution, with $ r_h $ varying from $ 10^{-3} $ to $ 10^2 $. Note that at $ r_h = 1 $ we switch to displaying $ \frac{m}{r_h} $ instead of $ m $ (digital only).}
  \label{fig:BHs}
\end{figure}
We see in figure \ref{fig:BHs} that as $ r_h $ approaches infinity, the solution approaches a limiting solution (only shifted in $ \ln r $) with $ m  = \frac{r_h}{2} $ and $ S = 1 $, thus the limiting metric is Schwarzschild.
Since semi-isotropic coordinates have $ \tau \sim \ln r $ for large $ r $, we use them to write the equations under these assumptions.
We solve $ \dydx{r}{\tau} = r\sqrt{N} $ with $ r(0) = r_h $ to find $ r = r_h \cosh^2\left(\frac{\tau}{2}\right) $, so $ N = \tanh^2\left(\frac{\tau}{2}\right) $ and $ \alpha = \tanh\left(\frac{\tau}{2}\right) $. This explicit formula for the Schwarzschild family is another benefit of semi-isotropic coordinates.
We can then write equation (\ref{w stat mag}) (cf. (\ref{wtautau})) as
\begin{equation} \label{w Schw bg}
w_{\tau\tau} +\frac{2-\cosh\tau}{\sinh \tau}w_\tau = (w^2-1)w \;.
\end{equation}
This is now independent of $ r_h $, and we numerically find the solutions in a similar way.
We find a power series with respect to $ \tau $ of
\[ w(\tau) = b\left(1+\frac{b^2-1}{4}\tau^2 +\frac{(b^2-1)(9b^2+1)}{192}\tau^4 +\frac{(b^2-1)(105b^4-30b^2-7)}{11520}\tau^6 +O\mathopen{}\left(\tau^8\right)\mathclose{}\right), \]
where $ b = w(0) $, and use this to set initial data for equation (\ref{w Schw bg}). The results for the first five solutions are shown in table \ref{table:BHs}.

\begin{table}[!ht]
  \centering
  \begin{tabular}{cll}
    \toprule
    $ k $ & $ b_k $ \\
    \midrule
    0 & 1 \\
    1 & 0.267 949 192 431 122 706 472 553 \\
    2 & 0.044 629 014 377 715 818 892 432 \\
    3 & 0.007 280 146 238 139 316 138 429 \\
    4 & 0.001 186 925 823 102 705 120 978 \\
    5 & 0.000 193 508 789 851 997 045 407 \\
    \vdots & \vdots \\
    $ \infty $ & 0 \\
    \bottomrule
  \end{tabular}
  \caption{The calculated parameter $ b $ for the first 5 Schwarzschild background solutions.}
  \label{table:BHs}
\end{table}

This limiting system was considered by Bizon \cite{Bizon94} and as the Yang-Mills field on a background Schwarzschild solution by Boutaleb-Joutei, Chakrabarti, and Comtet, who found an analytic expression for the first solution \cite{B-JCC79}, which in semi-isotropic coordinates is
\[ w_1(\tau) = \frac{2+3\sqrt{3}-\cosh\tau}{4+3\sqrt{3}+\cosh\tau} \;. \]
This means that the value $ b_1 $ in table \ref{table:BHs} is exactly $ 2-\sqrt{3} $, which we see we have achieved numerically to 24 decimal places.
As with the purely magnetic regular origin parameters (table \ref{table:BKs}), these decay approximately by a factor of $ e^{-\frac{\pi}{\sqrt{3}}k} $.

\begin{figure}[!ht]
  \centering
  \includegraphics[scale=0.75]{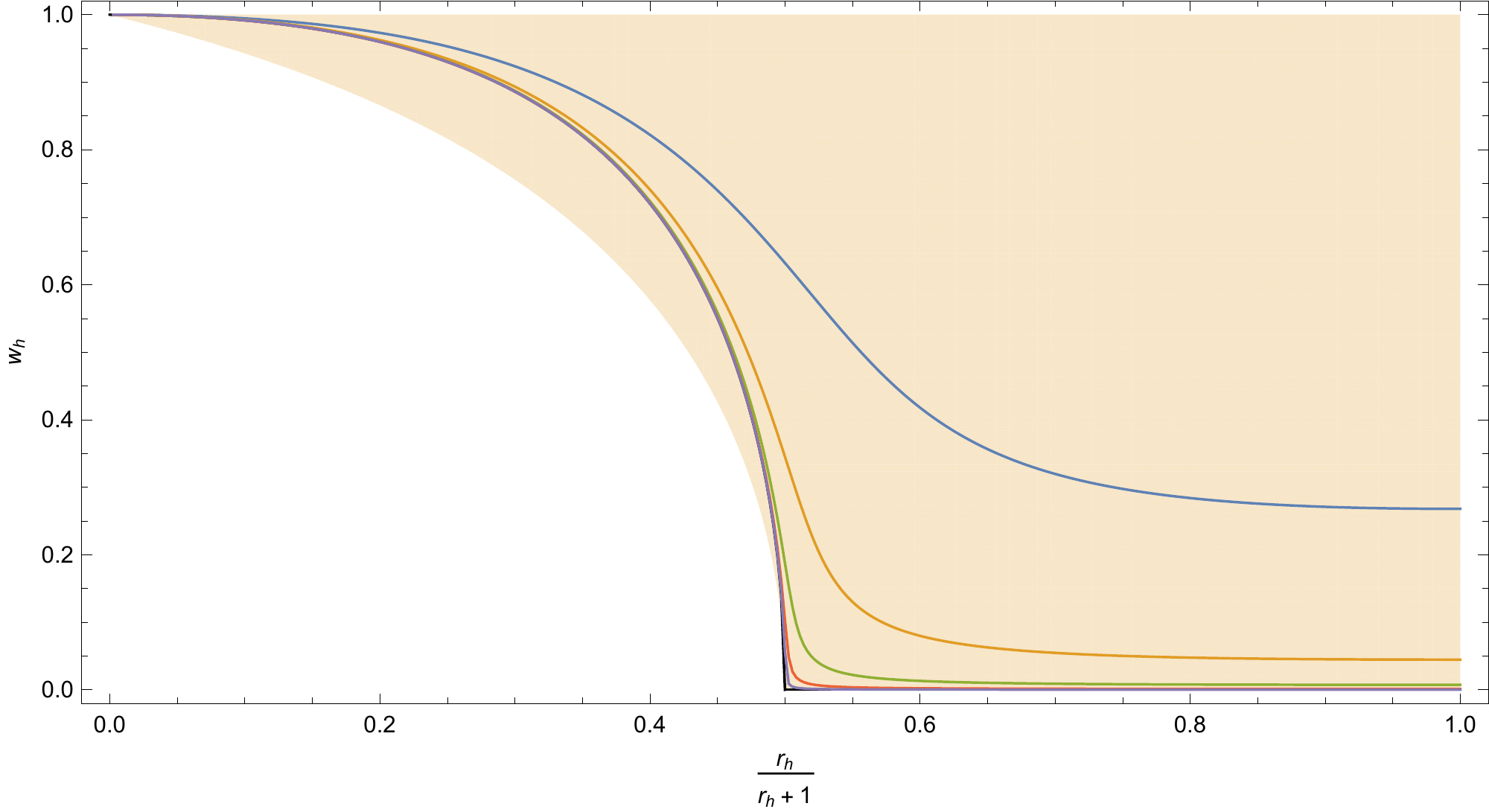}
  \caption{The $ (r_h, w_h) $ parameter space with the regular horizon ($ N'(r_h) > 0 $) part of the space shaded and the first five asymptotically flat black hole families marked, and the limiting solution marked in black.}
  \label{fig:rhwh}
\end{figure}

The solutions in the $ (r_h, w_h) $ parameter space are shown in figure \ref{fig:rhwh}. This view matches that shown in \cite{BFM94}, figure 3.

Similarly to the regular origin case, we use isothermal coordinates to find the values $ (w_h)_\infty $. Since $ R \to -\infty $ as the horizon is approached from outside, we find starting values for the equations (\ref{EYME stat TR}) by converting the values of the polar-areal functions at $ r = r_h + \rho $ to isothermal variables at $ R = R_0 $;
\begin{align*}
w(R_0) &= w(r_h+\rho) \;, & r(R_0) &= r_h+\rho \;, & \alpha(R_0) &= \left.S\sqrt{N}\right\vert_{r=r_h+\rho} \;, \\
\left.\dydx{w}{R}\right\vert_{R_0} &= \left.SNw'\right\vert_{r=r_h+\rho} \;, & \left.\dydx{r}{R}\right\vert_{R_0} &= \left.SN\right\vert_{r=r_h+\rho} \;, & \left.\dydx{\alpha}{R}\right\vert_{R_0} &= \left.SN(S\sqrt{N})'\right\vert_{r=r_h+\rho} \;.
\end{align*} 

The interiors of these black hole solutions have been analysed in the papers \cite{DGZ97} and \cite{BLM98}, and it was found that SL and RNL singularities can occur for discrete values of the parameters, while the generic behaviour is infinitely oscillating near the singularity.
We do not reproduce those results here.

\subsection{General solutions} \label{s:EH full}
When considering the solutions with a non-zero electric field, again the main known result is in \cite{BP92}; the only way that the regular horizon and asymptotically flat conditions can both be realised is in the essentially Abelian Reissner-Nordstr\"{o}m family of solutions.
So again here we investigate how the solutions fail to become asymptotically flat. We find that it is quite analogous to the regular origin case, in that two generic behaviours occur; an equator forms or the solution becomes long-lived with what appears to be infinite oscillations.

\begin{figure}[!ht]
  \centering
  \includegraphics[scale=1]{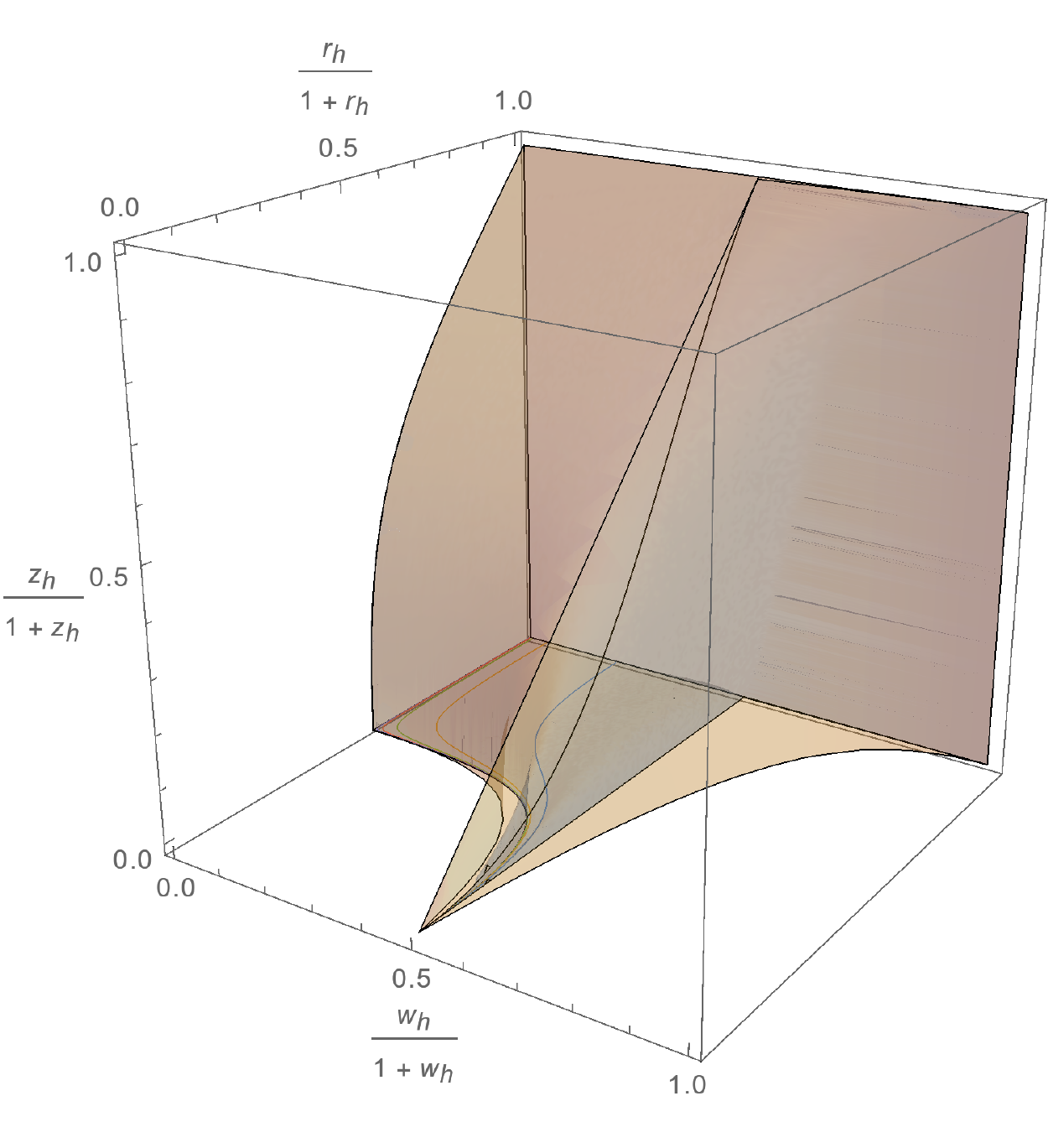}
  \caption{The boundary of the long-lived region in $ (r_h,w_h,z_h) $ space (blue) in the regular horizon part of the parameter space (yellow). The asymptotically flat solutions occur in the $ z_h = 0 $ and $ w_h = 0 $ planes.}
  \label{fig:stat elec BH space}
\end{figure}

We find numerically the boundary of the long-lived region, and this is plotted in figure \ref{fig:stat elec BH space}, where the parameter space has been compactified.
Analogously to the regular case, the boundary near the magnetic solutions (the lower boundary) is more difficult to determine than the other boundaries, where a bisection search can accurately determine the boundary by whether $ r' = 0 $ or $ N_R > 0 $ and $ N > \sqrt{w^2+\frac{w_R^2}{\bar{a}^2}} $ (the amplitude of $ w $).
Near the magnetic solutions, many oscillations could occur before eventually $ N \to 0 $.
The behaviour of the boundary solutions themselves are considered in section \ref{sec:s:ABR}.

In figure \ref{fig:stat elec BH space} the Reissner-Nordstr\"{o}m family ($ w_h = 0 $) are on the left side, with the small $ r_h $ boundary corresponding to the extremal Reissner-Nordstr\"{o}m family. As $ w_h \to 0 $, the long-lived region expands to include the entire $ (r_h,z_h) $ space of this family.
The Schwarzschild solutions occur for $ w_h = 1 $, $ z_h = 0 $, with the $ r_h \to 0 $ limit point representing both Minkowski space and all of the Bartnik-McKinnon solutions.
In the $ w_h = 1 $ plane, the $ r_h = z_h $ boundary of the regular horizon region as well as the upper $ z_h $ boundary of the long-lived region are shown with black lines.
As $ r_h \to \infty $, the long-lived region expands to include the entire $ (w_h,z_h) $ parameter space. Note that this expansion is much more abrupt for small $ z_h $.

The lower boundary of the long-lived region when $ w_h < 1 $ varies between $ 0 \leqslant z_h \lesssim 0.04 $ and extends continuously to the magnetic solutions (figure \ref{fig:rhwh}) at $ z_h = 0 $. Its maximum occurs between the Schwarzschild and first black hole solution with $ r_h \approx 2 $.

Let us consider the asymptotic system in the limit $ r_h \to \infty $.
As in the magnetic case, we have $ m = \frac{r_h}{2} $ and $ S = 1 $, so this is equivalent to considering the full Yang-Mills field on a Schwarzschild background. As before, we have $ r = r_h \cosh^2\left(\frac{\tau}{2}\right) $ and equations (\ref{wtautau},\ref{atautau}) become
\begin{subequations} \label{YM infinite rh}
\begin{align}
w_{\tau\tau} +\frac{2-\cosh\tau}{\sinh\tau}w_\tau &= \left((w^2-1)-\frac{r_h^2(\cosh\tau+1)^3}{4(\cosh\tau-1)}\bar{a}^2\right)w \;, \label{w infinite rh} \\
\bar{a}_{\tau\tau} -\frac{2-\cosh\tau}{\sinh\tau}\bar{a}_\tau &= 2w^2 \bar{a} \;.
\end{align}
\end{subequations}

We find it convenient to write these equations in terms of $ \bar\phi = \frac{r}{\alpha}\bar{a} $, in which case they are written
\begin{subequations} \label{YM Schwarzschild bg}
\begin{align}
w_{\tau\tau} +\frac{2-\cosh\tau}{\sinh\tau}w_\tau &= \left((w^2-1)-\bar\phi^2\right)w \;, \\
\bar\phi_{\tau\tau} +\frac{2-\cosh\tau}{\sinh\tau}\bar\phi_\tau &= \left(2w^2 -\frac{1-2\cosh\tau}{\sinh^2\tau}\right) \bar\phi \;,
\end{align}
\end{subequations}
and we see we have removed all dependency on $ r_h $ (the choice of $ \bar\phi = r\bar{a} $ would also be sufficient for our purposes).
These are singular at $ \tau = 0 $ and we find the power series
\begin{align*}
w(\tau) &= b\left(1+\frac{b^2-1}{4}\tau^2 +\frac{(b^2-1)(9b^2+1)-12d^2}{192}\tau^4 +O\mathopen{}\left(\tau^6\right)\mathclose{}\right), \\
\bar\phi(\tau) &= d\left(\tau +\frac{3b^2+2}{12}\tau^3 +\frac{15b^4+2}{240}\tau^5 +\frac{1155b^6-420b^4-231b^2+16-420d^2}{80640}\tau^7 +O\mathopen{}\left(\tau^{9}\right)\mathclose{}\right),
\end{align*}
where $ b = w(0) $ and $ d = \bar\phi_\tau(0) $.

We find numerically that solutions to equations (\ref{YM Schwarzschild bg}) exist for all $ (b,d) $ with $ d \neq 0 $. This corresponds to the observed sudden expansion of the long-lived region in figure \ref{fig:stat elec BH space} at large horizon radius, particularly noticeable for $ w_h > 1 $ and small $ z_h $.
This can be explained by considering the additional term in equation (\ref{w infinite rh}). In the magnetic case, $ |w| > 1 $ implies $ w \to \infty $. The additional $ -\bar{a}^2 $ term, where $ |\bar{a}| $ is monotonically increasing, demands that for large $ |w| $, $ w_{\tau\tau} $ must obtain the opposite sign to $ w $, thus causing $ w $ to oscillate indefinitely.

We now consider the asymptotics as $ \tau \to \infty $. Since the Schwarzschild background is asymptotically flat, we can consider the asymptotics of the flat-space Yang-Mills equations
\begin{subequations} \label{YM flat bg}
\begin{align}
w_{\tau\tau} -w_\tau &= \left((w^2-1)-\bar\phi^2\right)w \;, \label{w flat bg} \\
\bar\phi_{\tau\tau} -\bar\phi_\tau &= 2w^2 \bar\phi \;. \label{phi flat bg}
\end{align}
\end{subequations}
However, we can use the analysis of section \ref{sec:s:full} with the semi-isotropic coordinate to find the asymptotic behaviour on any background, including Schwarzschild and Minkowski space.
Note that for flat space $ r = e^\tau $ and $ \alpha = 1 $, while for the Schwarzschild background we have $ r = r_h \cosh^2\left(\frac{\tau}{2}\right) $ and $ \alpha = \tanh\frac{\tau}{2} $ (which approach the flat behaviour as $ \tau \to \infty $).
Equations (\ref{YME tau}) written in terms of $ \bar\phi $ are
\begin{subequations} \label{YME wphi tau}
\begin{align}
\frac{r}{\alpha}\left(\frac{\alpha}{r}w_\tau\right)_\tau &= \left((w^2-1)-\bar{\phi}^2\right)w \;, \label{wtautaua} \\
\left(\frac{r}{\alpha}\left(\frac{\alpha}{r}\bar\phi\right)_\tau\right)_\tau &= 2w^2\bar{\phi} \;. \label{phitautau}
\end{align}
\end{subequations}

We will assume as before that $ w \sim h\cos\lambda $ and $ \bar\phi \sim \bar{o} $. Inserting these ansatzes into equations (\ref{YME wphi tau}) with the assumption $ |w^2-1| \ll \bar{\phi}^2 $, we find
\begin{align*}
h &\sim \frac{A}{\sqrt{\bar{o}}}\sqrt{\frac{r}{\alpha}} \;, \\
\bar{o} &= A^2\frac{r}{\alpha}\left(\int\frac{\alpha}{r}\int\frac{r}{\alpha}\dx{\tau}\dx{\tau}+B\int\frac{\alpha}{r}\dx{\tau}+C\right), \\
\lambda &= A^2\left(\int\frac{r}{\alpha}\int\frac{\alpha}{r}\int\frac{r}{\alpha}\dx{\tau}\dx{\tau}\dx{\tau}+B\int \frac{r}{\alpha}\int\frac{\alpha}{r}\dx{\tau}\dx{\tau}+C\int \frac{r}{\alpha}\dx{\tau} +D\right),
\end{align*}
and so $ \bar{a} \sim A^2\left(\int\frac{\alpha}{r}\int\frac{r}{\alpha}\dx{\tau}\dx{\tau}+B\int\frac{\alpha}{r}\dx{\tau}+C\right) $.
While there is a nice symmetry to the integrals in these coordinates, we can see the advantage of using isothermal coordinates, for which $ \int\frac{r}{\alpha}\dx{\tau} = R $.

We can evaluate the Yang-Mills asymptotics for a given background, starting with flat space where $ r = e^\tau $ and $ \alpha = 1 $.
We find $ \bar{o} = A^2e^\tau(\tau+C-Be^{-\tau}) $, and the remaining functions in both semi-isotropic and the more familiar polar-areal coordinates are
\begin{align*}
\bar{a} &\sim A^2(\tau+C-Be^{-\tau}) \;, & \bar{a} &\sim A^2\left(\ln r+C-\frac{B}{r}\right), \\
\lambda &\sim A^2\left((\tau+C-1)e^\tau-B\tau+D\right), & \lambda &\sim A^2\left(\left(\ln r+C-1\right)r-B\ln r+D\right), \\
h &\sim \frac{1}{\sqrt{\tau+C-Be^{-\tau}}} \;, & h &\sim \frac{1}{\sqrt{\ln r+C-\frac{B}{r}}} \;.
\end{align*}
We note that the controlling factor $ \frac{1}{\sqrt{\ln r}} $ of the amplitude of $ w $ is independent of the initial conditions, although for modest $ r $ the constants $ C $ and $ B $ are important for a close numerical match.

For the Schwarzschild background ($ r = r_h \cosh^2\left(\frac{\tau}{2}\right) $, $ \alpha = \tanh\frac{\tau}{2} $) this results in
\[ \bar{o} = A^2r_h\left(\sinh\tau\ln\sinh\left(\frac{\tau}{2}\right) +C\frac{\cosh^3\left(\frac{\tau}{2}\right)}{\sinh\left(\frac{\tau}{2}\right)} +\frac{1-\frac{B}{r_h}}{\tanh\left(\frac{\tau}{2}\right)}\right). \]
Then we have
\begin{equation} \label{Schw bg asy}
\begin{aligned}
\bar{a} &\sim A^2\left(2\tanh^2\left(\frac{\tau}{2}\right)\ln\sinh\left(\frac{\tau}{2}\right)+C+\frac{1-\frac{B}{r_h}}{\cosh^2\left(\frac{\tau}{2}\right)}\right), \\
\lambda &\sim A^2r_h\left(\left(2\ln\sinh\left(\frac{\tau}{2}\right)+C-1\right)\sinh^2\left(\frac{\tau}{2}\right)+2\left(C+1-\frac{B}{r_h}\right)\ln\sinh\left(\frac{\tau}{2}\right)+D\right), \\
h &\sim \frac{1}{\sqrt{2\tanh^2\left(\frac{\tau}{2}\right)\ln\sinh\left(\frac{\tau}{2}\right)+C+\frac{1-\frac{B}{r_h}}{\cosh^2\left(\frac{\tau}{2}\right)}}} \;,
\end{aligned}
\end{equation}
and in polar-areal coordinates,
\begin{align*}
\bar{a} &\sim A^2\left(\sqrt{1-\frac{r_h}{r}}\ln\left(\frac{r}{r_h}-1\right)+C+\frac{r_h-B}{r}\right), \\
\lambda &\sim A^2\left(\left(\ln\left(\frac{r}{r_h}-1\right)+C-1\right)\left(r-r_h\right)+\left((C+1)r_h-B\right)\ln\left(\frac{r}{r_h}-1\right)+Dr_h\right), \\
h &\sim \frac{1}{\sqrt{\sqrt{1-\frac{r_h}{r}}\ln\left(\frac{r}{r_h}-1\right)+C+\frac{r_h-B}{r}}} \;.
\end{align*}
We thus find the same controlling factors $ \bar{a} \sim A^2\ln(r) $ and $ w \sim \frac{1}{\sqrt{\ln(r)}}\cos\left(A^2r\ln(r)\right) $ for both backgrounds (we're using the symbol $ \sim $ loosely with regards to $ w $ as it is oscillatory and the zeros of the approximation do not match the zeroes of $ w $ exactly).
This is not surprising as the Schwarzschild solution is asymptotically flat.

We confirm the approximations (\ref{Schw bg asy}) by determining the four constants $ A $, $ B $, $ C $, and $ D $ for the numerical solution to  (\ref{YM Schwarzschild bg}) with $ b = w(0) = 1 $ and $ d = \bar{\phi}_\tau(0) = 1 $.
It would be difficult in general to find the four constants by minimising the error in the approximations to $ w $ and $ \bar\phi $.
Instead we are able to approximate $ A $ quite well for each numerical solution by plotting $ \sqrt{\frac{\sinh\left(\frac{\tau}{2}\right)}{\cosh^3\left(\frac{\tau}{2}\right)}\frac{w\indices{_\tau^2}+\bar\phi^2w^2}{\bar\phi}} $, which approaches a constant ($ \sqrt{r_h}A $) quite rapidly. We can then find $ C $ from the plot of $ \frac{2}{\sinh\tau}\left(\tanh\left(\frac{\tau}{2}\right)\frac{\bar\phi}{A^2r_h}\right)_\tau-2\ln\sinh\left(\frac{\tau}{2}\right)-1 $, and $ \frac{B}{r_h} $ from $ 1+\tanh\left(\frac{\tau}{2}\right)\left(\sinh\tau\ln\sinh\left(\frac{\tau}{2}\right) +C\frac{\cosh^3\left(\frac{\tau}{2}\right)}{\sinh\left(\frac{\tau}{2}\right)}-\frac{\bar\phi}{A^2r_h}\right) $. Finally $ D $ can be determined from the phase of $ w $.

\begin{figure}[!ht]
  \centering
  \includegraphics[scale=0.75]{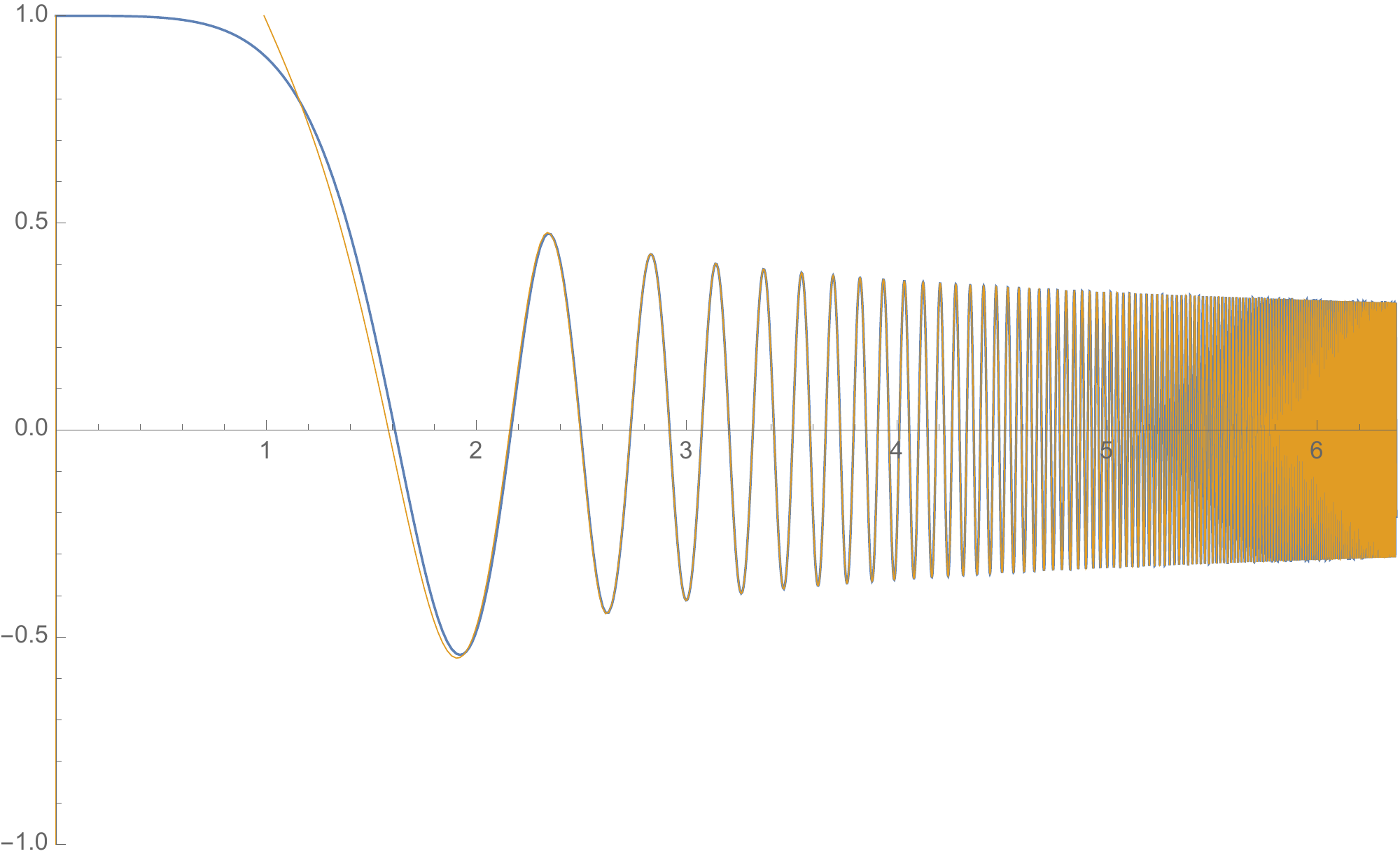}
  \caption{The numerical solution $ w(\tau) $ for $ (b,d) = (1,1) $ (blue) and the asymptotic approximation (yellow).}
  \label{fig:Schw bg w}
\end{figure}

\begin{figure}[!ht]
  \centering
  \includegraphics[scale=0.75]{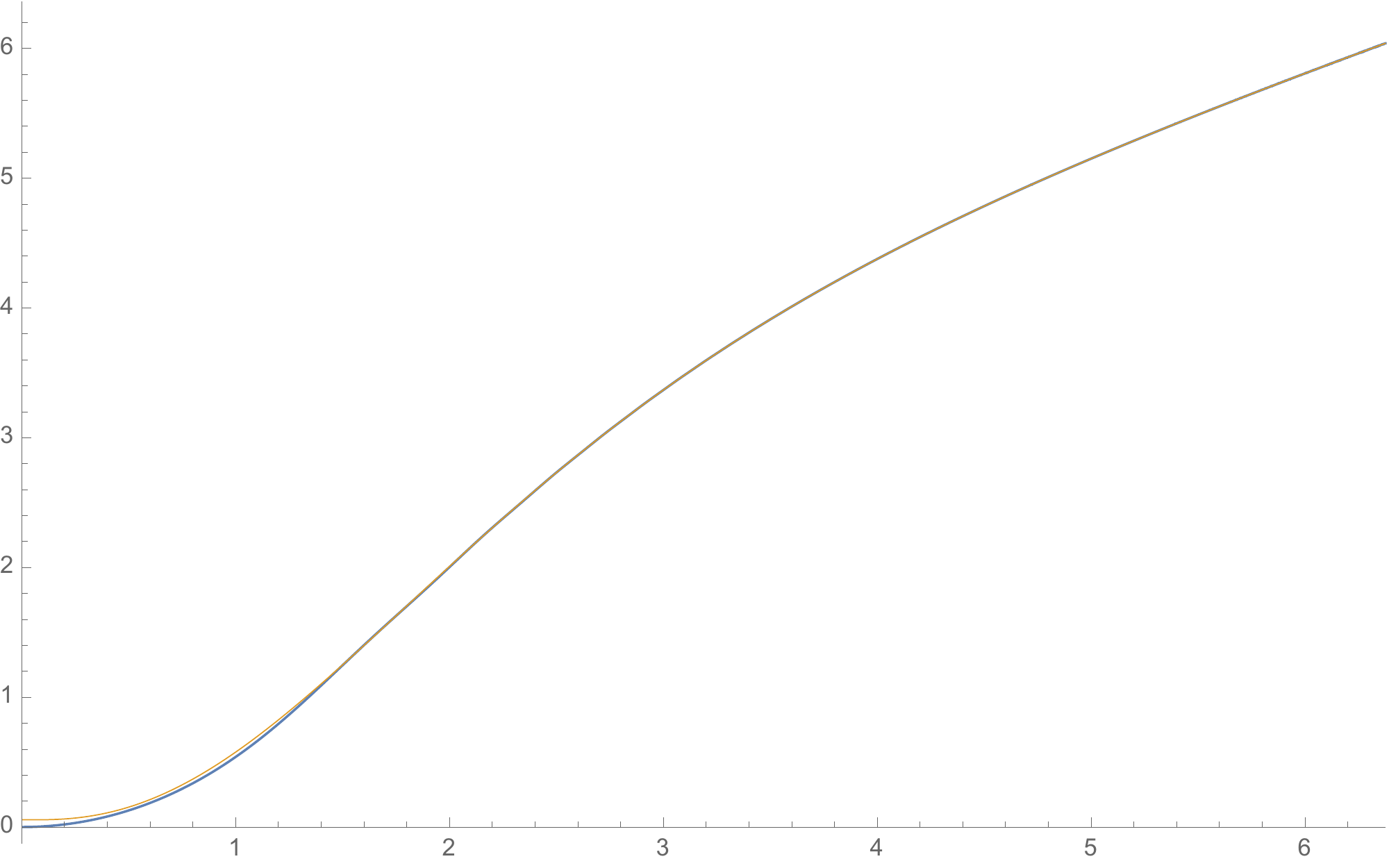}
  \caption{The numerical solution $ \bar{a}(\tau) $ for $ (b,d) = (1,1) $ and $ r_h = 1 $ (blue) and the asymptotic approximation (yellow).}
  \label{fig:Schw bg a}
\end{figure}

In figures \ref{fig:Schw bg w} and \ref{fig:Schw bg a} we plot the numerically determined $ w(\tau) $ and $ \bar{a}(\tau) $ for $ (b,d) = (1,1) $ as well as the asymptotic approximations given by $ w = \frac{A}{\sqrt{\bar{a}}}\cos\lambda $ and equations (\ref{Schw bg asy}).
The parameters found were $ A\sqrt{r_h} \approx 0.75519 $, $ C \approx 5.67345 $, $ \frac{B}{r_h} \approx 6.574 $ and $ D \approx -0.6 $.
They clearly give a very good fit, even for small values of $ \tau $.

We have this description of the asymptotics for an infinite horizon radius, which corresponds to the Yang-Mills equations on a Schwarzschild background. When $ r_h < \infty $, the asymptotics are the same as for the regular origin case.

\section{Equators} \label{sec:s:equators}
As polar-areal coordinates are singular on an equator, we use isothermal coordinates here.
First observing (\ref{rRR+}) we see that when $ r_R = 0 $, $ r_{RR} \leqslant 0 $, and so $ r $ can only ever attain a maximum; never a minimum.
Note that this does not discount a Schwarzschild-like throat; only that these coordinates can not see it. We also note that since $ N_R = N_T = 0 $, an equator has $ N = 0 $ but is not a marginally trapped surface.
The equator condition $ r_R = 0 $ in (\ref{ralpha1}) puts the following condition on the gauge fields at the equator.
\begin{equation} \label{equator}
1+\frac{2}{\alpha^2}\left(w\indices{_R^2}+\bar{a}^2w^2\right) = \frac{(w^2-1)^2}{r^2}+\frac{r^2\bar{a}\indices{_R^2}}{\alpha^4} \;.
\end{equation}
Also, while generically equation (\ref{rRR+}) can be used to specify $ \alpha_R $ at a point, with $ r_R = 0 $ this no longer occurs and $ \alpha_R $ becomes freely specifiable.

So at an equator ($ r_R(0) = 0 $) there are five essential free parameters that can be specified.
This means an equator can be expected to occur generically, which is indeed what we observe numerically when integrating through increasing $ r $ (for example, from a regular origin (section \ref{s:reg}), a singular origin (\ref{sec:s:stat sing}), or a regular horizon (\ref{s:EH})).
We find it most useful to specify $ b = w(0) $, $ c = \frac{w_R(0)}{\alpha(0)} $, $ d = \frac{r(0)^2}{\alpha(0)^2}\bar{a}_R(0) $, $ e = \frac{\bar{a}(0)}{\alpha(0)} $, so that (\ref{equator}) gives $ r(0)^2 = \frac{(b^2-1)^2+d^2}{1+2(c^2+b^2e^2)} $.
These variables are free as long as $ (b,d) \neq (\pm 1,0) $, which results in the singular solution $ r \equiv 0 $.
The final essential parameter is $ f = \alpha_R(0) $ and the trivial parameter is $ a = \alpha(0) $.
Note that a horizon would also have $ r_R = 0 $ in isothermal coordinates since $ N = 0 $, however a horizon can only occur as $ R \to -\infty $, and so the $ r_R(0) = 0 $ condition restricts to only equators or Bertotti-Robinson solutions.

A Bertotti-Robinson solution is achieved when $ b = c = 0 $, which gives a radius of $ r = \sqrt{1+d^2} $. The precise form it takes depends on the remaining parameters.
When $ f^2 < \frac{1}{r^2} $, the particular form of the Bertotti-Robinson solution this picks out is $ \alpha = \frac{r}{\cos(R)} $.
Precisely, we have $ \alpha(R) = \frac{Cr}{\cos(CR+D)} $ and $ \bar{a}(R) = Cd(\tan(CR+D)-\tan D)+e $ where $ C = a\sqrt{\frac{1}{r^2}-f^2} $ and $ D = \arcsin\left(fr\right) $.
When $ f^2 = \frac{1}{r^2} $, we find a Bertotti-Robinson solution in the $ \alpha = \frac{r}{R} $ family. Precisely, we have $ \alpha(R) = -\frac{1}{fR-\frac{1}{a}} $ and $ \bar{a}(R) = -\frac{d}{R-\frac{1}{af}}-afd+e $.
When $ f^2 > \frac{1}{r^2} $, we find a Bertotti-Robinson solution in the $ \alpha = \frac{r}{\sinh(R)} $ family. Precisely, we have $ \alpha(R) = -\frac{\sgn(f)Cr}{\sinh(CR+D)} $ and $ \bar{a}(R) = -Cd\left(\coth(CR+D)-\coth D\right)+e $ where $ C = -\sgn(f)a\sqrt{f^2-\frac{1}{r^2}} $ and $ D = \arcosh\left(|f|r\right) $.

When $ d = e = 0 $, $ \bar{a} $ is identically zero and we recover the purely magnetic case. Breitenlohner, Forg\'{a}cs and Maison proved in this case that an equator leads to a magnetic Reissner-Nordstr\"{o}m-like singularity \cite{BFM94}.
When $ b $ or $ c $ are non-zero and $ d $ or $ e $ are non-zero, we have (either at $ R = 0 $ or at a small, possibly negative $ R $) $ \bar{a} \neq 0 $ and $ |\bar{a}| $ increasing as $ |R| $ increases. By equation (\ref{aRR}), $ |\bar{a}| $ increases without bound, while at the same time $ r $ decreases to zero.
We find numerically that at least one side of an equator will be a Reissner-Nordstr\"{o}m-like singularity. 
From numerical integrations we see that this is typically the case on both sides, however as we have seen in sections \ref{sec:s:full} and \ref{s:EH full} it is also possible for $ |\bar{a}| $ to decrease to zero and produce a regular origin or horizon.
Such cases are not seen generically because they have codimensions three and two respectively. We can now interpret figure \ref{fig:stat elec S3 typ} as showing a solution with regular origin, equator, and a Reissner-Nordstr\"{o}m singularity.

\section{Solutions on the boundary of the long-lived region} \label{sec:s:ABR}
In this section we will consider the behaviour of solutions on the boundary of the long-lived regions shown in figure \ref{fig:stat elec space} and \ref{fig:stat elec BH space}.
We will use the semi-isotropic coordinates, and generalise the analysis used in \cite{BFM94} for the purely magnetic case.
In that paper, the authors note three possible cases when integrating the equations with increasing $ \tau $:
\begin{enumerate}
\item $ N(\tau) $ has a zero at $ \tau = \tau_0 $,
\item $ N(\tau) > 0 $ for all $ \tau $ and $ r(\tau) $ tends to infinity,
\item $ N(\tau) > 0 $ for all $ \tau $ and $ r(\tau) $ remains bounded.
\end{enumerate}
The first case indicates an equator, and the second the long-lived solutions.
It is the third case to which we now turn.

We begin by writing the Yang-Mills equations (\ref{wtautau},\ref{atautau}) in the following form, 
\begin{subequations} \label{FYM}
\begin{align}
w_{\tau\tau} +\left(\frac{\alpha_\tau}{\alpha}-\frac{r_\tau}{r}\right)w_\tau +\left(\bar{\phi}^2-(w^2-1)\right)w &= 0 \;, \\
\bar{\phi}_{\tau\tau} +\left(\frac{\alpha_\tau}{\alpha}-\frac{r_\tau}{r}\right)\bar{\phi}_\tau -\left(2w^2-\left(\frac{\alpha_\tau}{\alpha}-\frac{r_\tau}{r}\right)_\tau\right)\bar{\phi} &= 0 \;, \label{FYM phi}
\end{align}
\end{subequations}
where we find it convenient, as in section \ref{s:EH full}, to use the new variable $ \bar{\phi} = \frac{r}{\alpha}\bar{a} $ (cf. equations (\ref{YME wphi tau})).

Now considering case (iii), $ r_\tau(\tau) > 0 $ and $ r(\tau) $ being bounded implies $ r_\tau \to 0 $ and $ r $ has a limit, say $ r_0 $.
Following \cite{BFM94}, we will determine the behaviour of $ r $ and $ \alpha $ through the intermediate variables $ \kappa = \frac{r_\tau}{r} +\frac{\alpha_\tau}{\alpha} $ and $ N = \frac{r\indices{_\tau^2}}{r^2} $. Differentiating $ \kappa $, we find
\begin{equation} \label{kappa}
\kappa_\tau = 1 -\kappa^2 +\frac{2}{r^2}\left(w\indices{_\tau^2}+\bar\phi^2w^2\right).
\end{equation}
Lemma 10 in \cite{BFM94} generalises immediately to the general case, and so we know that, for any $ \epsilon > 0 $, $ 1 \leqslant \kappa < 2+\epsilon $ for sufficiently large $ \tau $.
Generalising the ``energy" from \cite{BFM94} using (\ref{alphatau}), we have
\begin{align}
E &= \frac{r^2}{4}\left(-1+\frac{r\indices{_\tau^2}}{r^2} +2\frac{r_\tau\alpha_\tau}{r\alpha}\right) = -\frac{r^2}{4}\left(1+N-2\kappa\sqrt{N}\right) \nonumber \\
 &= \frac{1}{2}\left(w\indices{_\tau^2} +\bar\phi^2w^2\right) -\frac{1}{4}\left((w^2-1)^2 +\left(\bar\phi\indices{_\tau^2} +\left(\frac{\alpha_\tau}{\alpha}-\frac{r_\tau}{r}\right)\bar\phi\right)^2\right), \label{E stat}
\end{align}
and $ E_\tau = \left(w\indices{_\tau^2}+\bar\phi^2w^2\right)\left(\frac{r_\tau}{r}-\frac{\alpha_\tau}{\alpha}\right) = \left(w\indices{_\tau^2}+\bar\phi^2w^2\right)\left(2\sqrt{N}-\kappa\right) $.
Since $ N = \frac{r\indices{_\tau^2}}{r^2} $ approaches zero, $ E \to -\frac{r_0^2}{4} $, which implies that $ E_\tau $ is integrable and $ w\indices{_\tau^2}+\bar\phi^2w^2 \to 0 $. Then (\ref{kappa}) tells us that $ \kappa \to 1 $, that is, $ \frac{\alpha_\tau}{\alpha} \to 1 $, so $ \alpha \sim A e^\tau $.

We note that for asymptotically flat spacetimes, $ \frac{r_\tau}{r} \to 1 $ and $ \frac{\alpha_\tau}{\alpha} \to 0 $, and the asymptotics we have found give the flat space Yang-Mills equations, with the direction of the coordinate reversed (see (\ref{FYM})). We therefore first consider the solutions to the flat space Yang-Mills equations
\begin{align*}
w_{\tau\tau} -w_\tau +\left(\bar{\phi}^2-(w^2-1)\right)w &= 0 \;, \\
\bar{\phi}_{\tau\tau} -\bar{\phi}_\tau -2w^2\bar{\phi} &= 0 \;.
\end{align*}
With these coordinates and variables, the flat space Yang-Mills equations are autonomous, and we consider the phase space of points $ (w,w_\tau,\bar{\phi},\bar{\phi}_\tau) $, with vector field
\[ X = (w_\tau,((w^2-1)-\bar{\phi}^2)w+w_\tau,\bar{\phi}_\tau,2w^2\bar\phi+\bar\phi_\tau) \;. \]
There are two singular points $ (\pm 1,0,0,0) $ and a singular line $ (0,0,B,0) $ with $ B \in \mathbb{R} $.
Linearisation reveals the singular points are saddle point with eigenvalues $ \lambda = -1,-1,2,2 $, while each point on the singular line has eigenvalues $ \lambda = 0,1,\frac{1 \pm i\sqrt{3+4B^2}}{2} $.
The stable manifold at the two singular points can be written as
\begin{align*}
w &= \pm \left(1 +ce^{-\tau} +O(e^{-2\tau})\right), & \bar\phi &= (e)e^{-\tau} +O(e^{-2\tau}) \;,
\end{align*}
and the unstable manifold as
\begin{align*}
w &= \pm \left(1 +be^{2\tau} +O(e^{4\tau})\right), & \bar\phi &= d e^{2\tau} +O(e^{4\tau}) \;.
\end{align*}
The solution near the singular line is
\begin{align*}
w(\tau) &= Ce^{\frac{\tau}{2}}\sin\left(\frac{\sqrt{3+4B^2}}{2}\tau+\theta\right), & \bar\phi(\tau) &= B +De^\tau \;.
\end{align*}

Returning to our asymptotics by reversing the direction of $ \tau $, we expect the solution will approach a singular point. Since the isolated singular points are not compatible with negative $ E $, we assume the solution approaches a point on the singular line.
We therefore expect the asymptotic behaviour
\begin{align} \label{wphi asy}
w(\tau) &\sim Ce^{-\frac{\tau}{2}}\sin\left(\frac{\sqrt{3+4B^2}}{2}\tau+\theta\right), & \bar\phi(\tau) &\sim B +De^{-\tau} \;.
\end{align}
Inserting this behaviour into (\ref{E stat}) we find $ r_0^2 = 1+B^2 $.
We also find
\[ w\indices{_\tau^2}+\bar\phi^2w^2 \sim \frac{1}{2}C^2e^{-\tau}\left(1+2B^2 +\sqrt{1+B^2}\cos\left(\sqrt{3+4B^2}\tau +\arctan\sqrt{3+4B^2} +2\theta\right)\right). \]
Inserting this into (\ref{rtautau})
\[ \left(\frac{r_\tau}{r}\right)_\tau = \frac{r_\tau\alpha_\tau}{r\alpha} -\frac{2}{r^2}\left(w\indices{_\tau^2}+\bar\phi^2w^2\right), \]
and using the asymptotic behaviour $ r \sim \sqrt{1+B^2} $ and $ \frac{\alpha_\tau}{\alpha} \sim 1 $ we integrate for $ \frac{r_\tau}{r} $ and find
\begin{align*}
\frac{r_\tau}{r} &= A_1e^\tau +\frac{C^2}{1+B^2}e^{-\tau}\left(\frac{1+2B^2}{2}-\sqrt{\frac{1+B^2}{7+4B^2}}\cos\left(\sqrt{3+4B^2}\tau+\arctan\left(-\frac{3\sqrt{3+4B^2}}{1+4B^2}\right)+2\theta\right)\right),
\end{align*}
where clearly we require $ A_1 = 0 $ for a bounded $ r $.
Integrating a further time we find the asymptotic behaviour of $ r $ to be
\[ r \sim \sqrt{1+B^2}e^{-\frac{C^2}{1+B^2}e^{-\tau}\left(\frac{1+2B^2}{2}-\frac{1}{2\sqrt{7+4B^2}}\cos\left(\sqrt{3+4B^2}\tau+2\theta+\arctan\left(-\frac{(1-2B^2)\sqrt{3+4B^2}}{5+8B^2}\right)\right)\right)} \;. \]

We then find the next to leading order behaviour of $ \kappa $ by setting $ \kappa = 1 + \epsilon $ in (\ref{kappa});
\[ \kappa(\tau) \sim 1 + \frac{C^2}{1+B^2} e^{-\tau} \left(1+2B^2+\frac{1}{2}\cos\left(\sqrt{3+4B^2}\tau +2\theta\right)\right) +E e^{-2\tau} \;. \]
This can then be used to determine the lower order terms of $ \alpha $, and we find
\begin{align*}
\frac{\alpha_\tau}{\alpha} &\sim 1+\frac{C^2}{2(1+B^2)}e^{-\tau}\left(1+2B^2+\sqrt{\frac{13+16B^2}{7+4B^2}}\cos\left(\sqrt{3+4B^2}\tau+2\theta+\arctan\left(-\frac{3\sqrt{3+4B^2}}{8(1+B^2)}\right)\right)\right) \\
 &\qquad+Ee^{-2\tau} \;, \\
\alpha &\sim Ae^{\tau-\frac{C^2}{2(1+B^2)}e^{-\tau}\left(1+2B^2+\frac{1}{2}\sqrt{\frac{13+16B^2}{(7+4B^2)(1+B^2)}}\cos\left(\sqrt{3+4B^2}\tau+2\theta+\arctan\left(\frac{(5+8B^2)\sqrt{3+4B^2}}{17+20B^2}\right)\right)\right) -\frac{E}{2}e^{-2\tau}} \;.
\end{align*}
Using this behaviour of $ r $ and $ \alpha $, we re-evaluate the Yang-Mills variables. The dominant behaviour of $ w $ is unchanged from (\ref{wphi asy}), while the nonlinear terms in (\ref{FYM phi}) change the next-to-leading order behaviour of $ \bar\phi $ and we find
\begin{align*}
\bar\phi &\sim B -BC^2 e^{-\tau}\left(\tau +D \vphantom{\frac{\sqrt{43+59B^2+4B^4}}{2(1+B^2)\sqrt{(3+4B^2)(7+4B^2)}}}\right. \\
 &\left.-\frac{\sqrt{43+59B^2+4B^4}}{2(1+B^2)\sqrt{(3+4B^2)(7+4B^2)}}\cos\left(\sqrt{3+4B^2}\tau +2\theta +\arctan\left(\frac{(19+4B^2)\sqrt{3+4B^2}}{11+20B^2}\right)\right)\right).
\end{align*}

We see also that these solutions have free parameters $ C $, $ \theta $ and $ E $ (and trivially $ A $) as in the magnetic case, plus $ B $ and $ D $. This is codimension 1, which is the correct dimensionality for this boundary.
The asymptotics of $ w = 0 $, constant $ \bar\phi $, constant $ r $ and $ \alpha \sim Ae^\tau $ match a Bertotti-Robinson solution. In particular, the solution which in isothermal coordinates is
\begin{align*}
w &= 0 \;, & \bar{a} &= Q\tan R \;, & r &= \sqrt{1+Q^2} \;, & \alpha &= \frac{\sqrt{1+Q^2}}{\cos R} \;,
\end{align*}
with domain $ -\frac{\pi}{2} < R < \frac{\pi}{2} $ becomes
\begin{align*}
w &= 0 \;, & \bar\phi &= Q\tanh\tau \;, & r &= \sqrt{1+Q^2} \;, & \alpha &= \sqrt{1+Q^2}\cosh\tau \;,
\end{align*}
in semi-isotropic coordinates, where $ \tau \in \mathbb{R} $. Recall that a simple coordinate transformation can rescale $ \alpha $ freely.
Thus the solutions on the boundaries of the long-lived regions shown in figures \ref{fig:stat elec space} and \ref{fig:stat elec BH space} all asymptote to Bertotti-Robinson solutions (other than the points on the boundaries corresponding to the purely magnetic Minkowsi space, BK solitons, Schwarzschild family, or the EYM black holes).

The only way to approach the purely magnetic Bertotti-Robinson solution asymptotically is if $ \bar\phi \equiv 0 $.
For a solution with a regular origin this means $ d = 0 $ and the only limit point of the long-lived solutions on this line (the BK solitons) is $ (b_\infty,0) $. This is asymptotically Bertotti-Robinson, as was found in \cite{BFM94}.
For a solution with a regular horizon this means $ z_h = 0 $ and the limit points of the long-lived solutions in this plane (the EYM black holes) are the black lines in figure \ref{fig:rhwh}. For $ r_h \geqslant 1 $ these correspond to Reissner-Nordstr\"{o}m solutions, while for $ 0 < r_h < 1 $ the solutions are asymptotic to the magnetic Bertotti-Robinson solution, as found in \cite{BFM94}.

\section{Asymptotically flat solutions analytic at spatial infinity} \label{s:AF}
In this section we consider the solutions that have a power series in $ \frac{1}{r} $ at $ r = \infty $.
The polar-areal coordinates are well-suited to describing the asymptotically flat solutions, and we shall use them throughout this section.

The asymptotically flat condition (\ref{BCs inf pa}) tells us $ \lim\limits_{r \to \infty} \bar{a}w = 0 $. We first consider the case $ w(\infty) = 0 $.
Generalising the idea used in a number of papers \cite{GE89,EG90,Zhou92} to the case of $ \bar{a} \not\equiv 0 $, we multiply (\ref{w stat}) by $ w^2-1 $ and integrate by parts to find
\[ (1-w^2)SNw'|_{r_0} = \int_{r_0}^\infty \left(2SNw'^2+S\frac{(w^2-1)^2}{r^2}+\frac{(1-w^2)\bar{a}^2}{SN}\right) w \dx{r} \;, \]
since $ w'(\infty) = 0 $.
Assuming $ w^2 \leqslant 1 $ and $ w $ doesn't change sign, we can take $ r_0 = 0 $ and conclude $ w \equiv 0 $.
If $ w^2 > 1 $ or $ w $ does change sign, we let $ r_0 $ be the largest point where $ w^2 = 1 $ or $ w' = 0 $ and we can conclude $ w \equiv 0 $ on $ [r_0,\infty) $, and therefore everywhere.
The case $ w \equiv 0 $ results in the essentially Abelian potential (\ref{potentialAbel}) with $ k = 1 $, whose solutions with an asymptotically flat region are the Reissner-Nordstr\"{o}m family (\ref{RN pa}).
Thus non-trivial solutions must have $ \bar{a}(\infty) = 0 $.
The most general such power series is
\begin{equation} \label{stat series inf}
\begin{aligned}
w(r) &= \pm\left(1 + \frac{c}{r} + \frac{3c^2-e^2+6cM}{4r^2} + \frac{11c^3-9ce^2+42c^2M-16Me^2+48cM^2}{20r^3} + O\left(\frac{1}{r^4}\right)\right), \\
\bar{a}(r) &= S_\infty e\left(\frac{1}{r^2} + \frac{c+M}{r^3} + \frac{9c^2-e^2+22cM+12M^2}{10r^4} + O\left(\frac{1}{r^5}\right)\right), \\
m(r) &= M - (c^2+e^2)\left(\frac{1}{r^3} + \frac{4c+5M}{2r^4} + \frac{3(37c^2-3e^2+100cM+68M^2)}{40r^5} + O\left(\frac{1}{r^6}\right)\right), \\
S(r) &= S_\infty\left(1 - (c^2+e^2)\left(\frac{1}{2r^4} + \frac{6(c+2M)}{5r^5} + O\left(\frac{1}{r^6}\right)\right)\right).
\end{aligned}
\end{equation}

There are three essential parameters; the ADM mass $ M = m(\infty) $, $ c = \mp r^2 w'(\infty) $, and $ e = r^2\bar{a}(\infty) $. The parameter $ S_\infty = S(\infty) $ again corresponds to the freedom to scale the time coordinate, and is chosen to be one for the purpose of numerical integration.

The Bartnik-McKinnon solutions (equation (\ref{seriesO stat}), table \ref{table:BKs}, and figure \ref{fig:BKs}) are asymptotically flat and we can calculate the parameter values for the expansions at infinity (\ref{stat series inf}), shown in table \ref{table:BKs inf}.
\begin{table}[!ht]
  \centering
  \begin{tabular}{clll}
    \toprule
    $ k $ & $ c_k $ & $ M_k $ & $ S_k $ \\
    \midrule
    0 & 0 & 0 & 1 \\
    1 & $ -0.893 382 587 $ & 0.828 646 982 112 440 562 649 254 & $ 7.909 430 661 212 136 750 839 34 $ \\
    2 & $ -8.863 910 44 $ & 0.971 345 494 316 017 225 547 032 & $ 4.7865 361 169 412 807 645 844 9 \times 10^1 $ \\
    3 & $ -5.89325538 \times 10^1 $ & 0.995 316 472 184 260 542 978 824 & $ 2.94792 934 496 198 456 634 74 \times 10^2 $ \\
    4 & $ -3.66334899 \times 10^2 $ & 0.999 236 192 794 180 794 848 655 & $ 1.810123 348 388 125 387 023 5 \times 10^3 $ \\
    5 & $ -2.251908 \times 10^3 $ & 0.999 875 468 061 823 570 554 977 & $ 1.1 104883 863 971 957 913 59 \times 10^4 $ \\
    6 & $ -1.38174758 \times 10^4 $ & 0.999 979 696 967 825 527 674 876 & $ 6.8 116258 317 884 713 152 \times 10^4 $  \\
    7 & $ -8.4757283 \times 10^4 $ & 0.999 996 689 920 921 630 940 935 & $ 4.17 807354 324 768 064 12 \times 10^5 $ \\
    8 & $ -5.198813 \times 10^5 $ & 0.999 999 460 346 001 720 942 571 & $ 2.562 710 220 964 698 593 \times 10^6 $ \\
    9 & $ -3.188 804 7 \times 10^6 $ & 0.999 999 912 018 298 273 648 824 & $ 1.571 891 681 994 513 832 \times 10^7 $ \\
    10 & $ -1.955 920 0 \times 10^7 $ & 0.999 999 985 656 032 093 345 011 & $ 9.641 523 867 412 406 2 \times 10^7 $ \\
    11 & $ -1.199 704 \times 10^8 $ & 0.999 999 997 661 452 206 828 142 & $ 5.913 828 656 871 30 \times 10^8 $ \\ 
    12 & $ -7.358 634 \times 10^8 $ & 0.999 999 999 618 738 286 877 288 & $ 3.627 369 465 343 980 \times 10^9 $ \\
    13 & $ -4.513 571 \times 10^9 $ & 0.999 999 999 937 841 555 216 072 & $ 2.224 922 295 689 314 \times 10^{10} $ \\
    14 & $ -2.768 49 \times 10^{10} $ & 0.999 999 999 989 866 089 027 827 & $ 1.364 702 236 358 54 \times 10^{11} $ \\
    15 & $ -1.698 11 \times 10^{11} $ & 0.999 999 999 998 347 832 672 638 & $ 8.370 684 214 481 \times 10^{11} $ \\
    16 & $ -1.041 57 \times 10^{12} $ & 0.999 999 999 999 730 641 320 503 & $ 5.134 332 776 16 \times 10^{12} $ \\
    17 & $ -6.388 \times 10^{12} $ & 0.999 999 999 999 956 085 502 346 & $ 3.149 249 497 524 \times 10^{13} $ \\
    18 & $ -3.918 6 \times 10^{13} $ & 0.999 999 999 999 992 840 464 217 & $ 1.931 657 496 70 \times 10^{14} $ \\
    19 & $ -2.403 \times 10^{14} $ & 0.999 999 999 999 998 832 755 573 & $ 1.184 822 189 3 \times 10^{15} $ \\
    20 & $ -1.474 \times 10^{15} $ & 0.999 999 999 999 999 809 700 015 & $ 7.267 352 63 \times 10^{15} $ \\
    \bottomrule
  \end{tabular}
  \caption{The calculated parameters at infinity for the first 20 Bartnik-McKinnon solutions. The decimal expansions are truncated, not rounded.}
  \label{table:BKs inf}
\end{table}
The asymptotic parameters were found by matching the numerical Bartnik-McKinnon solutions at large $ r $ to the asymptotic expansions (\ref{stat series inf}). The precision was found by doing this for two values of $ b $ that bracket the true solution, and only taking the digits of the answers that agree.

We can confirm the asymptotic behaviour of $ M_k $ and $ c_k $ found in \cite{BFM94} with extra precision, as well as the similar behaviour of $ S_k $:
\begin{align*}
c_k &= -0.2595 e^{1.81379 k} \;, & M_k &= 1-1.08118461 e^{-1.813799364 k} \;, & S_k &= 1.2791313 e^{1.8137 k} \;.
\end{align*}
The coefficient of the exponents again are completely consistent with the expected $ -\frac{\pi}{\sqrt{3}} $.

\begin{figure}[!ht]
  \centering
  \includegraphics[scale=0.75]{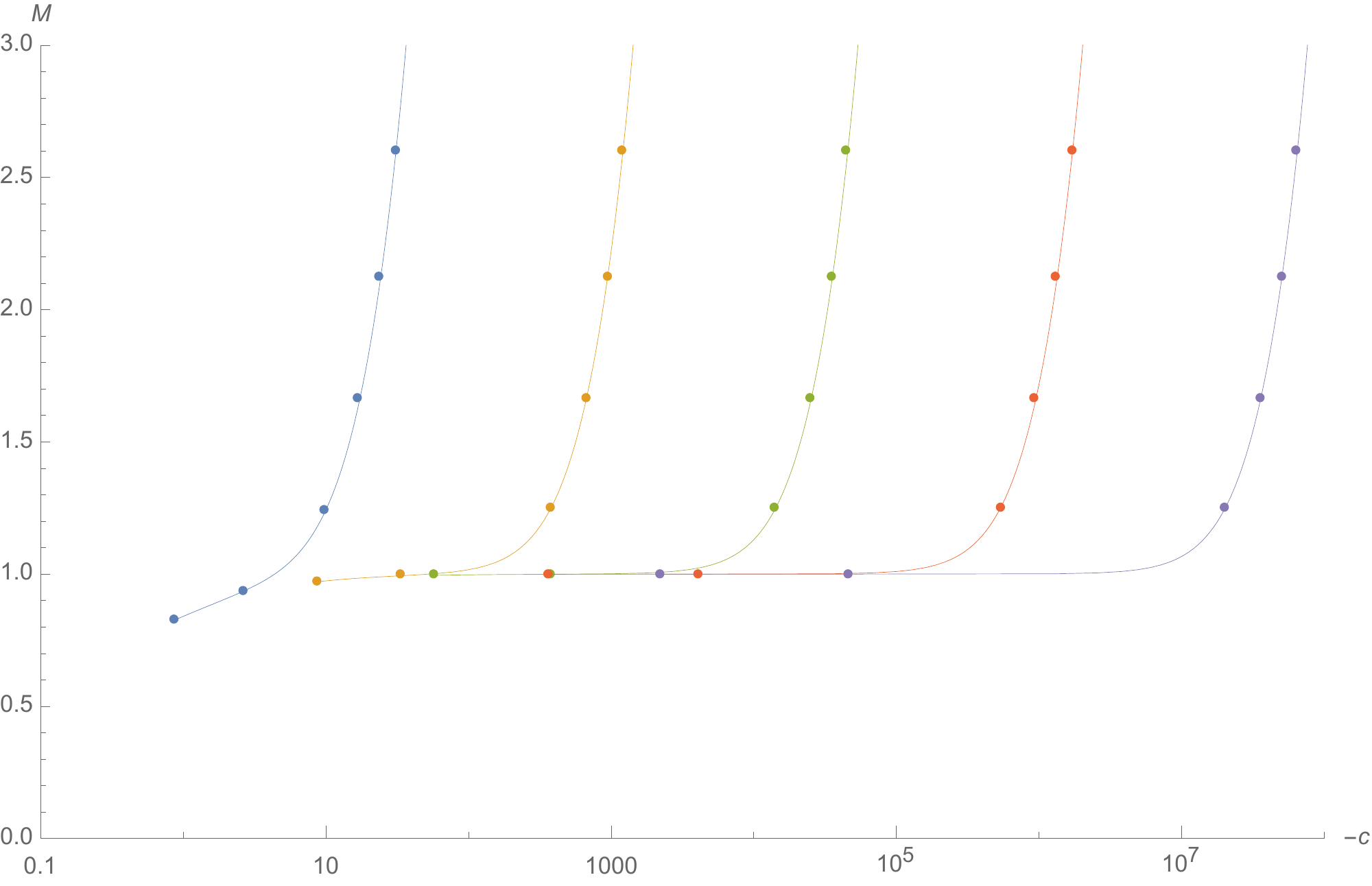}
  \caption{A plot of the generic static, asymptotically flat parameter space $ (c,M,e=0) $, with the first five Bartnik-McKinnon and EYM black hole solutions marked (blue, yellow, green, orange, purple respectively). The solutions with horizon radii 0 (the BK solitons), 1, 2, 3, 4, and 5 are marked with dots.}
  \label{fig:AF cMe}
\end{figure}

The EYM black holes are also asymptotically flat, and hence can be matched to the expansions (\ref{stat series inf}) at large $ r $. They and the Bartnik-McKinnon solutions all occur in the $ e = 0 $ plane and this is shown in figure \ref{fig:AF cMe}, which presents a precise confirmation of the schematic figure 8 of \cite{SW98}.
We see in this figure that the black hole mass remains less than one for all solutions with horizon radius less than or equal to 1, while for larger horizon radii it appears to grow linearly with $ r_h $.
The parameter $ c $ grows exponentially with $ k $ (the number of zero crossings), but transitions from a growth factor of $ \frac{\pi}{\sqrt{3}} $ at $ r_h = 0 $ to $ \frac{2\pi}{\sqrt{3}} $ for $ r_h \gtrsim 2 $.
We also see that the solutions with small horizon radius and large $ k $ are very close together around the $ M = 1 $ line.

We observe that the black hole mass is well-approximated by
\begin{equation*}
M \approx \begin{cases} 1 & \text{if } r_h \leqslant 1 \\ \frac{1}{2}\left(r_h+\frac{1}{r_h}\right) & \text{if } r_h > 1 \end{cases} \;,
\end{equation*}
and that these values are approached as $ k $ increases.
This was proved by Smoller and Wasserman \cite{SW96}.
This is consistent with our observation in section \ref{s:EH:m} that in the limit $ r_h \to \infty $, the ADM mass $ M $ approaches $ \frac{r_h}{2} $.
The limit for $ r_h > 1 $ corresponds to the magnetic Reissner-Nordstr\"{o}m solution with horizon radius $ r_h $, and the $ r_h = 1 $ limit is the extremal magnetic Reissner-Nordstr\"{o}m solution.

Not shown on the logarithmic plot (\ref{fig:AF cMe}) is the line representing the Schwarzschild solutions on $ c = 0 $. It extends for all $ M > 0 $ and terminates at $ M = 0 $, Minkowski space. The black hole mass of course corresponds to $ \frac{r_h}{2} $. The EYM black holes can be seen as generalisations of this behaviour, with the Bartnik-McKinnon solutions generalising the Minkowski space terminus. For each successive $ k $ the EYM solutions become less stable.

As was shown by Smoller and Wasserman \cite{SW95,SW97,SW98,Wasserman00}, a purely magnetic, asymptotically flat solution to (\ref{EYME stat}) exists for all $ r > 0 $, and is either a Bartnik-McKinnon solution (figure \ref{fig:BKs}), a black hole solution (figure \ref{fig:BHs}), or a Reissner-Nordstr\"{o}m-like (RNL) solution (figure \ref{fig:RNLs}).
Thus the remaining magnetic asymptotically flat solutions all have a naked singularity; they are Reissner-Nordstr\"{o}m-like (including negative-mass Schwarzschild for $ c = 0, M < 0 $).

We observe numerically that all asymptotically flat solutions with $ e \neq 0 $ extend to $ r = 0 $ and have a Reissner-Nordstr\"{o}m-like naked singularity.
Due to the number of parameters this RNL behaviour is expected to be generic, and this is what we have found. Note that, if as expected the long-lived solutions indeed extend to $ r = \infty $, then asymptotic flatness is not the generic behaviour for solutions defined in the far field when $ \bar{a} \not\equiv 0 $.

\section{Summary}
In this chapter we have added to the known results concerning the static spherically symmetric SU(2) EYM equations. Numerically integrating the equations in the general case has shown new behaviour, in particular long-lived solutions that appear to oscillate indefinitely as the areal radius goes to infinity.
We have determined the asymptotics of such solutions to the SU(2) Yang-Mills equations on an arbitrary background, and have obtained a corresponding partial explanation of the asymptotics for the coupled EYM system.
We have numerically found the boundary of the long-lived region in the parameter spaces of solutions with a regular origin and those with a regular horizon. We found that the boundary solutions that are not purely magnetic asymptote to Bertotti-Robinson spacetimes.
Considering solutions with a singular origin we found no essentially new series expansions at $ r = 0 $; the three known series for purely magnetic solutions all generalise.
We found numerically that solutions that are asymptotically flat and analytic at spatial infinity generically have a Reissner-Nordstr\"{o}m-like singularity.
We have also re-affirmed previous results in the magnetic sector with additional precision, in particular using arbitrary-precision arithmetic to find the Bartnik-McKinnon solutions with a large (greater than ten) number of zero crossings.
In this section we collect our results concerning the static equations to form an overview of the solutions to the static spherically symmetric SU(2) Einstein-Yang-Mills equations.

We have seen that the only generic points are the Ressiner-Nordstr\"{o}m-like singular origins and the equators.
Numerically, this is also what we see when generic initial data is evolved with (\ref{EYME stat TR}); an RNL singularity with decreasing $ R $, and an equator followed by an RNL singularity with increasing $ R $.

Given either a regular origin or a regular horizon, the solutions follow similar behaviour to the magnetic case \cite{BFM94}.
That is, there are three behaviours that occur depending on the initial parameters. Two behaviours occur generically; the solution either forms an equator and returns to $ r = 0 $ with an RNL singularity, or extends apparently indefinitely with infinite oscillations and fails to become asymptotically flat. When the electric field is zero the long-lived solutions exist for all $ r $, are no longer generic, and become asymptotically flat. These are the Bartnik-McKinnon \cite{BK88} and the black hole \cite{VG89,KM90,Bizon90} solutions.
Occurring on the boundary between these two generic regions, the third behaviour is that the solution approaches a Bertotti-Robinson solution, with $ w $ exhibiting infinite oscillations while decaying to zero, and $ \bar{a} $, if non-zero, diverging.

Solutions that are analytic and asymptotically flat behave similarly to the magnetic case, again with three possibilities as $ r $ decreases, depending on the initial parameters. Two cases; the solution continuing to a regular origin or a regular horizon, only occur when the electric field is zero. The third is generic; an RNL singularity.

With the above results we can clearly see what effect the gravitational field has on solutions to the (flat-space) SU(2) Yang-Mills equations (\ref{YM flat bg}).
In the purely magnetic case $ \bar\phi \equiv 0 $, the solution space is described in \cite{BFM94}.
If we assume the solutions are regular at $ r = 0 $ (so that $ w \to \pm 1 $ as $ \tau \to -\infty $), then there are the trivial solutions $ w = \pm 1 $ and solutions such that $ w \to \pm\infty $ in finite $ \tau $.
There are analogous solutions on the Schwarzschild background (\ref{w Schw bg}), as well as additional solutions that are characterised by the number of zero crossings \cite{Bizon94}.
In the EYM system, $ w \equiv \pm 1 $ gives the Schwarzschild family of solutions with one parameter, the ADM mass $ M > 0 $, and Minkowski spacetime for $ M = 0 $.
As on the Schwarzschild background, there remain solutions with $ w $ going from $ \pm 1 $ to $ \pm 1 $ crossing zero an integer number of times; the Bartnik-McKinnon and EYM black hole solutions. Again, for each integer, these are a one-parameter family depending on $ r_h $.
The solutions with $ w \to \pm\infty $ are regularised by the gravitational field; as $ |w| $ and $ |w_R| $ increase, the Misner-Sharp mass $ m $ also increases (\ref{mR}) until an equator forms, after which $ r $ decreases to zero and $ w $ remains bounded.
There is an additional solution in the EYM system; the magnetic Bertotti-Robinson solution.

In the flat-space Yang-Mills system, there are also solutions with $ w \to \pm\infty $ for finite decreasing $ \tau $ (that, is the solution becomes singular before $ r = 0 $ is reached from above).
Again, the gravitational field seems to create more regular solutions as in the EYM system there are no solutions that cannot be extended down to $ r = 0 $, nor solutions with unbounded $ w $.

The full system with non-zero electric field is perhaps simpler.
The solutions to (\ref{YM flat bg}) with $ \bar\phi \not\equiv 0 $ come in two families.
If $ w \equiv 0 $, $ \bar\phi = Ae^\tau+B $ and so $ \bar{a} = A + \frac{B}{r} $ for some constants $ A $ and $ B $.
On the Schwarzschild background (\ref{YM Schwarzschild bg}) the corresponding solution is $ w \equiv 0 $ and $ \bar\phi = A\frac{\cosh^3\frac{\tau}{2}}{\sinh\frac{\tau}{2}} +B\frac{\cosh\frac{\tau}{2}}{\sinh\frac{\tau}{2}} $, $ \bar{a} = \frac{A}{r_h}+\frac{B}{r} $, and in the EYM system we have the Reissner-Nordstr\"{o}m family of solutions with $ \bar{a} = \frac{A}{r_h} + \frac{B}{r} $, which again have an additional parameter of the ADM mass $ M $.
There is also the Bertotti-Robinson family with $ r = \text{constant} $ and $ \bar{a} = A+\frac{B}{R} $, this time with no additional parameter.

In the generic case where $ w $ is not identically zero, the solutions to (\ref{YM flat bg}) have $ w $ oscillating and $ |\bar\phi| $ increasing indefinitely. There are corresponding solutions on the Schwarzschild background, and again this is what we see in the full EYM case.
Since $ w $ is already bounded and the solutions already exist for all $ \tau $, we do not see entirely new behaviour in the EYM system.
However, we do see that for some parameter choices the mass $ m $ increases too quickly (\ref{mR}) so that the solutions do not extend to $ r = \infty $, but rather an equator forms after a typically small number of oscillations before the areal radius $ r $ returns to zero.
Note that in figures (\ref{fig:stat elec space}) and (\ref{fig:stat elec BH space}), the long-lived region remains close to the asymptotically flat solutions, and it is as the parameters increase away from these that the solution forms an equator.

\chapter{The code to solve the dynamic equations} \label{ch:code}
In this chapter we describe in detail the code we developed to numerically solve the spherically symmetric Einstein-Yang-Mills equations.
There are a few key challenges such a code must meet.
The assumption of spherical symmetry reduces the dimensionality of the problem as well as significantly reducing the number of variables, but the coordinate systems that take full advantage of this simplification will be singular at the centre of symmetry, the line $ r = 0 $. This coordinate singularity will result in expressions that are formally singular, and any numerical code will have to treat this region extra carefully to avoid small numerical errors from later blowing up there.
The natural domain of the problem is infinite in both space and time, and it is desirable to capture as much of this region as possible. We also wish to avoid introducing an artificial spatial boundary due to the difficulty of choosing boundary conditions there that will not interfere with the evolution in the interior.
As with any numerical code, it needs to accurately approximate a true solution of the equations and converge to the corresponding true solution as the resolution is increased (we only consider finite difference schemes). We wish to make the code as efficient as possible due to the desire to run a large number of evolutions to explore the solution space.
We know that the EYM equations can produce self-similar behaviour that shows variation on vastly different length (space and time) scales. Before evolving it is not clear where the extra resolution will be required thus we choose to use some kind of adaptive mesh refinement (AMR).

We choose to use double-null coordinates for this investigation for a number of reasons.
For one, the equations themselves are relatively simple when written in double-null coordinates (compare (\ref{EEuv},\ref{YMEuv}) with (\ref{EEtr},\ref{YMEtr})).
For another, we desire a coordinate system that can penetrate into the interior of any black hole that forms. Polar-areal coordinates, while extremely useful and relatively simple for regular solutions, cannot cross the boundary of a black hole (a marginally trapped tube) because they become singular when $ N = 0 $. Other choices of the time-slicing can penetrate inside an MTT (such as the maximal slicing used in \cite{CHM99} and the hyperboloidal slicing used in \cite{RM13}) however these give up some of the simplicity in the equations.
An MTT is indicated in double-null coordinates simply by $ f < 0 $ and $ g = 0 $ (as defined in (\ref{dn first})).
While it is somewhat natural to specify initial conditions on a spacelike slice of constant time and evolve to future time-slices, the long-time behaviour of the system often involves radiation moving out to infinity, and this method involves using significant computer resources tracking the outgoing waves.
One way to address this problem is to introduce an artificial outer boundary, with boundary conditions that attempt to only allow outgoing radiation through, but with nonlinear equations it does not seem possible to practically reduce the reflection completely.
One can use a compactified spatial coordinate to avoid an outer boundary and reach spatial infinity, however this inevitably results in a decrease in resolution of the outgoing waves and hence spurious effects.

In contrast, double-null coordinates will automatically follow outgoing radiation, and as the outgoing waveform settles down, less and less resources are required to describe the waveform because the relevant functions approach constants in these coordinates.
Furthermore, no outer boundary is required because compactifying the double-null coordinates in a natural way brings null infinity to a finite coordinate value, and hence the complete domain of dependence of the initial data can be found with finite resources.
As an added benefit, the functions can be determined actually on null infinity, and then interpreted physically as the radiation that could be observed from an isolated object.
The most natural way to set ``initial" conditions in this case is to specify appropriate data on an outgoing null ray -- this is then known as a characteristic initial value problem. While perhaps less intuitive this specification presents no practical problem and in fact we will see (section \ref{s:IBCs}) that satisfying the constraints is even simpler than in the time-slicing case.
We also mention that hyperboloidal slices, which are spacelike and approach future null infinity, share many of the same advantages at the expense of slightly more complicated equations.

Given a numerical solution in double-null coordinates, which naturally cover the domain of dependence of the initial data, we are able to transform the solution into any other desired coordinate system relatively simply. The details for numerically transforming into polar-areal or isothermal coordinates are given in appendix \ref{sec:A:nct}.

For these reasons we believe -- at least in the spherically symmetric case -- double-null coordinates to be the simplest and best option available for numerical evolution.

\section{Hamad\'{e} and Stewart's method} \label{s:HS}
There have been a number of codes written to solve Einstein's equations in spherical symmetry using double-null coordinates, for example \cite{HS96,PL04,Thornburg11}. Each of them used a version of adaptive mesh refinement based on Berger and Oliger's work \cite{BO84}, as we will also.
Pretorius and Lehner \cite{PL04} preferred to use a single null coordinate with the areal coordinate $ (v,r) $. Thornburg's code \cite{Thornburg11} did not evolve near the $ r = 0 $ singularity.
Another comparable code was shown in \cite{PHA05}, in this case with Bondi coordinates $ (u,r) $, which would not penetrate an event horizon. They also use a simpler form of mesh refinement tailored to the critical collapse of the massless scalar field.
We base our method on Hamad\'{e} and Stewart's, so we describe this as it is presented in \cite{HS96} in some detail here in order to highlight the changes we make. 

Hamad\'{e} and Stewart were concerned with the closely related Einstein-scalar field system, which also exhibits self-similar critical behaviour and is the first system where this behaviour was observed \cite{Choptuik93}. It is somewhat simpler than the EYM system in that there is only one matter variable (the massless scalar field $ \phi $) and there are no globally regular or regular black hole static solutions to the equations, meaning only ``type II" critical behaviour (see section \ref{s:critical}) is present.

With the equations written in double-null coordinates, Hamad\'{e} and Stewart introduced new variables for all the $ u $ and $ v $ derivatives of the metric and scalar functions (in our notation $ \alpha $, $ r $, and $ \phi $). This resulted in 9 unknown functions and 14 first-order equations. One function was discarded as it played no part in quantities of geometrical significance or the equations of those that did. Two of the functions only had evolution equations with $ u $ derivatives, while the remaining 6 were chosen to be evolved with their $ v $-derivative equations.
Thus $ u $ and $ v $ were treated as the evolving and constraining directions respectively, analogous to time and space in time-slicing evolution.
Note that this is in harmony with the assertion in \cite{GP97} that a stable numerical code for the spherically symmetric Einstein equations in double-null coordinates requires using a maximum number of equations containing only $ v $-derivatives (is ``fully constrained").

Their equations were written schematically as
\begin{align} \label{dne schem}
y_u &= F(y,z) \;, & z_v &=G(y,z) \;.
\end{align}
where $ y $ denotes two variables and $ z $ six.

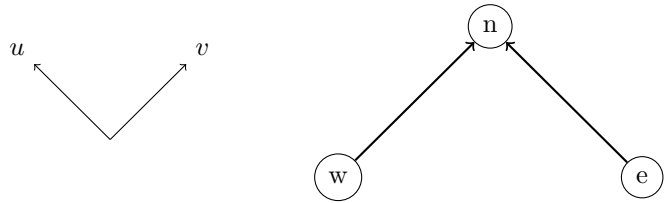
\begin{figure}[!ht]
  \centering
  \begin{tikzpicture}
    \draw[->] (0,0)--(-1,1) node[above left]{$ u $};
    \draw[->] (0,0)--(1,1) node[above right]{$ v $};
    \node[shape=circle,draw] (w) at (3,-0.5) {w};
    \node[shape=circle,draw] (n) at (5,1.5) {n};
    \node[shape=circle,draw] (e) at (7,-0.5) {e};
    \draw[->,thick] (w) -- (n);
    \draw[->,thick] (e) -- (n);
  \end{tikzpicture}
  \caption{The basic computational cell for the Hamad\'{e} and Stewart algorithm (based on \cite{HS96} figure 1).}
  \label{fig:wne}
\end{figure}

Their stated algorithm on a basic computational cell (figure \ref{fig:wne}) proceeded via a predictor-corrector method as follows.
An Euler step in the $ u $ direction followed by a trapezoidal integration step in the $ v $ direction gave the predicted values
\begin{subequations} \label{HS evolution}
\begin{align}
\hat{y}_n &= y_e +hF(y_e,z_e) \;,\\
\hat{z}_n &= z_w +\frac{h}{2}\left(G(y_w,z_w)+G(\hat{y}_n,\hat{z}_n)\right) \;, \label{HS2}
\end{align}
where $ h $ is both the change in $ u $ from e to n and the change in $ v $ from w to n.
An important part of their method that we will preserve is that the new variables and the order of the $ v $ integration can be chosen so that each $ v $ integration is linear in the unknown quantity. In this way this apparently implicit step (\ref{HS2}) can be made explicit.
This is why $ q $ and $ y $ are chosen as shown in (\ref{dn first}).
The corrected values are in both cases achieved by averaging the predicted values with the result of a backwards Euler step, which effectively gives a modified Euler method for $ y $.
\begin{align}
y_n &= \frac{1}{2}\left(\hat{y}_n +y_e+hF(\hat{y}_n,\hat{z}_n)\right) \;, \\
z_n &= \frac{1}{2}\left(\hat{z}_n +z_w+hG(\hat{y}_n,\hat{z}_n)\right) \;. \label{HS4}
\end{align}

The above computation occurred on every cell in a grid determined by an adaptive mesh refinement method based on the Berger and Oliger algorithm \cite{BO84}.
Adaptive refinement of the mesh was useful for this problem because regions typically evolved where the solution would change much more rapidly than in other regions.
This behaviour was particularly strong when evolving a near-critical solution that exhibits self-similarity; where the same structure is repeated on finer and finer scales. The adaptive mesh then would place grid points only where they were needed, allowing the majority of the spacetime to be evolved efficiently with few points. The method proceeds without knowledge of where refinement will be needed, which is typically not known from the initial conditions.
We first describe briefly the central ideas of the original Berger and Oliger algorithm.

The algorithm was designed to solve hyperbolic partial differential equations with one time and two spatial dimensions. Let the unknown variable be $ u(x,t) $ where $ x $ has two components $ x_i $. The method began by numerically evolving on a grid with spacings $ \Delta t = k $ and $ \Delta x_i = h $ using an explicit difference method $ Q $ of order $ q $ in time and space.
Using the assumption of a smooth solution, the local truncation error is $ u(x,t+k)-Qu(x,t) = \tau + k\:O(k^{q+1}+h^{q+1}) $. The leading term $ \tau = O(k^{q+1}+h^{q+1}) $ could be approximated by repeating the evolution with $ Q_{2h} $ acting on a grid with double the spacings in space and time. Comparing the result with two evolutions by $ Q $ gave the following estimation of the local truncation error:
\[ \frac{Q^2 u(x,t) -Q_{2h}u(x,t)}{2^{q+1}-2} = \tau +O(h^{q+2}) \;. \]

Where the estimated local truncation error was above a given tolerance a refined (``child") grid was introduced, where the spacings of the new grid were decreased by a predetermined integral factor (Berger and Oliger preferred four) in both space and time, which means any satisfied CFL (Courant-Friedrichs-Lewy) condition is maintained.
For the 2+1 dimensional problem Berger and Oliger introduced multiple rotated rectangles to cover the region requiring refinement, while in the 1+1 case ($ x $ has only one component), no orientation was required for the single spatial dimension. In either case a buffer zone was added either side of the points needing refinement, so that any fine structure would be captured on that child grid for some time into the future.
An example mesh for one spatial dimension is shown in figure \ref{fig:mesh eg}.

\begin{figure}[!ht]
  \centering
  \begin{tikzpicture}
  \draw[->] (0,0)--(1,0) node[right]{$ x $};
  \draw[->] (0,0)--(0,1)node[above]{$ t $};
  \foreach \x in {5.5,5.5625,...,6.75}{
    \draw[MMA2] (\x,0)--(\x,0.5);
    \draw[MMA2] (\x+0.25,0.5)--(\x+0.25,0.75);
    \draw[MMA2] (\x+0.5,0.75)--(\x+0.5,1);
    \draw[MMA2] (\x+0.75,1)--(\x+0.75,1.25);
    \draw[MMA2] (\x+1,1.25)--(\x+1,1.75);
    \draw[MMA2] (\x+1.25,1.75)--(\x+1.25,2);
  }
  \foreach \t in {0,0.0625,...,0.5}{
    \draw[MMA2] (5.5,\t)--(6.75,\t);
    \draw[MMA2] (5.5+1,\t+1.25)--(6.75+1,\t+1.25);
  }
  \foreach \t in {0,0.0625,...,0.25}{
    \draw[MMA2] (5.5+0.25,\t+0.5)--(6.75+0.25,\t+0.5);
    \draw[MMA2] (5.5+0.5,\t+0.75)--(6.75+0.5,\t+0.75);
    \draw[MMA2] (5.5+0.75,\t+1)--(6.75+0.75,\t+1);
    \draw[MMA2] (5.5+1.25,\t+1.75)--(6.75+1.25,\t+1.75);
  }
  \foreach \x in {4,4.25,...,8}{
    \draw[MMA1] (\x,0)--(\x,1);
    \draw[MMA1] (\x+1,1)--(\x+1,2);
  }
  \foreach \t in {0,0.25,...,1}{
    \draw[MMA1] (4,\t)--(8,\t);
    \draw[MMA1] (5,\t+1)--(9,\t+1);
  }
  \foreach \x in {2,3,...,10}
    \draw (\x,0)--(\x,2);
  \foreach \t in {0,1,2}
    \draw (2,\t)--(10,\t);
  \end{tikzpicture}
  \caption{An example adaptive mesh in 1+1 dimensions, with refinement ratio four.}
  \label{fig:mesh eg}
\end{figure}
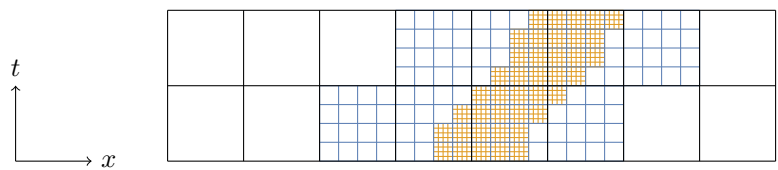

The boundary values for the newly created refined grid were set via interpolation from the coarser (``parent") grid the new grid was contained in. In the 2+1 case this involved interpolation in both time and space, while in the 1+1 case only interpolation in time was required.
As the child grid was evolved to the final time of the parent grid, the points which coincided with points in the parent grid were ``injected" onto the coarse grid.
The above evolution and refining processes were implemented recursively, so that as many refinement levels were introduced as was necessary to accurately follow the solution.

We now note how the Berger and Oliger algorithm was adapted for use by Hamad\'{e} and Stewart. With only two independent variables, they adapted the simpler 1+1 algorithm, with $ u $ treated as the time coordinate and $ v $ as the spatial coordinate. Since these are null coordinates, the required boundary conditions changed slightly. When a child grid was introduced in the range $ v_0 \leqslant v \leqslant v_1 $ say, boundary values were only required on the $ v_0 $ side rather than on both sides. Correspondingly, the parent level was no longer accurate for $ v > v_1 $, and thus should be updated via additional integration in the $ v $ direction as the child grid evolves. This was done by Hamad\'{e} and Stewart, but only justified as a solution to extraneous noise.
Rather than introduce an arbitrary number of refined child grids at a particular level, Hamad\'{e} and Stewart restrict to allow only one child grid for each parent, which they found to give identical results to the full Berger and Oliger algorithm. They also used a mesh refinement ratio of four.

\section{Ensuring second-order accuracy}
We have found it necessary to adjust the Hamad\'{e} and Stewart algorithm for our purposes, so we now describe the issues encountered and the changes made to address them.
Firstly, we analyse the behaviour of the evolution scheme (\ref{HS evolution}) for the equations (\ref{dne schem}). We expect a (global) second-order scheme, and thus a local truncation error of order $ h^3 $.
Considering $ z $ first, we insert a power series for $ z $ about the point $ \text{w} = (u,v) $ and find that the truncation error for the predicted $ \hat{z}_n $ is
\[ z(u,v+h)-\hat{z}_n = -\frac{h^3}{12}z_{vvv} +O(h^4) \;, \]
where $ z_{vvv} $ is evaluated at an unknown point between the points w and n.
Obviously the predictor step for $ z_n $ is already second-order. According to \cite{HS96} it was then averaged with a first-order backwards Euler step, which makes the resulting $ v $ integration only first-order. This seems to be an oversight, and we find it better to repeat the trapezoidal step with the corrected value of $ y_n $ to retain the $ h^3 $ accuracy in the truncation error, so using
\begin{equation} \label{zcorr}
z_n = z_w +\frac{h}{2}\left(G(y_w,z_w)+G(y_n,z_n)\right)
\end{equation}
\end{subequations}
instead of (\ref{HS4}).

Now turning our consideration to $ y $, we see that since the truncation error for $ \hat{z}_n $ is $ O(h^3) $, the error in the prediction for $ z $ does not affect the calculation for the truncation error for $ y_n $. Inserting a power series for $ y $ about $ \text{e} = (u,v) $ we find
\[ y(u+h,v)-y_n = -\frac{h^3}{12}\left(y_{uuu}-3F_y y_{uu}\right) + O(h^4) \;. \]
The above is indeed $ O(h^3) $ if the coefficient of $ h^3 $ is bounded, but for our system of equations (\ref{F4}), $ F_y $ is $ O\mathopen{}\left(\frac{1}{r}\right)\mathclose{} $ as $ r \to 0 $ (while $ y $ and its derivatives are well-behaved). Since $ r = O(h) $ for points near the origin line, this effectively decreases the order of the method by one, and we observe this in numeric results using this method.
In fact, the system considered in \cite{HS96} also had $ F_y = O\mathopen{}\left(\frac{1}{r}\right)\mathclose{} $ (see equation F8 in the system (2.6)) but this issue was not addressed there.

We proceed by making the predictor second-order. We might try a Taylor's method that makes use of $ y_{uu} $ at the point $ e $, and this worked well in the purely magnetic case. In the full case however, $ y_{uu} $ requires an explicit equation for $ a_u $ which is not included in the system of equations (\ref{F4},\ref{G11},\ref{H7}).
So we instead use the explicit multistep Adams-Bashforth method (\ref{my evolution y}), which uses data from a previous point $ ee $ to calculate $ y_n $.:
\begin{subequations} \label{my evolution}
\begin{align}
y_n &= y_e +h\left(\left(1+\frac{h}{2H}\right)F(y_e,z_e)-\frac{h}{2H}F(y_{ee},z_{ee})\right), \label{my evolution y} \\
z_n &= z_w +\frac{h}{2}\left(G(y_w,z_w)+G(y_n,z_n)\right) \;. \label{my evolution z}
\end{align}
Due to the adaptive nature of the algorithm, the distance $ H $ from ee to e will not be $ h $ in general (figure \ref{fig:wnee}), so we present the method for an arbitrary previous step size $ H $.

\begin{figure}[!ht]
  \centering
  \begin{tikzpicture}
    \draw[->] (0,0)--(-1,1) node[above left]{$ u $};
    \draw[->] (0,0)--(1,1) node[above right]{$ v $};
    \node[shape=circle,draw] (w) at (3,-0.5) {w};
    \node[shape=circle,draw] (n) at (5,1.5) {n};
    \node[shape=circle,draw] (e) at (7,-0.5) {e};
    \node[shape=circle,draw] (ee) at (8,-1.5) {ee};
    \draw[->,thick] (w) -- (n);
    \draw[->,thick] (e) -- (n);
    \draw[->,thick] (ee) to[bend left=20] (n);
    \draw[|<->|] ($(n)+(0.5,0.5)$) --node[above right]{$ h $} ($(e)+(0.5,0.5)$);
    \draw[|<->|] ($(e)+(0.5,0.5)$) --node[above right]{$ H $} ($(ee)+(0.5,0.5)$);
  \end{tikzpicture}
  \caption{The basic computational cell for our algorithm.}
  \label{fig:wnee}
\end{figure}
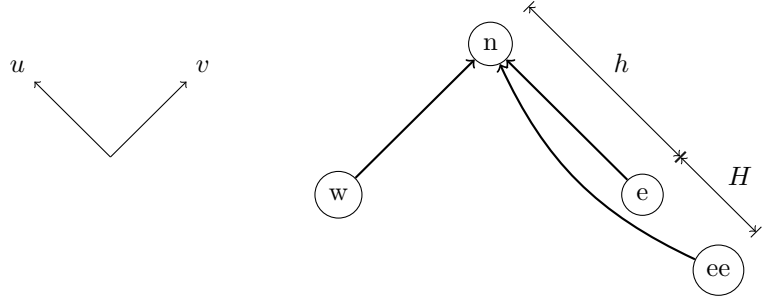

The truncation error for the $ y_n $ of (\ref{my evolution y}) is
\[ y(u+h,v)-y_n = \left(\frac{h^3}{6}+\frac{h^2H}{4}\right)y_{uuu} +O(h^4) \;. \]
We are able to order the $ v $ integration (see (\ref{G11})) so that (\ref{my evolution z}) is linear in the unknown quantity and can be made explicit, so we have then an explicit second-order scheme, and find that correction to these values are no longer necessary for either accuracy or stability. Thus instead of (\ref{HS evolution}) we will use (\ref{my evolution}).

Since we use a two-step method in the $ u $ direction, we require a special method to evolve the second row of constant $ u $. For this row, and only this row, we use the modified Euler method as Hamad\'{e} and Stewart do:
\begin{equation} \label{my evolution y1}
\begin{aligned}
\hat{y}_n &= y_e +hF(y_e,z_e) \;,\\
y_n &= \frac{1}{2}\left(\hat{y}_n +y_e+hF(\hat{y}_n,\hat{z}_n)\right) \;.
\end{aligned}
\end{equation}
\end{subequations}
For the $ v $ direction, we use the trapezoidal method for the prediction and the correction; (\ref{HS2}) and (\ref{zcorr}).
Since (\ref{my evolution y1}) is only used once, the error due to the $ F_y $ term in the truncation error does not affect the global order of the method.

\section{Ensuring regularity at the origin and null infinity} \label{s:final variables}
Having determined a second-order accurate method for solving a system of first-order equations in double-null coordinates, we now consider how to apply this to equations (\ref{F4},\ref{G11}). The equations cannot be immediately used due to the appearances of $ r $ in the denominators of the quotients and the fact that many of the variables go to infinity at future null infinity (for example $ r $ and $ g $). We here detail how to change variables to resolve these problems.

Our assumption of regularity at the origin implies that all of the quotients containing $ r $ in the denominator remain finite at the origin, and can indeed be evaluated at $ r = 0 $ with the use of l'H\^{o}pital's rule. We will explicitly do so once we have our final variables.
We will use the remaining coordinate freedom to bring future null infinity to the finite coordinate value $ v = 1 $.
To proceed we assume that our solutions are not only asymptotically flat, but also that they are analytic in $ \frac{1}{r} $ and $ \frac{1}{1-v} $ near null infinity.

We use up the remaining coordinate freedom of the double-null coordinates by the specification of $ \gamma = \frac{\alpha_v}{\alpha} $ on the initial outgoing null ray $ u = 0 $, as well as ensuring the (regular) origin remains on the line $ u = v $. We choose
\[ \gamma(0,v) = \frac{1}{1-v} \;, \]
so that future null infinity occurs at $ v = 1 $.

Let $ \gamma(u_0,v) = \frac{1}{1-v} +O(1-v) $ for some $ u_0 $ (the order of the second term will be justified shortly). We will determine the behaviour of the remaining metric functions as $ v \to 1 $.
Solving (\ref{alphav}) we find $ \alpha(u_0,v) = \frac{\alpha_1}{1-v} +O(1-v) $, where the subscript $ 1 $ on $ \alpha $ and subsequent variables indicates a constant value.
Writing (\ref{rv},\ref{gv}) as a single equation
\begin{equation} \label{rvv1}
r_{vv} = 2\gamma r_v -2(q^2+y^2)r \;,
\end{equation}
we use (\ref{BCs inf dn}) and assume for the moment that $ q^2+y^2 $ is analytic at $ v = 1 $.
Equation (\ref{rvv1}) is then a linear homogeneous ordinary differential equation with a regular singular point at $ v = 1 $.
Solving by the Frobenius method we find $ r = \frac{r_1}{1-v} +O(1) $ and $ g = \frac{r_1}{(1-v)^2} +O(1) $. This result along with the boundary conditions (\ref{BCs inf dn}) confirms the assumption on $ q^2+y^2 $.
Now considering (\ref{fv}), we see that the last term is of order $ (1-v) $ and can be ignored relative to the others. We can solve (\ref{fv}) by the method of dominant balance \cite{B&O} and find $ f = -\frac{\alpha_1^2}{r_1} +f_1(1-v) +O\mathopen{}\left((1-v)^2\right)\mathclose{} $.
We can now consider (\ref{gau}) to see if this behaviour is preserved to later $ u $. We find $ \gamma_u = \frac{f_1}{a_1}(1-v) +O\left((1-v)^2\right) $, so we see that it is only the second term in $ \gamma $ that is affected, which is why it was included.

We use this behaviour near $ v = 1 $ to introduce new variables (cf. (\ref{dn first})) that remain finite (unless a true singularity occurs):
\begin{equation} \label{dn first final}
\begin{aligned}
\gamma &= \frac{1}{1-v} +(1-v)\tilde\gamma \;,\qquad & w &= 1-\frac{\tilde{r}^2\tilde{W}}{(1-u)^2} \;,\qquad & f &= \frac{\tilde{f}}{(1-u)^2} \;, \\
\alpha &= \frac{\tilde\alpha}{(1-u)(1-v)} \;, & d &= \frac{\tilde{r}\tilde{D}}{1-u} \;, & \beta &= \frac{\tilde\beta}{1-u} \;, \\
r &= \frac{\tilde{r}}{(1-u)(1-v)} \;, & q &= (1-v)\tilde{q} \;, \\
g &= \frac{\tilde{r}+(1-v)\tilde\alpha^2\tilde{G}}{(1-u)(1-v)^2} \;, & y &= (1-v)\tilde{y} \;, \\
 && z &= \frac{\tilde{r}^2\tilde{Z}}{(1-u)^2} \;.
\end{aligned}
\end{equation}
We use symmetry and numerical experiments to determine the factors of $ (1-u) $ to include in the metric functions to keep them finite over the domain $ 0 \leqslant u \leqslant 1, u \leqslant v \leqslant 1 $.
The behaviour of the $ w $ and $ d $ at a regular origin as given in (\ref{regular tr}) and $ z $ as given in its definition  (\ref{dn first}) suggests the new regularising variables $ W $, $ D $, and $ Z $ given by:
\begin{align} \label{WDZ}
w &= 1-r^2W \;, & d &= rD \;, & z &= r^2Z \;.
\end{align}
The first of these was also favoured for numerical calculations in \cite{Rinne14}.
These choices improve the behaviour of the numerical solution at the origin, but also lose resolution near null infinity as $ r \to \infty $. Using the knowledge of $ r $ as $ v \to 1 $ this is easy to take into account and results in the final choice of variables $ \tilde{W} $, $ \tilde{D} $, and $ \tilde{Z} $.
Taking into account the behaviour of $ r $ at $ \mathscr{I}^+ $ also gives the final choices $ \tilde{q} $ and $ \tilde{y} $, while $ p $ and $ x $ are unaffected.
The final choice of the new $ g $ variable $ \tilde{G} $ ensures that the $ r $ and $ g $ equations (\ref{rv},\ref{gv}) do not include denominators with $ (1-v) $, by exploiting the fact that $ (1-v)g-r = O(1) $.
A capital is used because of the factor of $ \tilde\alpha^2 $ that is taken out to simplify the resulting equation further, in much the same way that $ G = \frac{g}{\alpha^2} $ simplifies (\ref{gv}).

With these variables we have
\begin{align} \label{Nm final}
N &= -\frac{\tilde{f}(\tilde{r}+(1-v)\tilde\alpha^2\tilde{G})}{(1-u)\tilde\alpha^2} \;, & m &= \frac{\tilde{r}\left((1-u)\tilde\alpha^2+\tilde{f}(\tilde{r}+(1-v)\tilde\alpha^2\tilde{G})\right)}{2(1-u)^2(1-v)\tilde\alpha^2} \;,
\end{align}
and the Bondi mass is given by $ \lim\limits_{v \to 1} m = -\frac{\tilde{r}^2\tilde{f}_v}{2(1-u)^2\tilde\alpha^2} $.
By taking the limit as $ r \to \infty $ of (\ref{tau dn}), we obtain the Bondi time $ \tau_B $ on future null infinity, which idealises an asymptotic observer \cite{PHA05,Frauendiener00} (also used by \cite{RM13,Rinne14}).
We find
\[ \tau_B = \int_{u_0}^u \frac{2\tilde\alpha^2}{(1-u)\tilde{r}} \dx{u} \;. \]

\section{The first-order equations}
We now write the final form of the first-order equations. The $ u $-direction equations (\ref{F4}) are
\begin{subequations} \label{F final}
\begin{align}
\tilde{q}_u &= -\frac{(1-v)\tilde{f}\tilde{q}+\tilde\alpha^2(-2\tilde{W}+\tilde{D}^2)}{(1-u)\tilde{r}} -\frac{\tilde\alpha^2}{(1-u)^2}\left(\tilde{D}\tilde{Z}+\frac{\tilde{r}\tilde{W}}{1-u}\left(3\tilde{W}-\tilde{D}^2-\frac{\tilde{r}^2\tilde{W}^2}{(1-u)^2}\right)\right) -a\tilde{y} \;, \label{qu final} \\
\tilde{y}_u &= -\frac{(1-v)\tilde{f}\tilde{y}-\tilde\alpha^2\tilde{Z}}{(1-u)\tilde{r}} -\frac{\tilde\alpha^2}{(1-u)^2}\left(\tilde{D}\left(-2\tilde{W}+\tilde{D}^2+\frac{\tilde{r^2}\tilde{W}^2}{(1-u)^2}\right)+\frac{\tilde{r}\tilde{W}\tilde{Z}}{1-u}\right) +a\tilde{q} \;, \\
\tilde\gamma_u &= \frac{\tilde{f}(\tilde{r}+(1-v)\tilde\alpha^2\tilde{G})+(1-u)\tilde\alpha^2}{(1-u)(1-v)\tilde{r}^2} -\frac{2(1-v)\tilde\alpha^2}{(1-u)^2}\left(\left(-2\tilde{W}+\tilde{D}^2+\frac{\tilde{r}^2\tilde{W}^2}{(1-u)^2}\right)^2+\tilde{Z}^2\right) \;, \label{gau final} \\
b_u &= -\frac{\tilde\alpha^2\tilde{Z}}{(1-u)^2} \;.
\end{align}
\end{subequations}
Note that it may be tempting to simplify the above equations for $ \tilde{q} $ and $ \tilde{y} $ by taking all the $ \tilde{q} $s and $ \tilde{y} $s to the left-hand side, introducing factors of $ \tilde{r} $ inside the derivative. For example, $ \left(\frac{\tilde{r}\tilde{q}}{1-u}\right)_u $. However, using (\ref{my evolution y}) on this would put a factor of $ \tilde{r}_n $ in front of the unknown $ \tilde{q}_n $, which would nullify this term at the origin, leaving $ \tilde{q}_n $ incalculable there by this approach.

Although (\ref{qu final}-\ref{gau final}) contain singular terms at $ \tilde{r} = 0 $ and $ u = 1 $, this is not a problem because the right-hand sides are not evaluated at the larger $ u $ point in our method (\ref{my evolution y}). However, since a modified Euler step (\ref{my evolution y1}) is used to begin at $ u = 0 $, for the very first step we require the following limit,
\[ \lim_{\tilde{r} \to 0} \tilde\gamma_u = -(1-u)\left(\tilde{q}^2+\tilde{y}^2\right), \]
which is determined by l'H\^{o}pital's rule.
The limits for $ \tilde{q}_u $ and $ \tilde{y}_u $ are not required because $ \tilde{q} $ and $ \tilde{y} $ on $ \tilde{r} = 0 $ are set by extrapolation, as detailed in section \ref{s:IBCs}.
Since (\ref{gau final}) contains a term that is singular at $ v = 1 $, we use l'H\^{o}pital's rule to find
\[ \lim_{v \to 1} \tilde\gamma_u = -\frac{\tilde{f}_v}{(1-u)\tilde{r}} \;, \]
where the right-hand side is evaluated at $ v = 1 $. Since we will shortly see that $ \tilde{f}_v $ can not be simply evaluated at $ v = 1 $, we use a second-order accurate finite difference approximation to calculate it.

The $ v $-direction equations (\ref{G11}) are
\begin{subequations} \label{G final}
\begin{align}
(\ln\tilde\alpha)_v &= (1-v)\tilde\gamma \;, \label{alphav final} \\
\tilde{r}_v &= \tilde\alpha^2\tilde{G} \;, \label{rv final} \\
\tilde{G}_v &= 2\left(\tilde\gamma-(1-v)^2(\tilde{q}^2+\tilde{y}^2)\right)\frac{\tilde{r}}{\tilde\alpha^2} \;, \label{Gv final} \\
\tilde{W}_v &= -\frac{2\tilde\alpha^2\tilde{G}\tilde{W}+(1-u)(\tilde{q}-b\tilde{D})}{\tilde{r}} \;, \label{Wv final} \\
\tilde{D}_v &= -\frac{\tilde\alpha^2\tilde{G}\tilde{D}-(1-u)b}{\tilde{r}} +\tilde{y} -\frac{\tilde{r}\tilde{W}b}{1-u} \;, \label{Dv final} \\
\tilde{Z}_v &= -2\frac{\tilde\alpha^2\tilde{G}\tilde{Z}+(1-u)\tilde{y}}{\tilde{r}} +2\tilde{D}\tilde{q} +\frac{2\tilde{r}\tilde{W}\tilde{y}}{1-u} \;, \label{Zv final} \\
a_v &= \frac{\tilde\alpha^2\tilde{Z}}{(1-u)^2} \;, \label{av final}
\end{align}
\begin{align}
\tilde{f}_v &= -\frac{\tilde{f}(\tilde{r}+(1-v)\tilde\alpha^2\tilde{G})+(1-u)\tilde\alpha^2}{(1-v)\tilde{r}} +\frac{(1-v)\tilde{r}\tilde\alpha^2}{1-u}\left(\left(-2\tilde{W}+\tilde{D}^2+\frac{\tilde{r}^2\tilde{W}^2}{(1-u)^2}\right)^2+\tilde{Z}^2\right), \label{fv final} \\
p_v &= -bx -\frac{\tilde\alpha^2}{(1-u)^2}\left(\left(1-\frac{\tilde{r}^2\tilde{W}}{(1-u)^2}\right)\left(-2\tilde{W}+\tilde{D}^2+\frac{\tilde{r}^2\tilde{W}^2}{(1-u)^2}\right)-\frac{\tilde{r}\tilde{D}\tilde{Z}}{1-u}\right), \label{pv final} \\
x_v &= bp -\frac{\tilde\alpha^2}{(1-u)^2}\left(\frac{\tilde{r}\tilde{D}}{1-u}\left(-2\tilde{W}+\tilde{D}^2+\frac{\tilde{r}^2\tilde{W}^2}{(1-u)^2}\right)+\left(1-\frac{\tilde{r}^2\tilde{W}}{(1-u)^2}\right)\tilde{Z}\right), \label{xv final} \\
\tilde\beta_v &=\frac{\tilde{f}(\tilde{r}+(1-v)\tilde\alpha^2\tilde{G})+(1-u)\tilde\alpha^2}{\tilde{r}^2} -\frac{2(1-v)^2\tilde\alpha^2}{1-u}\left(\left(-2\tilde{W}+\tilde{D}^2+\frac{\tilde{r}^2\tilde{W}^2}{(1-u)^2}\right)^2+\tilde{Z}^2\right). \label{bev final}
\end{align}
\end{subequations}
Again, it may be tempting to simplify the equations for each of $ \tilde{W} $, $ \tilde{D} $, and $ \tilde{Z} $ by bringing that variable to the left-hand side. For $ \tilde{W} $, for example, this would result in $ \left(\tilde{r}^2\tilde{W}\right)_v = -(1-u)\tilde{r}(\tilde{q}-b\tilde{D}) $. Not only is this simpler, but the right-hand side is no longer singular at $ \tilde{r} = 0 $. Furthermore, since integration in the $ v $ direction is away from the regular origin $ \tilde{r} = 0 $, the unknown is not nullified as above. However, we can determine the truncation error for $ \tilde{W} $ when (\ref{my evolution z}) is applied to this equation and find it to be
\begin{align*}
\tilde{W}(u,v+\Delta v)-\tilde{W}_n &= \left(-\frac{\tilde{W}_{vvv}}{12} -\frac{\tilde{r}_v}{2\tilde{r}}\tilde{W}_{vv} -\frac{1}{2}\left(\frac{\tilde{r}_v^2}{\tilde{r}^2}+\frac{\tilde{r}_{vv}}{\tilde{r}}\right)\tilde{W}_v -\frac{1}{2}\left(\frac{\tilde{r}_v\tilde{r}_{vv}}{\tilde{r}^2}+\frac{\tilde{r}_{vvv}}{3\tilde{r}}\right)\tilde{W}\right)(\Delta v)^3 \\
&\quad+ O\left((\Delta v)^4\right).
\end{align*}
The additional terms involving $ \tilde{r} $ are clearly unbounded at the origin and cause large errors in the calculation of $ \tilde{W} $, which in tests result in much spurious refinement at the origin. Therefore, the more complicated equations for $ \tilde{W} $, $ \tilde{D} $, and $ \tilde{Z} $ are used, and to evaluate the formally singular right-hand sides at $ \tilde{r} = 0 $, l'H\^{o}pital's rule is used;
\begin{equation} \label{lH}
\begin{aligned}
\lim_{\tilde{r} \to 0} \tilde{W}_v &= -\frac{1}{6}\left(\frac{\tilde\alpha\tilde{D}\tilde{Z}}{1-u}+2(1-u)\tilde\gamma(4\tilde{W}+\tilde{D}^2)+\frac{(1-u)(2\tilde{q}_v-3b_v\tilde{D})}{\tilde\alpha}\right), \\
\lim_{\tilde{r} \to 0} \tilde{D}_v &= -\frac{1}{2}\left(\frac{\tilde\alpha\tilde{Z}}{1-u}+2(1-u)\tilde\gamma\tilde{D}-\frac{(1-u)b_v}{\tilde\alpha}\right), \\
\lim_{\tilde{r} \to 0} \tilde{Z}_v &= -\frac{2}{3}\left(\frac{\tilde\alpha\tilde{D}(2\tilde{W}-\tilde{D}^2)}{1-u}+2(1-u)\tilde\gamma\tilde{Z}+\frac{(1-u)\tilde{y}_v}{\tilde\alpha}\right),
\end{aligned}
\end{equation}
where the right-hand side functions are evaluated at $ \tilde{r} = 0 $.

Attempting to simplify the equation for $ \tilde{f} $ results in
\[ \left(\frac{\tilde{r}\tilde{f}}{1-v}\right)_v = -\frac{(1-u)\tilde\alpha^2}{(1-v)^2} + \frac{\tilde{r}^2\tilde\alpha^2}{1-u}\left(\left(-2\tilde{W}+\tilde{D}^2+\frac{\tilde{r}^2\tilde{W}^2}{(1-u)^2}\right)^2+\tilde{Z}^2\right), \]
which contains a singular term as $ v \to 1 $. This then results in spurious refinement near $ v = 1 $, so (\ref{fv final}) is used in preference.
l'H\^{o}pital's rule gives $ \lim\limits_{\tilde{r} \to 0} \tilde{f}_v = 0 $ but does not produce a result for $ \tilde{f}_v $ at $ v = 1 $, it merely shows that (\ref{fv final}) is consistent. So instead, we multiply (\ref{fv final}) by $ (1-v) $ and evaluate at $ v = 1 $ to find
\begin{equation*} \label{fI}
\tilde{f} = -\frac{(1-u)\tilde\alpha^2}{\tilde{r}}
\end{equation*}
at $ v = 1 $.

There is an additional complication for this system as compared to the scalar field case.
The terms on the right hand side of (\ref{fv final}) are each $ O(v-u) $ at the origin, but this highest order behaviour exactly cancels to create a right hand side that is $ O\mathopen{}\left((v-u)^2\right)\mathclose{} $.
In the scalar field case the right hand side consists of the first term only.
This cancellation causes small numerical errors in the $ \tilde{f} $ calculation, which also affect the $ \tilde\gamma $ calculation since this involves the same first term. These errors are typically only seen as small oscillations when the (first term of the) right hand side of (\ref{fv final}) is calculated near $ r = 0 $.
However these errors seem to be amplified (only) when $ \tilde{q} $ and $ \tilde\gamma $ are interpolated with respect to $ v $ at the origin in order to decrease the $ u $ step size there (detailed in section \ref{sec:code:AMR}).
We found a solution to this issue by introducing a new variable $ \tilde{F} $ that removes the $ \tilde{r} $ from the denominator of the first term in (\ref{fv final});
\begin{align} \label{fF}
\tilde{f} &= \tilde\alpha^2\frac{(1-v)\tilde{r}\tilde{F}-(1-u)}{\tilde{r}+(1-v)\tilde\alpha^2\tilde{G}} \;, & \tilde{F} &= \frac{(1-u)+\frac{\tilde{f}}{\tilde\alpha^2}(\tilde{r}+(1-v)\tilde\alpha^2\tilde{G})}{(1-v)\tilde{r}} \;.
\end{align}
This substitution results in the equation
\begin{align}
\tilde{F}_v &= -2\left(\frac{\tilde\alpha^2\tilde{G}}{\tilde{r}}+\frac{(1-v)^3(\tilde{q}^2+\tilde{y}^2)\tilde{r}}{\tilde{r}+(1-v)\tilde\alpha^2\tilde{G}}\right)\tilde{F} \nonumber \\
 &\quad +\frac{2(1-u)(1-v)^2(\tilde{q}^2+\tilde{y}^2)}{\tilde{r}+(1-v)\tilde\alpha^2\tilde{G}} +\frac{\tilde{r}+(1-v)\tilde\alpha^2\tilde{G}}{1-u}\left(\left(-2\tilde{W}+\tilde{D}^2+\frac{\tilde{r}^2\tilde{W}^2}{(1-u)^2}\right)^2+\tilde{Z}^2\right). \label{Fv final}
\end{align}
We also use $ \tilde{F} = O(\tilde{r}) $ as $ \tilde{r} \to 0 $ and $ \lim\limits_{\tilde{r} \to 0} \tilde{F}_v = \frac{(1-u)^2}{\tilde\alpha}\left(\tilde{q}^2+\tilde{y}^2\right) $.
This then removes the errors in $ \tilde{f}_v $ and the undesired refinement.
Note that due to (\ref{fF}) and (\ref{Nm final}), writing $ \tilde{f} $ in terms of $ \tilde{F} $ results in a $ \frac{0}{0} $ on a future marginally trapped surface ($ r_v = 0 $, that is, $ \tilde{r}+(1-v)\tilde\alpha^2\tilde{G} = 0 $). For this reason we do not use $ \tilde{F} $ instead of $ \tilde{f} $ everywhere, but rather merely integrate (\ref{Fv final}) for any points near $ r = 0 $, and use (\ref{fF}) to set the values of $ \tilde{f} $.
For regular spacetimes $ \tilde{F} $ would certainly be a suitable variable and because (\ref{Fv final}) doesn't contain any $ (1-v) $ terms in the denominators, it results in a simpler calculation at future null infinity as well as a simpler formula for the mass (\ref{Nm final}) that can be evaluated directly on $ \mathscr{I}^+ $; $ m = \frac{\tilde{r}^2\tilde{F}}{2(1-u)^2} $.

Finally, considering equation (\ref{bev final}), we see it also contains an $ \tilde{r} $ in the denominator, so we use $ \lim\limits_{\tilde{r} \to 0} \tilde\beta_v = -(1-u)^3\left(\tilde{q}^2+\tilde{y}^2\right) $.

The equations that can be used for checking the numerical solution are
\begin{subequations} \label{check final}
\begin{align}
\tilde\alpha_u +\frac{1-\tilde\beta}{1-u}\tilde\alpha &= 0 \;, \\
\tilde{r}_u +\frac{\tilde{r}-(1-v)\tilde{f}}{1-u} &= 0 \;, \\
\tilde{f}_u +\frac{2(1-\tilde\beta)\tilde{f}}{1-u} +\frac{2(1-u)^3(1-v)(p^2+x^2)}{\tilde{r}} &= 0 \;, \\
\tilde{W}_u +\frac{(1-u)^2p +\frac{2(1-v)\tilde{r}\tilde{f}}{1-u}\tilde{W}-(1-u)\tilde{r}a\tilde{D}}{\tilde{r}^2} &= 0 \;,
\end{align}
\begin{align}
\tilde{D}_u -\frac{(1-u)x-\frac{(1-v)\tilde{f}}{1-u}\tilde{D}+(1-u)a}{\tilde{r}} +\frac{\tilde{r}a\tilde{W}}{1-u} &= 0 \;, \\
\tilde{Z}_u -2\frac{(1-u)^2x-\frac{(1-v)\tilde{r}\tilde{f}}{1-u}\tilde{Z}-(1-u)\tilde{r}\tilde{D}p}{\tilde{r}^2} +2\tilde{W}x &= 0 \;, \\
\tilde{G}_u +\frac{(2\beta-1)G+\frac{\tilde{f}(2\tilde{r}+(1-v)\tilde\alpha^2\tilde{G})+(1-u)\tilde\alpha^2}{\tilde{r}\tilde\alpha^2}}{1-u} & \\
-\frac{(1-v)^2\tilde{r}}{(1-u)^2}\left(\left(-2\tilde{W}+\tilde{D}^2+\frac{\tilde{r}\tilde{W}^2}{(1-u)^2}\right)^2+\tilde{Z}^2\right) &= 0 \;.
\end{align}
\end{subequations}

In the purely magnetic case we have $ \tilde{y} \equiv b \equiv \tilde{D} \equiv \tilde{Z} \equiv x \equiv a \equiv 0 $ and the equations dramatically simplify. Our code proceeds without calculating these variables, resulting in a quicker evolution and less required memory for the final data.

We previously claimed that the $ v $-direction equations could be ordered so that the unknown variable appears linearly, making the nominally implicit numerical method explicit in practice.
We now see that the order of equations (\ref{G final}) does indeed allow (\ref{my evolution z}) to be made explicit for each variable.
Given $ \tilde{q} $, $ \tilde{y} $, $ \tilde\gamma $, and $ b $ on a row of constant $ u $, equation (\ref{alphav final}) can be solved for $ \tilde\alpha $, and equations (\ref{rv final},\ref{Gv final}) can be solved simultaneously for $ \tilde{r} $ and $ \tilde{G} $. Equations (\ref{Wv final},\ref{Dv final}) can then be solved simultaneously for $ \tilde{W} $ and $ \tilde{D} $, followed by equation (\ref{Zv final}) for $ \tilde{Z} $. Equations (\ref{av final}) and (\ref{fv final}) can be solved for $ a $ and $ \tilde{f} $ respectively, and equations (\ref{pv final},\ref{xv final}) simultaneously for $ p $ and $ x $. Finally equation (\ref{bev final}) gives $ \tilde\beta $.
The paired equations may be solved simultaneously with (\ref{my evolution z}) by solving three $ 2 \times 2 $ linear systems, see appendix \ref{sec:A:num} for more details.
Note that since $ p $, $ x $, and $ \tilde\beta $ are not required to evolve the remaining equations, we could ignore their calculation as Hamad\'{e} and Stewart did for the variable we call $ \beta $.
However, they are necessary to calculate all of the checking equations (\ref{check final}), and $ p $ and $ x $ will appear in some quantities of interest, so we include these equations in the code, and calculate them when desired.

\section{Initial and boundary conditions, final coordinate and gauge choices} \label{s:IBCs}
We need to specify $ \tilde{q}(0,v) $, $ \tilde{y}(0,v) $, $ \tilde\gamma(0,v) $, and $ b(0,v) $ as initial conditions for the system (\ref{F final}), where $ \tilde\gamma(0,v) $ along with $ \tilde\alpha(0,0) = \alpha(0,0) $ fix the coordinates and $ b(0,v) $ (equivalently $ \lambda_v(0,v) $) fixes the gauge.
As stated in section \ref{ss:dn choices} we wish to separate the initial conditions from the coordinate choice and so we will specify $ W_0(r) $ and $ D_0(r) $ on the line $ u = 0 $.
Note that there are no restrictions on the parity of these functions; we only require them to be smooth and their behaviour at infinity should be $ W_0(r) = O\mathopen{}\left(\frac{1}{r^2}\right)\mathclose{} $ and $ D_0(r) = O\mathopen{}\left(\frac{1}{r}\right)\mathclose{} $.
To compute (using (\ref{dn first},\ref{dn first final})) the resulting
\begin{align*}
\tilde{q}(0,v) &= -\frac{1}{1-v}\left(\frac{\tilde{r}}{1-v}\right)_v\left(\frac{\tilde{r}W_0'}{1-v}+2W_0\right) +\frac{bD_0}{1-v} \;, \\
\tilde{y}(0,v) &= \frac{\left(\frac{\tilde{r}}{1-v}\right)_vD_0-b}{\tilde{r}} +\frac{1}{1-v}\left(\frac{\tilde{r}}{1-v}\right)_vD_0' +\frac{\tilde{r}bW_0}{(1-v)^2} \;,
\end{align*}
we first compute $ \tilde{r}(0,v) $ from (\ref{rv final},\ref{Gv final}):
\[ \tilde{r}_{vv} = 2(1-v)\tilde\gamma \tilde{r}_v +2\left(\tilde\gamma-(1-v)^2\left(\tilde{q}^2+\tilde{y}^2\right)\right)\tilde{r} \;. \]
We choose the gauge
\begin{subequations}
\begin{equation} \label{dn gauge final}
b(0,v) = r_vD_0
\end{equation}
to simplify this calculation.
Typically we also choose $ \tilde\gamma(0,v) = 0 $ to further simplify the calculation, but for some initial data will desire $ \tilde\gamma(0,v) $ non-zero, so the final coordinate choice can be specified when the program is called:
\begin{align} \label{dn coord final}
\tilde\gamma(0,v) &= \tilde\gamma_0(v) \;, \\
\tilde\alpha(0,0) &= \alpha_0 \;.
\end{align}
\end{subequations}
Therefore we numerically solve
\begin{align} \label{r0}
\tilde{r}_{vv} &= 2\tilde\gamma_0\left(\tilde{r}+(1-v)\tilde{r}_v\right) -2\tilde{r}\left(\left(\frac{\tilde{r}}{1-v}\right)_v\right)^2\left(\left(\frac{\tilde{r}W_0'}{1-v}+2W_0-D_0^2\right)^2+\left(D_0'+\frac{\tilde{r}W_0D_0}{1-v}\right)^2\right), \nonumber \\
\tilde{r}_v(0,0) &= \alpha_0 \;, \\
\tilde{r}(0,0) &= 0 \;, \nonumber
\end{align}
where $ W_0 $, $ D_0 $, and their derivatives with respect to $ r $ are evaluated at $ \frac{\tilde{r}}{1-v} $.

We note that this choice for $ b(0,v) $ keeps $ \tilde{y} $ finite at the origin, but may blow up at $ v = 1 $ if $ D_0(r) = O\mathopen{}\left(\frac{1}{r}\right)\mathclose{} $ as $ r \to \infty $. This could be avoided by adding the term $ rD_0'r_v $ so that $ b(0,v) $ is simply $ d_v $, but unfortunately this gauge choice is incompatible with $ \lambda(0,0) = 0 $.
We could make a more complicated gauge choice but we choose to keep (\ref{dn gauge final}) and simply demand $ D_0(r) = O\mathopen{}\left(\frac{1}{r^2}\right)\mathclose{} $ as $ r \to \infty $.

Thus we freely specify $ \alpha_0 $, $ \tilde\gamma_0(v) $, $ W_0(r), $ and $ D_0(r) $ (both $ O\mathopen{}\left(\frac{1}{r^2}\right)\mathclose{} $) and use (\ref{r0}) to find $ \tilde{r}(0,v) $. We then set
\begin{equation} \label{ICs}
\begin{aligned}
\tilde{q}(0,v) &= -\frac{1}{1-v}\left(\frac{\tilde{r}}{1-v}\right)_v\left(\frac{\tilde{r}W_0'}{1-v}+2W_0-D_0^2\right), \\
\tilde{y}(0,v) &= \frac{1}{1-v}\left(\frac{\tilde{r}}{1-v}\right)_v\left(D_0' +\frac{\tilde{r}W_0D_0}{1-v}\right), \\
\tilde\gamma(0,v) &= 0 \;, \\
b(0,v) &= \left(\frac{\tilde{r}}{1-v}\right)_vD_0 \;.
\end{aligned}
\end{equation}
We note that we have the following limits:
\begin{align*}
\lim_{v \to 1} \tilde{q}(0,v) &= -\frac{1}{\tilde{r}(0,1)^2}\lim_{r \to \infty} r^4W_0'+2r^3W_0 \;, \\
\lim_{v \to 1} \tilde{y}(0,v) &= \frac{1}{\tilde{r}(0,1)^2}\lim_{r \to \infty} r^2D_0(r^2W_0-2) \;, \\
\lim_{v \to 1} b(0,v) &= \frac{1}{\tilde{r}(0,1)}\lim_{r \to \infty} r^2D_0 \;.
\end{align*}
This completes the initial condition specification.
Note that $ \alpha_0 $ and $ \tilde\gamma(0,v) $ determine how quickly $ r $ increases with $ v $ and so in practice are chosen so that the variations in $ w $ and $ d $ are most spread out along $ 0 \leqslant v \leqslant 1 $. This decreases the amount of $ v $ refinement required to accurately trace the initial conditions.

We note that the above choices make it somewhat easier to specify simple bump initial data than using $ w $ or using a spatial slicing. On a spatial slice, $ w $ and $ W $ must be even functions of $ r $, and further, $ w $ must be 1 at $ r = 0 $, meaning any Gaussian in $ w $ must decay very quickly. In our set-up on an outgoing null slice, $ W $ is allowed to have non-zero odd derivatives at the origin, and there is no restriction on its value at the origin. Thus we can choose Gaussians with much wider profiles.
These bumps in $ W_0(r) $ then correspond to bumps in $ w_0(r) = 1-r^2W_0(r) $, however note that they are no longer symmetric about their maximums due to the $ r^2 $ factor.

The boundary conditions outlined in section \ref{ss:dn choices} are slightly modified for the new variables (\ref{dn first final}).
We first note that at the origin, $ \tilde{q} $, $ \tilde{y} $, and $ b $ should satisfy
\begin{subequations} \label{BCs final all}
\begin{align} \label{BCs 0 dn final}
\tilde{q} &= -\frac{\tilde\alpha(2\tilde{W}-\tilde{D}^2)}{1-u} \;, & \tilde{y} &= -\frac{\tilde\alpha\tilde{Z}}{1-u} \;, & b &= \frac{\tilde\alpha\tilde{D}}{1-u} \;,
\end{align}
due to their appearances in fractions over $ \tilde{r} $ in (\ref{F final},\ref{G final}). These equations could be used to set the values of $ \tilde{W} $, $ \tilde{D} $ and $ \tilde{Z} $ on the origin $ u = v $ however we instead set them the same way as $ \tilde\alpha $; using the fact that $ \alpha $, $ W $, $ D $ and $ Z $ are even functions of $ r $, thus their derivatives in the direction $ \pypx{}{v}-\pypx{}{u} $ are zero. The results with either method are equivalent.
To calculate the derivative, we use points on the closest rows in the direction of increasing $ v-u $, using a second-order approximation to the derivative wherever possible.
On the initial row, we set $ \tilde{W}(0,0) = W_0(0) $, $ \tilde{D}(0,0) = D_0(0) $, and $ \tilde{Z}(0,0) = -D_0'(0) $, where the last equation follows from evaluating $ \tilde{y}(0,0) $ using (\ref{ICs}) and (\ref{BCs 0 dn final}).

If it is reached, $ u = 1 $ is a single point corresponding to timelike infinity. The values of $ \tilde{W} $, $ \tilde{D} $, and $ \tilde{Z} $ are all set to zero, while $ \tilde\alpha $ is approximated by its final value on the previous row.
On $ u = v $ we specify the remaining $ v $-direction variables (\ref{G final}) by using (\ref{BCs 0 dn}) expressed in the new variables;
\begin{equation} \label{BCs final}
\begin{aligned}
\tilde{r} &= 0 \;, & p &= 0 \;, \\
\tilde{G} &= \frac{1}{\tilde\alpha} \;, & x &= 0 \;, \\
\tilde{f} &= -\tilde\alpha \;,\qquad & \tilde\beta &= 1+(1-u)^2\tilde\gamma \;, \\
 && a &= -\frac{\tilde\alpha\tilde{D}}{1-u} \;.
\end{aligned}
\end{equation}
\end{subequations}

\section{The AMR implementation} \label{sec:code:AMR}
Here we describe the algorithm in some detail so as to highlight the changes we make to the basic adaptive mesh refinement described in section \ref{s:HS}.
For example, we alter the starting procedure, the refinement ratio, incorporate dissipation, and specify the excision used to avoid the singularities of a black hole collapse.

One major adjustment we make is to take advantage of the double-null set-up and refine the $ u $ and $ v $ directions separately. In the original AMR algorithm it is necessary to refine the time and spatial directions together in order to preserve the CFL condition. That is, a too-large $ \Delta t $ step will take the new point out of the domain of dependence of the points used in the calculation.
In double-null coordinates however, a new point given by any $ \Delta u $ and $ \Delta  v $ will remain in the domain of dependence of the points used in the calculation, and thus we are free to refine $ u $ and $ v $ separately.

Unless stated otherwise, all interpolation used in the code is achieved using cubic splines, with not-a-knot end conditions.

\subsection{Input and output}
The main program takes the following input:

\begin{center}
\begin{tabular}{rl}
\verb_IC_, & the initial conditions $ \alpha_0 $, $ \tilde\gamma_0(v) $, $ W_0(r) $, $ D_0(r) $, \\
\verb_ns_, & the initial number of equally spaced steps on $ u = 0 $, $ 0 \leqslant v \leqslant 1 $, \\
\verb_epsu_, & the accuracy requirement $ \epsilon_u $ for the truncation error estimate in the $ u $ direction $ \text{TE}_u $, \\
\verb_epsv_, & the accuracy requirement $ \epsilon_v $ for the truncation error estimate in the $ v $ direction $ \text{TE}_v $, \\
\verb_MRLu_, & the maximum refinement level in the $ u $ direction, \\
\verb_MRLv_, & the maximum refinement level in the $ v $ direction. \\
\end{tabular}
\end{center}

There are also a number of switches that can be used to improve the efficiency of the algorithm, particularly when multiple evolutions are being performed in a bisection search:

\begin{center}
\begin{tabular}{rp{13cm}}
\verb_ICcheck_, & after $ u = 0 $ has been integrated, plot relevant functions and prompt whether to continue, \\
\verb_BHcontinue_, & after a black hole is detected, continue with the evolution, \\
\verb_lite_, & save memory by only retaining the rows necessary to continue the evolution, \\
\verb_out_, & create additional output, saving variables at the origin and future null infinity, and if an MTT forms, also on the MTT and near future timelike infinity, \\
\verb_pxbe_, & evolve the variables $ p $, $ x $, and $ \tilde\beta $, \\
\verb_timeout_, & limit any evolution to a predetermined time, \\
\verb_memout_, & limit any evolution to a predetermined amount of memory. \\
\end{tabular}
\end{center}

The functional initial conditions are entered as strings (to be interpreted by Mathematica).
We always choose a power of two for \verb_ns_ (typically at least 512) so that any amount of refinement will result in points that are represented exactly as machine numbers.
For $ \epsilon_u $ and $ \epsilon_v $, we usually choose a power of $ \frac{1}{8} $, since a further factor of $ \frac{1}{8} $ results in a rough doubling of the resolution (see below). We find the best results are when $ \epsilon_v $ has around two factors of $ \frac{1}{8} $ more than $ \epsilon_u $.
For regular spacetimes, \verb_MRLu_ and \verb_MRLv_ are simply set large enough to complete the evolution. It is important that the refinement levels are limited when a black hole may form, because such a true singularity will ask for more refinement indefinitely.
Note the maximum refinement level is fundamentally limited by machine resolution, so neither can be set larger than $ 53-\log_2(\verb_ns_) $.

The functions plotted when \verb_ICcheck = true_ are $ w $, $ d $, $ z $, $ \sqrt{Q^2+P^2} $, $ N $, and $ m $.
When \verb_lite = true_, the rows that are kept after each evolution of three rows are the most recent row of each level before the first row.
While the main program is written in MATLAB \cite{MATLAB} to take advantage of its fast vector numerical calculation abilities, we pass the initial conditions to Mathematica, which solves (\ref{r0}) and returns the values of (\ref{ICs}) to MATLAB. We do this to ensure high (at least machine precision) accuracy of the initial conditions.

The output consists essentially of a structure array containing constant-$ u $ rows of data.
Like Hamad\'{e} and Stewart we allow only a single child grid when refining in the $ u $ direction, so each row is uniquely identified by its $ u $ value, and the size of the structure array corresponds to the number of rows, which are ordered by $ u $.
The fields of the structure array include row vectors for storing each of the fifteen first order variables as well as $ v $ since the spacing between points on the constant-$ u $ row is allowed to vary. There is also a scalar storing the value of $ u $, and on the appropriate rows the truncation error estimates $ \text{TE}_u $ and $ \text{TE}_v $ are stored.
This is more complicated than the traditional storage method for an adaptive mesh refinement in that a single array contains non-constant spacing (in $ v $). However, one advantage is there is no doubled-up data, and thus no injection step necessary.

There is also an optional output of a cell array with up to four entries when \verb_out = true_, which is particularly useful to extract information from grids evolved with \verb_lite = true_. The first entry is a 5-row array containing the data calculated at the origin; $ u $, $ \tilde\alpha $, $ \tilde{W} $, $ \tilde{D} $, and $ \tilde{Z} $. The second is a 7-row array containing the data at future null infinity; $ u $, $ \tilde\alpha $, $ \tilde{r} $, $ \tilde{W} $, $ \tilde{D} $, $ \tilde{Z} $, and $ m $. The third and fourth entries are only created if a black hole forms; the third then is a 7-row array containing the MTT data $ u $, $ v $, $ \tilde\alpha $, $ \tilde{r} $, $ \tilde{W} $, $ \tilde{D} $, and $ \tilde{Z} $ calculated using cubic spline interpolation, and the fourth is the entire final $ u $-row created before excision begins.

\subsection{Initial evolution}
It is on the initial row ($ u = 0 $) that refinement in $ v $ begins. With the initial conditions (\ref{ICs}) set on the ($ \verb_ns_ + 1 $) equally-spaced points, the equations (\ref{G final}) are integrated using the trapezoidal method (\ref{my evolution z}).
The estimated truncation error $ \text{TE}_v $ is determined by recalculating these variables using only every second point and
\begin{equation} \label{TEv}
\text{TE}_v = \sqrt{\sum \left(\frac{z_{\Delta v}-z_{2\Delta v}}{2^3-2}\right)^2} \;,
\end{equation}
where the sum is over all the $ v $-direction variables $ z $, $ z_{\Delta v} $ are the values from the original integration and $ z_{2\Delta v} $ are the values calculated at half the resolution.
Note that the \verb_pxbe_ switch will affect the value of $ \text{TE}_v $ by determining if three terms are included.
The calculation of $ z_{2\Delta v} $ can be done for the entire row at once (without a for loop), in contrast to the original $ v $ integration, since for each step of $ 2\Delta v $ we begin not at the result of the previous step, but at the more accurate result calculated at $ \Delta v $ resolution. Therefore this is much quicker than the initial integration which must be done in increasing $ v $ order.
Also note that because the $ v $ integration requires using (\ref{lH}) at $ r = 0 $ while (\ref{TEv}) assumes the straightforward application of (\ref{my evolution z}) to (\ref{G final}), we do not calculate $ \text{TE}_v $ for the steps beginning at $ r = 0 $. Instead we copy the subsequent value of $ \text{TE}_v $ to indicate the magnitude of the truncation error there.

Wherever $ \text{TE}_v > \epsilon_v $, this indicates that the two previous steps are too large, and we refine by inserting two points symmetrically to make this four steps, so we use a refinement ratio of two in the $ v $ direction.
We do not insert a buffer zone because we use double-null coordinates and because we re-determine the $ v $-spacing at almost every opportunity.
Multiple clusters of $ v $-refinement are allowed, as long as there are a minimum of four coarse steps between them.
The call to Mathematica to set the initial conditions on the chosen points, the integration, and the truncation error estimation are then repeated until either $ \text{TE}_v \leqslant \epsilon_v $ everywhere or \verb_MRLv_ refinements occur.
Thus we see that \verb_ns_ does not so much set the initial resolution as determine how many levels of refinement are used. The resolution of the initial conditions is practically determined by $ \epsilon_v $.

Of the functions plotted when \verb_ICcheck = true_, $ N $ and $ m $ are particularly useful as their behaviour can be difficult to predict from the initial conditions, but are critically important to how the initial conditions evolve.
For example, if $ m $ increases too quickly and passes $ \frac{r}{2} $, then $ N $ becomes negative.
This does not indicate a black hole however, because $ g $ remains positive. Instead, $ f $ becomes positive and it indicates a past trapped region (a white hole) in the initial data.
Also, we generally find that a black hole will not evolve unless the Bondi mass of the initial data is at least of a magnitude close to one.

We use cubic spline interpolation to set the boundary conditions on new $ u $-rows that begin away from the origin. Since this requires at least four rows before $ u $ can be refined at this accuracy, we always begin with four steps of size $ \Delta u = h = \frac{1}{\text{ns}} $ in the $ u $ direction.
The truncation error estimation $ \text{TE}_u $ is calculated by repeating the evolution with double the step size $ 2h $ and
\begin{equation} \label{TEu}
\text{TE}_u = \sqrt{\sum\left(\frac{2h+3H}{9(h+H)}(y_h-y_{2h})\right)^2} \;,
\end{equation}
and we store it on the middle row of the three it is calculated over, because this remains a unique position even as further refinements are made.
The sum is over the $ u $-direction variables $ y $.
Although the steps are initially equispaced, we have here presented the more general calculation which can handle a previous step of size $ H $, allowed to be different to the current step size $ h $.
If $ h < H $, we scale $ \text{TE}_u $ by $ \frac{5h}{2h+3H} $ so that it scales with the resolution determined by $ h $ and is not affected by changes in $ H $.

Recall we use (\ref{my evolution y1}) initially, and since the formula (\ref{TEu}) assumes the evolution is given by (\ref{my evolution y}), we only begin to calculate $ \text{TE}_u $ after the third row.
If $ \text{TE}_u > \epsilon_u $ on the fourth row, then two new $ u $-rows are created at $ u = \frac{\Delta u}{2} $ and $ \frac{3\Delta u}{2} $ (equally spaced between the first, second, and third rows), so our refinement ratio in the $ u $ direction is also two. In general newly created rows will have a $ v $ extent determined by the positions where $ \text{TE}_u > \epsilon_u $, but initially the new rows have the full extent $ u \leqslant v \leqslant 1 $.
In order to calculate the boundary values for these rows without interpolation we check if the new $ \Delta u $ is smaller than the initial $ \Delta v $ and if so recalculate the initial conditions with $ \Delta v = \Delta u $ for the first eight steps.
The evolution and $ \text{TE}_u $ calculation is repeated on these new five rows, and this process continues until either $ \text{TE}_u \leqslant \epsilon_u $ everywhere on the fourth row or \verb_MRLu_ levels of refinement are implemented.
We see that \verb_ns_ also affects the amount of initial $ u $-refinement, as the initial $ u $-spacing is $ \Delta u = \frac{1}{\mathtt{ns}} $.

We illustrate this starting algorithm with an example initial refinement in figure \ref{fig:evo initial}.
\begin{figure}[!ht]
  \centering
  \begin{animateinline}[step,scale=0.9,
                        begin={\begin{tikzpicture} \useasboundingbox (-0.5,-1.5) rectangle (16.5,5);},end={\end{tikzpicture}}]{}
    \draw[->] (0,0)--(0,1.5) node[above]{$ u $};
    \draw[->] (0,0)--(1.5,0) node[right]{$ v $};
    \draw[dashed] (0,0)--(4.5,4.5) node[above right]{$ r = 0 $};
    \fill[MMA3] (0,0) circle (1.5pt);
    \foreach \v in {1,...,16}
      \fill[MMA2] (\v,0) circle (1.5pt);
    \foreach \u in {1,...,4}
      \foreach \v in {\u,...,16}
        \fill[MMA1] (\v,\u) circle (1.5pt);
    \node at (8,-1) {First row contains $ y $.};
  \newframe
    \draw[->] (0,0)--(0,1.5) node[above]{$ u $};
    \draw[->] (0,0)--(1.5,0) node[right]{$ v $};
    \draw[dashed] (0,0)--(4.5,4.5) node[above right]{$ r = 0 $};
    \foreach \v in {0,...,16}
      \fill[MMA3] (\v,0) circle (1.5pt);
    \foreach \v in {6,8,14,16}
      \draw[MMA4] (\v,0) circle (3pt);
    \foreach \u in {1,...,4}
      \foreach \v in {\u,...,16}
        \fill[MMA1] (\v,\u) circle (1.5pt);
    \node at (8,-1) {$ z $ integrated; $ \text{TE}_v $ calculated and exceeds $ \epsilon_v $ at four points.};
  \newframe
    \draw[->] (0,0)--(0,1.5) node[above]{$ u $};
    \draw[->] (0,0)--(1.5,0) node[right]{$ v $};
    \draw[dashed] (0,0)--(4.5,4.5) node[above right]{$ r = 0 $};
    \foreach \v in {0,...,4}
      \fill[MMA3] (\v,0) circle (1.5pt);
    \foreach \v in {4.5,5,...,8,9,10,...,12,12.5,13,...,16}
      \fill[MMA2] (\v,0) circle (1.5pt);
    \foreach \u in {1,...,4}
      \foreach \v in {\u,...,16}
        \fill[MMA1] (\v,\u) circle (1.5pt);
    \node at (8,-1) {Additional points inserted and $ y $ set via initial conditions.};
  \newframe
    \draw[->] (0,0)--(0,1.5) node[above]{$ u $};
    \draw[->] (0,0)--(1.5,0) node[right]{$ v $};
    \draw[dashed] (0,0)--(4.5,4.5) node[above right]{$ r = 0 $};
    \foreach \v in {0,...,4,4.5,5,...,8,9,10,...,12,12.5,13,...,16}
      \fill[MMA3] (\v,0) circle (1.5pt);
    \foreach \u in {1,...,4}
      \foreach \v in {\u,...,16}
        \fill[MMA1] (\v,\u) circle (1.5pt);
    \node at (8,-1) {First row integrated to within set tolerance.};
  \newframe
    \draw[->] (0,0)--(0,1.5) node[above]{$ u $};
    \draw[->] (0,0)--(1.5,0) node[right]{$ v $};
    \draw[dashed] (0,0)--(4.5,4.5) node[above right]{$ r = 0 $};
    \foreach \v in {0,...,4,4.5,5,...,8,9,10,...,12,12.5,13,...,16}
      \fill[MMA3] (\v,0) circle (1.5pt);
    \foreach \v in {1,...,4,4.5,5,...,8,9,10,...,12,12.5,13,...,16}
      \fill[MMA3] (\v,1) circle (1.5pt);
    \foreach \v in {2,...,4,4.5,5,...,8,9,10,...,12,12.5,13,...,16}
      \fill[MMA3] (\v,2) circle (1.5pt);
    \foreach \u in {3,...,4}
      \foreach \v in {\u,...,16}
        \fill[MMA1] (\v,\u) circle (1.5pt);
    \node at (8,-1) {Next two rows evolved ($ \text{TE}_v $ is calculated on row 3 and within the tolerance $ \epsilon_v $).};
  \newframe
    \draw[->] (0,0)--(0,1.5) node[above]{$ u $};
    \draw[->] (0,0)--(1.5,0) node[right]{$ v $};
    \draw[dashed] (0,0)--(4.5,4.5) node[above right]{$ r = 0 $};
    \foreach \v in {0,...,4,4.5,5,...,8,9,10,...,12,12.5,13,...,16}
      \fill[MMA3] (\v,0) circle (1.5pt);
    \foreach \v in {1,...,4,4.5,5,...,8,9,10,...,12,12.5,13,...,16}
      \fill[MMA3] (\v,1) circle (1.5pt);
    \foreach \v in {2,...,4,4.5,5,...,8,9,10,...,12,12.5,13,...,16}
      \fill[MMA3] (\v,2) circle (1.5pt);
    \foreach \v in {3,...,4,4.5,5,...,8,9,10,...,12,12.5,13,...,16}
      \fill[MMA3] (\v,3) circle (1.5pt);
    \foreach \v in {4,4.5,5,...,8,9,10,...,12,12.5,13,...,16}
      \fill[MMA3] (\v,4) circle (1.5pt);
    \foreach \v in {7,7.5,8,9,10}
      \draw[MMA4] (\v,4) circle (3pt);
    \node at (8,-1) {Rows 4 and 5 evolved, $ \text{TE}_u $ calculated for the row 3 to row 5 steps and exceeds $ \epsilon_u $ at five points.};
  \newframe
    \draw[->] (0,0)--(0,1.5) node[above]{$ u $};
    \draw[->] (0,0)--(1.5,0) node[right]{$ v $};
    \draw[dashed] (0,0)--(4.5,4.5) node[above right]{$ r = 0 $};
    \fill[MMA3] (0,0) circle (1.5pt);
    \foreach \v in {0.5,1,...,8,9,10,...,12,12.5,13,...,16}
      \fill[MMA2] (\v,0) circle (1.5pt);
    \foreach \v in {0.5,1,...,8,9,10,...,12,12.5,13,...,16}
      \fill[MMA1] (\v,0.5) circle (1.5pt);
    \foreach \v in {1,1.5,...,8,9,10,...,12,12.5,13,...,16}
      \fill[MMA1] (\v,1) circle (1.5pt);
    \foreach \v in {1.5,2,...,8,9,10,...,12,12.5,13,...,16}
      \fill[MMA1] (\v,1.5) circle (1.5pt);
    \foreach \v in {2,2.5,...,8,9,10,...,12,12.5,13,...,16}
      \fill[MMA1] (\v,2) circle (1.5pt);
    \foreach \v in {3,...,4,4.5,5,...,8,9,10,...,12,12.5,13,...,16}
      \fill[MMA1] (\v,3) circle (1.5pt);
    \foreach \v in {4,4.5,5,...,8,9,10,...,12,12.5,13,...,16}
      \fill[MMA1] (\v,4) circle (1.5pt);
    \node at (8,-1) {To be able to set boundary conditions on the new rows, the first row is refined near $ r = 0 $.};
  \newframe
    \draw[->] (0,0)--(0,1.5) node[above]{$ u $};
    \draw[->] (0,0)--(1.5,0) node[right]{$ v $};
    \draw[dashed] (0,0)--(4.5,4.5) node[above right]{$ r = 0 $};
    \fill[MMA3] (0,0) circle (1.5pt);
    \foreach \v in {0.5,1,...,8,9,10,...,12,12.5,13,...,16}
      \fill[MMA3] (\v,0) circle (1.5pt);
    \foreach \v in {0.5,1,...,8,9,10,...,12,12.5,13,...,16}
      \fill[MMA1] (\v,0.5) circle (1.5pt);
    \foreach \v in {1,1.5,...,8,9,10,...,12,12.5,13,...,16}
      \fill[MMA1] (\v,1) circle (1.5pt);
    \foreach \v in {1.5,2,...,8,9,10,...,12,12.5,13,...,16}
      \fill[MMA1] (\v,1.5) circle (1.5pt);
    \foreach \v in {2,2.5,...,8,9,10,...,12,12.5,13,...,16}
      \fill[MMA1] (\v,2) circle (1.5pt);
    \foreach \v in {3,...,4,4.5,5,...,8,9,10,...,12,12.5,13,...,16}
      \fill[MMA1] (\v,3) circle (1.5pt);
    \foreach \v in {4,4.5,5,...,8,9,10,...,12,12.5,13,...,16}
      \fill[MMA1] (\v,4) circle (1.5pt);
    \node at (8,-1) {First row integrated.};
  \newframe
    \draw[->] (0,0)--(0,1.5) node[above]{$ u $};
    \draw[->] (0,0)--(1.5,0) node[right]{$ v $};
    \draw[dashed] (0,0)--(4.5,4.5) node[above right]{$ r = 0 $};
    \fill[MMA3] (0,0) circle (1.5pt);
    \foreach \v in {0.5,1,...,8,9,10,...,12,12.5,13,...,16}
      \fill[MMA3] (\v,0) circle (1.5pt);
    \foreach \v in {0.5,1,...,8,9,10,...,12,12.5,13,...,16}
      \fill[MMA3] (\v,0.5) circle (1.5pt);
    \foreach \v in {1,1.5,...,8,9,10,...,12,12.5,13,...,16}
      \fill[MMA3] (\v,1) circle (1.5pt);
    \foreach \v in {1.5,2,...,8,9,10,...,12,12.5,13,...,16}
      \fill[MMA1] (\v,1.5) circle (1.5pt);
    \foreach \v in {2,2.5,...,8,9,10,...,12,12.5,13,...,16}
      \fill[MMA1] (\v,2) circle (1.5pt);
    \foreach \v in {3,...,4,4.5,5,...,8,9,10,...,12,12.5,13,...,16}
      \fill[MMA1] (\v,3) circle (1.5pt);
    \foreach \v in {4,4.5,5,...,8,9,10,...,12,12.5,13,...,16}
      \fill[MMA1] (\v,4) circle (1.5pt);
    \node at (8,-1) {Next two rows evolved.};
  \newframe
    \draw[->] (0,0)--(0,1.5) node[above]{$ u $};
    \draw[->] (0,0)--(1.5,0) node[right]{$ v $};
    \draw[dashed] (0,0)--(4.5,4.5) node[above right]{$ r = 0 $};
    \fill[MMA3] (0,0) circle (1.5pt);
    \foreach \v in {0.5,1,...,8,9,10,...,12,12.5,13,...,16}
      \fill[MMA3] (\v,0) circle (1.5pt);
    \foreach \v in {0.5,1,...,8,9,10,...,12,12.5,13,...,16}
      \fill[MMA3] (\v,0.5) circle (1.5pt);
    \foreach \v in {1,1.5,...,8,9,10,...,12,12.5,13,...,16}
      \fill[MMA3] (\v,1) circle (1.5pt);
    \foreach \v in {1.5,2,...,8,9,10,...,12,12.5,13,...,16}
      \fill[MMA3] (\v,1.5) circle (1.5pt);
    \foreach \v in {2,2.5,...,8,9,10,...,12,12.5,13,...,16}
      \fill[MMA3] (\v,2) circle (1.5pt);
    \foreach \v in {3,...,4,4.5,5,...,8,9,10,...,12,12.5,13,...,16}
      \fill[MMA1] (\v,3) circle (1.5pt);
    \foreach \v in {4,4.5,5,...,8,9,10,...,12,12.5,13,...,16}
      \fill[MMA1] (\v,4) circle (1.5pt);
    \node at (8,-1) {Rows 4 and 5 evolved; $ \text{TE}_u $ calculated for the row 3 to row 5 steps and within $ \epsilon_u $ everywhere.};
  \end{animateinline}
  \caption{An example initial refinement. Yellow indicates $ u $-direction variables $ y $ have been calculated at these points and green indicates the $ v $-direction variables $ z $ have also been calculated there. Red circles indicate the truncation error estimate for the previous two steps is above the specified tolerance. Click to step through (digital only).}
  \label{fig:evo initial}
\end{figure}
The result of the initial refinement is five rows successfully evolved to within the specified tolerances, perhaps with later rows that will be overwritten by the recursive algorithm.

\subsection{The recursive algorithm}
Finally we come to describe the main recursive refinement algorithm.
The recursion consists of two basic processes; evolving two steps in the $ u $ direction, and using the values of $ \text{TE}_u $ to determine whether to refine and insert new rows.
If there are no existing future rows ready to be evolved, the algorithm creates two new rows at the coarsest resolution in $ u $, $ \Delta u = \frac{1}{\mathtt{ns}} $, and with exactly the same resolution in $ v $ as the current final row.
The end condition for the recursion is either the solution is calculated up to $ u = 1 $ (timelike infinity) or the current row is reduced to a single point (at $ u = v $), which occurs if a black hole forms (discussed in section \ref{s:AMR:s}).

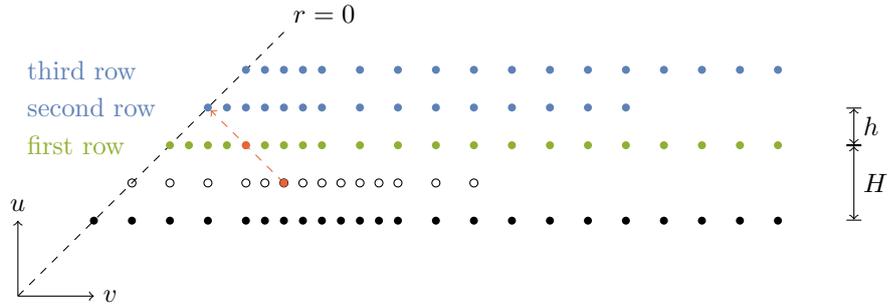
\begin{figure}[!ht]
  \centering
  \begin{tikzpicture}
    \draw[->] (1,-2)--(1,-1) node[above]{$ u $};
    \draw[->] (1,-2)--(2,-2) node[right]{$ v $};
    \draw[dashed] (1,-2)--(4.5,1.5) node[above right]{$ r = 0 $};
    \draw[MMA4,->,dashed,shorten >=1.5pt] (4.5,-0.5)--(3.5,0.5);
    \draw[MMA3] (1,0) node[right]{first row};
    \foreach \v in {3,3.25,...,5,5.5,6,...,11}
      \fill[MMA3] (\v,0) circle (1.5pt);
    \draw[MMA1] (1,0.5) node[right]{second row};
    \foreach \v in {3.5,3.75,...,5,5.5,6,...,9}
      \fill[MMA1] (\v,0.5) circle (1.5pt);
    \draw[MMA1] (1,1) node[right]{third row};
    \foreach \v in {4,4.25,...,5,5.5,6,...,11}
      \fill[MMA1] (\v,1) circle (1.5pt);
    \foreach \v in {2,2.5,...,4,4.25,4.5,...,6,6.5,7,...,11}
      \fill (\v,-1) circle (1.5pt);
    \foreach \v in {2.5,3,...,4,4.25,4.5,...,6,6.5,7}
      \draw[thin] (\v,-0.5) circle (1.5pt);
    \fill[MMA4] (4.5,-0.5) circle (1.5pt);
    \fill[MMA4] (4,0) circle (1.5pt);
    \draw[|<->|] (12,0) --node[right]{$ h $} (12,0.5);
    \draw[|<->|] (12,-1) --node[right]{$ H $} (12,0);
  \end{tikzpicture}
  \caption{An example block including $ r = 0 $ to be evolved two steps in the $ u $ direction (blue points), with the data on the first row already known (green), the previous row for the Adams-Bashforth method (solid black) and the points used for calculating the boundary conditions on the second row (red).}
  \label{fig:block}
\end{figure}

We evolve in blocks of three $ u $-rows (two steps), as this is the smallest block on which we can calculate $ \text{TE}_u $.
The boundary conditions for new rows are set in one of two ways, depending upon whether the rows reach $ r = 0 $ or not.
When the rows are formed by refinement in the $ u $ direction away from the origin, then the boundary values are set by interpolation in the $ u $ direction.
In the case when new rows are required to reach $ r = 0 $, the boundary values are calculated via (\ref{BCs final}) and by a second-order approximation to derivatives in the $ \pypx{}{v}-\pypx{}{u} $ direction for $ \tilde{\alpha} $, $ \tilde{W} $, $ \tilde{D} $, and $ \tilde{Z} $.
An example block for the second case is shown in figure \ref{fig:block}.

We see that the point (lower red point in figure \ref{fig:block}) previous to the first row (green) that is used to calculate the boundary condition for the middle row does not in general lie in the previous row used in the evolution (solid black). This is because the latter is the closest row that covers the $ v $-extent of the middle row of the block, while the former may be found in a closer row that is not as long as the middle row. If there is no previous row available (which occurs initially), then a first-order approximation for the derivative is used.
If the required points (red) do not already exist because the given rows are too coarse, the required variables are interpolated in the $ v $ direction onto those points. The points needed to set the boundary conditions on the third row are always found on rows one and two.

The $ v $-spacing on the latter two rows of the block is chosen to match the first row, but since (\ref{my evolution y}) requires a previous row which may have different spacing, some interpolation may be used on that row (solid black in figure \ref{fig:block}). No interpolation is required to evolve from rows one and two to the third row.
After the third row is evolved, $ \text{TE}_u $ (\ref{TEu}) is calculated at the $ v $-points of the middle row intersected with the third row, and only the values are stored on the middle row, not the points.
The evolution of the block ends with the integration of (\ref{G final}) along the final row.

The second basic step of the recursion is to either insert new rows or complete the integration, based on the values of $ \text{TE}_u $.
When $ \text{TE}_u > \epsilon_u $ we only take note of the minimum and maximum $ v $ where this inequality is satisfied, since we simply use one row for each value of $ u $.
We include a buffer zone in $ v $ of at least the current $ \Delta u $ to the left and right of the points in the row marked for refinement.
We further lengthen the new row if necessary by requiring that its boundaries occur on points that are on an imaginary $ 2\Delta v $ grid that begins at $ v = 0 $, where $ \Delta v $ is the $ v $ resolution at the beginning of the desired new row. This simplifies the calculation and interpretation of $ \text{TE}_v $.
This gives us the limits $ v_0 \leqslant v \leqslant v_1 $ for the two new rows, to be inserted symmetrically in $ u $ between the three rows of the block.
When $ v_0 = u $, we enforce $ \Delta v \leqslant \Delta u $ by interpolating and re-integrating in the $ v $ direction if necessary so that the boundary conditions can be calculated.
Otherwise, the most recent four rows to include $ v_0 $ are interpolated to set boundary conditions for the child blocks.
After new rows are created, we return to the first basic step and evolve the first new block, followed by the check of $ \text{TE}_u $.

When we find $ \text{TE}_u \leqslant \epsilon_u $ for a block, the final row is integrated to its end, and $ \text{TE}_v $ (\ref{TEv}) is calculated. As for the $ u = 0 $ row, a second integration is performed at half the resolution, and this occurs even where the resolution changes.
Points corresponding to $ \text{TE}_v > \epsilon_v $ are marked for refinement, while points that have $ \text{TE}_v < \frac{\epsilon_v}{8} $ are marked for coarsening.
Near $ r = 0 $, this criteria is strengthened to $ \text{TE}_v < \frac{\epsilon_v}{16} $.
We take care to ensure that any new coarse section begins and ends at points that correspond to a $ 2\Delta v $-spaced grid extending from $ v = 0 $ (where $ \Delta v $ is the coarser resolution).
Also, if a particular point corresponds to the boundary of a previous finer $ u $-row then the surrounding points that make it an even multiple of $ \Delta v $ are always kept.
Every row that begins at $ r = 0 $ with $ \Delta v = \Delta u $ and $ u $ being an odd multiple of $ \Delta u $ will have its first point not fall on an imaginary twice-coarser grid. We ensure consistency with all of the $ \text{TE}_v $ calculations by beginning the integration for $ z_{2\Delta v} $ in this case at the second point.
Once the new $ v $-points are determined, the evolution variables are interpolated onto these points, and the remaining variables are integrated. $ \text{TE}_v $ is calculated again and a warning is issued if it is not everywhere less than $ \epsilon_v $.
It is at this point that it is checked if a black hole is formed somewhere in that block.
The process we use when a black hole is detected, including excision, is described in section \ref{s:AMR:s}.

In figure \ref{fig:ref eg} we illustrate the algorithm with an example refinement.
\begin{figure}[!ht]
  \centering
  \begin{animateinline}[step,scale=0.9,
                        begin={\begin{tikzpicture} \useasboundingbox (-1.5,-1.5) rectangle (15.5,5);},end={\end{tikzpicture}}]{}
    \draw[->] (0,0)--(0,1.5) node[above]{$ u $};
    \draw[->] (0,0)--(1.5,0) node[right]{$ v $};
    \draw[dashed] (0,0)--(4.5,4.5) node[above right]{$ r = 0 $};
    \foreach \v in {2,3,4,4.5,5,5.5,6,6.25,...,14}
      \fill[MMA3] (\v,2) circle (1.5pt);
    \foreach \v in {3,4,4.5,5,5.5,6,6.25,...,14}
      \fill[MMA1] (\v,3) circle (1.5pt);
    \foreach \v in {4,4.5,5,5.5,6,6.25,...,14}
      \fill[MMA1] (\v,4) circle (1.5pt);
    \node at (7,-1) {First row contains $ z $.};
  \newframe
    \draw[->] (0,0)--(0,1.5) node[above]{$ u $};
    \draw[->] (0,0)--(1.5,0) node[right]{$ v $};
    \draw[dashed] (0,0)--(4.5,4.5) node[above right]{$ r = 0 $};
    \foreach \v in {2,3,4,4.5,5,5.5,6,6.25,...,14}
      \fill[MMA3] (\v,2) circle (1.5pt);
    \foreach \v in {3,4,4.5,5,5.5,6,6.25,...,14}
      \fill[MMA3] (\v,3) circle (1.5pt);
    \foreach \v in {4,4.5,5,5.5,6,6.25,...,7.75}
      \fill[MMA3] (\v,4) circle (1.5pt);
    \foreach \v in {8,8.25,...,14}
      \fill[MMA2] (\v,4) circle (1.5pt);
    \foreach \v in {7.75,8,8.25,10.25,10.5}
      \draw[MMA4] (\v,4) circle (3pt);
    \node at (7,-1) {Evolved two rows, $ \text{TE}_u > \epsilon_u $ on five points.};
  \newframe
    \draw[->] (0,0)--(0,1.5) node[above]{$ u $};
    \draw[->] (0,0)--(1.5,0) node[right]{$ v $};
    \draw[dashed] (0,0)--(4.5,4.5) node[above right]{$ r = 0 $};
    \foreach \v in {2,3,4,4.5,5,5.5,6,6.25,...,14}
      \fill[MMA3] (\v,2) circle (1.5pt);
    \foreach \v in {3,4,4.5,5,5.5,6,6.25,...,14}
      \fill[MMA3] (\v,3) circle (1.5pt);
    \foreach \v in {4,4.5,5,5.5,6,6.25,6.5}
      \fill[MMA3] (\v,4) circle (1.5pt);
    \foreach \v in {6.75,7,...,14}
      \fill[MMA2] (\v,4) circle (1.5pt);
    \foreach \v in {8,8.25,10.25}
      \draw[MMA4] (\v,4) circle (3pt);
    \node[inner sep=0pt,minimum size=6pt,circle,draw,MMA4] (n1) at (7.75,4) {};
    \node[inner sep=0pt,minimum size=6pt,circle,draw,MMA4] (n2) at (6.75,4) {};
    \node[inner sep=0pt,minimum size=6pt,circle,draw,MMA4] (n3) at (6.5,4) {};
    \draw[->,MMA4] (n1) to[out=135,in=45] node[above]{$ \Delta u $} (n2);
    \draw[->,MMA4] (n2) to[out=-110,in=-45] node[below]{$ 2\Delta v $ alignment} (n3);
    \node[inner sep=0pt,minimum size=6pt,circle,draw,MMA4] (n4) at (10.5,4) {};
    \node[inner sep=0pt,minimum size=6pt,circle,draw,MMA4] (n5) at (11.5,4) {};
    \draw[->,MMA4] (n4) to[out=45,in=135] node[above]{$ \Delta u $} (n5);
    \node at (7,-1) {Buffer of $ \Delta u $ and alignment with imaginary coarser $ v $ grid added.};
  \newframe
    \draw[->] (0,0)--(0,1.5) node[above]{$ u $};
    \draw[->] (0,0)--(1.5,0) node[right]{$ v $};
    \draw[dashed] (0,0)--(4.5,4.5) node[above right]{$ r = 0 $};
    \draw[dashed,MMA4] (6.5,1)--(6.5,4);
    \foreach \u in {1,2,3,4}
      \fill[MMA4] (6.5,\u) circle (1.5pt);
    \foreach \v in {2,3,4,4.5,5,5.5,6,6.25,6.75,7,...,14}
      \fill[MMA3] (\v,2) circle (1.5pt);
    \fill[MMA3] (6.5,2.5) circle (1.5pt);
    \foreach \v in {6.75,7,...,11.5}
      \fill[MMA1] (\v,2.5) circle (1.5pt);
    \foreach \v in {3,4,4.5,5,5.5,6,6.25}
      \fill[MMA3] (\v,3) circle (1.5pt);
    \foreach \v in {6.75,7,...,11.5}
      \fill[MMA1] (\v,3) circle (1.5pt);
    \foreach \v in {11.75,12,...,14}
      \fill[MMA2] (\v,3) circle (1.5pt);
    \fill[MMA3] (6.5,3.5) circle (1.5pt);
    \foreach \v in {6.75,7,...,11.5}
      \fill[MMA1] (\v,3.5) circle (1.5pt);
    \foreach \v in {4,4.5,5,5.5,6,6.25}
      \fill[MMA3] (\v,4) circle (1.5pt);
    \foreach \v in {6.75,7,...,11.5}
      \fill[MMA1] (\v,4) circle (1.5pt);
    \foreach \v in {11.75,12,...,14}
      \fill[MMA2] (\v,4) circle (1.5pt);
    \node at (7,-1) {New rows added, boundary conditions set by interpolation of four most recent rows.};
  \newframe
    \draw[->] (0,0)--(0,1.5) node[above]{$ u $};
    \draw[->] (0,0)--(1.5,0) node[right]{$ v $};
    \draw[dashed] (0,0)--(4.5,4.5) node[above right]{$ r = 0 $};
    \foreach \v in {2,3,4,4.5,5,5.5,6,6.25,...,14}
      \fill[MMA3] (\v,2) circle (1.5pt);
    \foreach \v in {6.5,6.75,7,...,11.5}
      \fill[MMA3] (\v,2.5) circle (1.5pt);
    \foreach \v in {3,4,4.5,5,5.5,6,6.25,6.5}
      \fill[MMA3] (\v,3) circle (1.5pt);
    \foreach \v in {6.75,7,...,14}
      \fill[MMA2] (\v,3) circle (1.5pt);
    \fill[MMA3] (6.5,3.5) circle (1.5pt);
    \foreach \v in {6.75,7,...,11.5}
      \fill[MMA1] (\v,3.5) circle (1.5pt);
    \foreach \v in {4,4.5,5,5.5,6,6.25,6.5}
      \fill[MMA3] (\v,4) circle (1.5pt);
    \foreach \v in {6.75,7,...,11.5}
      \fill[MMA1] (\v,4) circle (1.5pt);
    \foreach \v in {11.75,12,...,14}
      \fill[MMA2] (\v,4) circle (1.5pt);
    \node at (7,-1) {Two rows of first child block evolved, $ \text{TE}_u $ satisfactory.};
  \newframe
    \draw[->] (0,0)--(0,1.5) node[above]{$ u $};
    \draw[->] (0,0)--(1.5,0) node[right]{$ v $};
    \draw[dashed] (0,0)--(4.5,4.5) node[above right]{$ r = 0 $};
    \foreach \v in {2,3,4,4.5,5,5.5,6,6.25,...,14}
      \fill[MMA3] (\v,2) circle (1.5pt);
    \foreach \v in {6.5,6.75,7,...,11.5}
      \fill[MMA3] (\v,2.5) circle (1.5pt);
    \foreach \v in {3,4,4.5,5,5.5,6,6.25,6.5,...,14}
      \fill[MMA3] (\v,3) circle (1.5pt);
    \fill[MMA3] (6.5,3.5) circle (1.5pt);
    \foreach \v in {6.75,7,...,11.5}
      \fill[MMA1] (\v,3.5) circle (1.5pt);
    \foreach \v in {4,4.5,5,5.5,6,6.25,6.5}
      \fill[MMA3] (\v,4) circle (1.5pt);
    \foreach \v in {6.75,7,...,11.5}
      \fill[MMA1] (\v,4) circle (1.5pt);
    \foreach \v in {11.75,12,...,14}
      \fill[MMA2] (\v,4) circle (1.5pt);
    \foreach \v in {7.5,8,8.5,9,9.5,10.5,11,...,12}
      \draw[MMA5] (\v,3) circle (3pt);
    \node[MMA5] (n6) at (1,4) {$ \text{TE}_v $ calculated from this point};
    \node[inner sep=0pt,minimum size=3pt,circle,fill,MMA3] (n7) at (4,3) {};
    \draw[->,MMA5] (n6) -- (n7);
    \node at (7,-1) {Nine points are tagged for potential coarsening in $ v $, and none require refinement.};
  \newframe
    \draw[->] (0,0)--(0,1.5) node[above]{$ u $};
    \draw[->] (0,0)--(1.5,0) node[right]{$ v $};
    \draw[dashed] (0,0)--(4.5,4.5) node[above right]{$ r = 0 $};
    \foreach \v in {2,3,4,4.5,5,5.5,6,6.25,...,14}
      \fill[MMA3] (\v,2) circle (1.5pt);
    \foreach \v in {6.5,6.75,7,...,11.5}
      \fill[MMA3] (\v,2.5) circle (1.5pt);
    \foreach \v in {3,4,4.5,5,5.5,6,6.25,6.5,...,14}
      \fill[MMA3] (\v,3) circle (1.5pt);
    \fill[MMA3] (6.5,3.5) circle (1.5pt);
    \foreach \v in {6.75,7,...,11.5}
      \fill[MMA1] (\v,3.5) circle (1.5pt);
    \foreach \v in {4,4.5,5,5.5,6,6.25,6.5}
      \fill[MMA3] (\v,4) circle (1.5pt);
    \foreach \v in {6.75,7,...,11.5}
      \fill[MMA1] (\v,4) circle (1.5pt);
    \foreach \v in {11.75,12,...,14}
      \fill[MMA2] (\v,4) circle (1.5pt);
    \foreach \v in {7.5,8,8.5,9,10.5,11}
      \draw[MMA5] (\v,3) circle (3pt);
    \node[inner sep=0pt,minimum size=6pt,circle,draw,MMA5!50!] (n8) at (9.5,3) {};
    \node[MMA5] (n9) at (6.75,1) {Coarse section must end on $ 2\Delta v $ point.};
    \draw[->,MMA5] (n9) -- (n8);
    \node[inner sep=0pt,minimum size=6pt,circle,draw,MMA5!50!] (n10) at (11.5,3) {};
    \node[inner sep=0pt,minimum size=6pt,circle,draw,MMA5!50!] (n11) at (12,3) {};
    \node[MMA5] (n12) at (12,0) {Points kept so finer $ u $-rows end on $ 2\Delta v $ points.};
    \draw[->,MMA5] (n12) -- (n10);
    \draw[->,MMA5] (n12) -- (n11);
    \node at (7,-1) {Potential coarsening points are reduced.};
  \newframe
    \draw[->] (0,0)--(0,1.5) node[above]{$ u $};
    \draw[->] (0,0)--(1.5,0) node[right]{$ v $};
    \draw[dashed] (0,0)--(4.5,4.5) node[above right]{$ r = 0 $};
    \foreach \v in {2,3,4,4.5,5,5.5,6,6.25,...,14}
      \fill[MMA3] (\v,2) circle (1.5pt);
    \foreach \v in {6.5,6.75,7,...,11.5}
      \fill[MMA3] (\v,2.5) circle (1.5pt);
    \foreach \v in {3,4,4.5,5,5.5,6,6.25,6.5,...,14}
      \fill[MMA3] (\v,3) circle (1.5pt);
    \fill[MMA3] (6.5,3.5) circle (1.5pt);
    \foreach \v in {6.75,7,...,11.5}
      \fill[MMA1] (\v,3.5) circle (1.5pt);
    \foreach \v in {4,4.5,5,5.5,6,6.25,6.5}
      \fill[MMA3] (\v,4) circle (1.5pt);
    \foreach \v in {6.75,7,...,11.5}
      \fill[MMA1] (\v,4) circle (1.5pt);
    \foreach \v in {11.75,12,...,14}
      \fill[MMA2] (\v,4) circle (1.5pt);
    \foreach \v in {7.5,8,8.5,9}
      \draw[MMA5] (\v,3) circle (3pt);
    \node[inner sep=0pt,minimum size=6pt,circle,draw,MMA5!50!] (n13) at (10.5,3) {};
    \node[inner sep=0pt,minimum size=6pt,circle,draw,MMA5!50!] (n14) at (11,3) {};
    \node[MMA5] (n18) at (9.75,1) {Coarse sections must be at least 4 steps long.};
    \draw[->,MMA5] (n18) -- (n13);
    \draw[->,MMA5] (n18) -- (n14);
    \node at (7,-1) {Final coarsening points found.};
  \newframe
    \draw[->] (0,0)--(0,1.5) node[above]{$ u $};
    \draw[->] (0,0)--(1.5,0) node[right]{$ v $};
    \draw[dashed] (0,0)--(4.5,4.5) node[above right]{$ r = 0 $};
    \foreach \v in {2,3,4,4.5,5,5.5,6,6.25,...,14}
      \fill[MMA3] (\v,2) circle (1.5pt);
    \foreach \v in {6.5,6.75,7,...,11.5}
      \fill[MMA3] (\v,2.5) circle (1.5pt);
    \foreach \v in {3,4,4.5,5,5.5,6,6.25,6.5,6.75}
      \fill[MMA3] (\v,3) circle (1.5pt);
    \foreach \v in {7,7.5,...,9,9.25,9.5,...,14}
      \fill[MMA2] (\v,3) circle (1.5pt);
    \fill[MMA3] (6.5,3.5) circle (1.5pt);
    \foreach \v in {6.75,7,7.5,8,8.5,9,9.25,9.5,...,11.5}
      \fill[MMA1] (\v,3.5) circle (1.5pt);
    \foreach \v in {4,4.5,5,5.5,6,6.25,6.5}
      \fill[MMA3] (\v,4) circle (1.5pt);
    \foreach \v in {6.75,7,7.5,8,8.5,9,9.25,9.5,...,11.5}
      \fill[MMA1] (\v,4) circle (1.5pt);
    \foreach \v in {11.75,12,...,14}
      \fill[MMA2] (\v,4) circle (1.5pt);
    \node at (7,-1) {Points removed.};
  \newframe
    \draw[->] (0,0)--(0,1.5) node[above]{$ u $};
    \draw[->] (0,0)--(1.5,0) node[right]{$ v $};
    \draw[dashed] (0,0)--(4.5,4.5) node[above right]{$ r = 0 $};
    \foreach \v in {2,3,4,4.5,5,5.5,6,6.25,...,14}
      \fill[MMA3] (\v,2) circle (1.5pt);
    \foreach \v in {6.5,6.75,7,...,11.5}
      \fill[MMA3] (\v,2.5) circle (1.5pt);
    \foreach \v in {3,4,4.5,5,5.5,6,6.25,6.5,6.75,7,7.5,...,9,9.25,9.5,...,14}
      \fill[MMA3] (\v,3) circle (1.5pt);
    \fill[MMA3] (6.5,3.5) circle (1.5pt);
    \foreach \v in {6.75,7,7.5,8,8.5,9,9.25,9.5,...,11.5}
      \fill[MMA1] (\v,3.5) circle (1.5pt);
    \foreach \v in {4,4.5,5,5.5,6,6.25,6.5}
      \fill[MMA3] (\v,4) circle (1.5pt);
    \foreach \v in {6.75,7,7.5,8,8.5,9,9.25,9.5,...,11.5}
      \fill[MMA1] (\v,4) circle (1.5pt);
    \foreach \v in {11.75,12,...,14}
      \fill[MMA2] (\v,4) circle (1.5pt);
    \node at (7,-1) {Row re-integrated, $ \text{TE}_v $ satisfactory.};
  \newframe
    \draw[->] (0,0)--(0,1.5) node[above]{$ u $};
    \draw[->] (0,0)--(1.5,0) node[right]{$ v $};
    \draw[dashed] (0,0)--(4.5,4.5) node[above right]{$ r = 0 $};
    \foreach \v in {2,3,4,4.5,5,5.5,6,6.25,...,14}
      \fill[MMA3] (\v,2) circle (1.5pt);
    \foreach \v in {6.5,6.75,7,...,11.5}
      \fill[MMA3] (\v,2.5) circle (1.5pt);
    \foreach \v in {3,4,4.5,5,5.5,6,6.25,6.5,6.75,7,7.5,...,9,9.25,9.5,...,14}
      \fill[MMA3] (\v,3) circle (1.5pt);
    \foreach \v in {6.5,6.75,7,7.5,8,8.5,9,9.25,9.5,...,11.5}
      \fill[MMA3] (\v,3.5) circle (1.5pt);
    \foreach \v in {4,4.5,5,5.5,6,6.25,6.5}
      \fill[MMA3] (\v,4) circle (1.5pt);
    \foreach \v in {6.75,7,7.5,8,8.5,9,9.25,9.5,...,14}
      \fill[MMA2] (\v,4) circle (1.5pt);
    \node at (7,-1) {Two further rows evolved, $ \text{TE}_u $ satisfactory.};
  \newframe
    \draw[->] (0,0)--(0,1.5) node[above]{$ u $};
    \draw[->] (0,0)--(1.5,0) node[right]{$ v $};
    \draw[dashed] (0,0)--(4.5,4.5) node[above right]{$ r = 0 $};
    \foreach \v in {2,3,4,4.5,5,5.5,6,6.25,...,14}
      \fill[MMA3] (\v,2) circle (1.5pt);
    \foreach \v in {6.5,6.75,7,...,11.5}
      \fill[MMA3] (\v,2.5) circle (1.5pt);
    \foreach \v in {3,4,4.5,5,5.5,6,6.25,6.5,6.75,7,7.5,...,9,9.25,9.5,...,14}
      \fill[MMA3] (\v,3) circle (1.5pt);
    \foreach \v in {6.5,6.75,7,7.5,8,8.5,9,9.25,9.5,...,11.5}
      \fill[MMA3] (\v,3.5) circle (1.5pt);
    \foreach \v in {4,4.5,5,5.5,6,6.25,6.5,6.75,7,7.5,8,8.5,9,9.25,9.5,...,14}
      \fill[MMA3] (\v,4) circle (1.5pt);
    \node at (7,-1) {Final row integrated, $ \text{TE}_v $ satisfactory.};
  \end{animateinline}
  \caption{An example refinement. Yellow indicates $ u $-direction variables $ y $ have been calculated at these points and green indicates the $ v $-direction variables $ z $ have also been calculated there. Red circles indicate a truncation error estimate for the previous two steps is above the specified tolerance and purple circles indicate it is low enough to allow coarsening. Click to step through (digital only).}
  \label{fig:ref eg}
\end{figure}

If $ \text{TE}_u > \epsilon_u $ and child blocks are created, the middle row of the parent block is updated, changing in general both the positions and values of the points.
This accounts for why we always perform an additional coarse evolution step rather than utilise a self-shadow hierarchy \cite{PL04} (where the previously performed coarse step is compared with the current fine steps to determine the truncation error estimate).
Any second child block will be evolved from values that have changed since the second step of its parent.
This change also means the position of the points $ \text{TE}_u $ was calculated at are no longer simply given by the intersection of the middle row and final row, however these points can still be found by considering the first row of the parent block.
Finally, when $ \text{TE}_u > \epsilon_u $, (\ref{G final}) is integrated on the final row of the block only far enough to set boundary conditions for the new rows. The remaining $ v $ integration is performed only after the $ u $-variables are finalised by a satisfactory truncation error estimate.

While changing the $ v $ resolution makes the algorithm more efficient, it also can introduce high frequency noise into the numerical solution as is noted in \cite{Choptuik89}. We therefore add some Kreiss-Oliger dissipation \cite{KO73} to the evolution of the $ u $-direction variables. 
Let $ y $ stand for a $ u $-direction variable, and let $ y_i $ indicate $ y(u,v+i\Delta v) $, where $ u $ corresponds to the row that is being evolved to. Then after (\ref{my evolution y}), we replace $ y_i $ according to
\[ y_i \mapsto y_i - \epsilon_{\text{KO}}\frac{y_{i-2}-4y_{i-1}+6y_{i}-4y_{i+1}+y_{i+2}}{16} \;. \]
This does not affect the order of the method as the additional term is proportional to the fourth-order $ v $-derivative, but serves to smooth out any quickly varying behaviour that may develop from a change in resolution. 
We find through numerical experiments that a coefficient of $ \epsilon_{\text{KO}} = 0.3 $ works very well, and it is performed wherever the $ v $-spacing $ \Delta v $ is constant for five consecutive points.

\begin{figure}[!ht]
  \centering
  \includegraphics[scale=0.9]{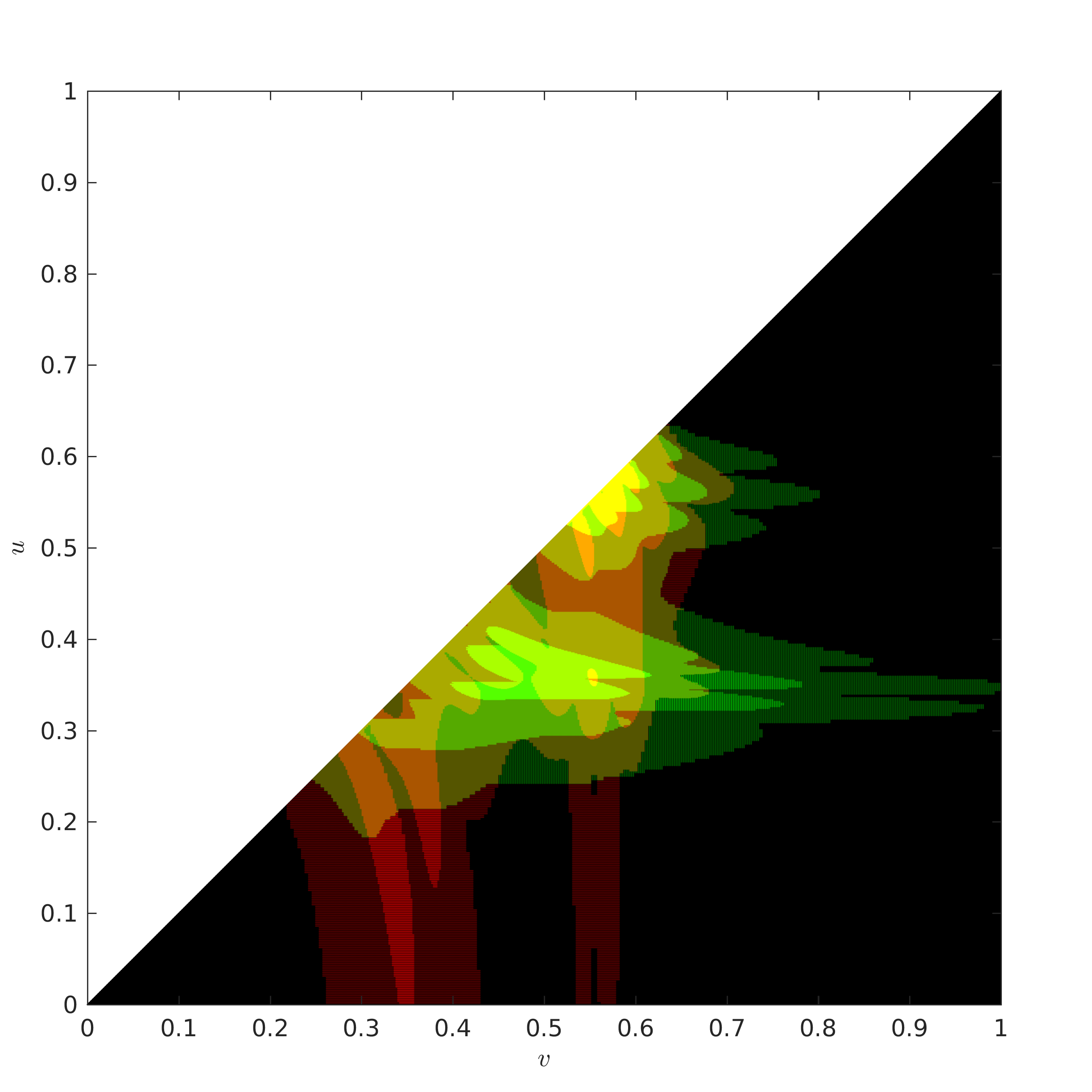}
  \caption{An example of the refinement created by this algorithm for a regular solution.}
  \label{fig:refinement eg}
\end{figure}

Figure \ref{fig:refinement eg} shows an example refinement, for initial data
\begin{equation} \label{ID ref eg}
\begin{aligned}
W_0(r) &= -0.034 e^{-(r-5)^2} \;, & \alpha_0 &= 10 \;, \\
D_0(r) &= 0.02 e^{-(r-10)^2} \;, & \tilde\gamma_0(v) &= 0 \;,
\end{aligned}
\end{equation}
and $ \verb_ns_ = 512 $, $ \epsilon_u = \frac{1}{8^5} $, $ \epsilon_v = \frac{1}{8^6} $.
We use colours to indicate the level of refinement. Black indicates the coarsest resolution, red increasing $ v $ resolution, and green increasing $ u $ resolution; thus regions equally refined in both directions show up as a shade of yellow. Full colour indicates the maximum refinement for that grid, not an absolute level of refinement.

We see in figure (\ref{fig:refinement eg}) the benefits of refining in the $ u $ and $ v $ directions separately; to retain the desired accuracy they require refinement in different regions in general. For example, $ v $ refinement is required initially to follow the initial conditions as they travel towards the origin, and no $ u $ refinement is necessary here.

For the majority of evolutions, with modest truncation error bounds, the resulting data structure is of the order of 100MB, and we keep all the data in memory.
When $ \epsilon_u $ and $ \epsilon_v $ are small and many (more than 20) levels of refinement are required, we set \verb_lite = true_.
This allows us to, for example, perform bisection searches with high accuracy without requiring unmanageable amounts of memory.

\subsection{Singularity formation} \label{s:AMR:s}
There are some aspects of the code particular to the case of a forming black hole, which we now describe.
Since there is a curvature singularity at the spacelike section of $ r = 0 $, the algorithm above will demand ever more refinement as this line is approached. Note also that this spacelike line extends from the regular section of $ r = 0 $ ($ u = v $) all the way to null infinity ($ v = 1 $).
As we evolve by increasing $ u $, this singular line will first be approached near null infinity.
There will come a point $ (u^*,v^*) $ when \verb_MRLu_ is already reached (the $ u $-step is already at the minimum we allow) and $ \text{TE}_u > \epsilon_u $. At this point we begin to remove points from the grid; all $ v \geqslant v^* $ are removed, and so $ v^* $ is never reached again for any $ u > u^* $.
Note that this excision begins before a black hole forms, as indicated by a (future) trapped surface ($ \tilde{r}+(1-v)\tilde\alpha^2\tilde{G} < 0 $), at $ (u_h,v_h) $ say.
In the cases where excision occurs but no trapped surface develops this indicates that more refinement is required to capture the evolution, and the evolution is restarted with larger \verb_MRLu_.
We then observe that the trapped region $ N < 0 $ reliably appears at approximately $ u_h \approx u^* + (u^*-u_\text{MRL}) $, where $ u_\text{MRL} $ is the first value of $ u $ for which the maximum refinement level is used.
Therefore, we see that only a little of the spacetime around timelike infinity is lost.

\begin{figure}[!ht]
  \centering
  \includegraphics[scale=0.9]{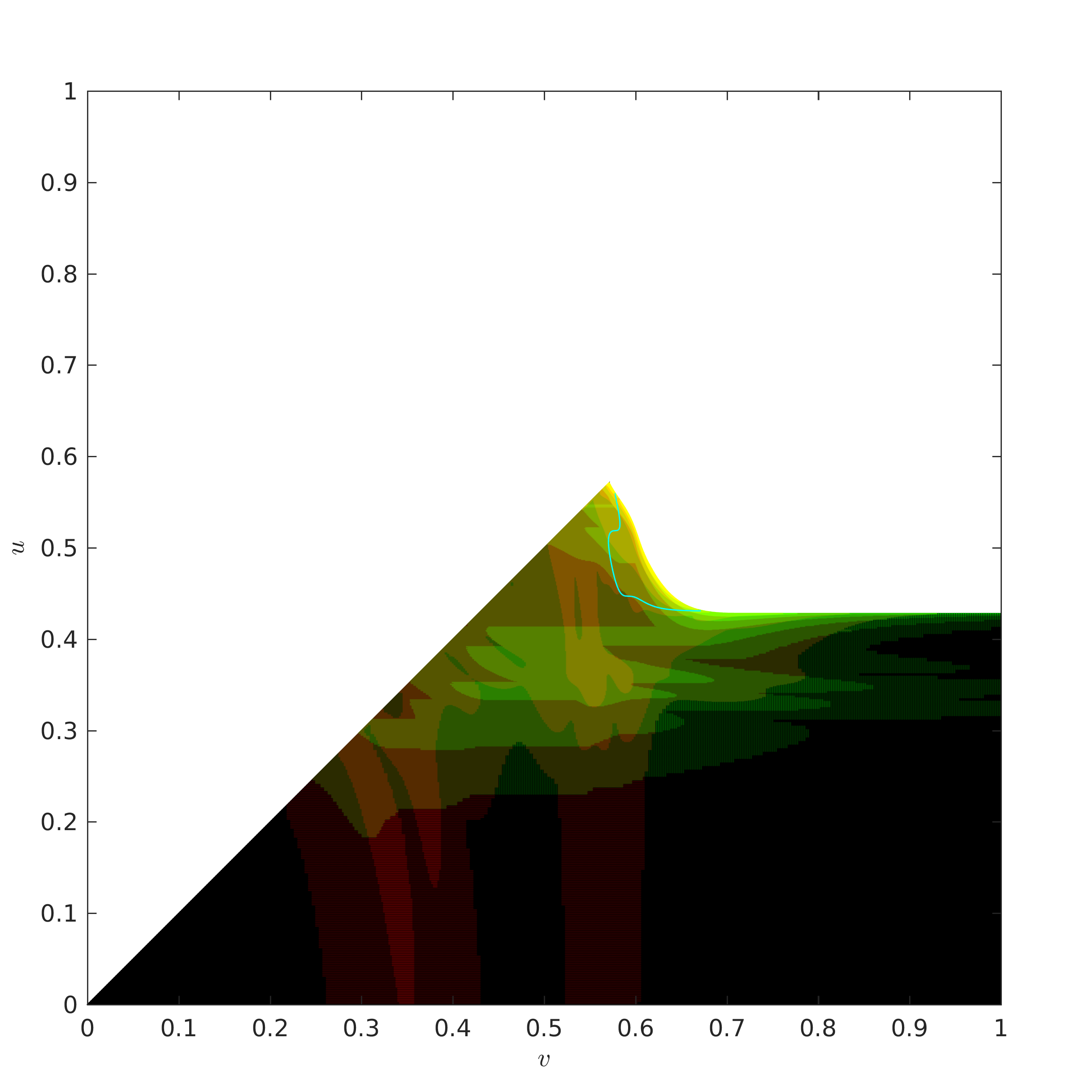}
  \caption{An example of the refinement created by this algorithm for a black hole solution, with MTT shown in cyan.}
  \label{fig:refinement BH eg}
\end{figure}

We want to use the refinements in particular to follow any self-similarities that arise during the evolution, and not to waste time and resources following the curve of a singularity.
After a trapped surface is detected at $ (u_h,v_h) $, we further limit the maximum refinement levels for $ u > u_h $ so that only modest resources are used to follow the singular and spacelike part of $ r = 0 $. Usually only the regular origin $ r = 0 $ or the MTT itself are specifically desired after $ u_h $. To ensure the future number of rows is limited by approximately $ 2^{12} $ we set $ \verb_MRLu_ = \verb_MRLv_ = \min(\verb_MRLu_,\left\lfloor 12-\log_2(\verb_ns_)-\log_2\left(\frac{v_h-u_h}{2}\right)\right\rfloor) $.
While the singular origin line is excised, the MTT is typically well captured until it approaches $ r = 0 $ on $ u = v $, where the spacetime around the meeting of the regular and singular origins is excised.
Figure \ref{fig:refinement BH eg} shows a typical refinement when a black hole forms during evolution, with the same initial data and tolerances as (\ref{ID ref eg}) except with the coefficient of $ D_0(r) $ changed to 0.04, and $ \verb_MRLu_ = \verb_MRLv_ = 6 $.

The creation of new, future rows at the coarsest evolution is subtly altered if a black hole is detected.
In that case, refinement in $ u $ will always be called for up to \verb_MRLu_ (at least near the singularity), and furthermore the evolution will stop before $ u = 1 $. We therefore decrease the step size of the new rows below $ \frac{1}{\mathtt{ns}} $, by measuring the minimum $ u $ refinement that is currently being used. This serves two purposes; it prevents repetitive refinement, and more importantly prevents the code from trying to evolve by an $ \frac{1}{\mathtt{ns}} $ step that takes it well past the singular endpoint. As the singular origin approaches the regular origin, the minimum $ u $ refinement approaches \verb_MRLu_, and this is the resolution at which the code ends.

We also calculate the position of the MTT. At every point we can of course calculate $ g $ by (\ref{dn first final}), and wish to approximate the points where this function is zero.
There is a first row (typically well after the event horizon) that contains both $ g > 0 $ and $ g < 0 $. For all such rows until the maximum $ u $, we calculate $ \tilde{r}+(1-v)\tilde\alpha^2\tilde{G} $ (the numerator of $ g $) on four points surrounding the change in sign, interpolate $ v $ with respect to this function, and evaluate it at zero. This gives the value of $ v $ for the MTS on this $ u $-row, which can then be used to evaluate any of the functions when they are interpolated.
For the previous $ u $-rows that are either entirely $ g > 0 $ or $ g < 0 $, we calculate $ \tilde{r}+(1-v)\tilde\alpha^2\tilde{G} $ on the four closest $ u $-points for each $ v $, and then interpolate $ u $ with respect to this function.

A final consideration for black hole spacetimes is that the final state of the black hole (the mass and also the sign of $ w $) is often desired, for which it is natural to consider the final point on future null infinity. However, due to the nature of the adaptive mesh refinement, this point is typically not well-resolved.
When the radiation is purely outgoing, the $ v $ derivatives are very small. Thus $ \tilde{q} $, $ \tilde{y} $, and $ \tilde\gamma $ are close to zero and there is therefore very little refinement in the $ u $ direction near $ \mathscr{I}^+ $. There is however maximal refinement at smaller $ v $ values just before excision begins at $ (u^*,v^*) $, and we use this information by integrating out in the $ v $ direction to $ v = 1 $.
This involves the extrapolation of the $ u $-direction variables, however this is reasonable because as discussed they are slowly changing in this region. Let $ (u',v') $ be the last point on the last completed row before $ (u^*,v^*) $. We hence, at each point $ v \in (v',1] $, take the two closest rows to $ u' $ and linear extrapolate $ \tilde{q} $, $ \tilde{y} $, $ \tilde\gamma $, and $ b $ to $ u' $. We then integrate the remaining variables along this extended row out to $ v = 1 $.
This row forms part of the additional output when \verb_out = true_.

\section{Convergence tests}
To confirm the code has the convergence properties we desire, we perform some simple tests.

The code was designed to be second-order accurate so that a doubling of the resolution results in the error decreasing by a factor of 4. To obtain a doubling of the resolution with this adaptive mesh code requires decreasing the truncation error tolerances.
We use the initial data (\ref{ID ref eg}) and vary \verb_ns_ and the truncation error tolerances as shown in figure \ref{fig:mesh convergence}.
\begin{figure}[!ht]
  \centering
  \animategraphics[scale=0.9,step]{}{mesh_convergence_}{256}{259}
  \caption{Plots of the automatically-generated meshes for the convergence test with successively double the resolution (Click to step through -- digital only).}
  \label{fig:mesh convergence}
\end{figure}
As seen in figure \ref{fig:mesh convergence}, doubling the initial resolution and reducing $ \epsilon_u $ and $ \epsilon_v $ by factors of 8 effectively doubles the resolution of the entire mesh.
Comparing the resulting meshes we see that the only change is that the edges of the levels become more defined as the resolution increases.
This indicates that the code is locally third-order as desired.

\begin{figure}[!ht]
  \centering
  \includegraphics[scale=0.9]{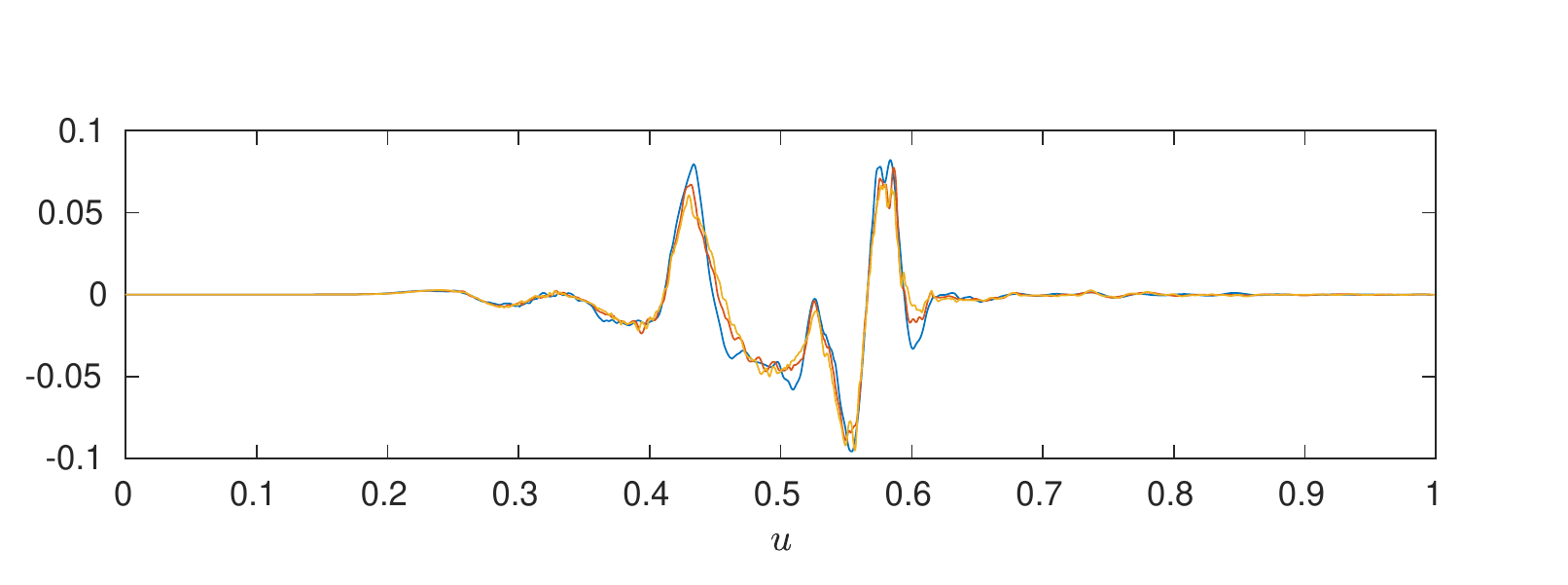}
  \includegraphics[scale=0.9]{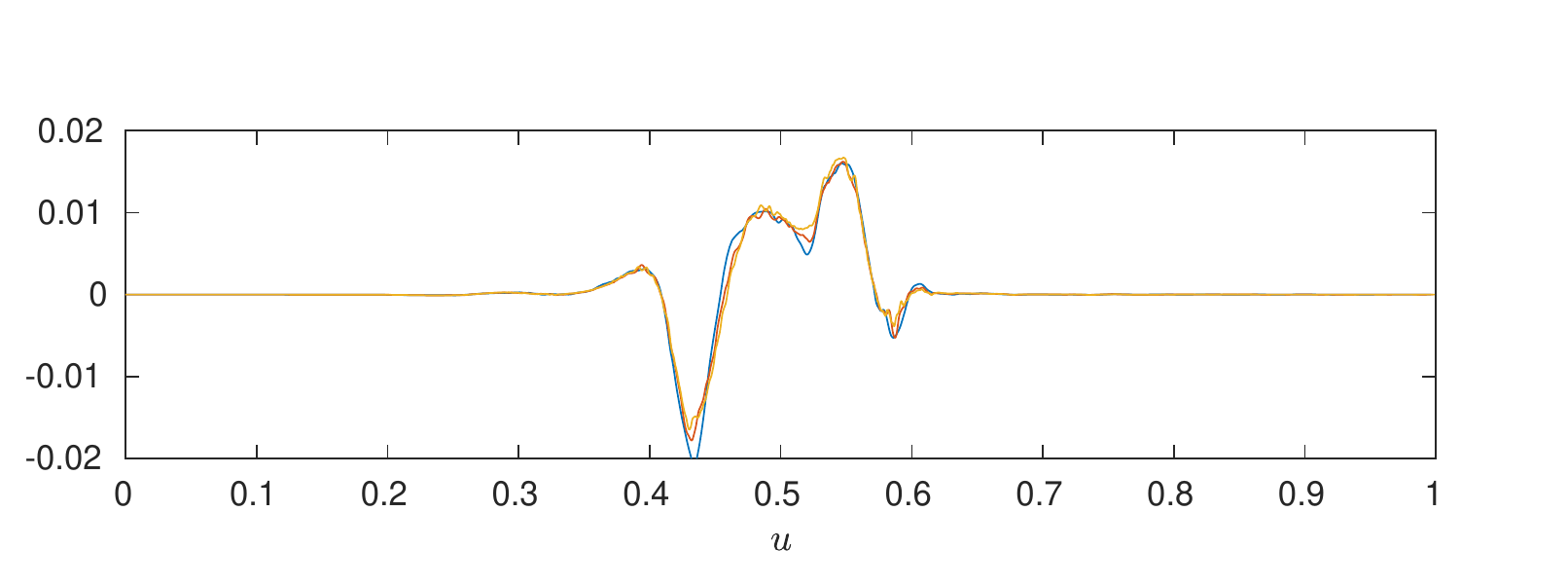}
  \caption{A plot of the scaled differences of $ \tilde{q} $ (top) and $ \tilde{W} $ (bottom) at the origin between evolutions with initial number of steps 256 and 512 (blue), 512 and 1024 (orange), and 1024 and 2048 (yellow).}
  \label{fig:convergence q}
\end{figure}

To test if the code is in fact globally second order, we consider the difference in the evolved quantities between the meshes with different resolutions, on the line of maximum $ u $, that is, the origin.
Figure \ref{fig:convergence q} shows the difference in $ \tilde{q} $ between consecutive grids of the four calculated above. The successive differences are multiplied successively by four and we see that they then align neatly with the first difference, indicating that the method is indeed globally second-order.
Figure \ref{fig:convergence q} also shows the same differences for an example $ v $-direction variable $ \tilde{W} $, which we see is also globally second-order convergent.

\begin{figure}[!ht]
  \centering
  \includegraphics[scale=0.9]{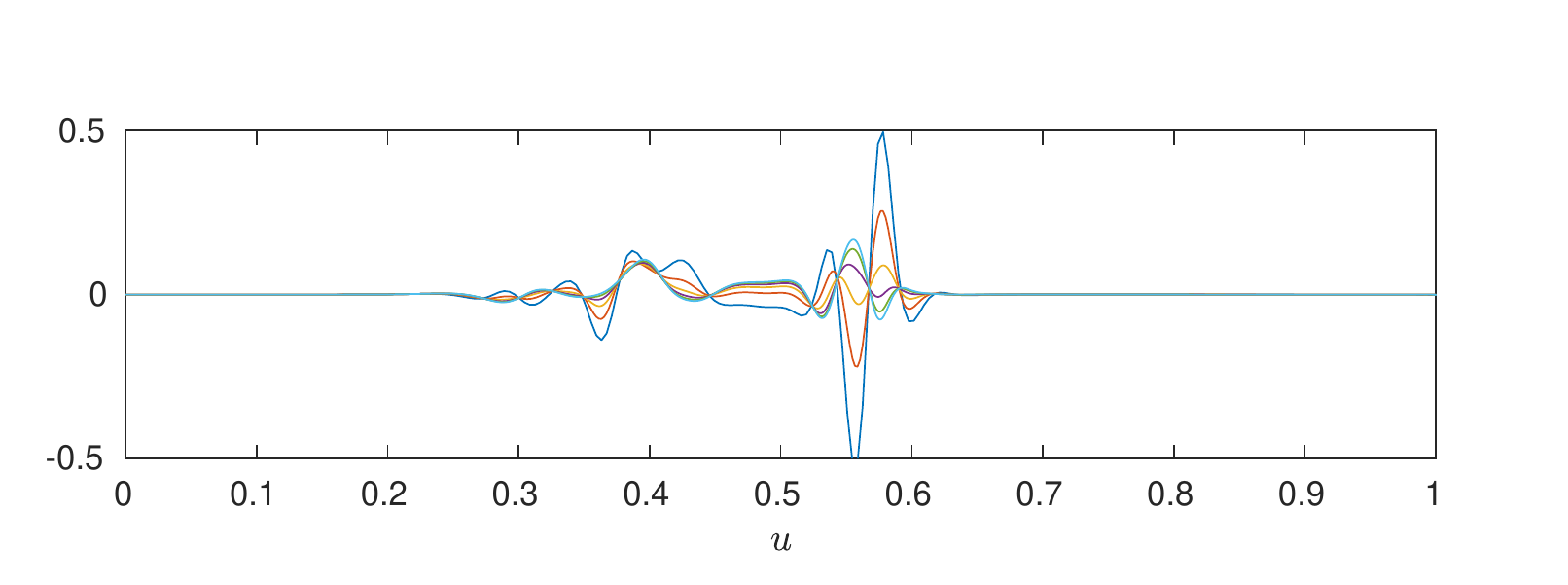}
  \includegraphics[scale=0.9]{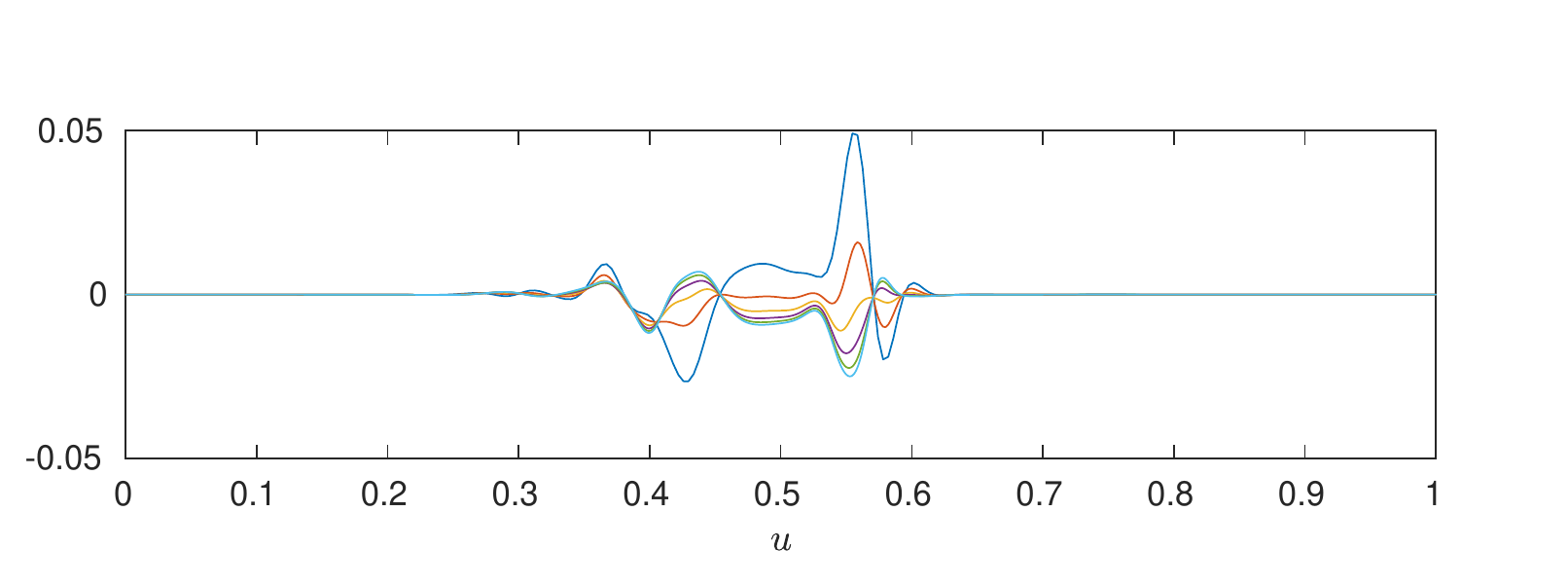}
  \caption{A plot of the scaled difference of $ \tilde{q} $ (top) and $ \tilde{W} $ (bottom) at the origin between evolutions with number of steps 256 and 512 (blue), 512 and 1024 (orange), and 1024 and 2048 (yellow), 2048 and 4096 (purple), 4096 and 8192 (green), 8192 and 16384 (cyan).}
  \label{fig:convergence qW unigrid}
\end{figure}

We perform the same test without any adaptive mesh refinement to observe the convergence of the scheme on a uniform grid and the results are shown in figure \ref{fig:convergence qW unigrid}.
We observe that the high-frequency fluctuations seen in figure \ref{fig:convergence q} are absent and thus are due to the AMR.
We can also see that much finer (initial) resolutions are required to capture the convergence of the scaled differences without the AMR.

\chapter{Purely magnetic dynamic solutions} \label{ch:Dm}
We begin the exploration of the dynamic solutions in the purely magnetic sector; $ a \equiv b \equiv d \equiv 0 $.
This sector has been previously explored by Zhou and Straumann in 1991-1992 \cite{ZS91,Zhou92}, Choptuik, Chmaj, and Bizo\'{n} in 1996 \cite{CCB96}, Choptuik, Hirschmann and Marsa in 1999 \cite{CHM99}, and Rinne in 2014 \cite{Rinne14}.
We will compare our numerical results with theirs, which will serve both as further confirmation of the code and confirmation of the previous numerical results as we produce them in significantly different ways.

The dynamic behaviours considered in this chapter are the instability of the static magnetic solutions (in particular the first Bartnik-McKinnon soliton BK1), and the different kinds of critical phenomena that can occur. We begin with a discussion of critical behaviours and how it can be realised numerically.

\section{Critical phenomena} \label{s:critical}
Critical phenomena refer to the various behaviours occurring when a family of parameterised initial data, which for different parameter values will evolve to two very different end states, is tuned to the boundary of these regions in the parameter space.
Typically the initial data is regular and the end states are either Minkowski space or a Schwarzschild black hole, so the evolution will determine if gravitational collapse occurs. It is on and near the threshold of black hole formation that the interesting behaviours occur, and a solution that is approached when the initial data is on this boundary is called a critical solution.
The critical behaviours of a universal (initial data family independent), self-similar critical solution, and supercritical black hole mass scaling with universal exponent were first observed by Choptuik in 1993 \cite{Choptuik93} in the context of the real massless scalar field, and have since been observed in many different scenarios \cite{GM-G07}.
In the Einstein-Yang-Mills system there is a richer breadth of critical phenomena, with ``type I" behaviour of a static critical solution and a supercritical black hole mass gap being distinguished from the ``type II" behaviour described above \cite{CCB96}. There is additionally ``Type III" behaviour, where the two end states are distinct Schwarzschild black holes and the critical solution is another static solution \cite{CHM99}.

In the relevant literature \cite{Choptuik93,HS96,CCB96,CHM99,Rinne14}, there is a parameter $ p $ in the initial data family that is tuned to the critical value $ p^* $. In each case it is claimed that this tuning is to machine precision. Before performing this tuning with our code we think it is important to expand on what this claim means.

While performing the bisection search for a critical parameter value with the code as it was being developed, it was noticed that slight changes in the algorithm (maintaining second-order convergence) would result in quite different critical parameters, only agreeing to the first few significant figures.
With the final version of the code there remain options (outside of the initial conditions) that ideally should not affect the critical parameter, but do: the choice of $ \alpha_0 $, and the choices of \verb_ns_, $ \epsilon_u $, and $\epsilon_v $.

So what could it mean to find the critical parameter $ p^* $ to machine precision? We believe that this is sensible if we consider it to depend not just on the initial condition family, but also on the final choice of algorithm, including all the options and hardware. Thus different algorithm/hardware combinations, given the same initial conditions, will find a different critical parameter.
This is not unreasonable because no numerical method can represent an arbitrary set of initial conditions with infinite accuracy. Rather, at least in the case of finite difference methods, the initial functions are represented by a finite set of points each with machine precision.

If we consider the (exact) initial conditions as a point $ f $ in an appropriate Banach space, then we can consider the numerically-defined initial conditions (at $ n $ points) as an equivalence class of all the functions in that space that pass through the $ n $ specified values at the determined points.
This will be the representative $ f $ (up to roundoff error) plus an infinite dimensional linear subspace (functions that are zero at the given points) so that it has codimension $ n $.

If the exact solution forms a continuous line in the Banach space, then the numerical solution defines a sequence of affine subspaces, which generically will not intersect this line, but should be close since the error at the evaluated points is assumed to be small. The AMR algorithm will periodically add points to (and remove points from) the numerical representation, which will increase (and decrease) the codimension of the affine subspace. So as finer grids are added, the codimension increases, and the solution becomes more precise. It is this refining (based on interpolation) that effectively further specifies the initial condition from the coarse initial equivalence class to a more precise equivalence class.
Thus in this sense the initial condition is determined not just from the initial specification, but also the way in which the AMR algorithm interpolates the functions, as well as when it does (the details of the algorithm).
So we see that it is sensible that the critical parameter depends not just on the nominal initial conditions, but also on the algorithm details.

This numerical issue is brought into focus in this particular scenario because on the actual boundary between dispersal and collapse, the system of partial differential equations is not well-posed (small changes in the initial data do not produce small changes in the solution). Far away from this boundary, it is not as critical where in the equivalence class the algorithm puts the effective initial conditions, since any choice will produce a similar solution.

One other consideration worth mentioning is that when performing a bisection search such as between dispersal and collapse, a result is always guaranteed because those are the only two possible outcomes. This does not mean, however, that the code will produce dispersal for all $ p < p^* $ and collapse for all $ p > p^* $ (for example).
In initial tests it was found that it took on the order of ten increments of machine precision ($ \approx 10^{-16} $) either side of $ p^* $ for the solution to be reliably dispersal or collapse for parameters further from $ p^* $.

Generic critical behaviour has been described in \cite{Gundlach97-2,Gundlach03,GM-G07}; we recall a few basic results when considering the linearisation around a critical solution.

If the critical solution is stationary (Type I) then the linearisation is
\begin{equation} \label{linearisation}
Z(\tau,x) = Z^*(x) + C(p) e^{\lambda \tau} \delta Z(x) \;,
\end{equation}
where $ (\tau,x) $ are coordinates adapted to the symmetry (such as polar-areal coordinates), $ Z $ is a representative variable, $ Z^* $ is the exactly critical solution, $ \delta Z(x) $ is a linear perturbation, $ C(p) $ is the perturbation amplitude which depends on the parameter $ p $, and $ \lambda > 0 $ is the eigenvalue of the one growing mode.
The lifetime $ T $ of the intermediate state is given by
\[ T = -\frac{1}{\lambda}\ln|p-p^*| + c \;, \]
where $ c $ is a constant.

If the critical solution is continuously self-similar (Type II) then the linearisation of a scale-invariant variable $ Z $ is
\begin{equation*}
Z(\tau,x) = Z^*(x) + C(p) e^{\lambda \tau} \delta Z(x) \;,
\end{equation*}
where $ (\tau,x) $ are coordinates adapted to the symmetry (such as $ x = \frac{r}{t^*-t} $, $ \tau = -\ln\left(\frac{t^*-t}{l}\right) $, where $ (t,r) $ are polar-areal coordinates and $ t^* $ and $ l $ are constants that depend on the family of initial data \cite{Gundlach03}).
The mass of the black hole $ M $ in the supercritical regime follows
\[ M \propto (p-p^*)^\frac{1}{\lambda} \;. \]
If the critical solution is discretely self-similar with period $ \Delta $ (Type II) then there is additional fine structure
\begin{equation} \label{m supercritical}
\ln M = \frac{1}{\lambda} \ln(p-p^*) + c + f\left(\frac{1}{\lambda}\ln(p-p^*) +c\right),
\end{equation}
where $ f $ is periodic with nominal period $ \Delta $, and only the constant $ c $ depends on the initial data family.
The observed period of $ f $ is $ \frac{\Delta}{2} $ because the mass depends quadratically on the Yang-Mills variables \cite{HP97}, see (\ref{mr}) or (\ref{mv}).

\section{Instability of static solutions} \label{s:PM instability}
An important difference between our code and the majority of codes used previously is that we set initial data on an outgoing null cone, rather than a spacelike Cauchy surface.
This means that we cannot set an unstable static solution as initial data and simply observe the time it takes for numerical errors to evolve the solution into either collapse or dispersal as a test of the accuracy of the method, as in \cite{ZS91,Zhou92}.
We can of course set the initial condition on $ u = 0 $ to be the first Bartnik-McKinnon solution (BK1, with gauge potential $ w_1 $), plus a small perturbation, whose sign should dictate which end-state is reached.

Since we require $ W_0(r) $ for the initial conditions (\ref{ICs}), we first calculate the solution to the static (magnetic) equations (\ref{EYME stat}) in terms of $ W $ instead of $ w $:
\begin{subequations} \label{EYME W stat}
\begin{align}
m' &= \frac{r^2W^2}{2}(2-r^2W)^2 + r(r-2m)(2W+r^2P)^2 \;, \\
S' &= 2rS(2W+r^2P)^2 \;, \\
W' &= rP \;, \\
P' &= -\frac{5P+W^2(3-8W)}{rN} +\left(\frac{8m}{r^2}+rW^2(2-r^2W)^2\right)\frac{P}{N} + \frac{rW^3}{N}\left(1-8W+2r^2W^2\right). \label{P stat}
\end{align}
\end{subequations}
Multiplying (\ref{P stat}) by $ r $ and evaluating at $ r = 0 $ gives the initial condition for $ P $. Also recall that $ m = O(r^3) $, and observe that $ S $, $ W $, and $ P $ are even functions of $ r $, so all the first derivatives are zero at the origin.
The initial conditions are
\begin{align*}
m(0) &= 0 \;, & S(0) &= 1 \;, & W(0) &= -b_1 \;, & P(0) &= -\frac{8b_1^3+3b_1^2}{5} \;,
\end{align*}
where $ b_1 $ is the free parameter calculated earlier and listed in table \ref{table:BKs}.
Only (\ref{P stat}) is formally singular at $ r = 0 $, but since it is straightforward to set $ P'(0) = 0 $, (\ref{EYME W stat}) can be solved in Mathematica from $ r = 0 $.
We confirm that repeating the previous bisection search for $ b_1 $ with this system gives an identical result.

Zhou and Straumann \cite{ZS91,Zhou92} found that a negative Gaussian added onto $ w_1 $ would result in collapse, while a positive Gaussian results in dispersal.
We can achieve similar initial conditions by adding or subtracting a Gaussian onto $ W_1 $, and we find a positive Gaussian would result in collapse, and a negative Gaussian would result in collapse.
Taking into account the sign in the definition of $ W $ (\ref{WDZ}) we see that the end-states are the same for these initial conditions on $ u = 0 $ as those for $ t = 0 $.

We can find the unstable eigenvalue and eigenmode ($ \lambda $ and $ Z(x) $ in (\ref{linearisation})) easily in polar-areal coordinates similarly to \cite{SZ90-1,Zhou92}. With the purely magnetic ansatz the EYM equations (\ref{EEtr},\ref{YMEtr}) simplify considerably.
\begin{subequations}
We rewrite (\ref{mt}) and (\ref{mr}) in terms of $ N $;
\begin{align}
\dot{N} &= -\frac{4N\dot{w}w'}{r} \;, \label{Nt} \\
N' &= -\frac{N}{r}\left(1+2\left(\frac{\dot{w}^2}{S^2N^2}+w'^2\right)\right) +\frac{1}{r}\left(1-\frac{(w^2-1)^2}{r^2}\right). \label{Nr}
\end{align}
We use the following combination of (\ref{mr}) and (\ref{Sr});
\begin{equation}
(SN)' = \frac{S}{r}\left(-N+1-\frac{(w^2-1)^2}{r^2}\right).
\end{equation}
The one remaining Yang-Mills equation (\ref{w}) can then be written with $ S $ removed from all terms bar the time derivatives:
\begin{equation} \label{wtt}
\frac{\ddot{w}}{S^2N} -\frac{(SN)^.\dot{w}}{S^3N^2} -Nw'' -\frac{1}{r}\left(-N+1-\frac{(w^2-1)^2}{r^2}\right)w' = -w\frac{w^2-1}{r^2} \;.
\end{equation}
\end{subequations}

Linearising (\ref{Nt}) around the static BK1 solution we obtain
\[ \dot{\delta N} = -\frac{4Nw'}{r}\dot{\delta w} \;, \]
and integrating with respect to $ t $ we find $ \delta N = -\frac{4Nw'}{r}\delta w + f(r) $ for some arbitrary function $ f $. Inserting this into the linearisation of (\ref{Nr}),
\[ \delta N' = -\frac{1+2w'^2}{r}\delta N -\frac{4Nw'}{r}\delta w' -\frac{4w(w^2-1)}{r^3}\delta w \;, \]
and using the static Yang-Mills equation (\ref{w stat}) produces
\[ f' +\frac{1+2w'^2}{r}f = 0 \;. \]
Since regularity requires $ \delta N = \delta w = 0 $ at $ r = 0 $, we desire $ f(0) = 0 $. The unique solution with this boundary condition is $ f(r) \equiv 0 $, and so we have
\begin{equation} \label{dN}
\delta N = -\frac{4Nw'}{r}\delta w \;.
\end{equation}

Linearising (\ref{wtt}) we obtain
\[ \frac{\ddot{\delta w}}{S^2N} -N\delta w'' -\frac{(SN)'}{S}\delta w' -w''\delta N +\frac{w'}{r}\delta N +\frac{4w(w^2-1)w'}{r^3}\delta w + \frac{3w^2-1}{r^2} \delta w = 0 \;. \]
Multiplying by $ S^2N $, using the static Yang-Mills equation (\ref{w stat}), and the formula for $ \delta N $ (\ref{dN}) this reduces to
\[ \ddot{\delta w} -SN(SN\delta w')' +S^2N\left(\frac{3w^2-1}{r^2} +\frac{8w(w^2-1)w'}{r^3} -\frac{4w'^2}{r^2}\left(1-\frac{(w^2-1)^2}{r^2}\right)\right)\delta w = 0 \;. \]

As in \cite{SZ90-1,Zhou92}, we find it convenient to convert to the isothermal coordinate $ R $ of the static solution, so $ S^2N = \alpha^2 $ and $ SN\dydx{}{r} = \dydx{}{R} $. We obtain
\[ \ddot{\delta w} -\delta w_{RR} +\alpha^2\left(\frac{3w^2-1}{r^2} +\frac{8w(w^2-1)w_R}{r^3r_R} -\frac{4w\indices{_R^2}}{r^2r\indices{_R^2}}\left(1-\frac{(w^2-1)^2}{r^2}\right)\right)\delta w = 0 \;. \]
Note that this is not the same as the equations that would result from performing the linearisation in isothermal coordinates, which would also have $ \delta r $ terms.

\begin{figure}[!ht]
  \centering
  \includegraphics[scale=0.8]{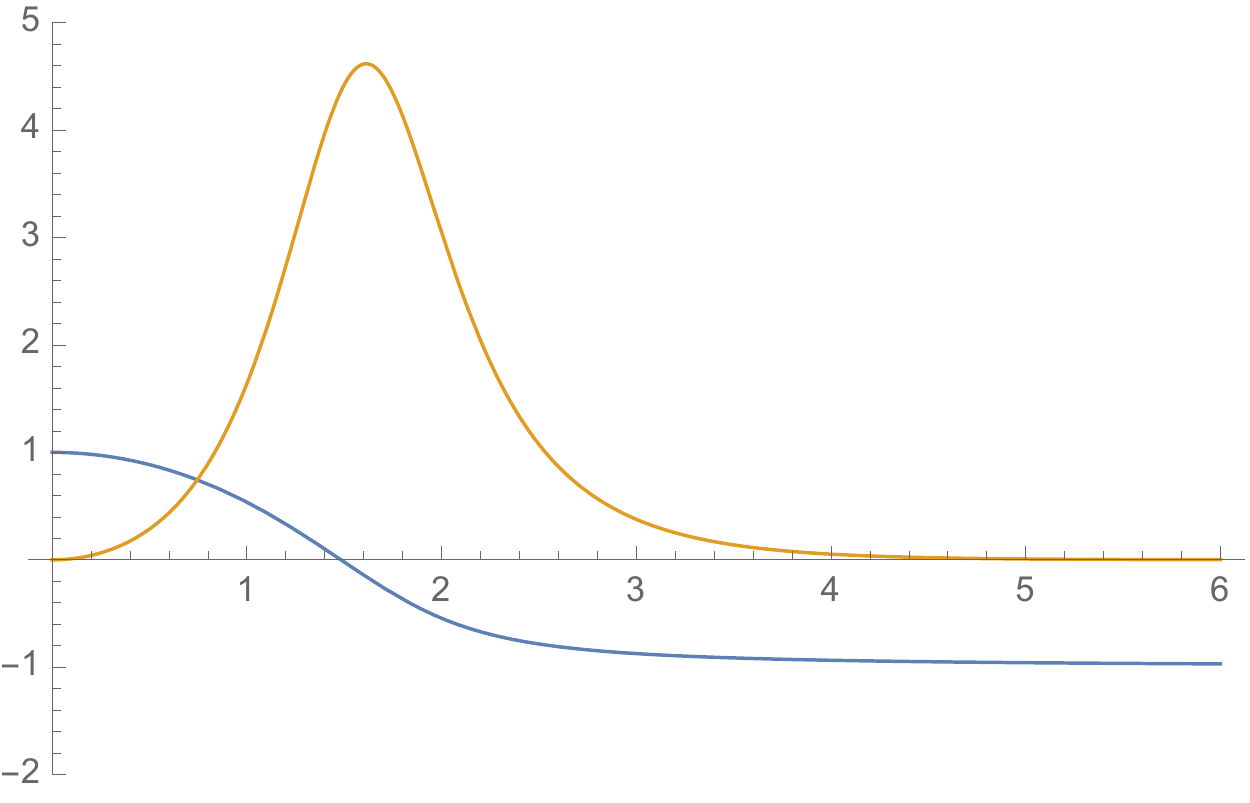}
  \caption{A plot of the first BK solution $ w(R) $ (blue) and the unstable eigenmode $ \xi(R) $ (yellow), with respect to the isothermal coordinate $ R $.}
  \label{fig:BK1 eigenmode}
\end{figure}

We insert the ansatz $ \delta w(t,R) = e^{\lambda t}\xi(R) $ and obtain
\begin{equation} \label{xiRR}
\xi_{RR}-\alpha^2\left(\frac{3w^2-1}{r^2} +\frac{8w(w^2-1)w_R}{r^3r_R} -\frac{4w\indices{_R^2}}{r^2r\indices{_R^2}}\left(1-\frac{(w^2-1)^2}{r^2}\right)\right)\xi = \lambda^2 \xi \;.
\end{equation}
We solve this eigenvalue problem numerically, shooting from $ r = 0 $ and varying $ \lambda $ so that the solution remains bounded.
We use a Taylor series for $ \xi $ to obtain initial conditions at $ r > 0 $;
\begin{equation} \label{xi}
\xi(R) = e\left(R^2 + \frac{6b_1+12b_1^2+\lambda^2}{10}R^4 + O(R^6)\right),
\end{equation}
where $ b_1 $ is the parameter found for the first Bartnik-McKinnon solution listed in table \ref{table:BKs}.
Because (\ref{xiRR}) is a linear homogeneous equation, there is a freedom to scale $ \xi $; we fix this by setting the coefficient of $ R^2 $ to be one ($ e = 1 $). The resulting eigenvalue is
\begin{equation} \label{lambdaI}
\lambda_\text{I} = 1.812 173 86 \;,
\end{equation}
where the subscript I denotes that this is the eigenvalue for the type I critical solution.
This is consistent with the result of \cite{SZ90-1,Zhou92} ($ \sigma^2 = -0.0525 $) who fix $ t $ to be the proper time at infinity. In our notation, $ \sigma^2 = -\frac{\lambda^2}{S_1^2} $, where $ S_1 $ is given in table \ref{table:BKs inf}.
The unstable mode of BK1 is plotted in figure \ref{fig:BK1 eigenmode}.

It is also possible to find an unstable mode with non-zero electric field \cite{VBLS95}. This is simple because for the purely magnetic BK1 solution, the two kinds of perturbations decouple. When the static solution is not purely magnetic this is no longer the case, so we do not have a linear stability analysis of the long-lived solutions.

Another important restriction of this code is that (just like \cite{HS96}) it assumes a regular origin, at least initially. This means that one cannot begin the evolution from a static (or perturbed static) black hole configuration.
Thus we cannot directly observe the evolutions of perturbations of Schwarzschild, Reissner-Nordstr\"{o}m, or EYM black holes. However, the latter two families are unstable and we will see that it is possible to evolve towards these configurations by appropriate tuning of the initial data. The Schwarzschild family is of course stable and a natural endpoint of evolution, along with Minkowski space.

\section{Purely magnetic critical phenomena}
In this section we use our code to investigate the critical phenomena occurring in the purely magnetic case.
We wish to make our analysis of the numerical results as independent of the coordinate choice as possible, thus wherever possible we will restrict to using data on the origin.
For all of the results in this chapter we set \verb_pxbe = false_ so we do not solve for $ p $, $ x $, or $ \tilde\beta $.

\subsection{Type I: the BK1 intermediate attractor}
Since the BK1 solution has $ w $ go smoothly from $ 1 $ at the origin to $ -1 $ at infinity, a common choice of an initial data family that may evolve to the BK1 solution after tuning is a modification of the $ \tanh(r) $ function. Such a choice has been used since the first work on this subject \cite{CCB96}. The function is essentially $ w_0(r) = \tanh\left(\frac{x-r}{s}\right) $ so that the tuneable parameters $ x $ and $ s $ correspond to the position and width of the step respectively. However it must be modified to satisfy the regularity requirements at the origin.
In the Cauchy problem, this is more difficult because $ w_0 $ is required to be an even function of $ r $.
In our characteristic initial value problem however we merely need to ensure $ w_0(0) = 1 $ and $ w_0'(0) = 0 $, which can be achieved with the adjustment detailed in \cite{CCB96}. Writing this explicitly in terms of $ W_0 $ we use
\[ W_\text{tanh}(r,x,s) = \begin{cases} \frac{1+\coth\left(\frac{x}{s}\right)^2-2\tanh\left(\frac{x}{s}\right)}{s^2} & \text{if } r = 0 \\
 \frac{1-\left(1+\left(\coth\left(\frac{x}{s}\right)-1\right)\left(1+\left(\coth\left(\frac{x}{s}\right)+1\right)\frac{r}{s}\right) e^{-2\left(\frac{r}{s}\right)^2}\right)\tanh\left(\frac{x-r}{s}\right)}{r^2} & \text{if } r > 0 \end{cases} \;. \]

We perform a convergence test similar to that of \cite{HLPTA00} in their appendix B. They consider the convergence of the critical parameter $ p^* $ as the number of grid points is increased (for the type II critical collapse in the SU(2) $ \sigma $ model coupled to gravity). We do what is analogous for our algorithm and observe how $ p^* $ converges as the tolerances $ \epsilon_u $ and $ \epsilon_v $ are decreased. Since a doubling of the resolution results in a local truncation error decrease by a factor of 8, we choose values of $ \epsilon_u $ and $ \epsilon_v $ that are powers of $ \frac{1}{8} $.

We here use the modified hyperbolic tangent function with position 5 and width given by the variable parameter $ p $:
\begin{equation} \label{ICs I}
\begin{aligned}
W_0(r) &= W_\text{tanh}(r,5,p) \;, & \alpha_0 &= 10 \;, \\
D_0(r) &= 0 \;, & \tilde\gamma_0(v) &= 0 \;.
\end{aligned}
\end{equation}
The results (all using \verb_ns = 512_) are shown in table \ref{table:pstarI}.
Our algorithm works best when the amount of refinement in the $ u $ and $ v $ directions is roughly the same. In particular, asking for much $ u $ refinement without any $ v $ refinement can cause spurious $ u $ refinement around the high $ \text{TE}_v $ values.
The values of $ p^* $ are found by performing a bisection search; given two separated values of $ p $ that cause the evolution of (\ref{ICs I}) to disperse and collapse respectively ($ p^* $ is bracketed), the mean value of $ p $ is tested and replaces the appropriate bound on $ p^* $, and this is repeated until the bounds for $ p^* $ are sufficiently close.
Here we only determine $ p^* $ to the resolution required to see the convergence of the algorithm as the tolerances go to zero, often it will be to machine precision.
Here and throughout we indicate a bracketed value by putting some of the differing digits in subscripts and superscripts, both of which are truncated, not rounded.

\begin{singlespace}
\begin{table}[!ht]
  \centering
  \renewcommand{\arraystretch}{1.5}
  \begin{tabular}{c|c|cccc|}
    \multicolumn{1}{c}{} & \multicolumn{1}{c}{} & \multicolumn{4}{c}{$ \epsilon_v $} \\
    \cline{3-6}
    \multicolumn{1}{c}{} & & $ \frac{1}{8^6} $ & $ \frac{1}{8^7} $ & $ \frac{1}{8^8} $ & $ \frac{1}{8^9} $ \\
    \cline{2-6}
    \multirow{4}{*}{$ \epsilon_u $} & $ \frac{1}{8^4} $ & $ 1.3872110_{54}^{84} $ & $ 1.3872206_{50}^{80} $ & &  \\
     & $ \frac{1}{8^5} $ & $ 1.387296_{289}^{319} $ & $ 1.3873050_{21}^{51} $ & $ 1.387291_{580}^{610} $ & \\
     & $ \frac{1}{8^6} $ & & $ 1.387528_{683}^{713} $ & $ 1.3875148_{53}^{83} $ & $ 1.3875092_{32}^{62} $ \\
     & $ \frac{1}{8^7} $ & & & $ 1.3876343_{54}^{83} $ & $ 1.3876294_{32}^{61} $ \\
    \cline{2-6}
  \end{tabular}
  \renewcommand{\arraystretch}{1}
  \caption{Bounds for $ p^* $ depending on the truncation error tolerances, for Type I critical behaviour.}
  \label{table:pstarI}
\end{table}
\end{singlespace}

We observe that $ \epsilon_v $ has little effect on the value of $ p^* $ other than a finer $ \Delta v $ allowing greater $ u $ resolution (without spurious refinement). Decreasing $ \epsilon_u $ by a factor of 8 appears to decrease the error in $ p^* $ by a factor of 2 (for small $ \epsilon_u $). We might have expected a factor of four here, so this is a bit disappointing.
This is somewhat surprising too because we would have expected a decrease in $ \epsilon_v $ to correspond to a more specific initial condition (and more precise evolution), and thus be changing the value of $ p^* $.

Observing the unstable eigenmode $ \xi(R) $ (\ref{xi}) we note that it has a non-zero $ R^2 $ term, which means that $ \delta W $ is non-zero at the origin.
This allows us to use $ W $ at $ r = 0 $ as a simple way of measuring the closeness of the numerical evolution to the static BK1 solution.
Note that this is conceptually much simpler than the method used in \cite{CCB96} where they measure the crossing of $ w $ passing consecutively higher values of $ r $ as the solution disperses. That method depends on the slicing choice (polar-areal) and only works for evolutions that disperse.

\begin{figure}[!ht]
  \centering
  \includegraphics[scale=0.9]{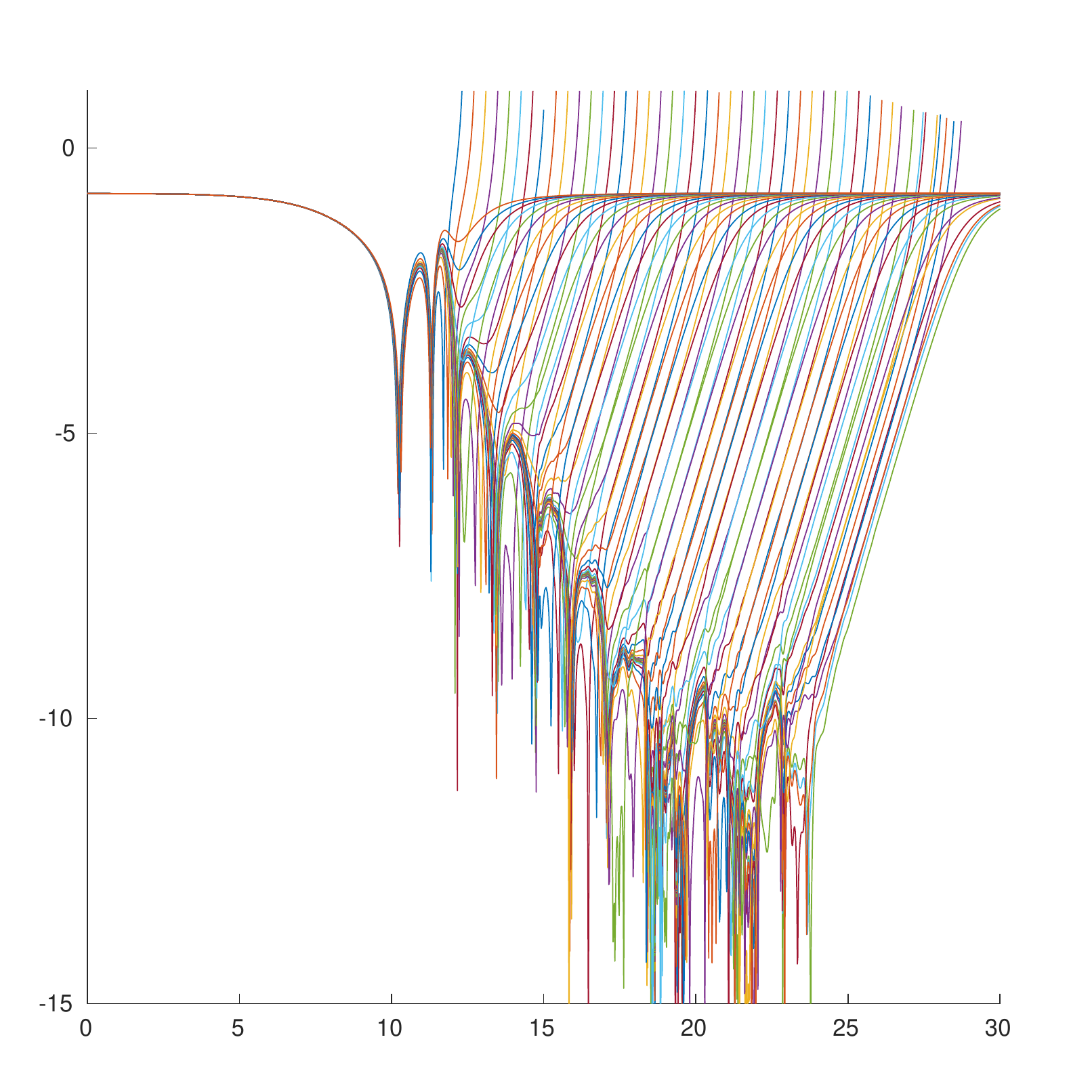}
  \caption{A plot of $ \ln|W(r=0)-b_1| $ versus proper time for a large number of evolutions with parameter $ p $ close to the critical parameter $ p^* $.}
  \label{fig:BK1 W(0)}
\end{figure}

We test the evolution by extracting $ \lambda_\text{I} $ from the numerical evolution data, using $ (\epsilon_u,\epsilon_v) = \left(\frac{1}{8^4},\frac{1}{8^6}\right) $, and comparing with (\ref{lambdaI}).
We do this by first finding $ p ^* $ to machine precision ($ p^* = 1.387211066176_{6998696}^{7000916} $) and then evolving the initial data (\ref{ICs I}) for a large number of values of $ p $ satisfying $ -\log_2|p-p^*| = \{7,8,\ldots,51,52\} $ (we use the upper value for $ p^* $ since that solution remained near BK1 for longer).
We then plot $ W $ at $ r = 0 $ as a function of the proper time (figure \ref{fig:BK1 W(0)}).
We see in figure \ref{fig:BK1 W(0)} that for near-critical evolutions, $ W $ at the origin oscillates around and approaches $ b_1 $ via a quasi-normal mode, then grows exponentially before the solution either decays ($ W \to 0 $ as $ \tau \to \infty $) or collapses ($ W \to \infty $ in finite proper time).
We also see from the equal spacings between the exponential (straight) sections that the time spent near the BK1 solution is proportional to $ \log_2|p-p^*| $.
Note that for 2 values of $ p $ either side but equidistant from $ p^* $, the solutions simultaneously converge towards the critical solution, evolve separately for a time, agree again for the exponentially growing phase, and finally diverge again.
Note the pattern is not so clear as $ p $ gets very close to $ p^* $ as here the small uncertainty in $ p^* $ ($ < 2^{-52} $) becomes significant when $ \log_2|p-p^*| $ is calculated.

\begin{figure}[!ht]
  \centering
  \includegraphics[scale=0.5]{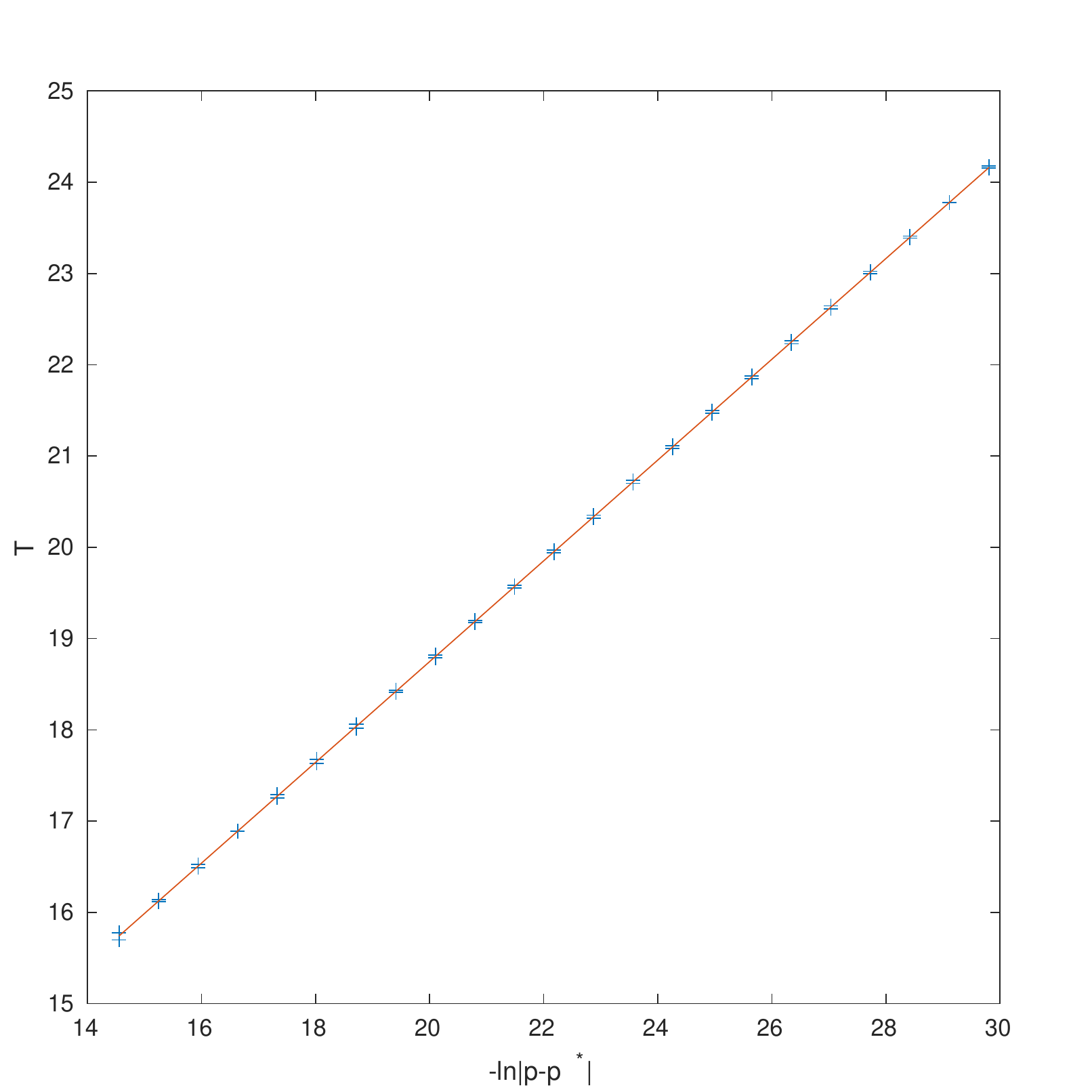}
  \caption{A plot of the proper time at the origin when $ W $ diverges from $ b_1 $ by 0.01 as a function of $ -\ln|p-p^*| $ and the line of best fit.}
  \label{fig:BK1 T}
\end{figure}

We use the region of the exponential growth where the evolutions of the two equidistant values of $ p $ approximately agree to determine $ \lambda_\text{I} $. We calculate the time $ T $ when the solutions cross $ |W-b_1| = 0.01 $ and plot that as a function of $ -\ln|p-p^*| $, shown in figure \ref{fig:BK1 T}.
The gradient of the fitted line is 0.5521 which compares reasonably to the expected value of $ \frac{1}{\lambda_\text{I}} = 0.551823 $ using (\ref{lambdaI}). 
This is more precise than the value obtained in \cite{CCB96} because we calculate only at $ r = 0 $ and can use supercritical evolutions as well as subcritical evolutions.

\begin{figure}[!ht]
  \centering
  \includegraphics[scale=0.5]{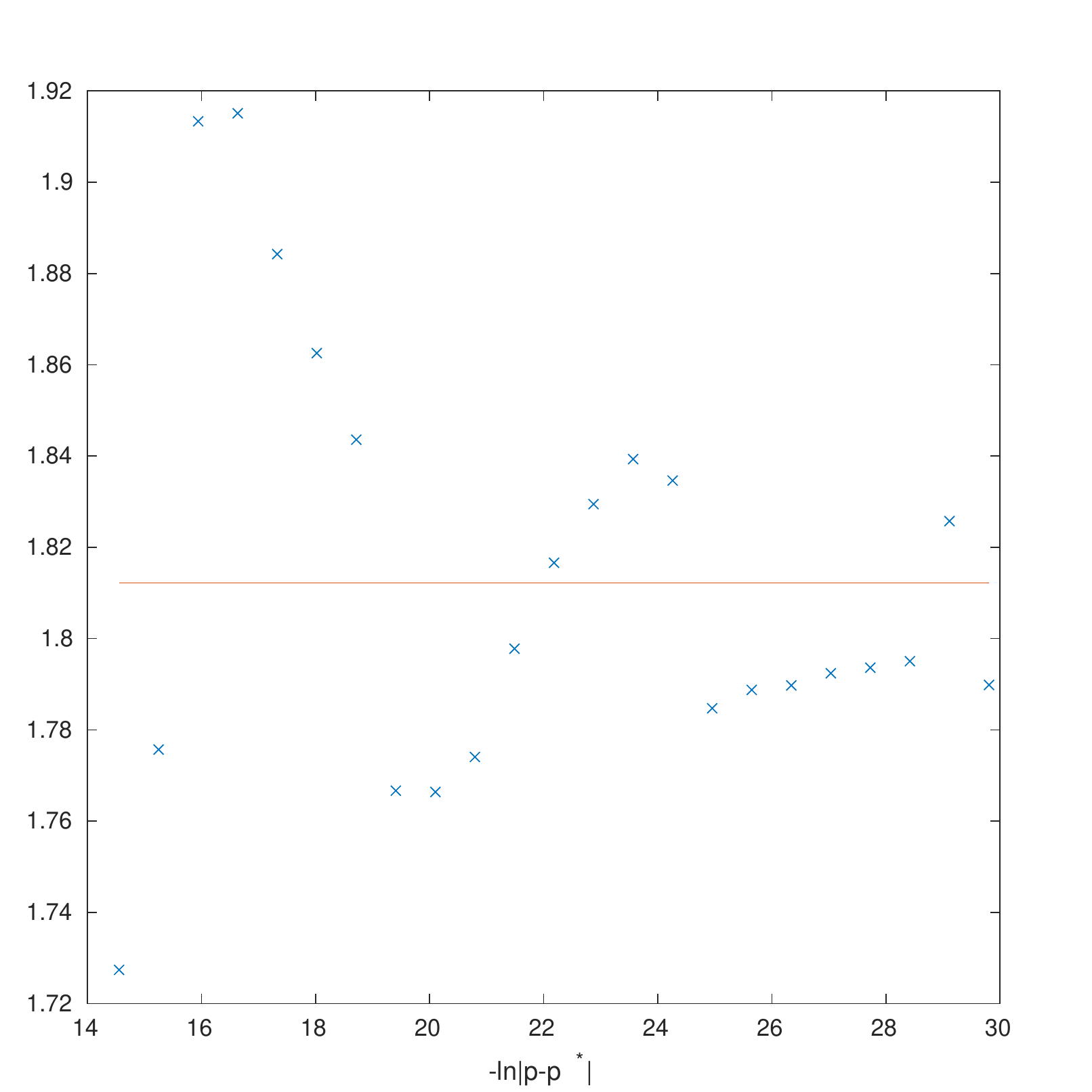}
  \caption{A plot of the gradient of the exponential growth of near-critical evolutions as a function of $ -\ln|p-p^*| $ (blue) with the theoretical value (orange).}
  \label{fig:BK1 grad}
\end{figure}

There is another way to obtain the value of $ \lambda_\text{I} $ from the data of figure \ref{fig:BK1 W(0)}, and that is to measure the gradient of the exponential sections, which should be $ \lambda_\text{I} $ itself. We do this by considering only the exponential section where the two solutions $ \ln(W-b_1) $ from $ p $ equidistant from $ p^* $ coincide within a tolerance of 0.1. We plot the gradient over these regions as a function of $ -\ln|p-p^*| $ in figure \ref{fig:BK1 grad}.
We see from this figure that the gradients are consistent with theory but not an accurate way to determine $ \lambda_\text{I} $.

\subsection{Type II: the self-similar solution}
We perform the same test as for the Type I critical phenomena to see the dependence of $ p^* $ on the error bounds. 
We take purely magnetic initial data:
\begin{equation} \label{ID m}
\begin{aligned}
W_0(r) &= -p e^{-(r-5)^2} \;, & \alpha_0 &= 10 \;, \\
D_0(r) &= 0 \;, & \tilde\gamma(v) &= 0 \;,
\end{aligned}
\end{equation}
(with $ \verb_ns_ = 512 $ and \verb_pxbe = false_) and tune $ p $ to the boundary of black hole formation using a bisection search as described above. We do this for various values of the tolerances $ \epsilon_u $ and $ \epsilon_v $.
Table \ref{table:pstar} shows the results.
Similarly to the Type I case, we observe that $ \epsilon_v $ has little effect on the value of $ p^* $, and decreasing $ \epsilon_u $ by a factor of 8 appears to decrease the error in $ p^* $ by a factor of 2 (for small $ \epsilon_u $).

\begin{singlespace}
\begin{table}[!ht]
  \centering
  \renewcommand{\arraystretch}{1.5}
  \begin{tabular}{c|c|cccccc|}
    \multicolumn{1}{c}{} & \multicolumn{1}{c}{} & \multicolumn{6}{c}{$ \epsilon_v $} \\
    \cline{3-8}
    \multicolumn{1}{c}{} & & $ \frac{1}{8^4} $ & $ \frac{1}{8^5} $ & $ \frac{1}{8^6} $ & $ \frac{1}{8^7} $ & $ \frac{1}{8^8} $ & $ \frac{1}{8^9} $ \\
    \cline{2-8}
    \multirow{6}{*}{$ \epsilon_u $} & $ \frac{1}{8^2} $ & $ 0.035312_{592}^{622} $ & $ 0.0353125_{63}^{92} $ & $ 0.0353105_{06}^{36} $ & & & \\
     & $ \frac{1}{8^3} $ & $ 0.0353188_{21}^{51} $ & $ 0.0353188_{51}^{81} $ & $ 0.0353167_{65}^{95} $ & $ 0.0353172_{12}^{42} $ & & \\
     & $ \frac{1}{8^4} $ & $ 0.0353198_{34}^{64} $ & $ 0.0353198_{34}^{64} $ & $ 0.035317_{599}^{629} $ & $ 0.0353181_{36}^{66} $ & $ 0.035318_{374}^{404} $ &  \\
     & $ \frac{1}{8^5} $ & & $ 0.0353153_{34}^{64} $ & $ 0.0353123_{54}^{85} $ & $ 0.0353127_{41}^{71} $ & $ 0.0353129_{50}^{80} $ & $ 0.0353130_{40}^{69} $ \\
     & $ \frac{1}{8^6} $ & & & $ 0.0353097_{02}^{31} $ & $ 0.035310_{089}^{119} $ & $ 0.0353103_{57}^{87} $ & $ 0.035310_{387}^{417} $ \\
     & $ \frac{1}{8^7} $ & & & & $ 0.0353089_{57}^{86} $ & $ 0.0353093_{14}^{44} $ & $ 0.0353093_{44}^{74} $ \\
    \cline{2-8}
  \end{tabular}
  \renewcommand{\arraystretch}{1}
  \caption{Bounds for $ p^* $ depending on the truncation error tolerances.}
  \label{table:pstar}
\end{table}
\end{singlespace}

To determine the instability eigenvalue $ \lambda_{\text{II}} $ and the echoing period $ \Delta $ we first find $ p^* $ precisely (with $ (\epsilon_u,\epsilon_v) = \left(\frac{1}{8^4},\frac{1}{8^6}\right) $):
\[ p^* = 0.0353176030432238_{094}^{163} \;. \]
We then observe $ W $ at the origin with respect to proper time for the evolution of the initial data with these two values of $ p $; it oscillates with increasing amplitude and frequency before either the evolution ends due to collapse or it approaches zero as $ \tau \to \infty $.

We find the local maxima of $ \log|W| $ and calculate the gradient (with respect to an index of the oscillation) for the $ p^* $ bounds above which gives us a value of
\[ \Delta = 0.73_{761}^{833} \;. \]
This compares favourably with the precise value obtained in \cite{Gundlach97-2} ($ \Delta = 0.73784 \pm 0.00002 $) and is more precise than the value in \cite{CCB96}.

We can determine $ \Delta $ another way, which involves finding the critical time $ \tau^* $. We find accurately the positions of the zero crossings of $ W $ (using linear interpolation) and since $ e^\Delta = \frac{\tau^*-\tau_n}{\tau^*-\tau_{n+1}} $ we calculate $ \tau^* = \frac{\tau_{n+1}^2-\tau_n\tau_{n+2}}{2\tau_{n+1}-\tau_n-\tau_{n+2}} $. We find
\[ \tau^* = 13.0382 \;. \]
We then look at the gradient (with respect to the oscillation index) of $ -\ln(\tau^*-\tau_n) $ and find
\[ \Delta = 0.73_{718}^{810} \;. \]

\begin{figure}[!ht]
  \centering
  \includegraphics[scale=0.9]{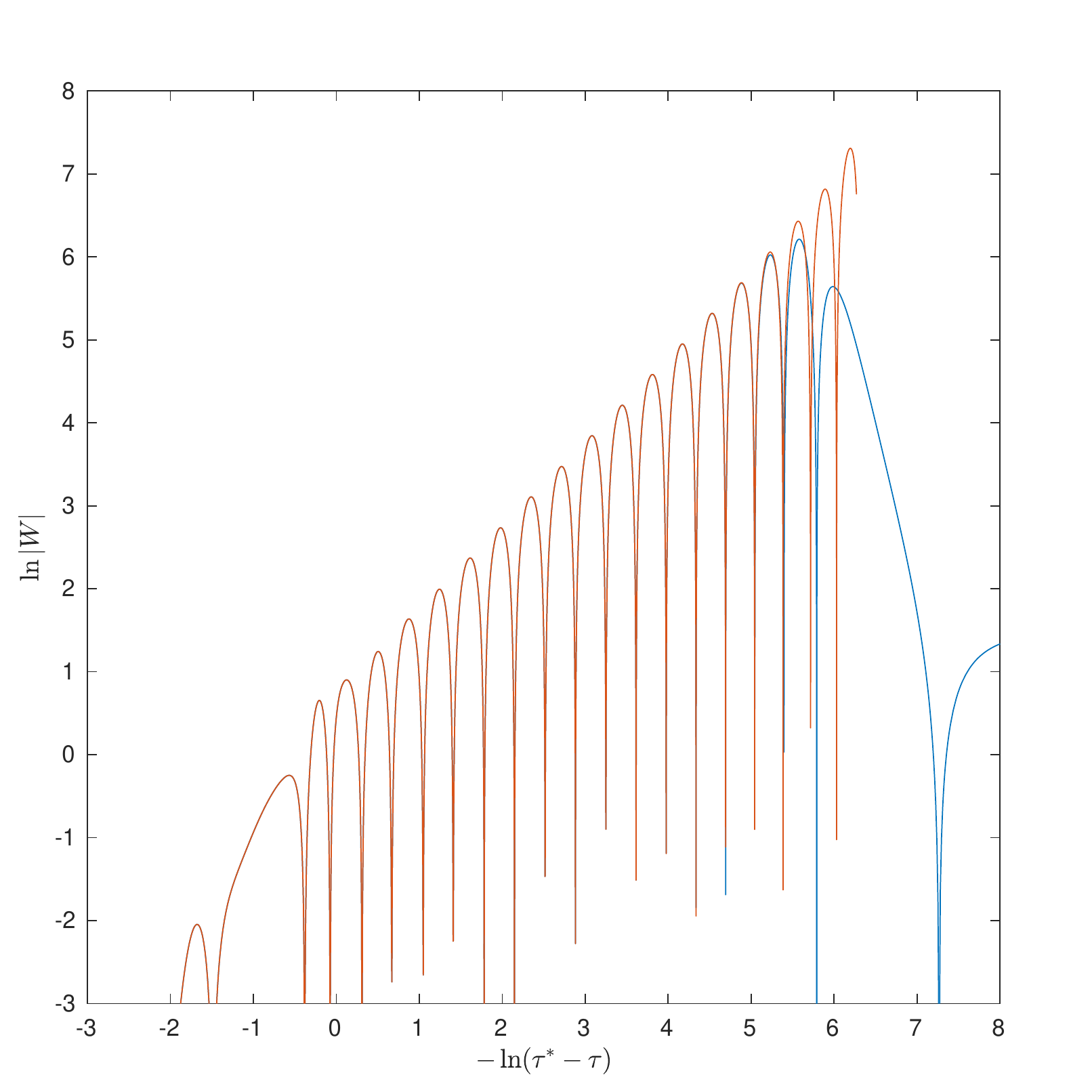}
  \caption{A plot of $ \ln|W| $ at the origin versus $ -\ln(\tau^*-\tau) $ for the just Type II subcritical (blue) and supercritical (orange) evolutions.}
  \label{fig:S-S W}
\end{figure}

Figure \ref{fig:S-S W} shows the behaviour of $ W $ at the origin for near critical evolutions. Note that the gradient of the maxima is 1 which corresponds to $ e^\Delta $ being both the growth factor for $ W $ and the reduction factor in $ \tau^*-\tau $.
These evolutions used 16 refinement levels (in $ v $) corresponding to an ultimate resolution of $ 2^{-25} $ or over 32 million points.

According to equation (\ref{m supercritical}), the instability eigenvalue $ \lambda_\text{II} $ manifests in the scaling of the mass of supercritical black holes.
Ideally we would find the black hole mass of supercritical evolutions by evaluating $ m $ at timelike infinity. However, in a black hole spacetime this is where future null infinity (and the MTT) meets the singular $ r = 0 $, and this will not be included in our data since we excise regions of the spacetime where the truncation error bounds are exceeded.
We have a few options to approximate $ m_{BH} $. Previous work investigating the mass scaling relation that did not evolve through the MTT \cite{Choptuik93,CCB96,PHA05} used the position $ r_h $ of the minimum of $ N $ to calculate the approximate black hole mass $ m_h = \frac{r_h}{2} $. This was because $ N $ could approach zero but never reach it in coordinates that don't penetrate the MTT.
The $ t = \text{constant} $ polar-areal slices approach an outgoing null direction near the MTT, so polar-areal and double-null coordinates should give similar results with this method.
Because analytically we expect the $ N = 0 $ MTT to begin at timelike infinity as $ u $ is increased, the results of finding the minimum of $ N $ at $ v < 1 $ come from the necessarily limited resolution.

\begin{figure}[!ht]
  \centering
  \includegraphics[scale=0.9]{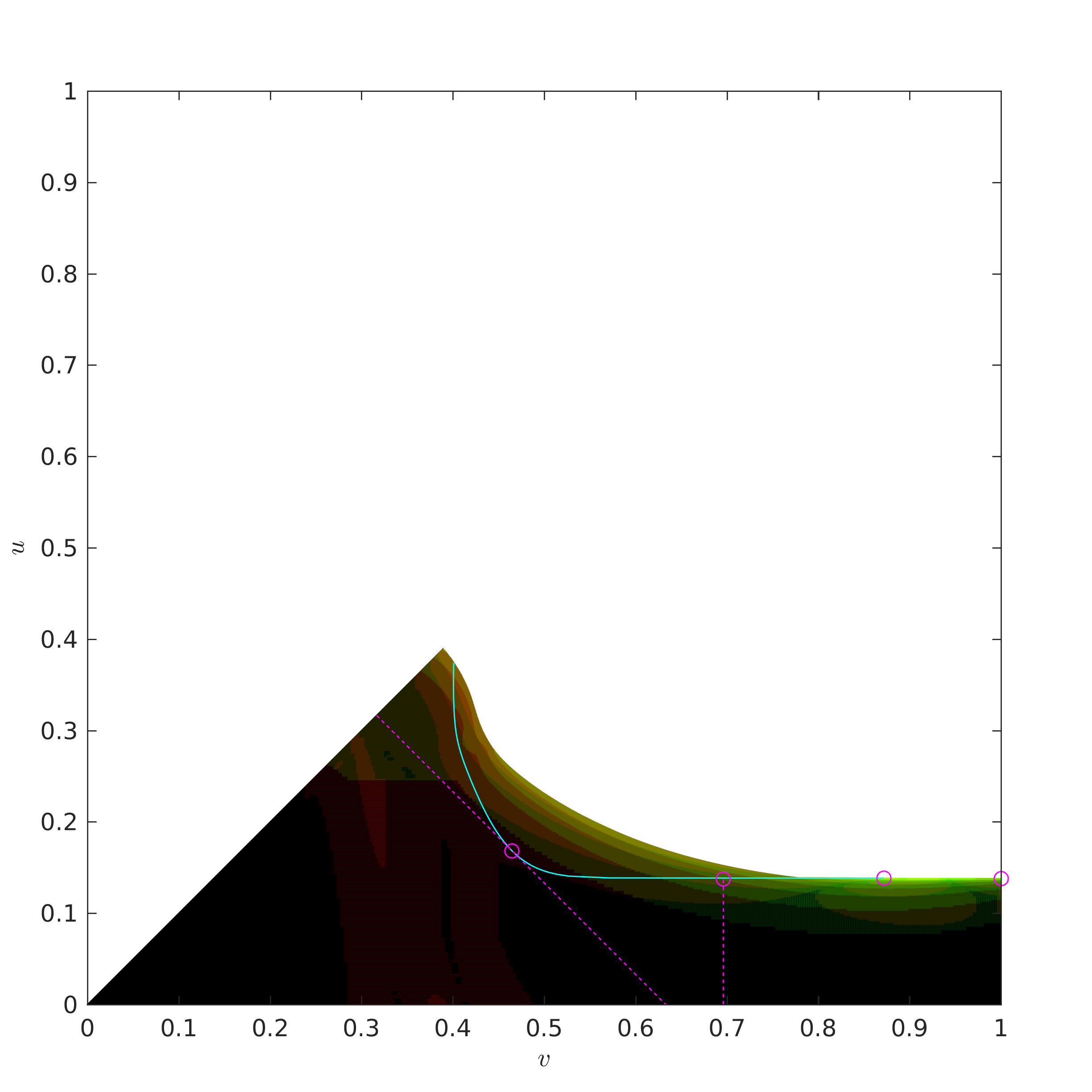}
  \caption{A plot of the mesh generated in a black hole evolution, (\ref{ID m}) with $ p = p^*+2^{-5} $, with the MTT marked in cyan. The maximum refinement levels in both directions are 8, and the four positions the mass of the black hole is calculated at are marked in magenta.}
  \label{fig:BH horizon}
\end{figure}

Since we actually evolve through the MTT, we use alternative methods to approximate the black hole mass.
Figure \ref{fig:BH horizon} shows the result of a (not near critical) black hole evolution; note that points near the singularity are excised due to the excessive refinement required to resolve them, including points near the regular origin and timelike infinity.
We approximate the black hole mass in four ways.
\begin{enumerate}
\item Calculate the apparent horizon mass $ m_h = \frac{r_h}{2} $ on the largest $ v $-point on the MTT,
\item Calculate the Bondi mass $ m_B = -\frac{\tilde{r}^2\tilde{f}_v}{2(1-u)^2\tilde\alpha^2} $ on the largest $ u $-point on future null infinity,
\item Calculate the apparent horizon mass $ m_h = \frac{r_h}{2} $ at a predetermined value of $ v = v_0 $, and
\item Calculate the apparent horizon mass $ m_h = \frac{r_h}{2} $ at the minimum value of $ u+v $ on the horizon.
\end{enumerate}
The first two methods take the values of $ m $ closest to timelike infinity, according to the maximum refinement levels chosen.
The Bondi mass in particular could however include mass contributions from late fall-in of matter rather than from the initial just-supercritical black hole formation, as pointed out in \cite{PHA05,GM-G07}. This will be tested below.
The third is the method used by \cite{HS96}.
The fourth is basically the method of following the $ N $ minimum described above for the horizon-penetrating isothermal coordinates.
The position of all these calculations are also shown on figure \ref{fig:BH horizon}.

We wish to compare these definitions of the black hole mass and see if the results like (\ref{m supercritical}) are dependent on the definition of $ m $.
Using $ (\epsilon_u,\epsilon_v) = \left(\frac{1}{8^2},\frac{1}{8^4}\right) $, we perform evolutions of (\ref{ID m}) with $ p = p^* + 2^{-n} $. With these tolerances we find $ p^* = 0.03531260825930455_{23}^{92} $, and we take $ n $ from $ 4.8 $ to $ 57 $ in steps of $ 0.2 $, noting that for $ n \leqslant 4.6 $, the initial conditions contain regions of $ N < 0 $ indicating past trapped regions.
We choose $ v_0 = 0.696 $ for (iii) so that the line intersects the MTT in each case.

\begin{figure}[!ht]
  \centering
  \includegraphics[scale=0.9]{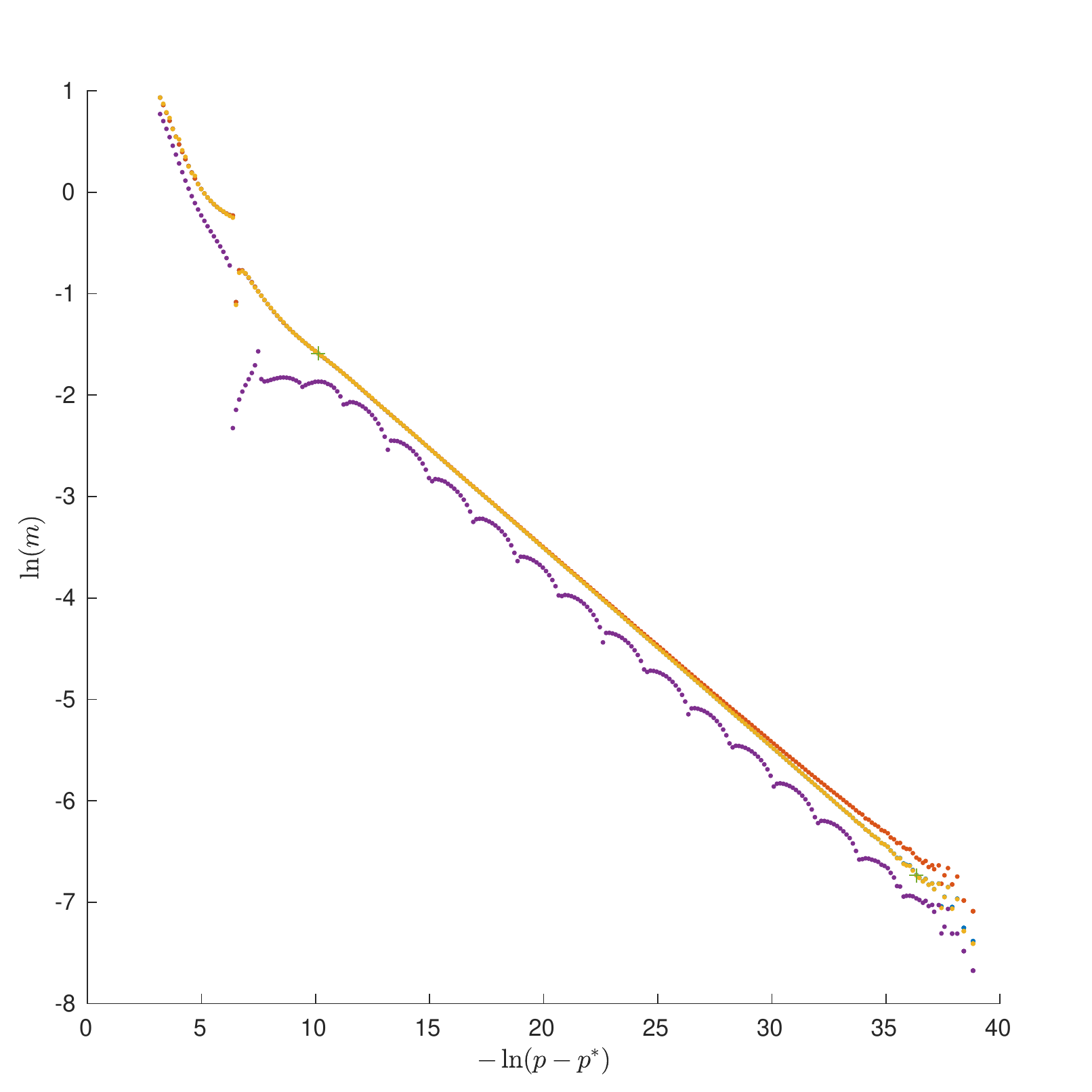}
  \caption{A plot of $ \ln(m) $ versus $ -\ln(p-p^*) $ for the four definitions of $ m $ given in the text; (i) blue, (ii) orange, (iii) yellow, (iv) purple. (i) and (iii) are almost coincident, and the green crosses indicate the first and last of these points used to find the line of best fit.}
  \label{fig:S-S m vs p 24}
\end{figure}

Figure \ref{fig:S-S m vs p 24} shows the four mass definitions plotted against $ -\ln(p-p^*) $.
We see that all four definitions of the mass have broadly the same behaviour, however (i) and (iii) show precisely the expected behaviour over a large parameter range.
When $ p \gg p^* $, the mass scaling relation (\ref{m supercritical}) is not yet observed, and when $ p $ is close to $ p^* $ numerical errors visibly affect the result.
From the data (i) and (iii) plotted in figure \ref{fig:S-S m vs p 24} we obtain a gradient of
\begin{equation} \label{gamma}
\gamma = \frac{1}{\lambda_\text{II}} = 0.196_{16}^{23} \;,
\end{equation}
which compares well with the precise value of $ \gamma = 0.1964 \pm 0.0007 $ found in \cite{Gundlach97-2} by considering the eigenvalue problem.

The behaviour of definition (ii) is quite sensitive to how close to timelike infinity the points on future null infinity reach.
We measure the final value of $ u $ reached on $ \mathscr{I}^+ $ and the first value of $ u $ on the MTT, corresponding to the positions the mass is calculated in definitions (ii) and (i) respectively. We find that as the critical parameter $ p^* $ is approached, the difference between these increases, which perhaps accounts for some of the artificially high value of $ m $ calculated by (ii); there is still energy to be radiated away that is not observed due to the finite resolution.
It was suggested in \cite{PHA05} that this final black hole mass can not go to zero, and there is necessarily some non-zero mass that falls in due to backscattering of the primary pulse. We will return to this idea when we consider explicit late in-falling matter below.

\begin{figure}[!ht]
  \centering
  \includegraphics[scale=0.9]{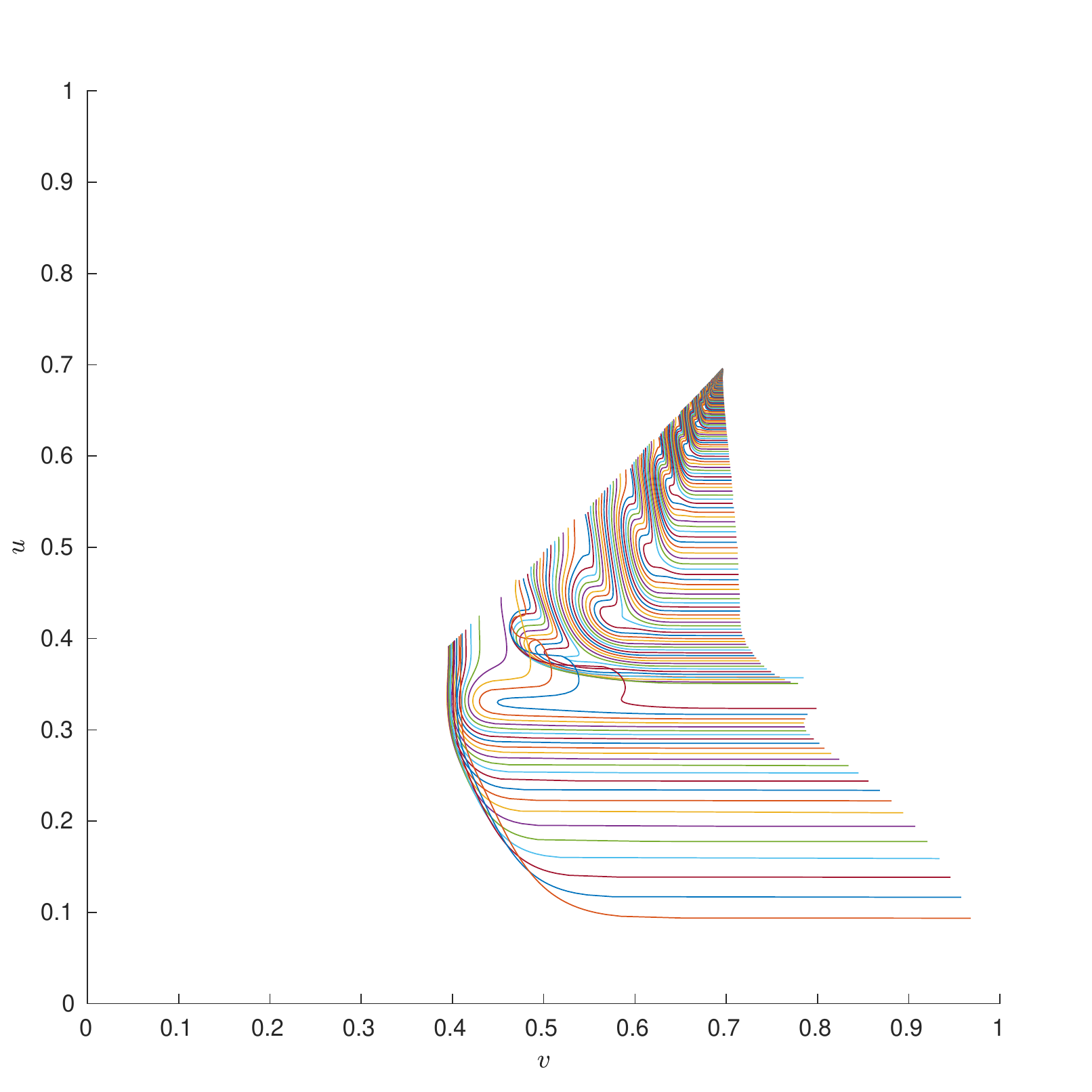}
  \caption{A plot of the MTTs for the evolutions corresponding to each value of $ p $.}
  \label{fig:S-S horizons 24}
\end{figure}

The initial apparent horizon mass in isothermal coordinates (iv) shows the same overall behaviour, however it contains periodic fluctuations; the fine structure is particularly prominent. In figure \ref{fig:S-S horizons 24} we plot the positions of the MTTs for each of the evolutions, and we see that their shape changes with this same period.
We also see in figure \ref{fig:S-S horizons 24} that for $ p $ near enough to $ p^* $ so that the solution expresses self-similarity, successive MTTs do not intersect each other.

A further thing we can note from figure \ref{fig:S-S horizons 24} is that we evolve MTTs that quite generically contain timelike sections.
We compare this to the massless scalar field case where these are impossible (\cite{Christodoulou93} prop. 1.3), and to the massive scalar field case where in \cite{BBGvdB06} section 4.4 the authors note they were unable to find the expected timelike membrane regions.

\begin{figure}[!ht]
  \centering
  \includegraphics[scale=0.8]{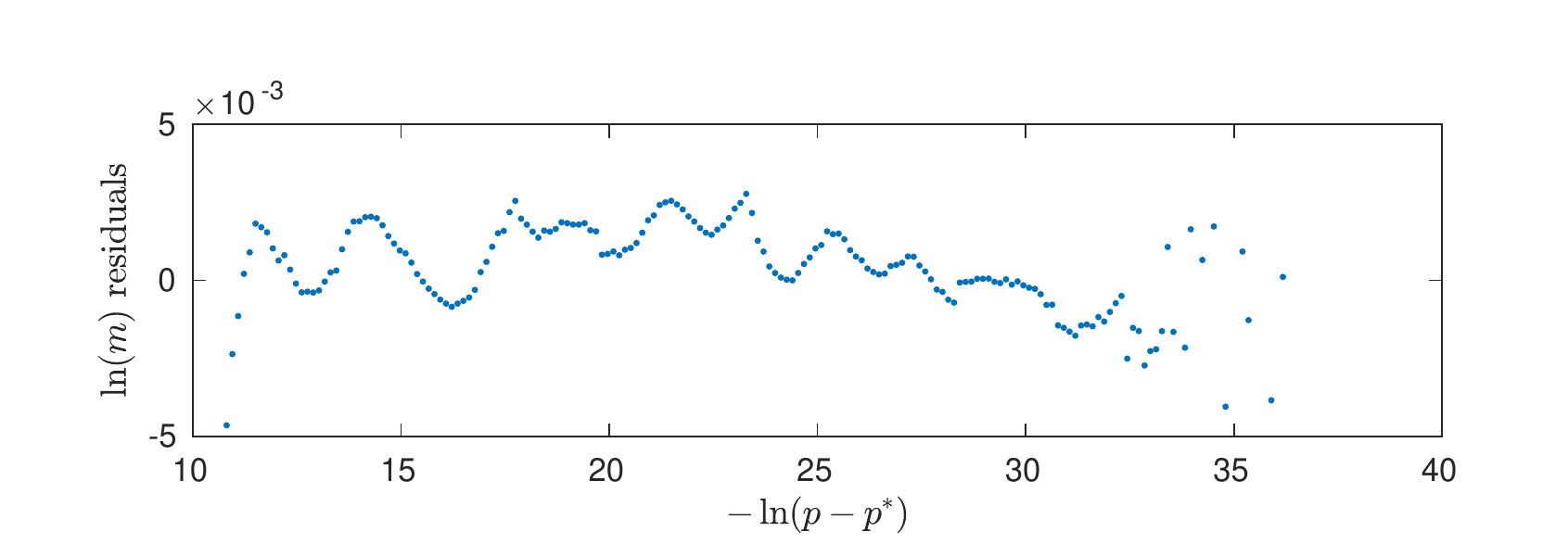}
  \includegraphics[scale=0.8]{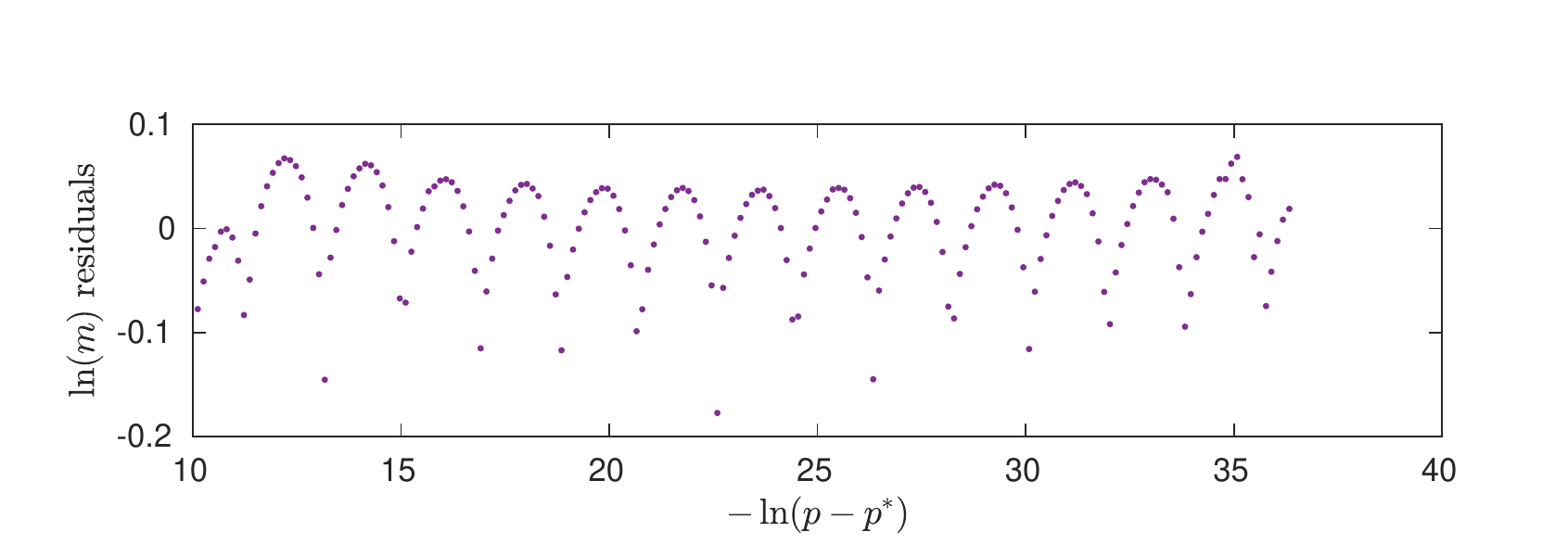}
  \caption{A plot of $ \ln(m) -\gamma\ln(p-p^*)-c $ versus $ -\ln(p-p^*) $ for the mass definitions (i) (blue) and (iv) (purple), where $ \gamma $ and $ c $ are given by the line of best fit.}
  \label{fig:S-S m fine 24}
\end{figure}

We plot mass (i) minus the line of best fit in figure \ref{fig:S-S m fine 24}.
We can observe the predicted fine structure wiggling here, which has not been observed before for the Einstein-Yang-Mills system.
It is predicted by the description of the system as a discretely self-similar solution \cite{GM-G07}, and has been observed in the Einstein scalar field case \cite{PHA05,Gundlach97,HP97,HP97-2}.
It is more apparent for definition (iv), also shown in figure \ref{fig:S-S m fine 24}.

\begin{figure}[!ht]
  \centering
  \includegraphics[scale=0.9]{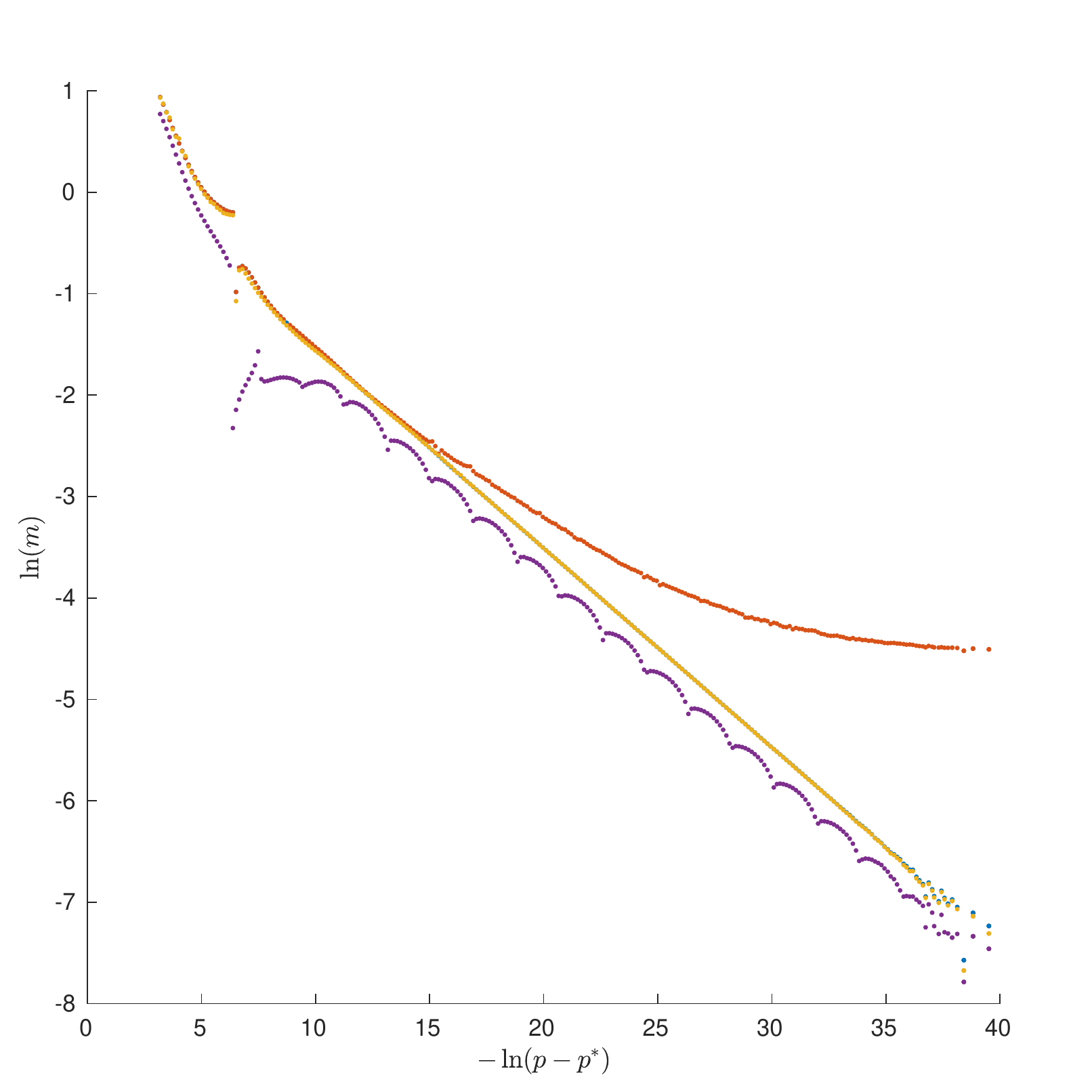}
  \caption{A plot of $ \ln(m) $ versus $ -\ln(p-p^*) $ for the four definitions of $ m $ given in the text, with late in-fall of matter; (i) blue, (ii) orange, (iii) yellow, (iv) purple.}
  \label{fig:S-S m vs p late}
\end{figure}

We repeat these tests with additional late in-falling YM matter that doesn't affect the initial black hole formation but does affect the final black hole mass and observe which of these mass definitions retains the scaling behaviour:
\begin{equation*} \label{ICs II late}
\begin{aligned}
W_0(r) &= -p e^{-(r-5)^2} +0.0004e^{-(r-17)^2} \;, & \alpha_0 &= 10 \;, \\
D_0(r) &= 0 \;, & \tilde\gamma_0(v) &= 0 \;.
\end{aligned}
\end{equation*}
The form of the additional Gaussian was chosen to produce a bump around $ v = 0.7 $ with an initial increase of mass of approximately 0.02.
We find the critical parameter in this case is slightly reduced; $ p^* = 0.0353126080426955_{09}^{16} $. For (iii) we continue to use $ v_0 = 0.696 $.
The results are shown in figure \ref{fig:S-S m vs p late}.
We see the late in-fall of matter increases the final black hole mass measured by (ii). The initial increase in mass was 0.0149 while the final black hole mass is approximately 0.011.
While on $ u = 0 $ the mass is constant after the additional bump, we observe that it is increasing on the final complete $ u $-row, which suggests that part of the final mass is due to backscattering -- the nonlinear interaction of the outgoing fields with themselves and the spacetime curvature. Our results are consistent with the conjecture of \cite{PHA05} that it is not possible to form arbitrarily small black holes. The primary cause of the increased mass (ii) in figure \ref{fig:S-S m vs p 24} is then probably due to backscattering.
With the additional pulse at large $ v $, and the numerically resolved MTTs much as in figure \ref{fig:S-S horizons 24}, we expect that for evolutions far from critical the mass definition (i) should include the increased mass from the second pulse. Indeed, this is what we observe in figure \ref{fig:S-S m vs p late}, where for large $ p $, the mass (i) aligns with (ii), while for $ p $ near $ p^* $ it aligns with (iii) as before.

In \cite{GD98} the authors predict (in the context of a spherically symmetric massless scalar field) and numerically observe another scaling relation, for subcritical evolutions, of the maximum of the Ricci scalar at the origin; $ |R|^\frac{1}{2} \sim (p^*-p)^{-\gamma} $, again with fine structure wiggling.
Since for the EYM system, $ R $ is identically zero, we instead consider other scalar quantities the Kretschmann scalar $ K = R^{\mu\nu\kappa\lambda}R_{\mu\nu\kappa\lambda} $ and $ L = R^{\mu\nu}R_{\mu\nu} $, which are predicted to follow the same scaling relation; $ |K|^\frac{1}{4} \sim (p^*-p)^{-\gamma} $ and $ |L|^\frac{1}{4} \sim (p^*-p)^{-\gamma} $.
In double-null coordinates $ K $ is
\[ K = \frac{4}{\alpha^4}\left(\left(\frac{\alpha^2+r_ur_v}{r^2}\right)^2 +((\ln\alpha)_{uv})^2 +\frac{\alpha^4}{r^2}\left(\frac{r_u}{\alpha^2}\right)_u\left(\frac{r_v}{\alpha^2}\right)_v +\frac{r\indices{_{uv}^2}}{r^2}\right). \]
Using the Einstein equations (\ref{EEuv}) and the first order variables (\ref{dn first}) this becomes
\begin{align*}
K &= \frac{4}{\alpha^4}\left(3\left(\frac{\alpha^2+fg}{r^2} -\frac{\alpha^2\left((w^2+d^2-1)^2+z^2\right)}{r^4}\right)^2\right. \\
&+\left.2\left(\frac{\alpha^2\left((w^2+d^2-1)^2+z^2\right)}{r^4}\right)^2 +\frac{4}{r^2}(p^2+x^2)(q^2+y^2)\right).
\end{align*}
With the numerical variables (\ref{dn first final}) this is
\begin{align*}
K &= \frac{4(1-u)^4(1-v)^4}{\tilde\alpha^4}\left(3\left(\frac{\tilde{f}(\tilde{r}+(1-v)\tilde\alpha^2\tilde{G})+(1-u)\tilde\alpha^2}{(1-u)\tilde{r}^2} \right.\right. \\
 & \qquad\left. -\frac{(1-v)^2\tilde\alpha^2}{(1-u)^2}\left(\left(-2\tilde{W}+\tilde{D}^2+\frac{\tilde{r}^2\tilde{W}^2}{(1-u)^2}\right)^2+\tilde{Z}^2\right)\right)^2 \\
 &\left. +2\left(\frac{(1-v)^2\tilde\alpha^2}{(1-u)^2}\left(\left(-2\tilde{W}+\tilde{D}^2+\frac{\tilde{r}^2\tilde{W}^2}{(1-u)^2}\right)^2+\tilde{Z}^2\right)\right)^2 +\frac{4(1-u)^2(1-v)^4}{\tilde{r}^2}(p^2+x^2)(\tilde{q}^2+\tilde{y}^2)\right).
\end{align*}
Evaluated at the origin $ \tilde{r} = 0 $ using l'H\^{o}pital's rule and the boundary conditions (\ref{BCs final all}) it is
\[ K = \frac{24(1-u)^{12}(\tilde{q}^2+\tilde{y}^2)^2}{\tilde\alpha^4} \;. \]
Using the regular variables this can be written as
\begin{equation} \label{K0}
K = 24\left((2W-D^2)^2+Z^2\right)^2,
\end{equation}
which for the magnetic case is simply $ K = 384W^4 $.

Similarly, we calculate $ L = R^{\mu\nu}R_{\mu\nu} $ in double-null coordinates:
\[ L = \frac{2}{\alpha^4}\left(\left(\frac{r_{uv}}{r}+\frac{\alpha^2+r_ur_v}{r^2}\right)^2+\left(\left(\ln\alpha\right)_{uv}+\frac{r_{uv}}{r}\right)^2+\frac{\alpha^4}{r^2}\left(\frac{r_u}{\alpha^2}\right)_u\left(\frac{r_v}{\alpha^2}\right)_v\right). \]
Using the Einstein equations (\ref{EEuv}) and the first order variables (\ref{dn first}) this becomes
\[ L = \frac{4\left((w^2+d^2-1)^2+z^2\right)^2}{r^8}+\frac{8}{r^2\alpha^4}(p^2+x^2)(q^2+y^2) \;. \]
With the numerical variables (\ref{dn first final}) this is
\[ L = 4(1-v)^8\left(\left(-2\tilde{W}+\tilde{D}^2+\frac{\tilde{r}^2\tilde{W}^2}{(1-u)^2}\right)^2+\tilde{Z}^2\right)^2 +\frac{8(1-u)^6(1-v)^8}{\tilde{r}^2\tilde\alpha^4}(p^2+x^2)(q^2+y^2) \;. \]
Evaluated at the origin $ \tilde{r} = 0 $ using l'H\^{o}pital's rule and the boundary conditions (\ref{BCs final all}) it is
\[ L = \frac{12(1-u)^{12}(\tilde{q}^2+\tilde{y}^2)^2}{\tilde\alpha^4} \;. \]
Using the regular variables this can be written
\begin{equation} \label{L0}
L = 12\left((2W-D^2)^2+Z^2\right)^2,
\end{equation}
which for the magnetic case is simply $ L = 96W^4 $.

\begin{figure}[!ht]
  \centering
  \includegraphics[scale=0.8]{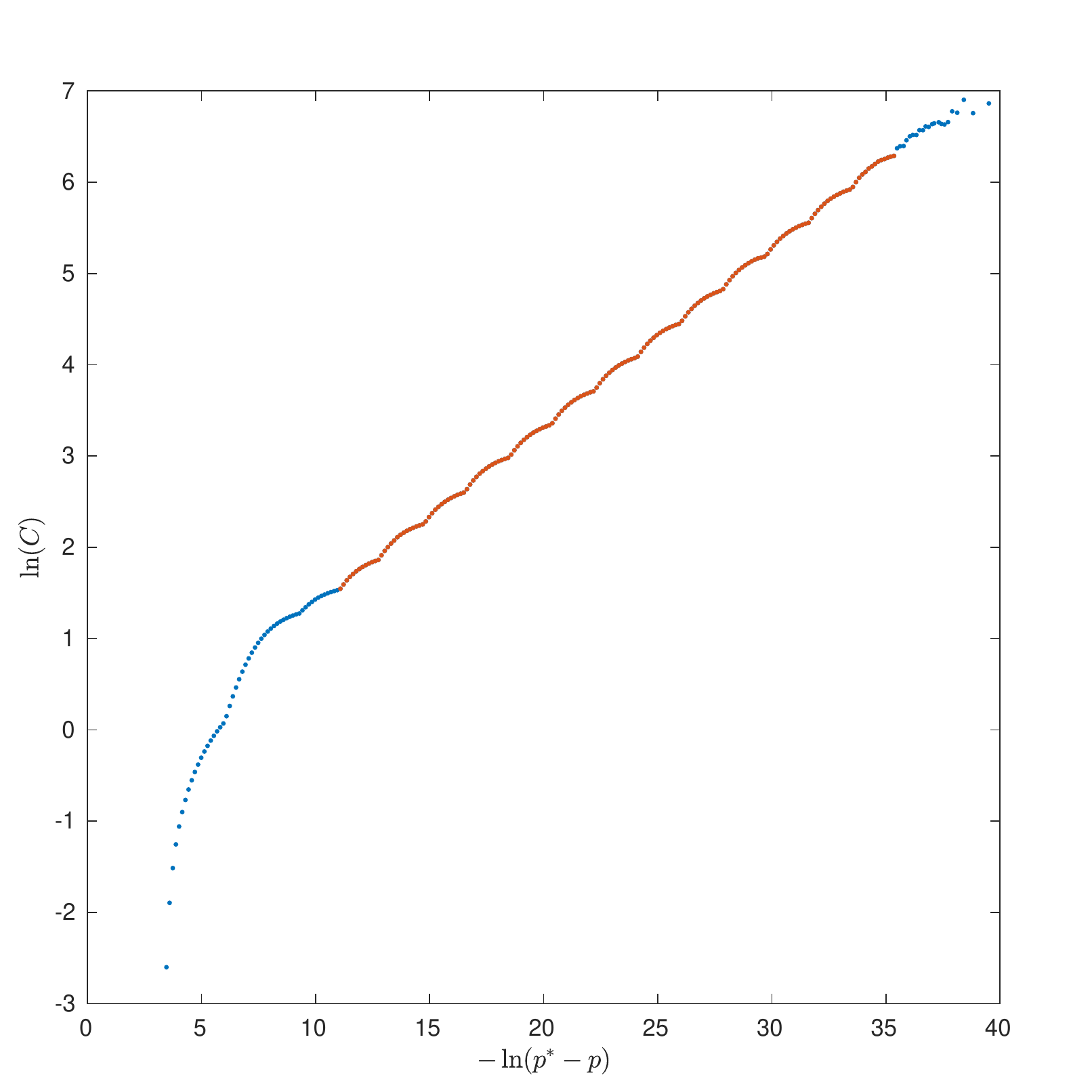}
  \caption{A plot of $ \ln C $ versus $ -\ln(p^*-p) $ for the maximum of $ |W| $ on the origin. The red points indicate those used to determine the line of best fit.}
  \label{fig:S-S W vs p}
\end{figure}

We hence define
\begin{equation} \label{C}
C := \sqrt{(2W-D^2)^2+Z^2}
\end{equation}
as a measure of the curvature at the origin, which is simply $ 2|W| $ in the purely magnetic case we have here.
This is shown in figure \ref{fig:S-S W vs p} as a function of the sub-critical parameter $ p $.
The line of best fit gives a value for $ \gamma = \frac{1}{\lambda_\text{II}} = 0.1960 $, which is consistent with the value we obtained from supercritical solutions (\ref{gamma}).

\begin{figure}[!ht]
  \centering
  \includegraphics[scale=0.9]{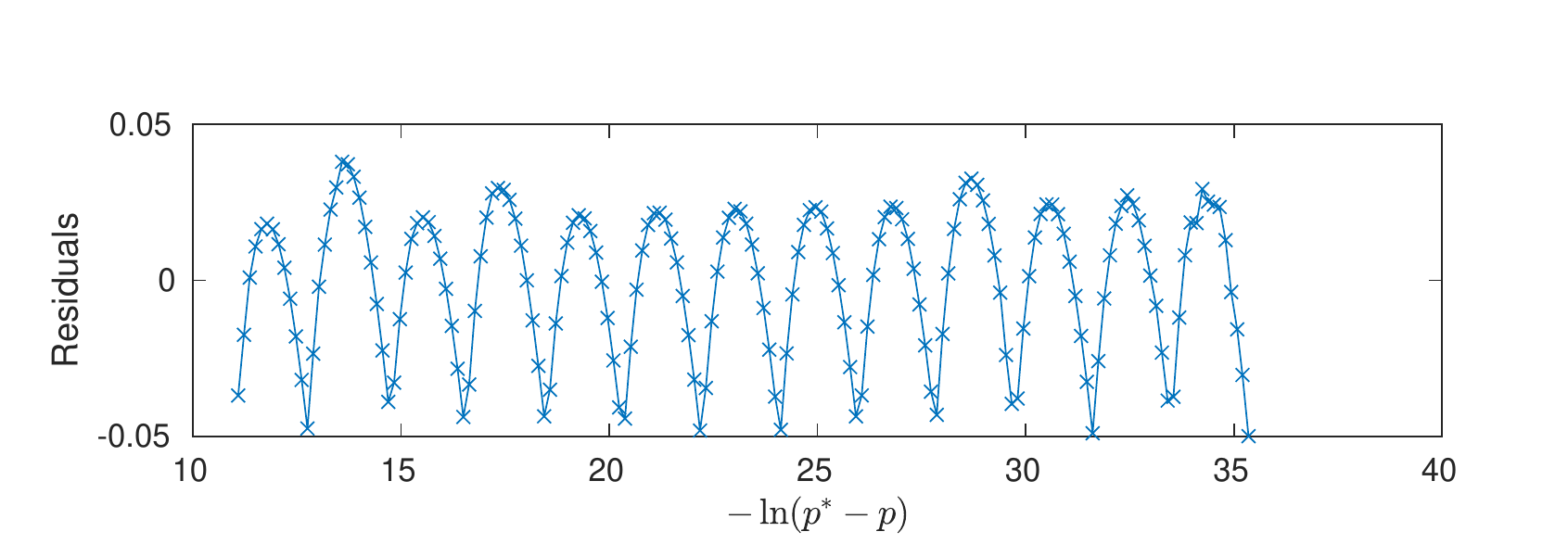}
  \caption{A plot of $ \ln C -2\gamma\ln(p^*-p)-c $ versus $ -\ln(p^*-p) $, where $ \gamma $ and $ c $ are given by the line of best fit.}
  \label{fig:S-S W fine 24}
\end{figure}

We also subtract the line of best fit and observe the fine-structure, see figure \ref{fig:S-S W fine 24}.
We can clearly observe the predicted fine structure in figure \ref{fig:S-S W fine 24}, with a measured period of approximately 1.883.
This compares well with the expected value of $ \frac{\Delta}{2\gamma} = 1.8784 $.
It is interesting that the critical behaviour can be captured accurately with relatively modest accuracy of the numerical scheme, and this will be exploited when evolving the full equations.

To create data for a comparison with the full case in section \ref{s:CB}, we also explore a two-parameter family in the purely magnetic ansatz:
\begin{align*}
W_0(r) &= -p_1 e^{-(r-5)^2} -p_2 e^{-(r-10)^2} \;, & \alpha_0 &= 10 \;, \\
D_0(r) &= 0 \;, & \tilde\gamma(0,v) &= 0 \;.
\end{align*}
The results in table \ref{table:p1star p2star} are for $ (\epsilon_u,\epsilon_v) = (\frac{1}{8^2},\frac{1}{8^4}) $ and give an indication of the uncertainties in the results.
\begin{singlespace}
\begin{table}[!ht]
  \centering
  \begin{tabular}{llc}
    \toprule
    $ p_1^* $ & $ p_2^* $ & $ \Delta $ \\
    \midrule
    0 & $ 0.00866624148373318_{18}^{35} $ & $ 0.73_{62}^{79} $ \\
    0.01 & $ 0.00844262440111723_{59}^{76} $ & $ 0.73_{65}^{78} $ \\
    0.02 & $ 0.00778753996683234_{46}^{55} $ & $ 0.7_{391}^{408} $ \\
    0.03 & $ 0.00572147016763677_{09}^{18} $ & $ 0.73_{59}^{77} $ \\
    $ 0.03531260825930455_{23}^{92} $ & 0 & $ 0.73_{68}^{80} $ \\
    \bottomrule
  \end{tabular}
  \caption{Calculations of a line of critical parameters $ p_1^* $ and $ p_2^* $.}
  \label{table:p1star p2star}
\end{table}
\end{singlespace}
Since all the $ \Delta $ values should be equal, we estimate the error for $ \Delta $ with tolerances of $ (\epsilon_u,\epsilon_v) = \left(\frac{1}{8^2},\frac{1}{8^4}\right) $ is about 0.008.

\subsection{Other purely magnetic critical phenomena}
There are two possible black hole end-states to evolution, Schwarzschild black holes with $ w = 1 $ or $ w = -1 $. In \cite{CHM99}, the authors found further critical behaviour, dubbed ``Type III" \cite{GM-G07}, where another co-dimension one attractor was evolved when the initial conditions were tuned to the boundary between these different end-states.
This critical solution was one of the static black hole solutions with one zero-crossing, described in section \ref{s:EH:m}.

The critical static black hole could have either $ \lim\limits_{r \to \infty} w = -1 $ (as in figure \ref{fig:BHs}) or 
$ \lim\limits_{r \to \infty} w = 1 $.
As investigated in \cite{Rinne14}, tuning a second parameter to the boundary between these regions produced a Reissner-Nordstr\"{o}m black hole, which is an approximate unstable attractor.
It is approximate because the EYM Reissner-Nordstr\"{o}m solution has a countably infinite number of unstable modes, however the third mode's eigenvalue is much smaller than the second mode's eigenvalue \cite{Rinne14}.
As pointed out in \cite{Rinne14}, it is not possible to search for critical behaviour separating $ w = 1 $ and $ w = -1 $ Minkowski end-states because regularity at the origin requires $ w $ to retain its initial value there.

We do not reproduce any of these magnetic critical phenomena here, but instead, having shown that our code can reproduce the relevant purely magnetic results, turn our attention to the general case.
In particular we consider what critical phenomena occur separating Minkowski and Schwarzschild end-states in the general case. Since the first Bartnik-McKinnon solution then has two unstable modes, we expect it to be like the magnetic Type II behaviour.

\chapter{General dynamic solutions} \label{ch:Df}
In this chapter we apply our code to the general case and investigate how the behaviour is different to the known magnetic case.
First, we observe the evolution of appropriately cut-off long-lived solutions, and begin to get an understanding of their (local) stability.
We then consider the most general critical behaviour, and find it to be substantially different to that in the magnetic case.

To interpret the solutions we find numerically, it will be important to view functions that do not depend on the choice of gauge, or the coordinates. For the metric functions, we view either the mass $ m $ or $ N $.
For a gauge-independent combination of the gauge variables, we use the Yang-Mills Lagrangian density $ |F|^2 $, which in double-null coordinates and using the first-order variables is
\[ |F|^2 = \frac{(w^2+d^2-1)^2-z^2}{r^4} -2\frac{pq+xy}{r\alpha^2} \;. \]

We can also use components of the stress-energy tensor $ T_{\mu\nu} $, as these will be gauge-invariant, however they are coordinate dependent.
The energy density (\ref{energy density dn}) also depends on the choice of orientation of the time direction (here chosen to be in the $ \pypx{}{u}+\pypx{}{v} $ direction).
We instead form coordinate-independent functions from the stress-energy tensor by choosing a non-coordinate basis $ \hat\theta^\alpha = e\indices{^\alpha_\mu} \dy{x}^\mu $ that diagonalises the stress-energy tensor $ T_{\mu\nu}\dy{x}^\mu\dy{x}^\nu = T_{\alpha\beta} \hat\theta^\alpha\hat\theta^\beta $.
Then the non-zero components of $ T_{\alpha\beta} $ will be along the diagonal and be both gauge- and coordinate-independent, representing the rest energy density $ \rho $ and the principal pressures $ p_i $ \cite{Wald}.

For a spherically symmetric stress-energy tensor in double-null coordinates that has non-zero components $ T_{uu} $, $ T_{uv} = T_{vu} $, $ T_{vv} $, and $ T_{\phi\phi} = T_{\theta\theta}\sin^2\theta $, the orthonormal basis that achieves this (up to sign choices) is
\begin{align*}
\hat\theta^0 &= \frac{\alpha}{\sqrt[4]{T_{uu}T_{vv}}}\left(\sqrt{T_{uu}}\dx{u}+\sqrt{T_{vv}}\dx{v}\right), \\
\hat\theta^1 &= \frac{\alpha}{\sqrt[4]{T_{uu}T_{vv}}}\left(-\sqrt{T_{uu}}\dx{u}+\sqrt{T_{vv}}\dx{v}\right), \\
\hat\theta^2 &= r\dy{\theta} \;, \\
\hat\theta^3 &= r\sin\theta\dy{\phi} \;.
\end{align*}
Note that for any matter satisfying the null energy condition, $ T_{uu} \geqslant 0 $ and $ T_{vv} \geqslant 0 $.

The resulting rest energy density and principal pressures are
\begin{align*}
\rho &= \frac{T_{uv}+\sqrt{T_{uu}T_{vv}}}{2\alpha^2} \;, \\
p_1 &= \frac{-T_{uv}+\sqrt{T_{uu}T_{vv}}}{2\alpha^2} \;, \\
p_2 &= \frac{T_{\theta\theta}}{r^2} = p_3 = \frac{T_{\phi\phi}}{r^2\sin^2\theta} \;.
\end{align*}
These results apply for any kind of matter in spherical symmetry. Inserting the Yang-Mills stress-energy tensor (\ref{T dn}) we find
\begin{align*}
8\pi\rho &= \frac{(w^2+d^2-1)^2+z^2}{r^4} +\frac{2}{r\alpha^2}\sqrt{p^2+x^2}\sqrt{q^2+y^2} \;, \\
8\pi p_1 &= -\frac{(w^2+d^2-1)^2+z^2}{r^4} +\frac{2}{r\alpha^2}\sqrt{p^2+x^2}\sqrt{q^2+y^2} \;, \\
8\pi p_2 &= 8\pi p_3 = \frac{(w^2+d^2-1)^2+z^2}{r^4} \;.
\end{align*}

Note that these functions can be evaluated at the origin using the regularising variables, and we find
\begin{align*}
8\pi p_2 &= (2W-D^2)^2+Z^2 \;, & 8\pi\rho &= 3\left((2W-D^2)^2+Z^2\right), & |F|^2 &= 3\left((2W-D^2)^2-Z^2\right).
\end{align*}
We thus can use $ \sqrt{\frac{8\pi\rho + |F|^2}{6}} = \left|2W-D^2\right| $ and $ \sqrt{\frac{8\pi\rho - |F|^2}{6}} = \left|Z\right| $ as gauge-independent functions at the origin.

Away from the origin, we find it useful to consider the following combinations:
\begin{subequations} \label{P123}
\begin{align}
P_1 &:= \frac{r^2}{2}\left(8\pi \rho-8\pi p_1\right) = r^2 8\pi p_2 = \frac{(w^2+d^2-1)^2+z^2}{r^2} \;, \\
P_2 &:= \frac{r^2}{2}(8\pi\rho+|F|^2) = \frac{(w^2+d^2-1)^2}{r^2} +\frac{r}{\alpha^2}\left(\sqrt{p^2+x^2}\sqrt{q^2+y^2}-(pq+xy)\right), \label{P2} \\
P_3 &:= \frac{r^2}{2}(8\pi\rho-|F|^2) = \frac{z^2}{r^2} +\frac{r}{\alpha^2}\left(\sqrt{p^2+x^2}\sqrt{q^2+y^2}+(pq+xy)\right).
\end{align}
\end{subequations}
These are all non-negative. This is because $ (xq-py)^2 \geqslant 0 $ shows that $ x^2q^2+p^2y^2 \geqslant 2pqxy $, which after adding $ p^2q^2+x^2y^2 $ to both sides and taking the square root, gives $ \sqrt{p^2+x^2}\sqrt{q^2+y^2} \geqslant |pq+xy| $.
In the purely magnetic case these simplify to:
\begin{align*}
P_1 &= \frac{(w^2-1)^2}{r^2} \;, \\
P_2 &= \frac{(w^2-1)^2}{r^2} +\frac{r}{\alpha^2}\left(|pq|-pq\right) \;, \\
P_3 &= \frac{r}{\alpha^2}\left(|pq|+pq\right) \;,
\end{align*}
which we see are manifestly non-negative, and that $ w $ and $ w_uw_v $ represent gauge- and coordinate-invariant terms.

Note that these functions require $ p $ and $ x $ to calculate, thus in the following sections we generally set \verb_pxbe = true_ (which affects the critical parameters in the sixth significant figure or so).

We will also consider the solutions in polar-areal coordinates by performing a numerical coordinate transformation (see Appendix \ref{sec:A:num}).
In these coordinates the Yang-Mills Lagrangian density is
\[ |F|^2 = \frac{(w^2+d^2-1)^2-z^2}{r^4} +\frac{2N}{r^2}\left(\hat{p}^2+\hat{x}^2-\hat{q}^2-\hat{y}^2\right). \]

The orthonormal basis that diagonalises the stress-energy tensor in polar coordinates with non-zero components $ T_{tt} $, $ T_{tr} = T_{rt} $, $ T_{rr} $, $ T_{\phi\phi} = T_{\theta\theta}\sin^2\theta $ is
\begin{align*}
\hat\theta^0 &= \frac{S\sqrt{N}}{\sqrt{2}}\sqrt{\frac{\frac{T_{tt}}{S^2N}+NT_{rr}}{\sqrt{\left(\frac{T_{tt}}{S^2N}+NT_{rr}\right)^2-\frac{4T_{tr}^2}{S^2}}}+1}\dx{t} +\frac{\sgn(T_{tr})}{\sqrt{2N}}\sqrt{\frac{\frac{T_{tt}}{S^2N}+NT_{rr}}{\sqrt{\left(\frac{T_{tt}}{S^2N}+NT_{rr}\right)^2-\frac{4T_{tr}^2}{S^2}}}-1}\dx{r} \;, \\
\hat\theta^1 &= \frac{\sgn(T_{tr})S\sqrt{N}}{\sqrt{2}}\sqrt{\frac{\frac{T_{tt}}{S^2N}+NT_{rr}}{\sqrt{\left(\frac{T_{tt}}{S^2N}+NT_{rr}\right)^2-\frac{4T_{tr}^2}{S^2}}}-1}\dx{t} +\frac{1}{\sqrt{2N}}\sqrt{\frac{\frac{T_{tt}}{S^2N}+NT_{rr}}{\sqrt{\left(\frac{T_{tt}}{S^2N}+NT_{rr}\right)^2-\frac{4T_{tr}^2}{S^2}}}+1}\dx{r} \;, \\
\hat\theta^2 &= r\dy{\theta} \;, \\
\hat\theta^3 &= r\sin\theta\dy{\phi} \;.
\end{align*}
This basis is written assuming $ N > 0 $, but the formulas work equally well when $ N < 0 $.

The resulting rest energy density and principal pressures are
\begin{align*}
\rho &= \frac{1}{2}\left(\frac{T_{tt}}{S^2N}-NT_{rr}+\sqrt{\left(\frac{T_{tt}}{S^2N}+NT_{rr}\right)^2-\frac{4T_{tr}^2}{S^2}}\right), \\
p_1 &= \frac{1}{2}\left(-\left(\frac{T_{tt}}{S^2N}-NT_{rr}\right)+\sqrt{\left(\frac{T_{tt}}{S^2N}+NT_{rr}\right)^2-\frac{4T_{tr}^2}{S^2}}\right), \\
p_2 &= \frac{T_{\theta\theta}}{r^2} = p_3 = \frac{T_{\phi\phi}}{r^2\sin^2\theta} \;.
\end{align*}
The null energy condition ensures $ \left|\frac{T_{tt}}{S^2N}+NT_{rr}\right| \geqslant \left|\frac{2T_{tr}}{S}\right| $.
These results apply for any kind of matter in spherical symmetry. Inserting the Yang-Mills stress-energy tensor (\ref{T pa}) we find
\begin{align*}
8\pi\rho &= \frac{(w^2+d^2-1)^2+z^2}{r^4} +\frac{2|N|}{r^2}\sqrt{(\hat{p}+\hat{q})^2+(\hat{x}+\hat{y})^2}\sqrt{(\hat{p}-\hat{q})^2+(\hat{x}-\hat{y})^2} \;, \\
8\pi p_1 &= -\frac{(w^2+d^2-1)^2+z^2}{r^4} +\frac{2|N|}{r^2}\sqrt{(\hat{p}+\hat{q})^2+(\hat{x}+\hat{y})^2}\sqrt{(\hat{p}-\hat{q})^2+(\hat{x}-\hat{y})^2} \;, \\
8\pi p_2 &= 8\pi p_3 = \frac{(w^2+d^2-1)^2+z^2}{r^4} \;.
\end{align*}

The useful combinations (\ref{P123}) in these coordinates are:
\begin{align*}
P_1 &= \frac{(w^2+d^2-1)^2+z^2}{r^2} \;, \\
P_2 &= \frac{(w^2+d^2-1)^2}{r^2} +|N|\sqrt{(\hat{p}+\hat{q})^2+(\hat{x}+\hat{y})^2}\sqrt{(\hat{p}-\hat{q})^2+(\hat{x}-\hat{y})^2}+N\left(\hat{p}^2+\hat{x}^2-\hat{q}^2-\hat{y}^2\right), \\
P_3 &= \frac{z^2}{r^2} +|N|\sqrt{(\hat{p}+\hat{q})^2+(\hat{x}+\hat{y})^2}\sqrt{(\hat{p}-\hat{q})^2+(\hat{x}-\hat{y})^2}-N\left(\hat{p}^2+\hat{x}^2-\hat{q}^2-\hat{y}^2\right),
\end{align*}
which again are all non-negative.
In the purely magnetic case these simplify to:
\begin{align*}
P_1 &= \frac{(w^2-1)^2}{r^2} \;, \\
P_2 &= \frac{(w^2-1)^2}{r^2} +\left|N\left(\hat{p}^2-\hat{q}^2\right)\right|+N\left(\hat{p}^2-\hat{q}^2\right), \\
P_3 &= \left|N\left(\hat{p}^2-\hat{q}^2\right)\right|-N\left(\hat{p}^2-\hat{q}^2\right),
\end{align*}
and we see that $ w $ and $ \dot{w}^2-w'^2 $ represent gauge- and coordinate-invariant terms.

As explained in section \ref{s:SS}, the (magnitudes of the) charges (\ref{charge}) are gauge-invariant, so we will also use these to interpret the solutions. In general they are coordinate dependent, however in all of the coordinate systems we choose (\ref{metric}), where $ \theta $ and $ \phi $ are coordinates on the orbit 2-spheres, they are well-defined.

\section{Evolution of long-lived solutions}
We evolve the long-lived solutions by setting the initial conditions on $ u = 0 $ to be solutions of the static equations (\ref{EYME stat TR}) with an appropriate cutoff.
We could use isothermal coordinates to calculate the solution because they have a simple transformation to double-null coordinates (\ref{uvTR}). However this then requires a further coordinate transformation to compactify the coordinates (\ref{dn coord final}).
So instead, we calculate the static solution in polar-areal coordinates. This also allows us to easily specify the initial conditions as functions of $ r $, and let the usual algorithm calculate the compactified $ v $ coordinate.
The transformation from polar-areal to double-null coordinates (with our gauge choice) requires a non-trivial gauge transformation, which we now describe.

Given a static solution in polar-areal coordinates $ w_s(r) $, $ \hat{a}_s(r) $, $ m_s(r) $, $ S_s(r) $, we can solve (\ref{pa to dn}) and find
\begin{align*}
\mu(r) &= \frac{1}{S\sqrt{N}} \;, & \nu(r) &= \frac{1}{S\sqrt{N}} \;.
\end{align*}
Thus (\ref{ab pa to dn}) gives
\begin{align*}
w(u,v) &= w_s(r) \;, & a(u,v) &= \hat{a}_s(r) \;, \\
d(u,v) &= 0 \;, & b(u,v) &= \hat{a}_s(r) \;,
\end{align*}
where $ r $ is a function of $ v-u $ given by $ \dydx{r}{R} = SN $ and $ R = v-u $. Thus $ r_u = -SN $ and $ r_v = SN $.
Our gauge choice (\ref{dn gauge},\ref{dn gauge final}) requires $ a_v + b_u = 0 $ and $ b(0,v) = \frac{r_v}{r}d(0,v) $. We see the first equation is satisfied, but the second is not. So we perform a gauge transformation (\ref{gauge transform}), choosing $ \lambda(0,v) $ such that (\ref{dn gauge final}) is satisfied for the transformed variables. Writing the equation in terms of the areal coordinate, we obtain
\begin{equation} \label{tr to uv l}
\dydx{\lambda_s}{r} = \frac{w_s\sin\lambda_s}{r}-\frac{\hat{a}_s}{S_sN_s} \;.
\end{equation}
Using the same approach as before, we insert a power series for $ \lambda $ into (\ref{tr to uv l}) to find
\[ \lambda_s(r) = gr -dr^2 +\frac{6bg-g^3}{12}r^3 -\frac{2\left(14b+16b^2+4d^2-5g^2\right)}{30}r^4 + O(r^5) \;, \]
and integrate (\ref{tr to uv l}) from a small positive $ r $ with the other static equations (\ref{EYME stat}).
We recall that we require $ \lambda(t,r) $ to be an odd function of $ r $, which means that on an outgoing null line the only restriction on $ \lambda_s(r) $ is $ \lambda_s(0) = 0 $.
There is then one free parameter $ g = \lambda'(0) $ (the two parameters $ b $ and $ d $ determine the static solution), which we set to zero.

Since we can neither integrate the static solutions to $ r = \infty $ nor set them as asymptotically flat initial data, we smoothly cut off the solutions at a finite radius $ r_c $.
The cutoff function is given by $ \co(x) = \int\limits_x^1 e^{-\frac{1}{1-x^2}} \dx{x} $ for $ x \in [-1,1] $, and we set
\[ w_c(r) = \begin{cases} w_s(r) & 0 \leqslant r \leqslant r_c \\ \frac{1}{\co(-1)}\co\left(2\left(\frac{r}{r_c}-1\right)-1\right)w_s(r) & r_c < r \leqslant 2r_c \\ 0 & 2r_c < r \end{cases} \;. \]
Note that $ w \nrightarrow 1 $ as $ r \to \infty $, so such a slice is not vacuum and Schwarzschild near infinity, but rather approaches a Reissner-Nordstr\"{o}m spacetime.

We can use these $ w_c $ and $ \lambda_s $ to set the initial conditions
\begin{align*}
W_0(r) &= \frac{1-w_c(r)\cos(\lambda_s(r))}{r^2} \;, & \alpha_0 &= 50 \;, \\
D_0(r) &= \frac{w_c(r)\sin(\lambda_s(r))}{r} \;, & \tilde\gamma_0(v) &= 0 \;.
\end{align*}
Since these expressions are singular at $ r = 0 $, we use the power series for small $ r $.

With initial data given by the choice of parameters $ (b,d) = \left(0,\frac{1}{40}\right) $, $ r_c = 50 $, and $ \left(\epsilon_u,\epsilon_v\right) $ given by $ \left(\frac{1}{8^4},\frac{1}{8^6}\right) $, we observe that the solution remains approximately static until the dynamics due to the cutoff travel in to the origin.
The function $ P_2 $ (\ref{P2}) in this case is shown in figure \ref{fig:LL (0,40)}.
\begin{figure}[!ht]
  \centering
  \includegraphics[scale=0.8]{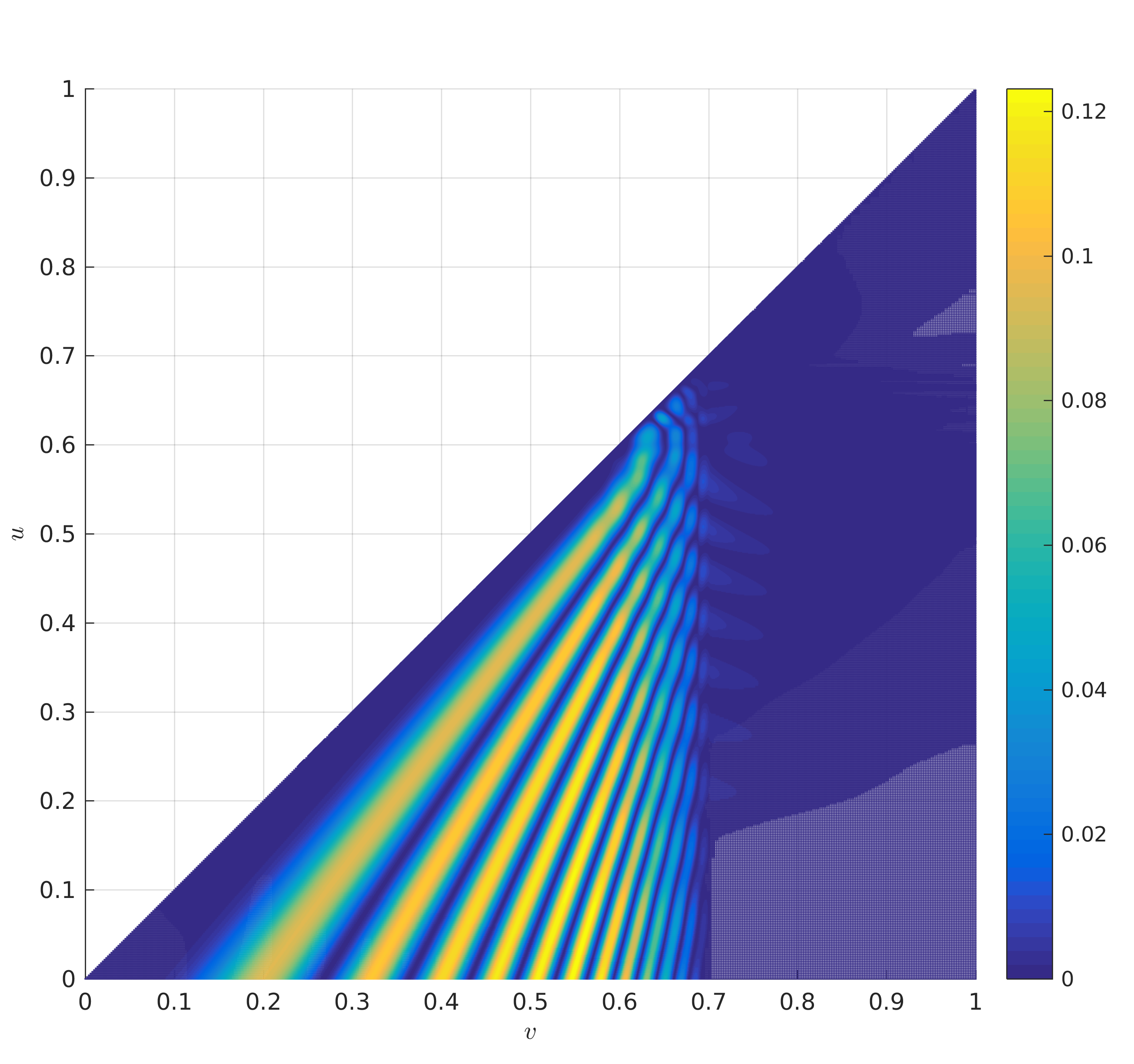}
  \caption{A plot of $ P_2 $ of the solution with initial data given by $ (b,d) = \left(0,\frac{1}{40}\right) $, $ r_c = 50 $.}
  \label{fig:LL (0,40)}
\end{figure}
The cutoff range $ r \in (50,100) $ is initially placed at $ v \in (0.53,0.71) $.
The change in lightness visible in this plot corresponds to a change in the density of points determined by the adaptive mesh.
Figure \ref{fig:LL (0,40) mPQI} shows the behaviour of some physical variables on future null infinity; the mass, magnetic charge, and electric charge (\ref{charge}).
\begin{figure}[!ht]
  \centering
  \includegraphics[scale=0.65]{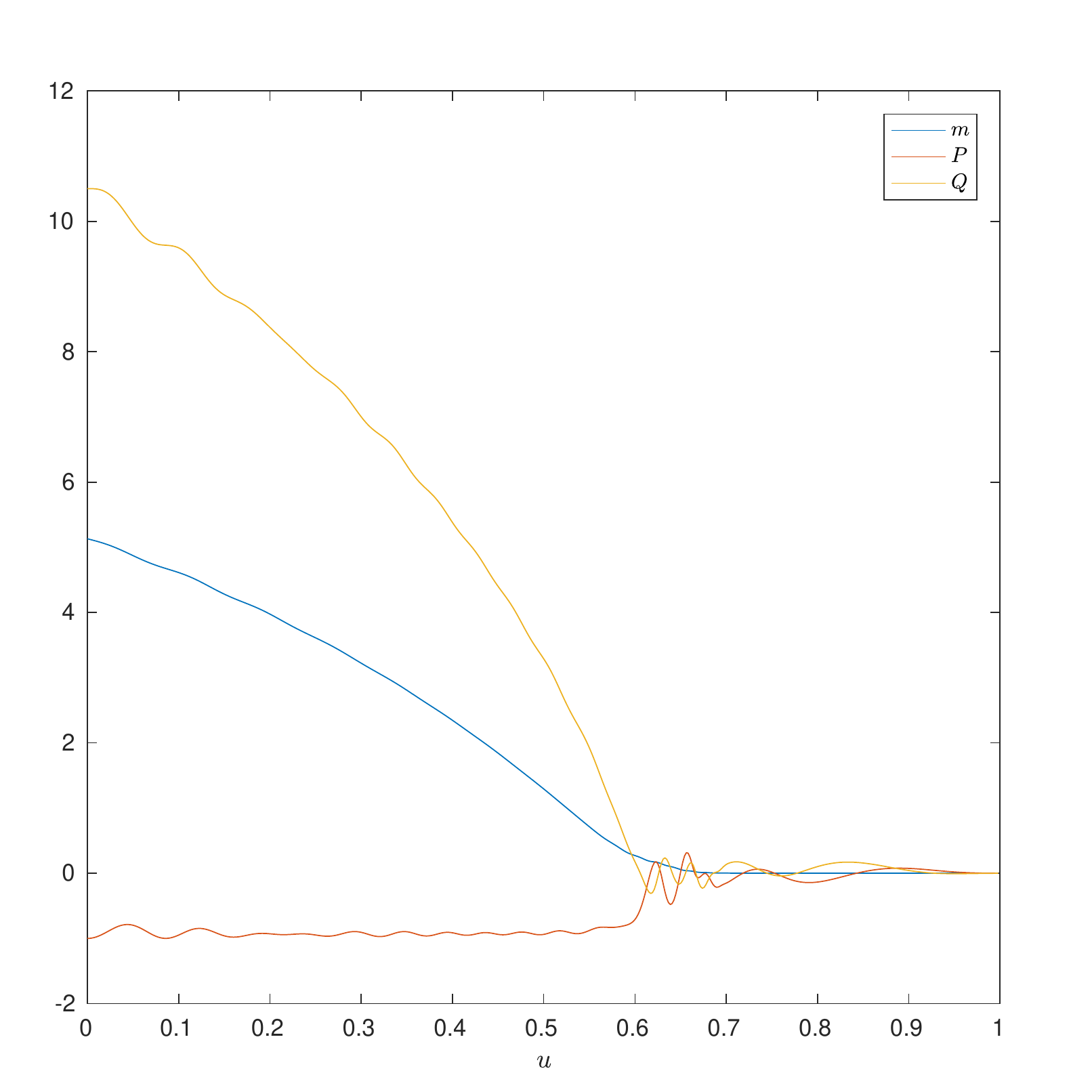}
  \caption{A plot of $ m $, $ P $, and $ Q $ on future null infinity of the solution with initial data given by $ (b,d) = \left(0,\frac{1}{40}\right) $, and $ r_c = 50 $.}
  \label{fig:LL (0,40) mPQI}
\end{figure}

If we add a small perturbation $ W_0(r) = -0.002 e^{-(r-5)^2} $ to the above initial data, we observe that the peaks of $ P_2 $ will oscillate around the static solution. This is shown in figure \ref{fig:LL (0,40) bump}.
\begin{figure}[!ht]
  \centering
  \animategraphics[scale=0.9]{10}{LL_0,40_bump_P2}{0001}{0351}
  \caption{An animated plot of $ P_2 $ with above initial data plus a small bump (orange), and the unperturbed (but cutoff) initial data (blue) -- digital only.}
  \label{fig:LL (0,40) bump}
\end{figure}
The details of how the data in polar-areal coordinates are determined are in Appendix \ref{sec:A:num}.
We see that the small bump at $ r = 5 $ quickly reflects through the origin and causes an oscillation in the peaks as it moves outward. Eventually, this is swamped by the dynamic effect of the cutoff moving inwards from $ r = 50 $.
A selection of the constant time slices for this solution, equally spaced in proper time at the origin, is shown in figure \ref{fig:LL (0,40) bump slice}. Note that the large-$ v $ regions of the ``initial data" on $ u = 0 $ are not reached in polar-areal coordinates until large $ t $.
\begin{figure}[!ht]
  \centering
  \includegraphics[scale=0.65]{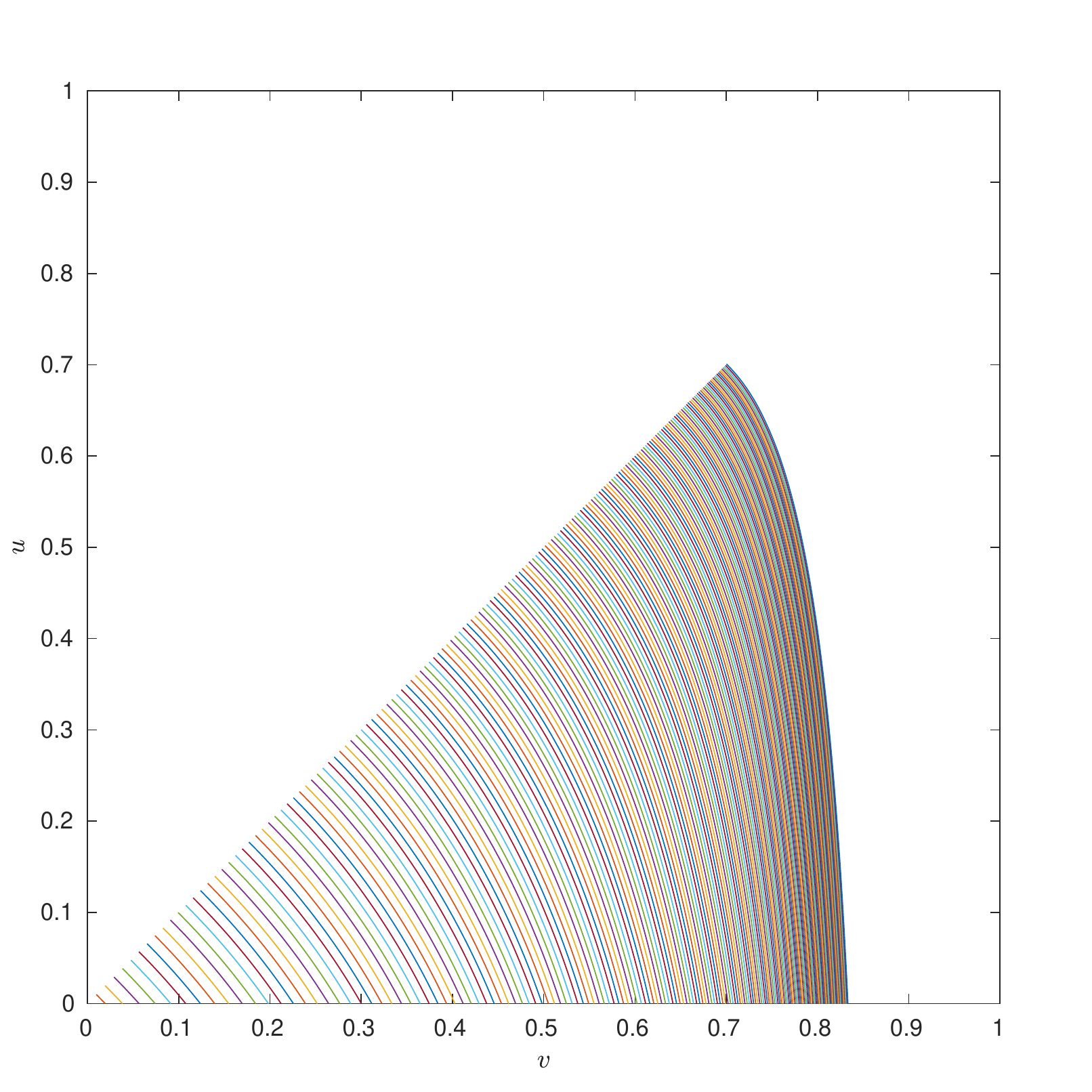}
  \caption{A plot of some equally-spaced polar-areal constant $ t $ slices for the solution with initial data given by $ (b,d) = \left(0,\frac{1}{40}\right) $, $ r_c = 50 $, and a small bump.}
  \label{fig:LL (0,40) bump slice}
\end{figure}

When evolving initial data based on other long-lived solutions, we find setting $ \tilde\gamma_0(v) $ as a non-zero function is useful.
This is because we desire both a reasonably small $ \alpha_0 $ so there is not a lot of $ u $ refinement initially, as well as $ r $ to increase rapidly as a function of $ v $ so that the many oscillations do not congregate near $ v = 1 $, resulting in much $ v $ refinement. Ideally we would like the initial required refinement to be reasonably well matched to the initial equispaced grid.
The choice of $ \tilde\gamma(0,v) = 10(1-v)^3 $ generally served this purpose well.
Note that we have $ \tilde\gamma(0,1) = 0 $ which prevents $ \tilde{r} $ from having large derivatives at $ v = 1 $.

Using $ \tilde\gamma(0,v) = 10(1-v)^3 $ to investigate the evolution of the long-lived solutions further from $ (b,d) = (0,0) $, we again find that in general a perturbation causes oscillations in the solution's peaks. For static solutions with a lower minimum of $ N $ we find that a black hole typically forms rather than dispersal. With the truncation error tolerances suitably low, this occurs as the dynamics due to the cutoff function travel in towards $ N $'s minimum.

\section{General critical phenomena} \label{s:CB}
In the general case, the first Bartnik-McKinnon solution gains a second unstable mode \cite{VG99} (this is sometimes referred to as in the sphaleron sector \cite{CHM99} due to the analogous unstable modes for sphalerons; they are both due to periodic topological vacua) and therefore cannot generically be a critical solution.
We therefore expect a generic critical solution to be asymptotically discretely self-similar, as in the magnetic case.

We begin our analysis by finding the critical parameter for the following two initial data families:
\begin{equation} \label{ID a}
\begin{aligned}
W_0(r) &= p e^{-(r-5)^2} \;, & \alpha_0 &= 10 \;, \\
D_0(r) &= 0.01 e^{-(r-5)^2} \;, & \tilde\gamma(0,v) &= 0 \;,
\end{aligned}
\end{equation}
and
\begin{equation} \label{ID e}
\begin{aligned}
W_0(r) &= 0 \;, & \alpha_0 &= 10 \; , \\
D_0(r) &= se^{-(r-5)^2} \;, & \tilde\gamma(0,v) &= 0 \;.
\end{aligned}
\end{equation}
With tolerances of $ (\epsilon_u,\epsilon_v) = \left(\frac{1}{8^2},\frac{1}{8^4}\right) $, we find the critical parameters to be $ p^* = 0.0350748450572620_{287}^{357} $ and $ s^* = 0.1691309336951535_{01}^{29} $.
Some of the following analysis was also performed with smaller tolerances with very similar results; however we use and present the relatively coarse tolerances of $ (\epsilon_u,\epsilon_v) = \left(\frac{1}{8^2},\frac{1}{8^4}\right) $ throughout the rest of this section as they result in quicker calculations, which is important when thousands of evolutions are required.

\begin{figure}[!ht]
  \centering
  \includegraphics[scale=0.65]{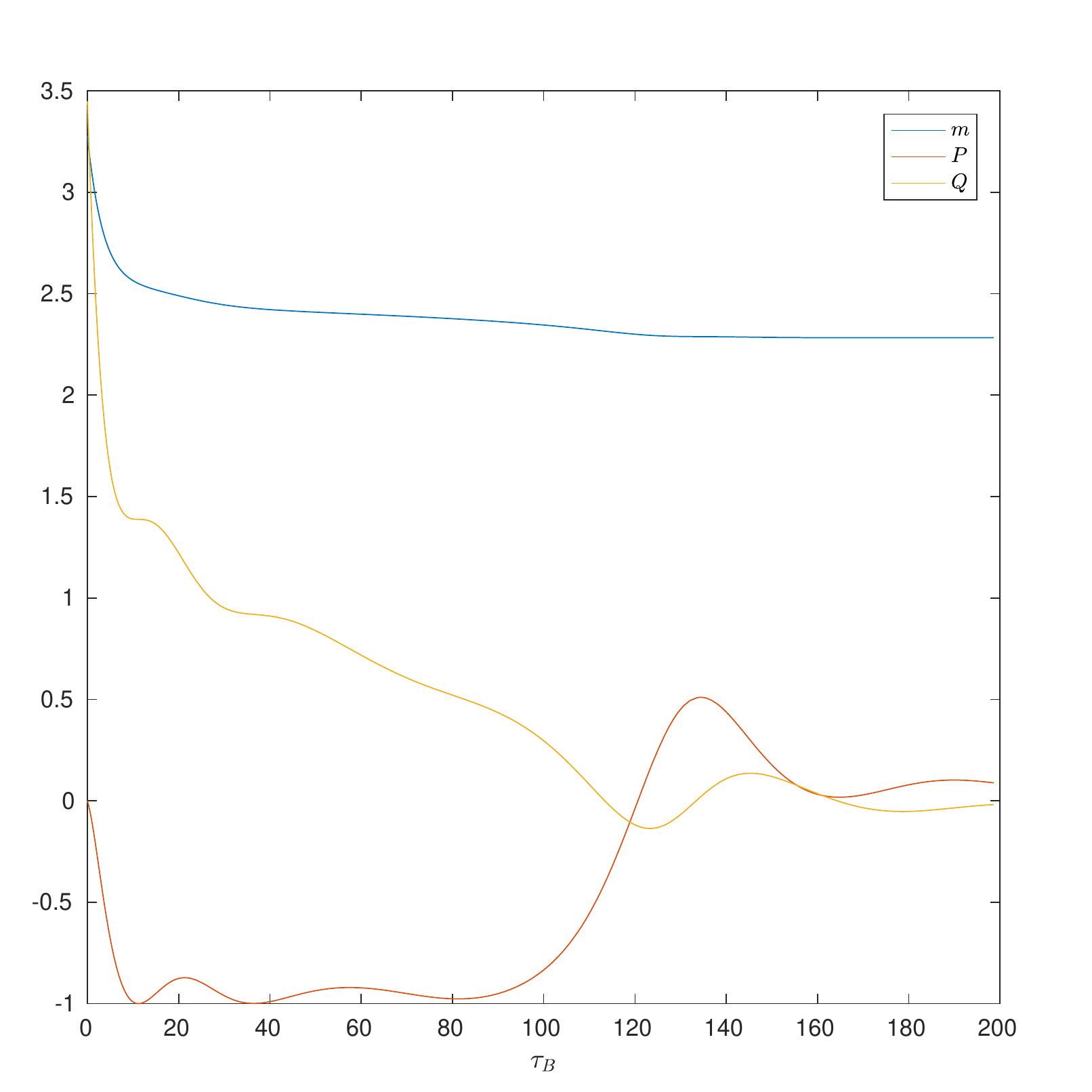}
  \caption{A plot of $ m $, $ P $, and $ Q $ on $ \mathscr{I}^+ $ with respect to Bondi time for the solution with initial data given by (\ref{ID e}), $ s = 0.35 $.}
  \label{fig:RN35}
\end{figure}

\begin{figure}[!ht]
  \centering
  \includegraphics[scale=0.65]{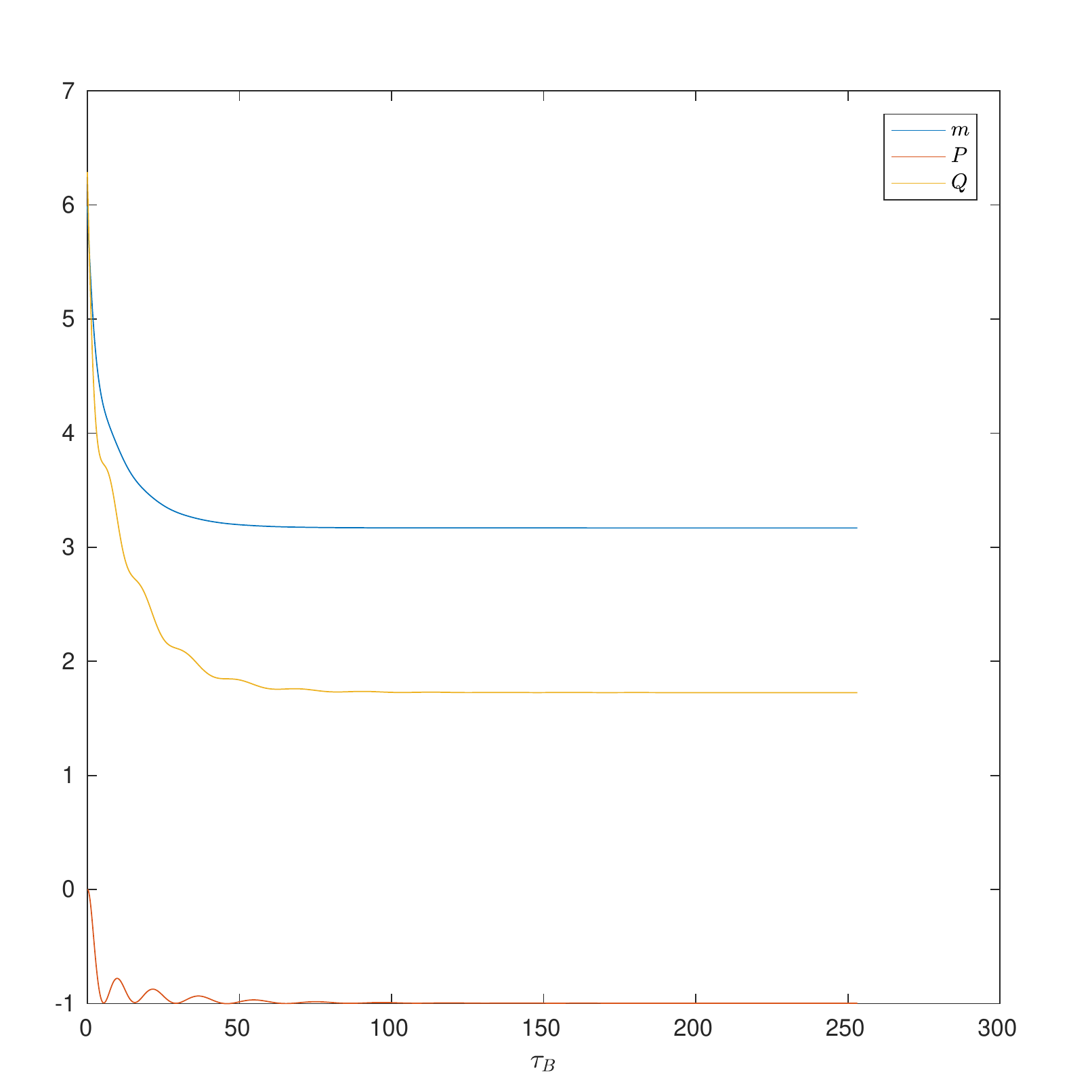}
  \caption{A plot of $ m $, $ P $, and $ Q $ on $ \mathscr{I}^+ $ with respect to Bondi time for the solution with initial data given by (\ref{ID e}), $ s = 0.45 $.}
  \label{fig:RN45}
\end{figure}

Before investigating the nature of the near-critical solutions, we notice some unexpected behaviour in the supercritical regime of the family (\ref{ID e}). For $ s \gtrsim 0.3 $ we observe that the solution approaches a Reissner-Nordstr\"{o}m black hole before decaying to a Schwarzschild black hole.
We plot the mass and charges on future null infinity in figure \ref{fig:RN35} for $ s = 0.35 $, and for $ s = 0.45 $ in figure \ref{fig:RN45}.

As $ s $ is increased, the time spent near the Reissner-Nordstr\"{o}m black hole also increases. Note that $ s $ cannot be increased past $ s = 0.4877 $ because a white hole is then introduced in the initial conditions.
The solution cannot remain near the Reissner-Nordstr\"{o}m black hole because it has infinitely many unstable modes. However, Rinne found that a Reissner-Nordstr\"{o}m black hole could be evolved in the purely magnetic case by tuning two parameters \cite{Rinne14}. It is surprising to find here that it can be achieved without any tuning.
In figure \ref{fig:RN45} we see that it remains near the Reissner-Nordstr\"{o}m black hole for as long as we are able to evolve into the future for a maximum refinement level of 30.

\begin{figure}[!ht]
  \centering
  \includegraphics[scale=0.9]{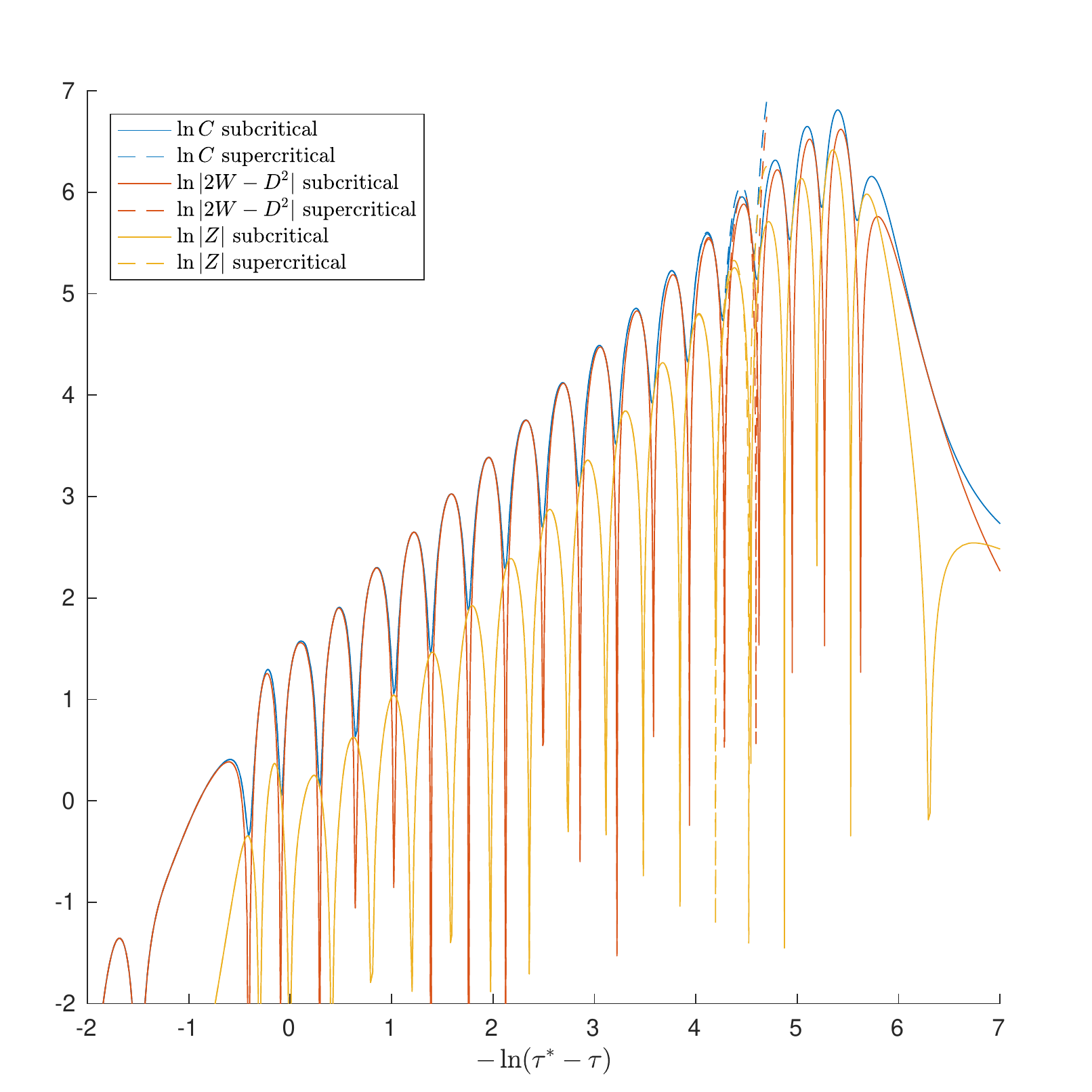}
  \caption{A plot of $ |2W-D^2| $, $ |Z| $, and $ C $ of the solution with initial data given by (\ref{ID a}).}
  \label{fig:S-S C vs t a}
\end{figure}

\begin{figure}[!ht]
  \centering
  \includegraphics[scale=0.9]{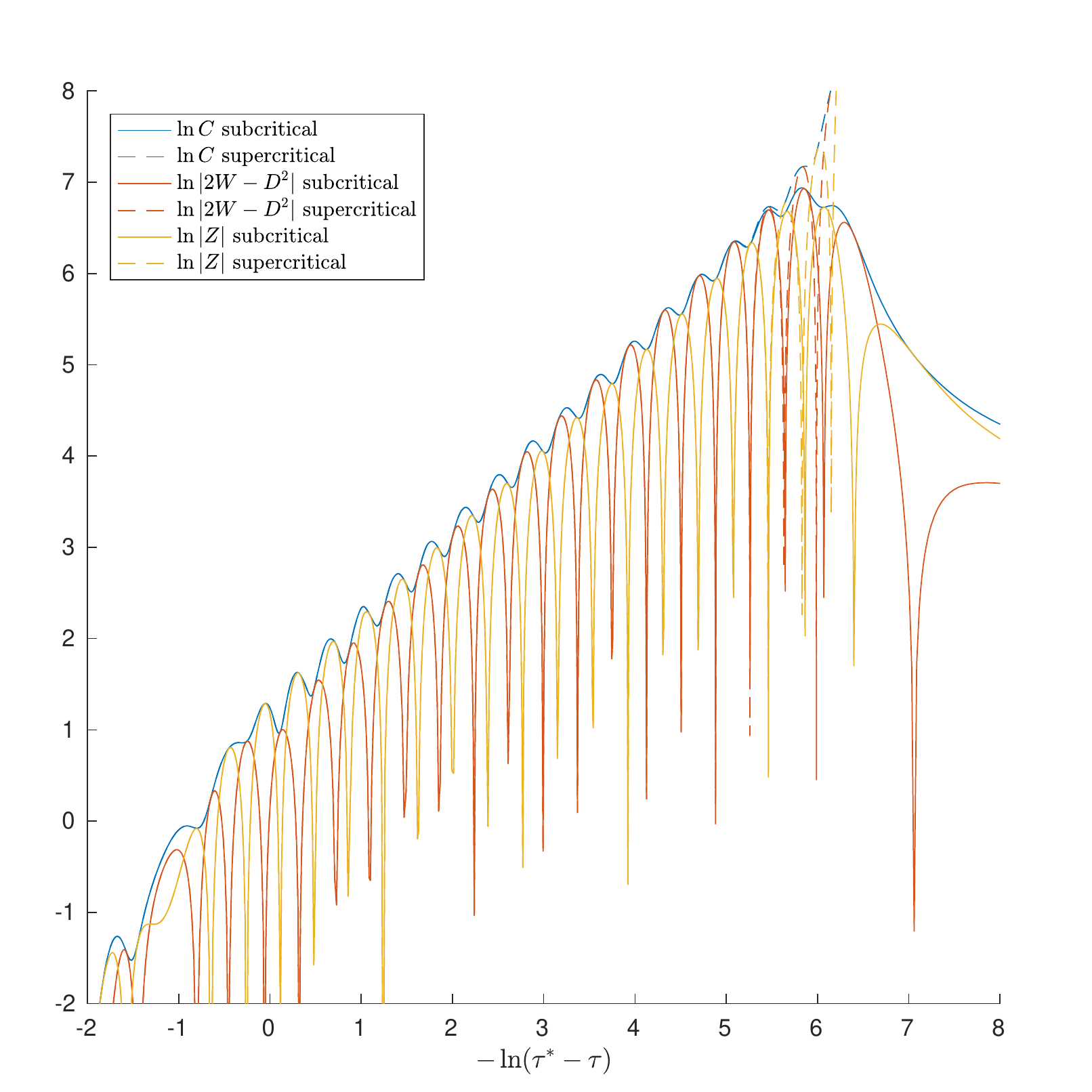}
  \caption{A plot of $ |2W-D^2| $, $ |Z| $, and $ C $ of the solution with initial data given by (\ref{ID e}).}
  \label{fig:S-S C vs t e}
\end{figure}

We now return to the critical phenomena.
We use zero crossings to estimate the critical time $ t^* $ and plot the following functions at the origin (figures \ref{fig:S-S C vs t a} and \ref{fig:S-S C vs t e});
\begin{align*}
&|2W-D^2| \;, &&|Z| \;, & C &= \sqrt{(2W-D^2)^2+Z^2} \;.
\end{align*}
Recall that these are gauge-independent functions, and that by equations (\ref{K0}-\ref{C}), $ C $ is a measure of the curvature as it is proportional to both $ |K|^\frac{1}{4} $ and $ |L|^\frac{1}{4} $ (see.
We use the maxima of $ C $ to determine the growth rate $ \Delta = 0.73_{34}^{67} $.
We immediately see a difference to the magnetic case (figure \ref{fig:S-S W}), in that the solution is not approximately discretely self-similar (although by itself $ C $ does appear approximately discretely self-similar).
In each case we observe the relative amplitudes of $ |2W-D^2| $ and $ |Z| $ change with $ \tau $.
Furthermore, there is no distinct period; the zeros of $ |Z| $ are closer together than the zeros of $ |2W-D^2| $.
Also, since the behaviour depends on the initial conditions, there is no longer a universal critical solution.

\begin{figure}[!ht]
  \centering
  \includegraphics[scale=0.8]{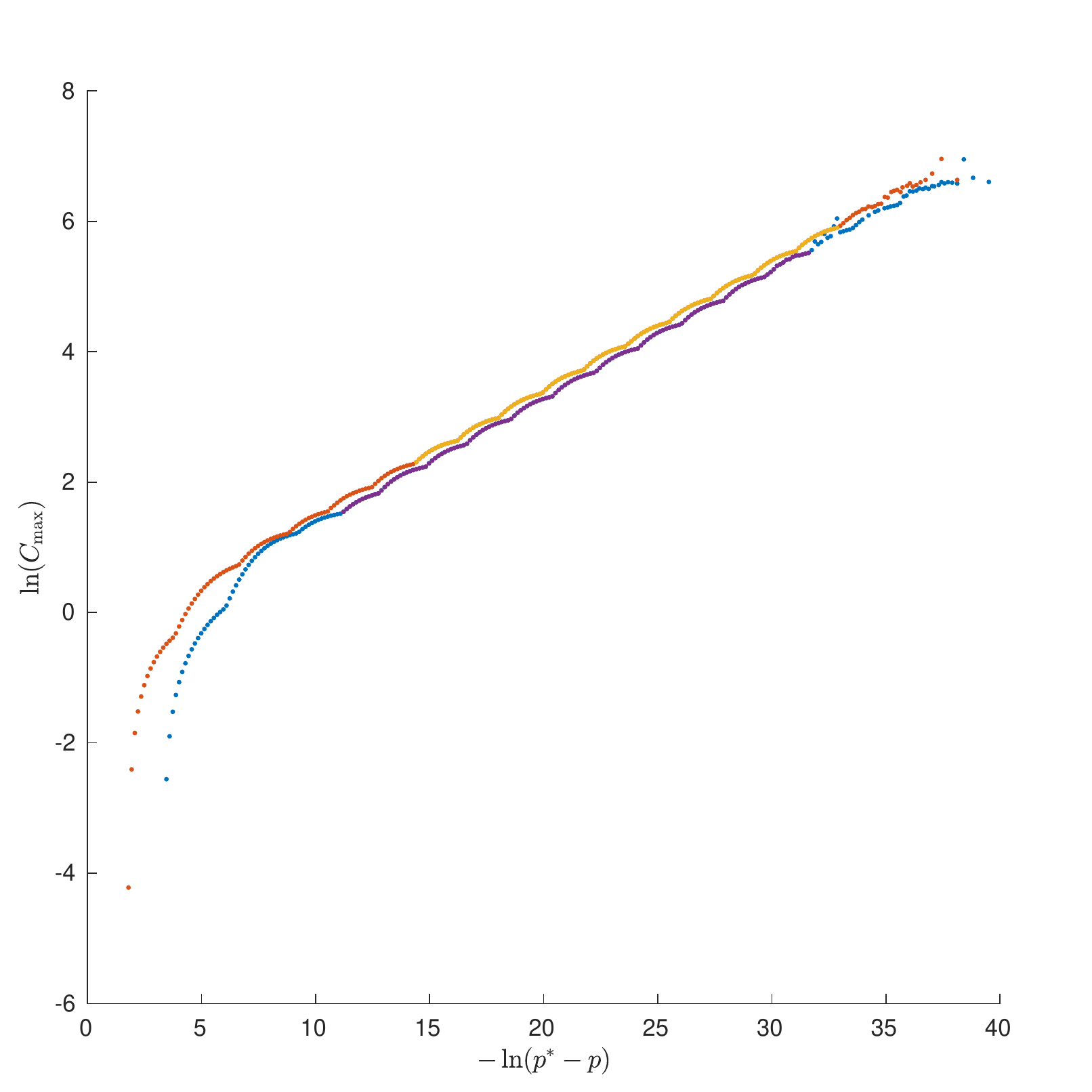}
  \caption{A plot of the maximum of $ C $ versus the parameter ($ p $ or $ s $) for the initial data families given by (\ref{ID a}) (blue) and (\ref{ID e}) (orange). The points used in the regression fit are marked in yellow and purple respectively.}
  \label{fig:S-S C vs p ae}
\end{figure}

We also evolve the initial data for many values of the parameters, approaching the critical parameter from below, and plot the maximum curvature $ C $ as a function of the parameter (figure \ref{fig:S-S C vs p ae}).
\begin{figure}[!ht]
  \centering
  \includegraphics[scale=0.8]{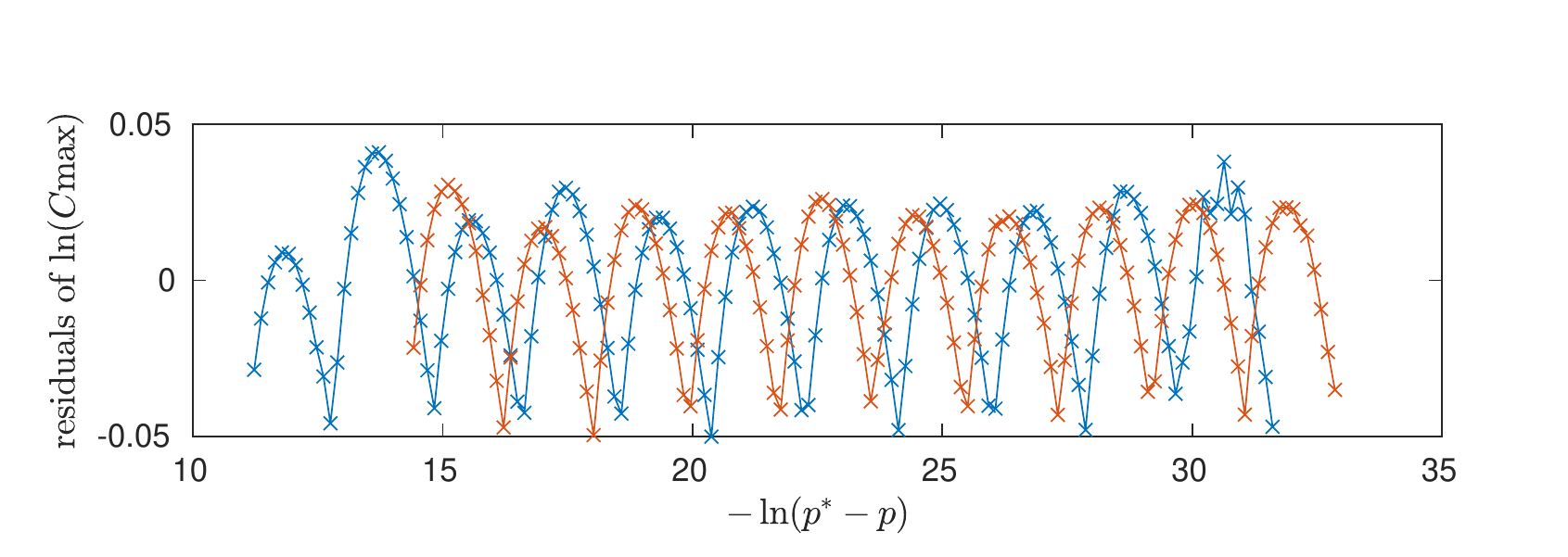}
  \caption{A plot of the residuals of $ \ln(C_\mathrm{max}) $ versus the parameter for the initial data family given by (\ref{ID a}) (blue) and (\ref{ID e}) (orange).}
  \label{fig:S-S resid vs p ae}
\end{figure}
The resulting value of $ \gamma $ is 0.195680 for family (\ref{ID a}) and 0.195688 for family (\ref{ID e}). These are equal up to our numerical error, which from the magnetic test case is at least 0.0001.
The fine structure appears as in the magnetic case for both families (figure \ref{fig:S-S resid vs p ae}).

\begin{figure}[!ht]
  \centering
  \includegraphics[scale=0.75]{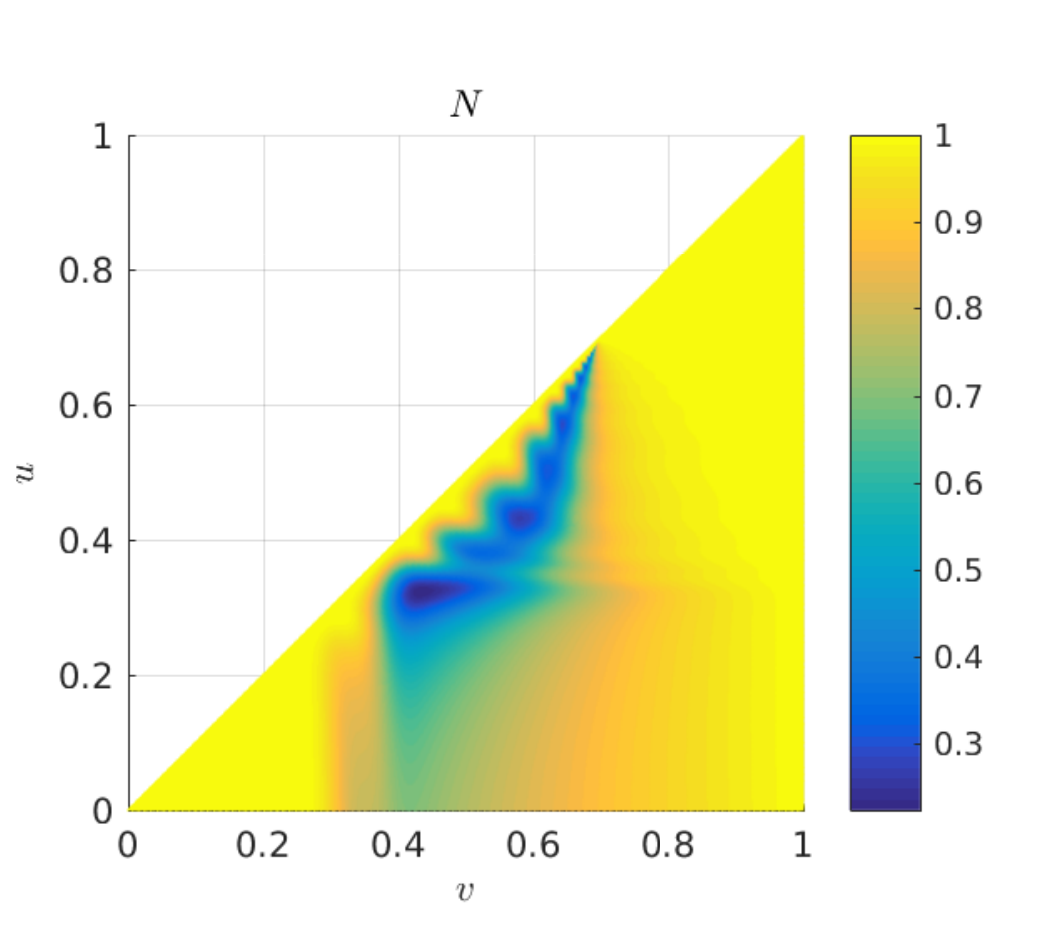}
  \includegraphics[scale=0.75]{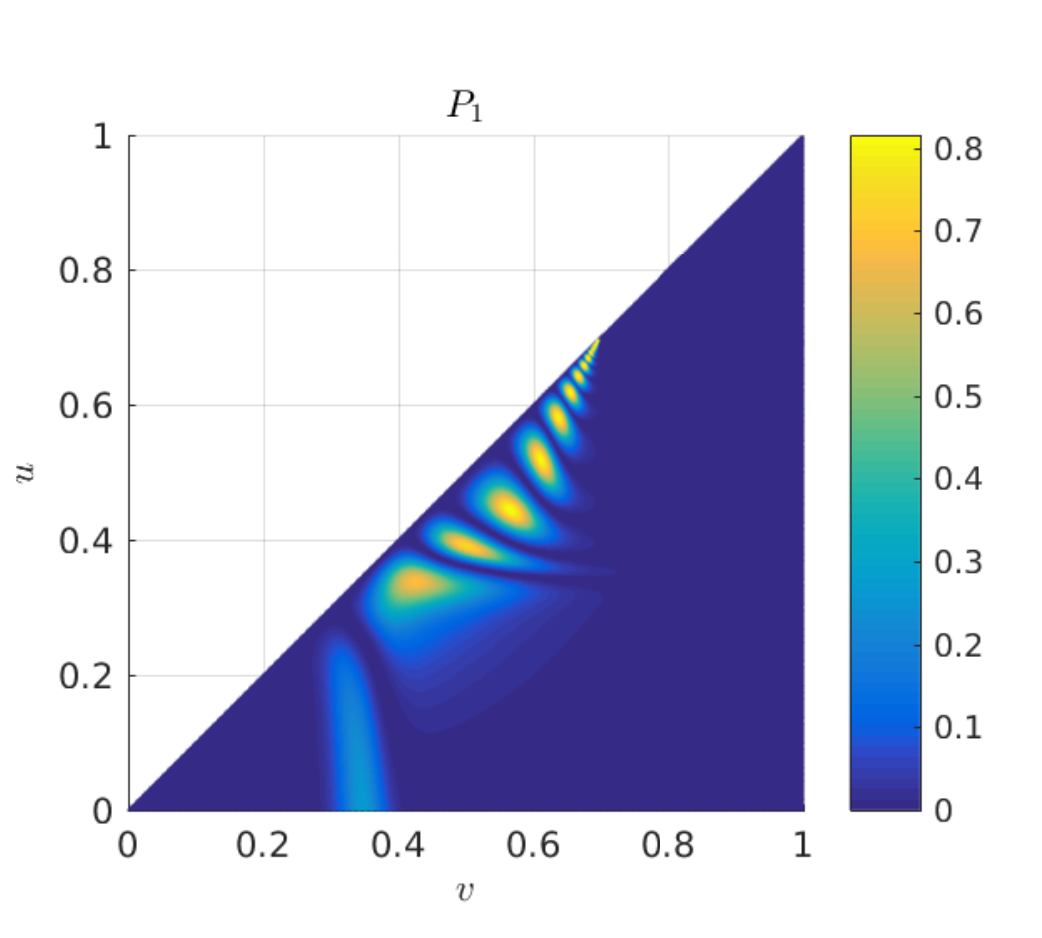}
  \includegraphics[scale=0.75]{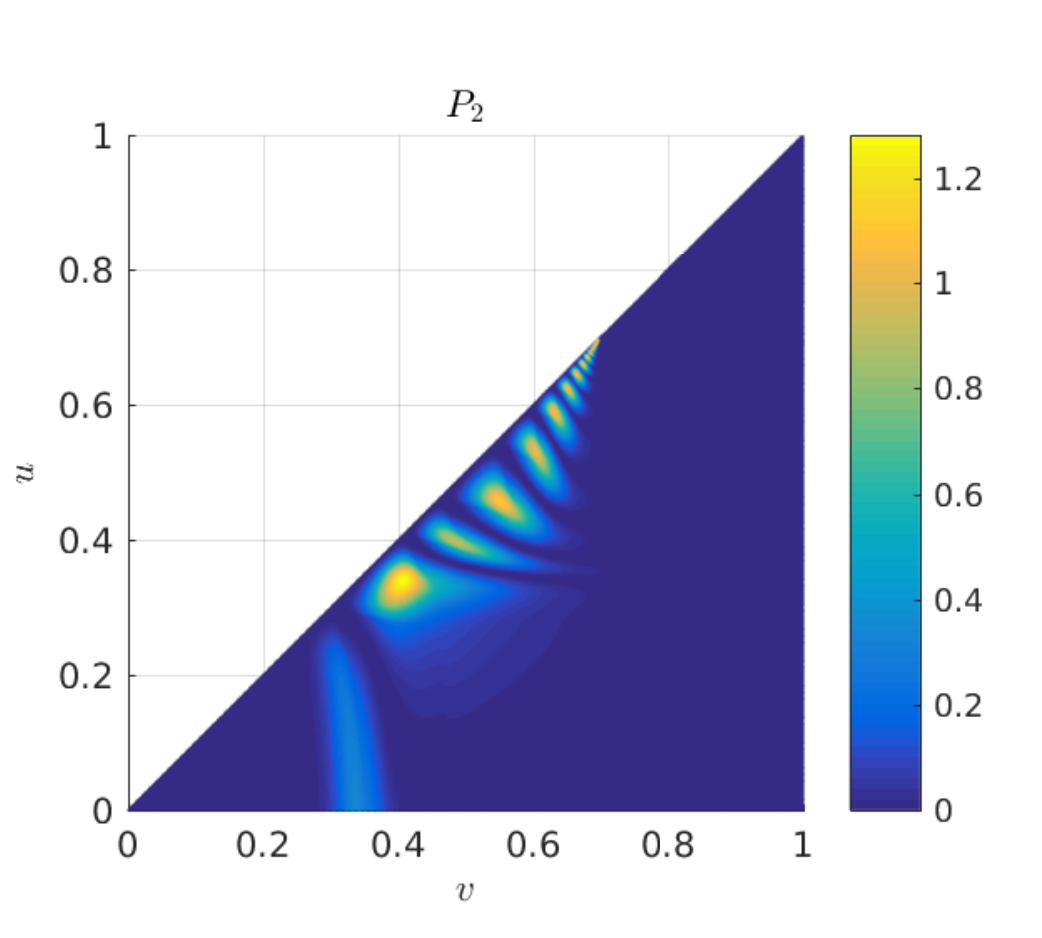}
  \includegraphics[scale=0.75]{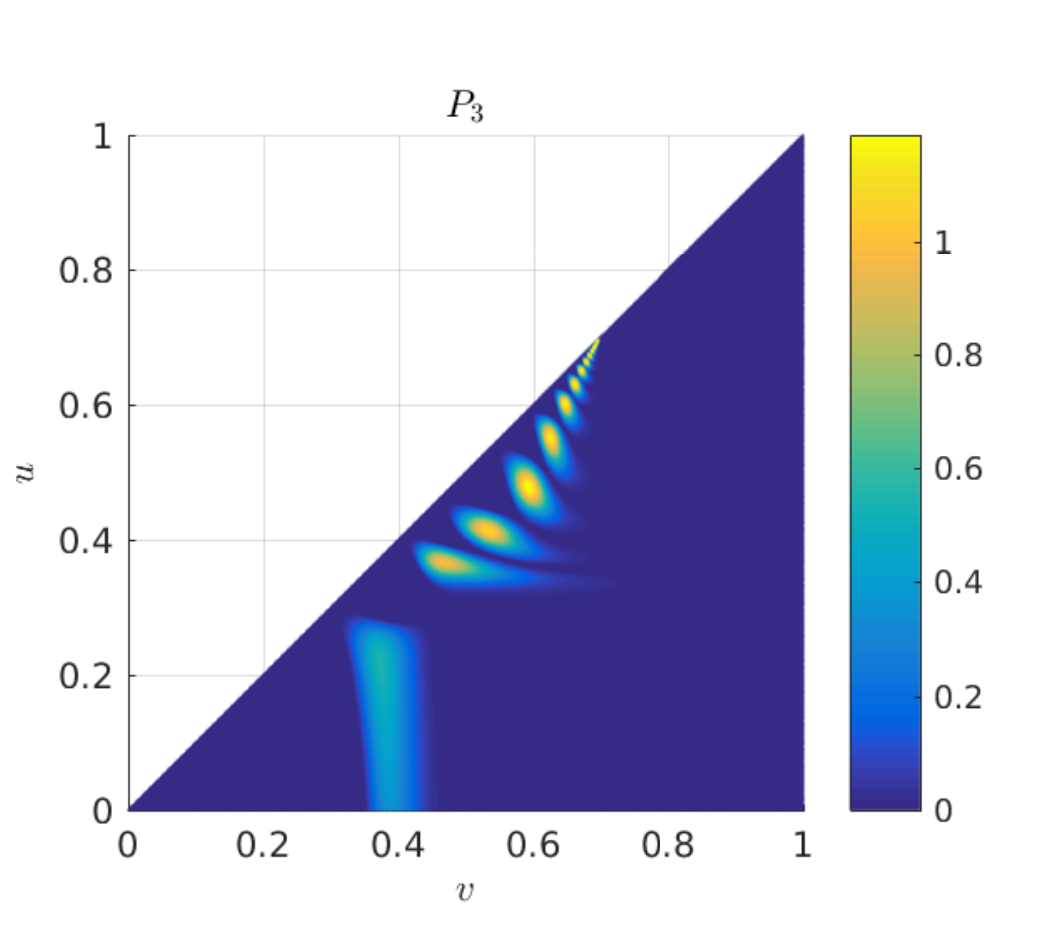}
  \caption{Plots of $ N $, $ P_1 $, $ P_2 $, and $ P_3 $ for the just subcritical evolution of (\ref{ID m}).}
  \label{fig:S-S NP123 m}
\end{figure}

\begin{figure}[!ht]
  \centering
  \includegraphics[scale=0.4]{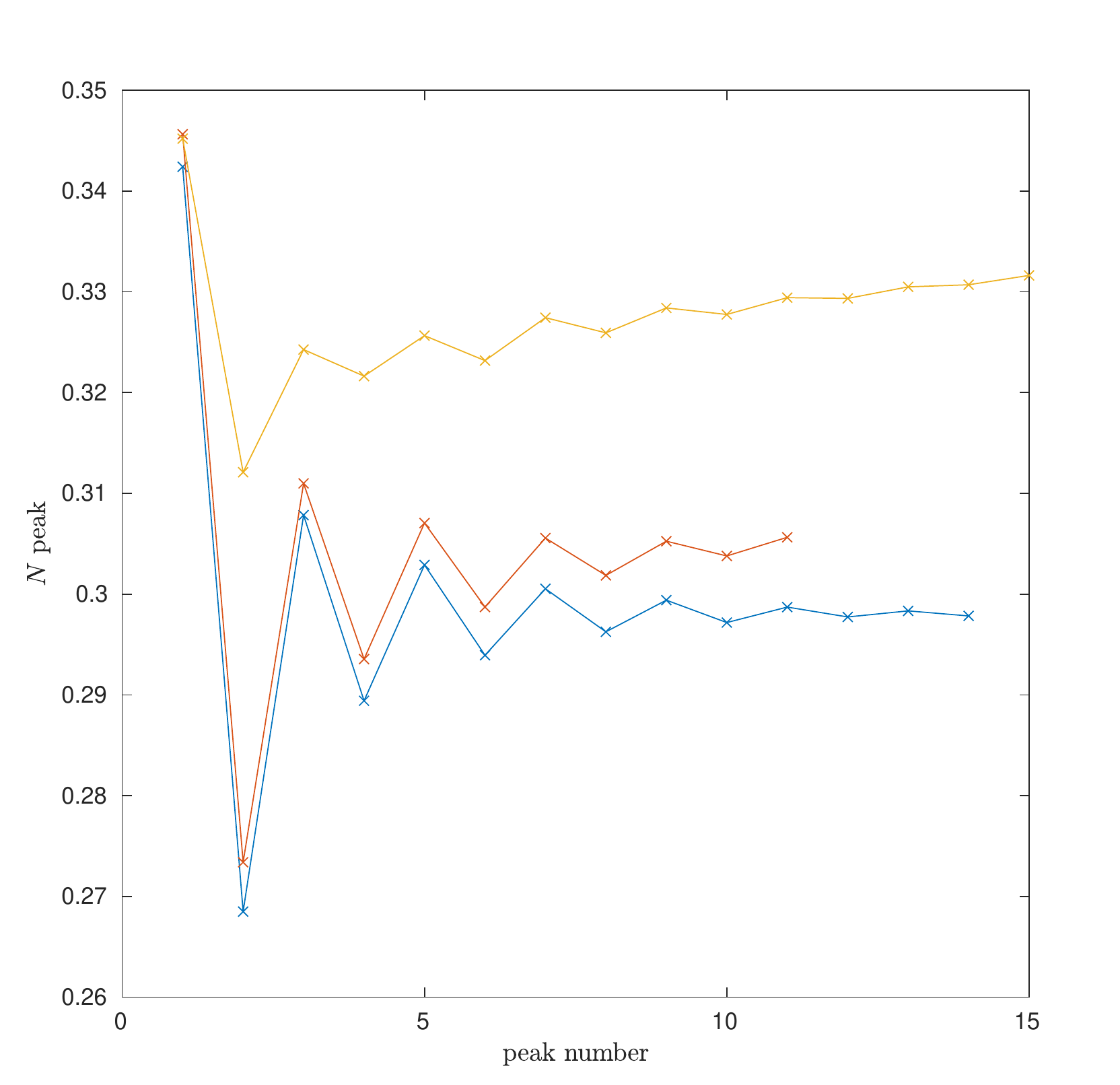}
  \includegraphics[scale=0.4]{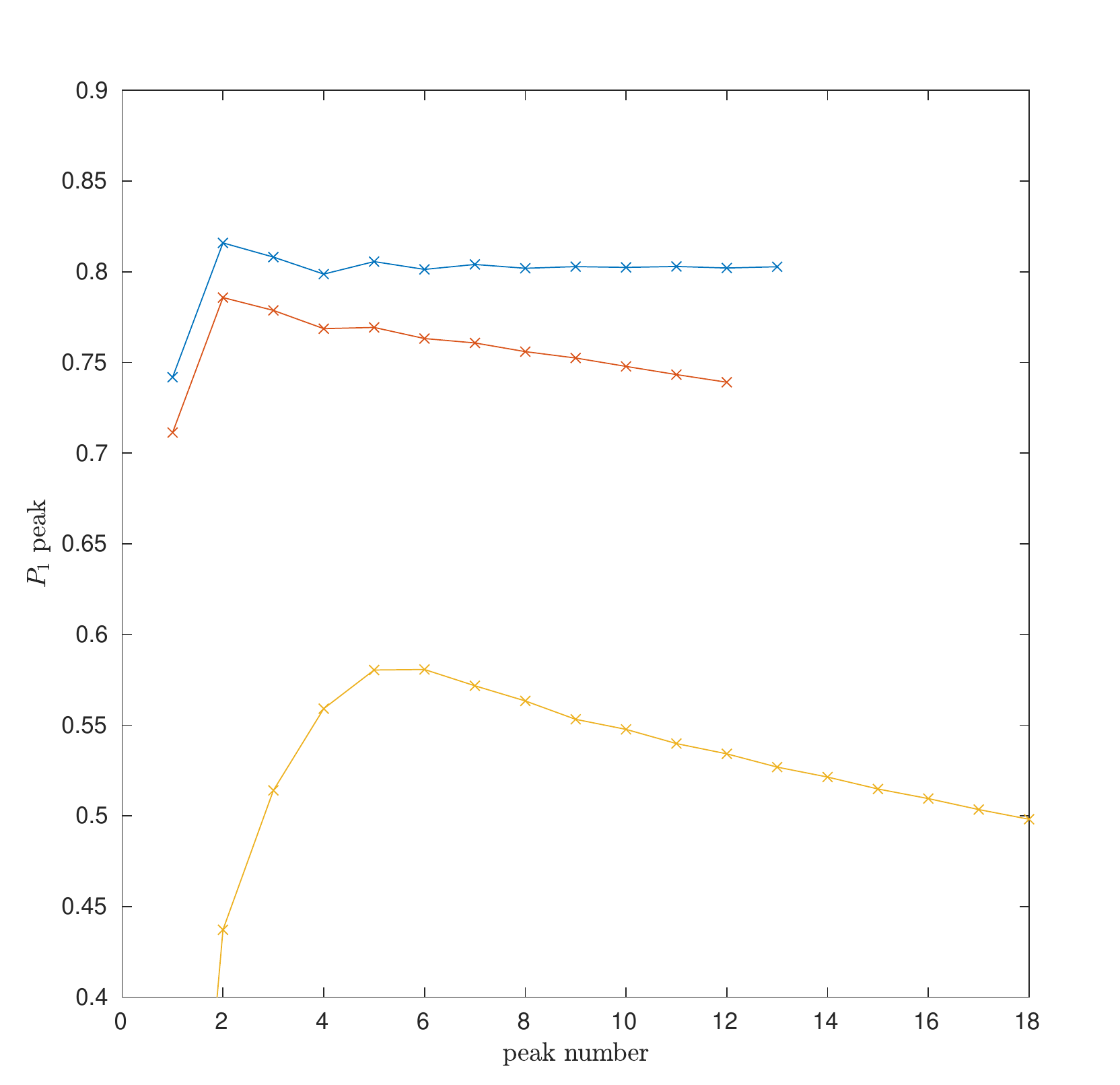}
  \includegraphics[scale=0.4]{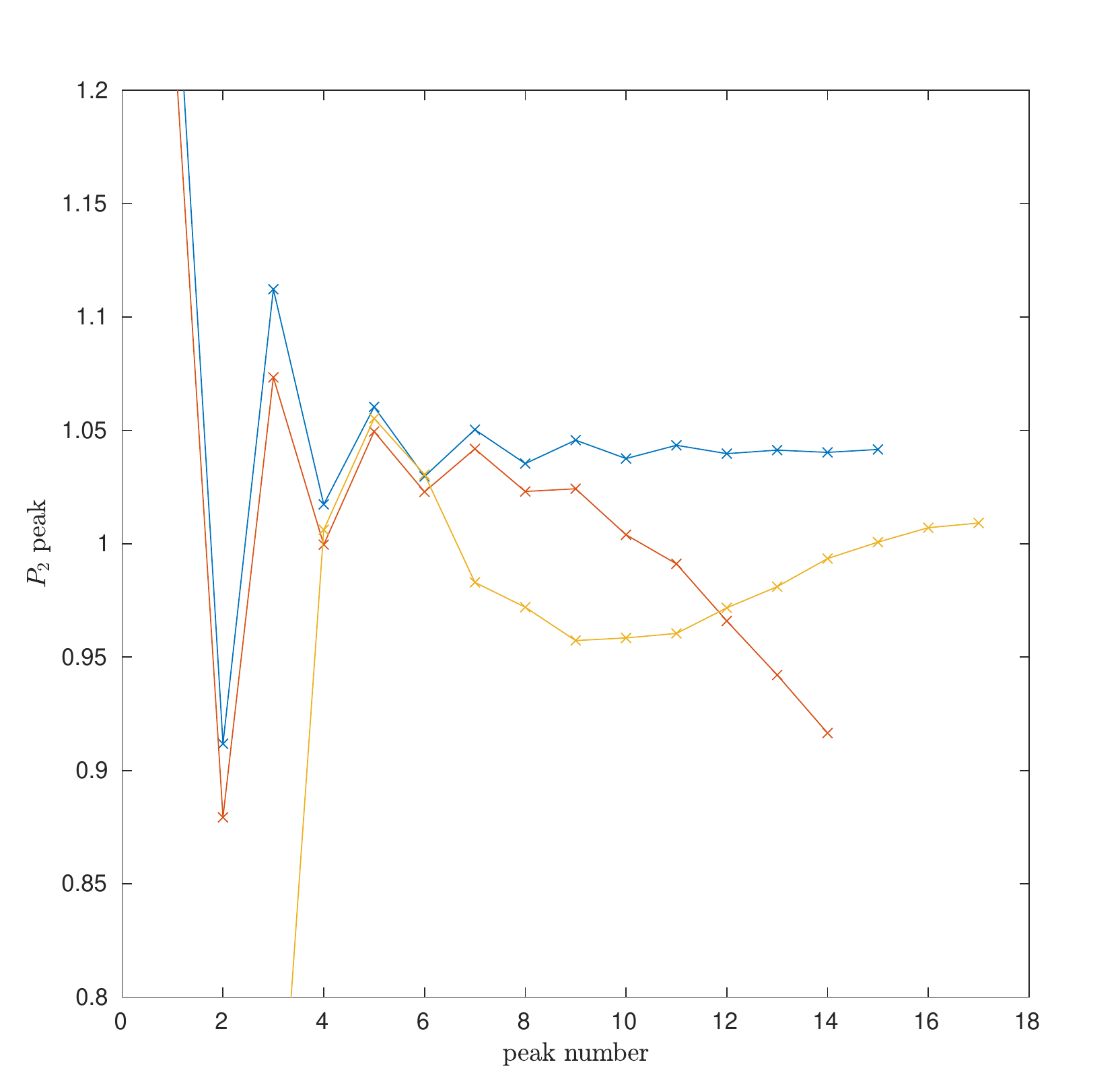}
  \includegraphics[scale=0.4]{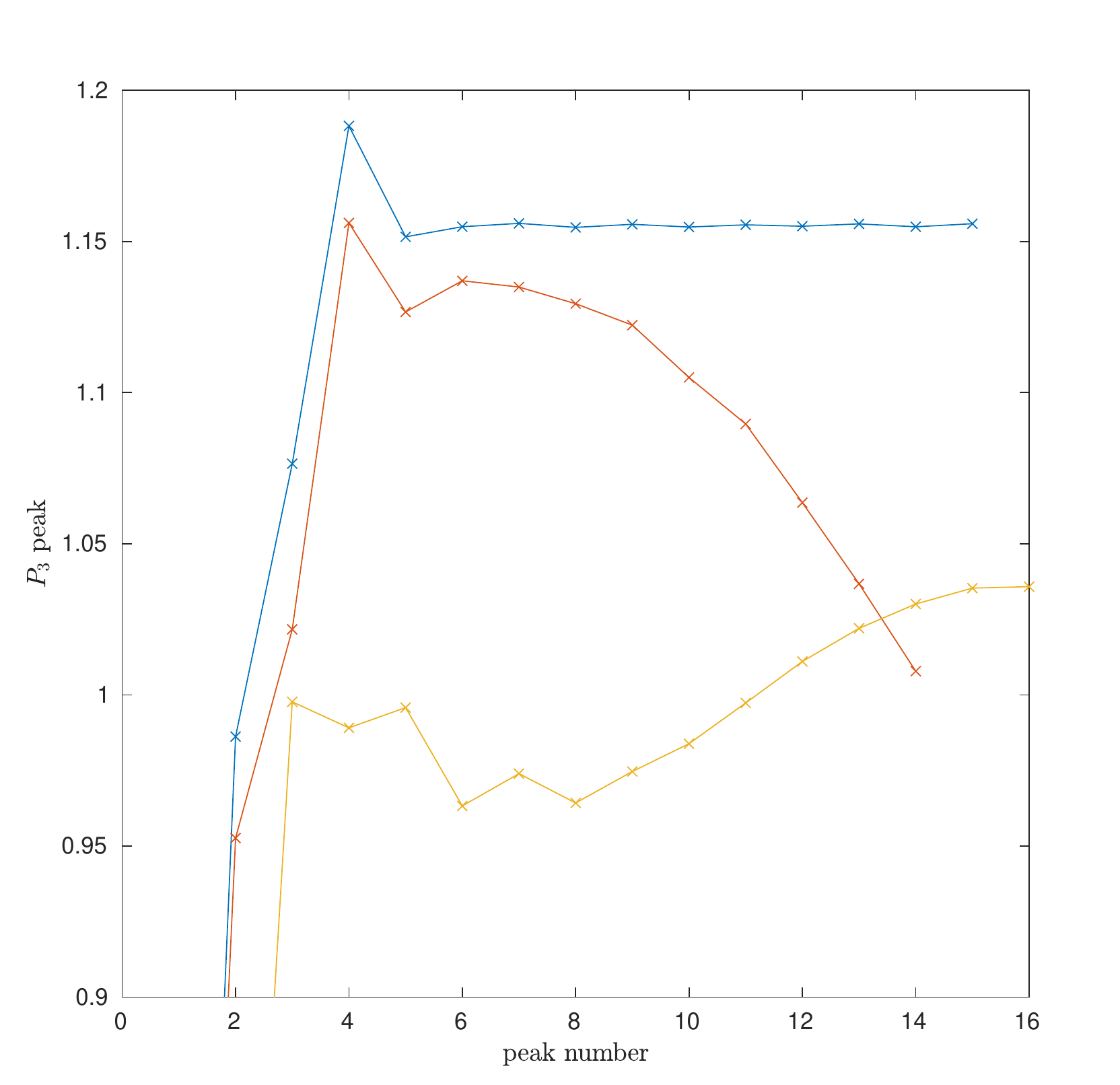}
  \caption{Plots of the peaks of $ N $, $ P_1 $, $ P_2 $, and $ P_3 $ for the evolutions of (\ref{ID m}) (blue), (\ref{ID a}) (orange), and (\ref{ID e}) (yellow).}
  \label{fig:S-S NP123peaks mae}
\end{figure}

To compare further the general critical solutions with the universal magnetic solution, we plot in figure \ref{fig:S-S NP123 m} some functions that have maxima away from the origin; $ N $ for the metric and $ P_1 $, $ P_2 $, and $ P_3 $ (\ref{P123}) for the gauge field.
The plots in figure \ref{fig:S-S NP123 m} are for a subcritical magnetic solution, and the full case looks generally very similar. However, there is an important difference in the behaviour of the peaks.
For each of the functions in the magnetic case, the peaks approach (from alternating sides) a constant value (the blue lines in figure \ref{fig:S-S NP123peaks mae}).
In the general case the peak amplitude continues to change after the alternating behaviour has settled down (the orange and yellow lines in figure \ref{fig:S-S NP123peaks mae}).
The values in figure \ref{fig:S-S NP123peaks mae} are found as follows.
Let $ e_{\text{sub},i} $ and $ e_{\text{super},i} $ be the extrema of any of $ \{N,P_1,P_2,P_3\} $ for the subcritical and supercritical evolution respectively, indexed by the peak number $ i $. We take the average $ e_i = \frac{e_{\text{sub},i}+e_{\text{super},i}}{2} $ to be the value of the extrema until the evolutions diverge. This is taken to be when their respective peak values differ by more than the difference between successive peaks; that is, when $ |e_{\text{super},i}-e_{\text{sub},i}| > |e_i-e_{i-1}| $.

We note the difficulty in exploring these trends any further; since the critical parameter was in each case determined to machine precision, this is as close as the subcritical and supercritical evolutions can be with this numerical method.
Nevertheless from these results we can see that some aspects of the critical behaviour in the magnetic case remain; the exponential scaling of the curvature at the origin and some kind of fine structure still appear.
The critical solutions still display approximate self-similarity, but no longer appear to approach true self-similarity asymptotically. Since their precise form now depends on the details of the initial data, it appears that there is no universal critical solution.

We investigate the critical solution for initial data (\ref{ID a}) a little further. We decrease the tolerances to $ (\epsilon_u,\epsilon_v) = \left(\frac{1}{8^3},\frac{1}{8^5}\right) $ and $ \left(\frac{1}{8^4},\frac{1}{8^6}\right) $, and determine the critical parameters with \verb_pxbe = false_ to save time and memory, as we will only consider the behaviour at the origin.
The critical parameters we find are shown in table \ref{table:peps}, and the equivalents of figure \ref{fig:S-S C vs t a} are shown together in figure \ref{fig:S-S C vs t a243546}.
\begin{table}[!ht]
  \centering
  \renewcommand{\arraystretch}{1.5}
  \begin{tabular}{ll}
    \toprule
    $ (\epsilon_u,\epsilon_v) $ & $ p^* $ \\
    \midrule
    $ \left(\frac{1}{8^2},\frac{1}{8^4}\right) $ & $ 0.03507483593605378_{17}^{86} $ \\
    $ \left(\frac{1}{8^3},\frac{1}{8^5}\right) $ & $ 0.03508189107275219_{01}^{70} $ \\
    $ \left(\frac{1}{8^4},\frac{1}{8^6}\right) $ & $ 0.03508136933654187_{18}^{88} $ \\
    \bottomrule
  \end{tabular}
  \renewcommand{\arraystretch}{1}
  \caption{The critical parameters for (\ref{ID a}) and \texttt{pxbe = false} for various tolerances.}
  \label{table:peps}
\end{table}

\begin{figure}[!ht]
  \centering
  \animategraphics[scale=0.9,step]{}{S-S_CWDZ_vs_tau_a_}{24}{26}
  \caption{A plot of $ |2W-D^2| $, $ |Z| $, and $ C $ of the solution with initial data given by (\ref{ID a}), with \texttt{pxbe = false}. Click to compare the results of different tolerances (digital only).}
  \label{fig:S-S C vs t a243546}
\end{figure}
We see in figure \ref{fig:S-S C vs t a243546} that the solutions with lower tolerances show the same behaviour, confirming again that we can accurately determine the critical behaviour with only modest tolerances.

\chapter{Conclusion} \label{ch:C}
In the context of the static, spherically symmetric SU(2) Einstein-Yang-Mills equations, we have found new solutions in the electric sector numerically. These long-lived solutions exhibit apparently infinite oscillations, have interesting, complicated asymptotics, and fail to be asymptotically flat.

We have developed a code to solve the SU(2) Einstein-Yang-Mills equations in spherical symmetry, based on double-null coordinates. It uses adaptive mesh refinement that exploits the double-null coordinates and results in the efficient calculation and storage of the entire future of the initial data, including future null infinity. It can evolve spacetimes that develop black holes and approach the singular line, and it can evolve spacetimes that exhibit self-similarity, requiring many orders of magnitude finer resolution in certain regions.

We used this code to study the critical behaviour in the general, not purely magnetic case. We found evidence suggesting that unlike the magnetic case, and unlike all critical behaviour found to date, there is not a single, universal critical solution, nor are the critical solutions (asymptotically) discretely self-similar. The introduction of the additional Yang-Mills variables apparently provides sufficient extra degrees of freedom for behaviour beyond the standard observed critical phenomena.

As noted in the introduction, the preprint \cite{MR18} was released as this thesis was being completed. The authors also considered the critical phenomena for the fully general spherically symmetric SU(2) Einstein-Yang-Mills system. They found similar subcritical scaling of the maximum curvature at the origin, but found that different families produced measurably different scaling exponents. They also produced evidence that the universality of the critical solution in the purely magnetic case is lost in the general case, consistent with our results.

\section{Future work}
It should be relatively easy to extend the code to solve the spherically symmetric Einstein equations with other matter models. The main property it relies on is that the first order equations can be ordered so that they are linear in their unknown variables. For each matter model it will be a case of choosing the first order variables so that this property, which can be achieved in the scalar field and EYM model presented here, is achieved.

While setting the initial data on an outgoing null slice is quite natural for a double-null code, it could be very useful to instead set the data on a Cauchy surface. This would allow, for example, black hole initial conditions to be set (with or without perturbations), as well as the entire spacetime (past and future) to be evolved from the Cauchy data.
The main issue in this case is that spatial infinity would become a point in the domain. While double-null coordinates could be chosen so that any point on past and future null infinity can be treated as a regular singular point, spatial infinity will be an irregular singular point when the mass is non-zero. One possible way of including spatial infinity would be to assume the initial data is precisely Schwarzschild for large enough $ r $; then a diamond of a Schwarzschild spacetime could be assumed around spatial infinity, and the remainder of the spacetime evolved as usual.

With Cauchy initial data implemented or not, we believe this code set-up could be useful as a teaching aid, to develop students' understanding and intuition about general relativity. The restricted dimension keeps it simple enough to easily compute and visualise, however there is still the interesting dynamics of dispersal, black hole formation, different kinds of critical phenomena, and self-similarity with this matter model.
Due to the spherical symmetry, there can be no binary systems or gravitational waves, but it is possible to compute any radiation at null infinity.
In particular, clearly seeing the possible behaviours of MTTs could be quite useful. With Cauchy data the relationship between white holes, black holes, and dispersal could be investigated.
The ability to numerically change coordinates after the evolution also enables a useful variety of perspectives.
We note that the hyperboloidal method used by \cite{RM13}, for example, is also a very promising approach to solving large regions of a spacetime. It uses space-like slices that intersect future null infinity, so cannot include spatial infinity, but it does generalise much easier beyond spherical symmetry.

The double-null coordinates we've used in this thesis brought many advantages to the numerical evolution of the equations, but we here note a difficulty that was encountered with these coordinates when evolving black hole spacetimes. In these cases, future timelike infinity is brought to a point $ (u^+,1) $ with $ u^+ < 1 $. This point is one end of the singular $ r = 0 $ line, and as such demands infinite refinement to reach.
It is of course desirable to evolve to later times in order to accurately determine the final black hole mass, but this proved difficult with our code. The nature of the adaptive refinement to evolve at a coarse resolution first then repeatedly at finer resolutions is inefficient here, and perhaps different coordinates or a method better able to account for the singularity would be preferable in this region.

In this work we considered a restrictive fall-off condition for the Yang-Mills field, such that $ w $ and $ d $ are bounded at infinity. This is stronger than what is required by finite ADM mass, and it remains to be seen if there is any new interesting behaviour without this assumption.

\newpage
\emergencystretch=1em
\printbibliography[heading=bibintoc]

\appendix
\chapter{Coordinate transformations}
\section{Polar-areal and isothermal (static)}
For a static spacetime that can be written in both polar-areal and isothermal coordinates (so $ N > 0 $), it is straightforward to relate the functions.
Assuming $ \pypx{}{T} $ is chosen to be the timelike hypersurface-orthogonal Killing vector, we can take $ T = t $ and $ \bar{a} = \hat{a} $.
The gauge function $ w $ is coordinate-independent, and the other gauge functions are trivial for a static spacetime (see chapter \ref{ch:S}).

Given a static solution in polar-areal coordinates (\ref{metric tr}) $ S(r) $ and $ m(r) $, we find $ R $ by integration and $ \alpha $ as follows;
\begin{align*}
\dydx{R}{r} &= \frac{1}{SN} \;, & \alpha &= S\sqrt{N} \;.
\end{align*}

Given a static solution in isothermal coordinates (\ref{metric TR}) $ r(R) $ and $ \alpha(R) $, we immediately have
\begin{align*}
N &= \frac{r'^2}{\alpha^2} \;, & S &= \frac{\alpha^2}{r'} \;.
\end{align*}

\section{Isothermal and double-null}
The double-null coordinates (\ref{metric dn}) relate straightforwardly to the isothermal coordinates (\ref{metric TR}):
\begin{subequations}
Let
\begin{align} \label{uvTR}
T &= v+u \;, & R &= v-u \;,
\end{align}
and note that $ \alpha $ has been kept consistent between the two coordinate charts.
The coordinate-dependent gauge functions relate by
\begin{align}
a &= \bar{a}-\bar{b} \;, & b &= \bar{a}+\bar{b} \;.
\end{align}
\end{subequations}

\section{Polar-areal and double-null} \label{ss:pa dn}
Given an EYM solution in double-null coordinates, we wish to use $ r $ as a coordinate and seek an orthogonal time coordinate $ t $.
We know $ \dy{r} = r_u\dy{u}+r_v\dy{v} $ and thus we use $ \dy{t} = -\frac{r_u}{c}\dy{u} +\frac{r_v}{c}\dy{v} $ where $ c $ is a to-be-determined positive function of $ u $ and $ v $. These relations allow us to transform the derivatives by
\begin{subequations}
\begin{align}
\pypx{}{t} &= -\frac{c}{2r_u}\pypx{}{u} +\frac{c}{2r_v}\pypx{}{v} \;, & \pypx{}{r} &= \frac{1}{2r_u}\pypx{}{u} +\frac{1}{2r_v}\pypx{}{v} \;, \label{pa derivs in dn} \\
\pypx{}{u} &= r_u\left(\pypx{}{r}-\frac{1}{c}\pypx{}{t}\right), & \pypx{}{v} &= r_v\left(\pypx{}{r}+\frac{1}{c}\pypx{}{t}\right). \label{dn derivs in pa}
\end{align}
\end{subequations}
Enforcing $ t_{uv} = t_{vu} $ gives us the following advection equation for $ c(u,v) $:
\begin{align} \label{c}
r_v c_u + r_u c_v &= 2c r_{uv} \;, & c(u,u) = 1 \;.
\end{align}
The initial condition completes the specification of the coordinates; here it is chosen so that $ t $ is the proper time at the origin.

We can then write the metric functions as
\begin{align} \label{SN from dn}
N &= -\frac{r_u r_v}{\alpha^2} \;, & S &= \frac{c}{N} \;.
\end{align}
The time coordinate can be found by solving $ t_v = c r_v $ or $ t_u = -c r_u $.
The gauge potentials are given by
\begin{align} \label{ab dn to pa}
\hat{a} &= \frac{c}{2}\left(\frac{b}{r_v} -\frac{a}{r_u}\right), & \hat{b} &= \frac{1}{2}\left(\frac{b}{r_v} +\frac{a}{r_u}\right).
\end{align}
Now using equations (\ref{SN from dn}), (\ref{ab dn to pa}), and (\ref{pa derivs in dn}) we see that the definitions of $ z $ (\ref{pa first},\ref{dn first}) are consistent.

Given an EYM solution in polar-areal coordinates, the null lines can be found from $ \dydx{t}{r} = \pm\frac{1}{SN} $.
We can convert (\ref{metric tr}) to (\ref{metric dn}) by finding the integrating factors $ \mu(t,r) $ and $ \nu(t,r) $ such that
\begin{align*}
\mu\left(S\sqrt{N}\dy{t}-\frac{1}{\sqrt{N}}\dy{r}\right) &= 2\dy{u} \;, & \nu\left(S\sqrt{N}\dy{t}+\frac{1}{\sqrt{N}}\dy{r}\right) &= 2\dy{v} \;.
\end{align*}
Then $ \alpha^2 = \frac{1}{\mu\nu} $, where $ \mu $ and $ \nu $ solve the advection equations:
\begin{subequations} \label{pa to dn}
\begin{align}
\dot\mu +SN\mu' &= \left(-S'N-\frac{SN'}{2}+\frac{\dot{N}}{2N}\right)\mu \;, \\
\dot\nu -SN\nu' &= \left(S'N+\frac{SN'}{2}+\frac{\dot{N}}{2N}\right)\nu \;,
\end{align}
with initial conditions
\begin{align}
\nu(t,0) &= 1 \;, & \mu(t,0) &= 1 \;,
\end{align}
for example, which use up the remaining coordinate freedom (\ref{coord freedom dn}) by specifying $ \alpha = 1 $ on the origin.
\end{subequations}
The gauge potentials are then given by
\begin{align} \label{ab pa to dn}
a &= \frac{\hat{a}-SN\hat{b}}{\mu S\sqrt{N}} \;, & b &= \frac{\hat{a}+SN\hat{b}}{\nu S\sqrt{N}} \;.
\end{align}

\chapter{Numerical details} \label{sec:A:num}
Applying the trapezoidal rule (\ref{my evolution z}) to the $ v $-direction equations (\ref{G final}) in a straightforward way requires an expensive ``for" loop, the expense of which is amplified by the repetition of the integration in the $ v $ direction. We here briefly explain how we are able to implement this algorithm in MATLAB without a for loop.

Since each of the equations (\ref{G final}) is linear in their unknown, we represent them by the equation $ z_v = Az+B $ where $ A $ and $ B $ are known functions of $ v $. Discretising and applying (\ref{my evolution z}) we find the recurrence relation
\[ z_k = \alpha_k z_{k-1} +\beta_k \;, \]
where $ \alpha_k = \frac{1+\frac{\Delta v}{2}A_{k-1}}{1-\frac{\Delta v}{2}A_k} $ and $ \beta_k = \frac{\beta_{k-1}+\beta_k}{\frac{2}{\Delta v}-A_k} $. This has solution
\[ z_n = \left(\prod_{j=1}^n \alpha_j\right)\left(z_0+\sum_{k=1}^n\frac{\beta_k}{\prod\limits_{j=1}^k \alpha_j}\right) \;. \]
We use the vectorised MATLAB functions \verb_cumsum_ and \verb_cumprod_ to calculate $ \prod_{j=1}^n \alpha_j $ and then the result for all $ v $-points at once.
While the formula generalises to systems of linear equations (with the appropriate order for the matrix multiplication), there is no built-in MATLAB function for performing the cumulative product of matrices. Therefore we use for loops for the pairs of equations that must be solved together; (\ref{rv final},\ref{Gv final}), as well as (\ref{Wv final},\ref{Dv final}) and (\ref{pv final},\ref{xv final}) in the general (not purely magnetic) case.
Nevertheless we find a significant increase in speed with this implementation.

\section{Numerical coordinate transformations} \label{sec:A:nct}
While double-null coordinates have proven to be very useful to numerically evolve the Einstein-Yang-Mills equations, it is sometimes preferable to view the resulting solutions in more familiar space plus time coordinates.
In this section we describe how we perform the coordinate transformations into polar-areal and isothermal coordinates numerically.

We begin with the output of the AMR algorithm; a grid of double-null coordinate values organised uniquely into rows of constant $ u $, with varying separation between the rows and varying $ v $-spacing along each row.
After the transformation we wish to have a new structure with data organised into rows of constant time of equal separation (as determined by the proper time along a line of constant $ r $). The spatial variable along each row will typically have variable separation.

In each case we begin by calculating the proper time at a specified radius $ r_0 $, which is typically taken to be $ 0 $.
A larger value is useful to see the exterior of black hole spacetimes, which requires $ r_0 $ to be chosen greater than the final horizon radius.
If $ r_0 > 0 $ then this will require interpolation. For this and all subsequent interpolations we use cubic splines with not-a-knot end conditions.
On a given row $ u = u_0 $, the value of $ v $ for which the function $ r(u_0,v) = \frac{\tilde{r}(u_0,v)}{(1-u_0)(1-v)} $ takes the value $ r_0 $ is found by interpolating $ v $ as a function of $ r(v) $.
Since the areal radius blows up as $ v \to 1 $, and is not even monotonic on a line of constant $ u $ if a black hole forms, we restrict to $ r < 500 $ and $ v < v(r_{max}) $ where $ r_{max} $ is the maximum value $ r $ takes on that row.
Thus for every row $ u_i $ that crosses $ r_0 $, we find a corresponding $ v_i $, onto which each double-null variable is interpolated.
The proper time $ \tau $ is then calculated by integrating (see (\ref{tau dn}))
\[ \tau(u) = 2\int_0^{u} \frac{\tilde\alpha(u,v(u))}{(1-u)}\sqrt{\frac{-\tilde{f}}{(1-u)(\tilde{r}+(1-v(u))\tilde\alpha^2\tilde{G})}} \dx{u} \;, \]
for $ u \in (0,u_\text{max}) $ where $ u_\text{max} $ is 1 for an evolution that disperses and is limited to less than one by the singularity if a black hole has formed.
This and other integrations are performed using the trapezoidal method.

The lines of constant time are then determined by integrating the relevant equation for $ \dydx{v}{u} $. These are given by 
$ \dydx{v}{u} = \frac{f}{g} = \frac{(1-v)^2\tilde{f}}{(1-u)(\tilde{r}+(1-v)\tilde\alpha^2\tilde{G})} $ for polar-areal coordinates and $ \dydx{v}{u} = -1 $ for isothermal coordinates.
The values of $ \dydx{v}{u} $ are interpolated onto the line of constant $ r = r_0 $ calculated above, and then each of $ u $, $ v $, $ \dydx{v}{u} $, and the double-null variables is interpolated along that line as a function of $ \tau $ onto equispaced values of $ \tau $.

We note that for polar-areal coordinates, $ \dydx{v}{u} $ is infinite on an MTT, and therefore does not provide a nice function to be interpolated when the rows include $ N \leqslant 0 $ points.
In this case we instead interpolate $ \dydx{u}{v} $, which merely goes to zero on the MTT.
We do not use $ \dydx{u}{v} $ in general because in these two cases it blows up at $ v = 1 $.

From these points, the lines of constant time are found by integrating $ \dydx{v}{u} $ to the nearest $ u $-rows, for $ u $ increasing up until the origin, and $ u $ decreasing to zero (or future null infinity as the may be the case for isothermal coordinates).
Using only the interpolated values of $ \dydx{v}{u} $, we then have for each constant time row the values of $ u $ (which correspond to a known double-null row) and $ v $ (which are typically in-between known points).
The double-null functions are then splined onto these points, so all that remains is to calculate the new metric functions and the coordinate-dependent gauge functions.

For polar-areal coordinates ($ t = \tau $ and $ r = \frac{\tilde{r}}{(1-u)(1-v)} $), we require $ S $ and $ m $. The mass is simply given by (\ref{Nm final}).
For $ S $ we first solve the advection equation (\ref{c}) for $ c $. Along the ``characteristics" (which are simply the lines of constant $ t $), we have
\begin{align*}
&\dydx{c}{u} = 2c\frac{g_u}{g} \\
 &= \frac{2(1-v)\tilde\alpha^2}{\tilde{r}+(1-v)\tilde\alpha^2\tilde{G}}\left(-\frac{1}{\tilde{r}}\left(1+\frac{\tilde{f}(\tilde{r}+(1-v)\tilde\alpha^2\tilde{G})}{(1-u)\tilde\alpha^2}\right)+\frac{(1-v)^2\tilde{r}}{(1-u)^2}\left(\left(-2\tilde{W}+\tilde{D}^2+\frac{\tilde{r}^2\tilde{W}^2}{(1-u)^2}\right)^2+\tilde{Z}^2\right)\right).
\end{align*}
We integrate this from the origin outwards and then set $ S = -\frac{c(1-u)\tilde\alpha^2}{\tilde{f}(\tilde{r}+(1-v)\tilde\alpha^2\tilde{G})} $.

While $ w $ and $ d $ are coordinate independent, we need to transform from $ a $ and $ b $ into $ \hat{a} $ and $ \hat{b} $ and further perform a gauge transformation into the polar gauge, which will then affect $ w $ and $ d $.
We calculate $ \hat{a} $ and $ \hat{b} $ using (\ref{ab dn to pa}) and then require
\[ \dydx{\lambda}{r} = -\frac{1}{2}\left(b\frac{(1-u)(1-v)^2}{\tilde{r}+(1-v)\tilde\alpha^2\tilde{G}} +a\frac{(1-u)^2}{\tilde{f}}\right). \]
We integrate this from $ \lambda(0) = 0 $ and then set (using (\ref{gauge transform}))
\begin{align}
w &= (1-\tilde{r}^2\tilde{W})\cos\lambda -\tilde{r}\tilde{D}\sin\lambda \;, \label{wdnpa} \\
d &= \tilde{r}\tilde{D}\cos\lambda +(1-\tilde{r}^2\tilde{W})\sin\lambda \;, \label{ddnpa} \\
\hat{a} &= \frac{c}{2}\left(b\frac{(1-u)(1-v)^2}{\tilde{r}+(1-v)\tilde\alpha^2\tilde{G}}-a\frac{(1-u)^2}{\tilde{f}}\right) +\dot{\lambda} \;, \nonumber
\end{align}
and note $ \hat{b} $ is transformed to zero.

Finally, using
\begin{align*}
\pypx{}{r} &= \frac{1}{2r_u}\pypx{}{u} + \frac{1}{2r_v}\pypx{}{v} \;, & \pypx{}{t} &= -\frac{c}{2r_u}\pypx{}{u} +\frac{c}{2r_v}\pypx{}{v} \;,
\end{align*}
and (\ref{ab dn to pa}) we set
\begin{align*}
\hat{p} &= \left(\frac{(1-u)^2p}{2\tilde{f}}+\frac{(1-v)^2\tilde{r}\tilde{q}}{2(\tilde{r}+(1-v)\tilde\alpha^2\tilde{G})}\right)\cos\lambda
          -\left(\frac{(1-u)^2x}{2\tilde{f}}+\frac{(1-v)^2\tilde{r}\tilde{y}}{2(\tilde{r}+(1-v)\tilde\alpha^2\tilde{G})}\right)\sin\lambda \;, \\
\hat{x} &= \left(\frac{(1-u)^2x}{2\tilde{f}}+\frac{(1-v)^2\tilde{r}\tilde{y}}{2(\tilde{r}+(1-v)\tilde\alpha^2\tilde{G})}\right)\cos\lambda
          +\left(\frac{(1-u)^2p}{2\tilde{f}}+\frac{(1-v)^2\tilde{r}\tilde{q}}{2(\tilde{r}+(1-v)\tilde\alpha^2\tilde{G})}\right)\sin\lambda  \;,\\
\hat{q} &= \left(-\frac{(1-u)^2p}{2\tilde{f}}+\frac{(1-v)^2\tilde{r}\tilde{q}}{2(\tilde{r}+(1-v)\tilde\alpha^2\tilde{G})}\right)\cos\lambda
          -\left(-\frac{(1-u)^2x}{2\tilde{f}}+\frac{(1-v)^2\tilde{r}\tilde{y}}{2(\tilde{r}+(1-v)\tilde\alpha^2\tilde{G})}\right)\sin\lambda \;, \\
\hat{y} &= \left(-\frac{(1-u)^2x}{2\tilde{f}}+\frac{(1-v)^2\tilde{r}\tilde{y}}{2(\tilde{r}+(1-v)\tilde\alpha^2\tilde{G})}\right)\cos\lambda
          +\left(-\frac{(1-u)^2p}{2\tilde{f}}+\frac{(1-v)^2\tilde{r}\tilde{q}}{2(\tilde{r}+(1-v)\tilde\alpha^2\tilde{G})}\right)\sin\lambda \;.
\end{align*}
The function $ z $ is the same in both coordinate systems, and is residual gauge invariant.

The transformation into isothermal coordinates is more straightforward. $ T $ and $ R $ are set by (\ref{uvTR}), and $ r $ and $ \alpha $ are unchanged; see (\ref{dn first final}).
Their derivatives are given by
\begin{align*}
r_T &= \frac{1}{2}\left(\frac{\tilde{r}+(1-v)\tilde\alpha^2G}{(1-u)(1-v)^2} +\frac{\tilde{f}}{(1-u)^2}\right) \;, & \alpha_T &= \frac{\tilde\alpha}{2(1-u)(1-v)}\left(\frac{1}{1-v}+(1-v)\tilde\gamma +\frac{\tilde\beta}{1-u}\right) \;, \\
r_R &= \frac{1}{2}\left(\frac{\tilde{r}+(1-v)\tilde\alpha^2G}{(1-u)(1-v)^2} -\frac{\tilde{f}}{(1-u)^2}\right) \;, & \alpha_R &= \frac{\tilde\alpha}{2(1-u)(1-v)}\left(\frac{1}{1-v}+(1-v)\tilde\gamma -\frac{\tilde\beta}{1-u}\right) \;.
\end{align*}
Note that in this case the proper time is not utilised; setting $ T = \tau $ would require a further coordinate transformation.

The gauge variables and their derivatives are found similarly to the polar-areal case. Again the polar gauge is implemented ($ \bar{b} \equiv 0 $) so $ \lambda(R) $ solves
\[ \dydx{\lambda}{R} = -\frac{b-a}{2} \;, \qquad \lambda(0) = 0\;, \]
$ w $ and $ d $ are again given by equations (\ref{wdnpa},\ref{ddnpa}), and $ \bar{a} = \frac{b+a}{2}+\dot\lambda $.
We can define appropriate first order variables as
\begin{align*}
\bar{p} &= w_R+\bar{b}d \;, & \bar{q} &= w_T+\bar{a}d \;, & \bar{x} &= d_R-\bar{b}w \;, & \bar{y} &= d_T-\bar{a}w \;, & z &= \frac{r^2\left(\bar{a}_R-\bar{b}_T\right)}{\alpha^2} \;,
\end{align*}
and then they are given by
\begin{align*}
\bar{p} &= \frac{1}{2}\left(\frac{\tilde{r}\tilde{q}}{1-u}-p\right)\cos\lambda -\frac{1}{2}\left(\frac{\tilde{r}\tilde{y}}{1-u}-x\right)\sin\lambda \;, \\
\bar{x} &= \frac{1}{2}\left(\frac{\tilde{r}\tilde{y}}{1-u}-x\right)\cos\lambda +\frac{1}{2}\left(\frac{\tilde{r}\tilde{q}}{1-u}-p\right)\sin\lambda \;, \\
\bar{q} &= \frac{1}{2}\left(\frac{\tilde{r}\tilde{q}}{1-u}+p\right)\cos\lambda -\frac{1}{2}\left(\frac{\tilde{r}\tilde{y}}{1-u}+x\right)\sin\lambda \;, \\
\bar{y} &= \frac{1}{2}\left(\frac{\tilde{r}\tilde{y}}{1-u}+x\right)\cos\lambda +\frac{1}{2}\left(\frac{\tilde{r}\tilde{q}}{1-u}+p\right)\sin\lambda \;,
\end{align*}
where $ z $ is again the same as in the other coordinates, and residual gauge invariant.

\end{document}